\documentclass[12pt]{uomthesis}
\usepackage{hyperref}
\usepackage{cleveref}
\usepackage{graphicx} 
\usepackage{dsfont}
\usepackage{amsmath}
\usepackage[singlelinecheck=false,justification=raggedright]{caption}
\usepackage{graphicx} 
\usepackage{tikz} 
\usetikzlibrary{backgrounds}
\usetikzlibrary{decorations.pathmorphing}
\usepackage{pgfplots}
\pgfplotsset{compat=1.18}
\usetikzlibrary{shadows.blur} 
\usetikzlibrary{shadows.blur} 
\usetikzlibrary{decorations.pathreplacing} 
\usetikzlibrary{decorations.pathreplacing} 
\usetikzlibrary{calc} 
\usetikzlibrary{arrows.meta} 
\usetikzlibrary{decorations.pathreplacing}

\ThesisTitle{Chaos, Entanglement and Measurement: Field-Theoretic Perspectives on Quantum Information Dynamics}
\ThesisAuthor{Anastasiia Tiutiakina}
\ThesisDegree{Doctor of Philosophy}
\ThesisDate{November 2025}
\ThesisDepartment{École Doctorale EM2PSI}
\ThesisUniversity{CY Cergy Paris Université,Laboratoire de Physique Théorique et Modélisation}
\ThesisUniversity{%
  CY Cergy Paris Université, Laboratoire de Physique Théorique et Modélisation\\[0.8em]
  \includegraphics[height=20mm]{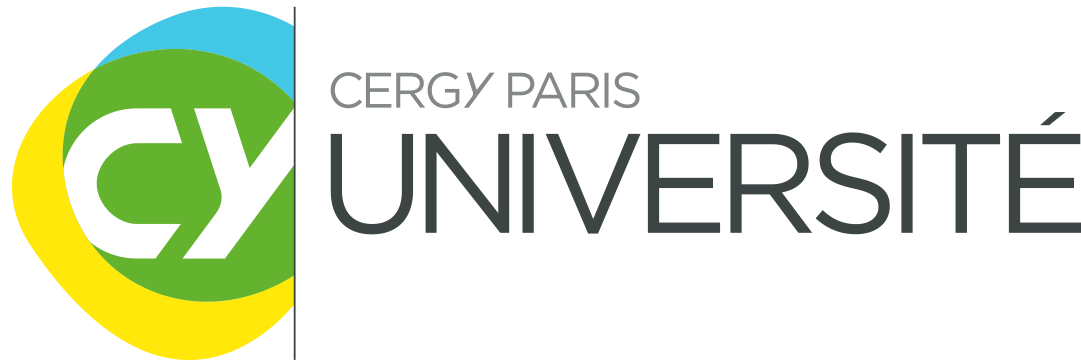}\\[2em]
  \noindent
  \begin{tabular*}{\textwidth}{@{}l@{\extracolsep{\fill}}r@{}}
    \begin{tabular}[t]{@{}l@{}}
      \textbf{Jury members}\\[0.5em]
      Philippe Lecheminant -- \textit{examinateur}\\
      Andreas Honecker -- \textit{examinateur}\\[0.3em]
      \textit{External members}\\
      Marco Schir\'o -- \textit{rapporteur, president}\\
      Laura Foini -- \textit{examinatrice}\\
      C\'ecile Monthus -- \textit{rapporteuse}
    \end{tabular}
    &
   \begin{tabular}[t]{@{}l@{}}
      \textbf{Supervisors}\\[0.5em]
      Jacopo De Nardis\\
      Andrea De Luca
    \end{tabular}
  \end{tabular*}%
}

\def\ii{\textbf{i}}
\def\jj{\textbf{j}}
\def\hchi{\hat \chi}
\def\uni{{\rm uni}}
\def\mon{{\rm mon}}
\def\hgamma{\hat \gamma}
\def\hhc{\mathfrak{h}}
\def\eec{\mathfrak{e}}
\def\mmu{\boldsymbol \mu}
\def\mnu{\boldsymbol \nu}
\def\mmu{\boldsymbol \mu}
\def\mnu{\boldsymbol \nu}

\def\hchi{\hat \chi}

\begin{document}

  \maketitle

  \frontmatter
  
  \chapter*{Abstract}
\addcontentsline{toc}{chapter}{Abstract} 

This thesis develops field-theoretic tools to understand how quantum information spreads, scrambles, and is reshaped by measurements in many-body systems. It is organized around three complementary projects. Project 1 — Scrambling and pseudorandomness in Brownian SYK. I quantify pseudorandomness using unitary \(k\)-designs and frame potentials, taking the Brownian SYK model as a strongly chaotic yet tractable test bed. Using Keldysh path integrals combined with replicas and disorder averaging, I obtain analytic control of the time-dependent approach to randomness and identify collective modes that delay convergence to Haar-like behavior. This yields design-time estimates as functions of model parameters and clarifies links between scrambling, complexity growth, and random-circuit phenomenology. I also outline randomized-measurement protocols that can access these predictions on current quantum hardware. Project 2 — Field theory for SYK clusters under weak measurements. Here I build a first-principles, field-theoretic description of interacting SYK clusters subject to gentle, frequent measurements. Starting from a system–ancilla picture, I pass to a continuum monitoring limit and represent the dynamics with fermionic coherent states. Using the replica method and disorder averaging, I derive a nonlinear sigma model that captures measurement back-action, the competition between interaction-induced scrambling and information extraction, and the resulting pattern of soft and massive collective modes. This framework predicts characteristic crossover scales in time and length, fluctuation spectra, and response signatures that distinguish weak-monitoring regimes from fully unitary evolution. Project 3 — Strong-disorder renormalization for measurement-only SYK clusters. I construct a strong-disorder RG tailored to measurement-only dynamics, starting from the \(SO(2n)\) replica algebra and deriving Dasgupta–Ma decimation rules. While the flow shows features reminiscent of infinite-randomness behavior, an order-of-limits subtlety in the replica treatment (\(n\!\to\!1\) vs.\ \(\Gamma'\!\to\!\infty\)) produces a simple pole in the induced coupling and makes the leading-order recursions non-robust. Consequently, the analytic evidence for an IRFP in the monitored setting is \emph{inconclusive} and requires further analysis (e.g., a replica-stable or replica-free formulation). Numerically, the \emph{average} second Rényi entropy follows the predicted logarithmic scaling. Across the three projects, the thesis offers a unified language—frame-potential and \(k\)-design diagnostics, Keldysh/replica techniques with a nonlinear sigma model, and disorder-based RG—to decide when many-body evolution generates operational randomness and how measurements redirect that flow. The results suggest concrete, testable signatures for near-term quantum simulators (superconducting qubits, neutral atoms, trapped ions).

  \chapter*{Declaration}
\addcontentsline{toc}{chapter}{Declaration}

I, \textbf{Anastasiia Tiutiakina}, declare that:
\begin{enumerate}[label=\roman*.]
  \item this thesis is the result of my own work carried out within the \textit{Doctorat de CY Cergy Paris Université (Spécialité: Physique)} except where otherwise indicated in the Preface;
  \item all sources are acknowledged and all quotations and ideas from other works are properly cited in the text;
  \item parts of this thesis that derive from co-authored publications are clearly identified, and my individual contribution is explicitly stated in the Preface.
\end{enumerate}

\vspace{1em}
\noindent\textit{Cergy, \rule{2cm}{0.15mm} \quad Signature:} \rule{4cm}{0.15mm}

\chapter*{Preface}
\addcontentsline{toc}{chapter}{Preface}

This dissertation was prepared within the Doctoral Program in Physics at CY Cergy Paris Université (École Doctorale EM2PSI, LPTM). It gathers original research conducted between \textit{2022} and \textit{2025}. The thesis consists of \textit{4} chapters and \textit{22} appendices. Below I detail authorship, prior/parallel submissions, and the use of editorial assistance.

\subsection*{Candidate's contributions to co-authored works}
Several chapters incorporate material from co-authored manuscripts. My specific contributions are identified for each item below; any text or figures reused or adapted from published or submitted work are clearly indicated at the beginning of the corresponding chapter.


\begin{description}[leftmargin=0pt, labelindent=\parindent, style=nextline]
\item[\textbf{Chapter 2 — Frame Potential of the Brownian SYK Model} \ (pp.~22–59)]
\textbf{Source:} Anastasiia Tiutiakina, Andrea De Luca, Jacopo De Nardis, “\textit{Frame potential of Brownian SYK model of Majorana and Dirac fermions}”, JHEP01(2024)115,2024.  
\textbf{My contributions:} conceived problem; developed Keldysh/replica formalism; analytic calculations; manuscript writing; Figs.~[1].

\item[\textbf{Chapter 3 — Field theory for monitored Brownian SYK clusters} \ (pp.~59–95)]
\textbf{Source:} Anastasiia Tiutiakina, Hugo Lóio, Guido Giachetti, Jacopo De Nardis, Andrea De Luca, “\textit{Field theory for monitored Brownian SYK clusters}”, q-2025-07-14-1794, 2025.  
\textbf{My contributions:} development of the coherent state formalism for generic number of replicas, considerationg of the $n=2$ case, calculations of the purity for two clusters, derivation of non-linear sigma model for the SYK-2 fermions and derivation of the field theory with permutation symmetry for interacting SYK-4. Fig[1,2] 

\item[\textbf{Chapter 4 — Strong Disorder Renormalisation Group approach for monitored fermions} \ (pp.~95-119)]
\textbf{Source:} Anastasiia Tiutiakina, Jacopo De Nardis, Andrea De Luca, “\textit{Strong Disorder Renormalisation Group approach for monitored fermions}”, unpublished work, 2025.  
\textbf{My contributions:} development of Dasgupta-Ma procedure for the chain of generators SO(2n), numerical analysis of the entanglement entropy for monitored fermions. 
\end{description}

\subsection*{Work submitted for other qualifications}
None of the work included here has been submitted towards any other degree or qualification.

\subsection*{Work carried out prior to enrolment}
 All research presented was conducted during my doctoral enrolment. \\

\subsection*{Digital editorial assistance}

\begin{itemize}
  \item Light language polishing tools (e.g.\ grammar/spell checking, reference managers) were used. No generative content was adopted without full verification; all scientific content, derivations, and conclusions are my own.
  \item No digital editorial assistance beyond standard typesetting (LaTeX) and reference management was used.
\end{itemize}

\subsection*{Third-party human assistance}
 No third-party editorial services were employed. Scientific guidance by my supervisors and co-authors is acknowledged in the Acknowledgments. \\

\subsection*{Publications included in the thesis}
\begin{itemize}
  \item \textbf{In preparation:} \textit{Strong disorder renormalisation group in monitored fermions},
  Anastasiia Tiutiakina, Jacopo De Nardis, and Andrea De Luca. 
  \item \textbf{Published on \textit{2025-07-14} by \textit{Quantum}}: \textit{Field theory for monitored Brownian SYK clusters}, 
  
  Anastasiia Tiutiakina, Hugo Lóio, Guido Giachetti, Jacopo De Nardis, and Andrea De Luca.
  
  \item \textbf{Published on \textit{2024-01-23} by \textit{JHEP}}: \textit{Frame potential of Brownian SYK model of Majorana
and Dirac fermions},

Anastasiia Tiutiakina, Andrea De Luca, and Jacopo De Nardis.
\end{itemize}

\vspace{1.0em}

\subsection*{Other publications by the author}
\begin{itemize}

  \item \textbf{Published on \textit{2023-05-25} by \textit{PRB}}: \textit{Adiabatic eigenstate deformations and weak integrability breaking of Heisenberg chain},
  
  Pavel Orlov, Anastasiia Tiutiakina, Rustem Sharipov, Elena Petrova,Vladimir Gritsev, and Denis V. Kurlov.
  
  \item \textbf{Published on \textit{2024-09-09} by \textit{SciPost}}: \textit{Adiabatic gauge potential and integrability
breaking with free fermions},

Balázs Pozsgay, Rustem Sharipov, Anastasiia Tiutiakina and István Vona.
  \item \textbf{Preprint on \textit{2024-11-18}}: \textit{Hilbert space geometry and quantum chaos},
  
  Rustem Sharipov, Anastasiia Tiutiakina, Alexander Gorsky, Vladimir Gritsev, and Anatoli Polkovnikov.
\end{itemize}

\vspace{1.0em}

  \chapter*{Dedication}
\addcontentsline{toc}{chapter}{Dedication}

\emph{To becoming--- kinder, braver, and ever so slightly more precise.}

  \chapter*{Acknowledgments}
\addcontentsline{toc}{chapter}{Acknowledgments}

\noindent\textbf{Jacopo De Nardis.} Thank you for being a steady presence---especially through the stretches when the work felt lonely---and for your scientific guidance and many clarifying discussions. Thank you as well for the opportunities you created for collaborations and travel. You taught me what responsibility in research means and the standards expected of a scientist---in rigor, reliability, and care for the community. When I set out to organize a workshop on integrability---and later helped run two wonderful editions in Bad Honnef and Budapest---it was your idea to propose me as an organizer. I am deeply grateful.

\noindent\textbf{Andrea De Luca.} Our scientific discussions were always insightful, and I was inspired by the spark of deep curiosity you bring to your work. I cherish your guidance and our conversations about the meaning and purpose of academic work. You also model a thoughtful balance between scientific work and personal life—an example I aspire to.

\noindent\textbf{At LPTM.} I am grateful to \textbf{Jean Avan} and \textbf{Genevi\`eve Rollet} for their care and steady support whenever problems arose, and for their unfailing readiness to help with any issue I encountered. I also thank all my LPTM colleagues for the warm atmosphere and the laughter during lunch breaks, which made the difficult days lighter.

\noindent\textbf{Tom Roussel.} Thank you for your love, patience, and unwavering support. You made space for this thesis in our everyday life---from late-night deadlines to weekends spent working---and kept me grounded when the work felt all-consuming. Thank you for believing in me when I doubted myself, for listening even when I got overly excited about physics, and for celebrating each small step forward. You kept asking what truly makes me happy---and because of you, I began asking myself the same question. I am grateful for that. You also showed me that I don't have to constantly prove myself to feel valued and loved---a revelation that changed me as a person. You brought perspective into my life, and a sense of the future I now cherish. At the same time, you embraced my competitive spirit---from the countless arm-wrestling matches we had (often alongside others thanked here) to the friendly challenges that kept things fun. Your kindness, humor, and steadiness are woven through these pages. I could not have done this without you.

\noindent\textbf{Sascha Gehrmann and Friedrich H\"ubner.} My co-organisers of the Integrability Workshop---Bad Honnef and Budapest editions. Thank you both for turning a first-time organising adventure into something we are proud of. With \textbf{Friedrich}, the silly moments and inside jokes we tried to keep under wraps during crunch time---and the pure fun of shaping the workshop together---made the work a joy; you always had my back when I needed help. \textbf{Sascha}, you taught us how to handle the audience, stay organised, and think through every contingency and urgency that conference organisation can bring. I am personally thankful for your deep understanding and for being there for me when no one else could.

\noindent\textbf{Leonardo Biagetti, Hugo L\'oio, and Guillaume C\'ecile.} From officemates to close friends—thank you for being the people I could count on, for good ideas and good company. Between deadline sprints and long blackboard sessions, your perspective and humor kept things in balance. There’s a kind of understanding that only those at the same stage of life can offer—and I felt that with you. The conference trips we shared—equal parts science and mini-holiday—are among my happiest memories of these years. Special thanks to Hugo and Leonardo for the darts skills I'll carry with me.

\noindent\textbf{Elena Petrova, Viktoriia Pinchenkova, and Sara Vanovac.} Fellow women in science and PhD students like me—you brought perspectives on being women in academia that mattered deeply, and your support meant a great deal. Thank you for the solidarity, the laughter, and the honest conversations. The joy we shared, the ability to rely on one another—and to be there for you in return—are things I will always hold dear.

\providecommand{\cp}{cp}
\providecommand{\bow}{bow}

  \tableofcontents
  \printglossaries
  
  \listoffigures     
  \listoftables  
  



  \mainmatter

  \chapter{Introduction}
\label{chap:introduction}
The history of physics in the 19-20th centuries is a testament to the power of experiment to shape theory. The birth of quantum mechanics arose not from abstract mathematical conjecture, but from experimental puzzles that demanded explanation.
In the late 19th century, physicists tried to account for the spectrum emitted by an ideal blackbody (an object in thermal equilibrium that absorbs all incident radiation). The spectral energy density of the black body $u(\nu,T)$ grows with frequency $\nu$, attains a maximum, and then decays approximately exponentially. And the peak position increases with temperature $T$ in accordance with Wien’s displacement law.
But classical physics predictions failed dramatically. The Rayleigh-Jeans law \cite{Jeans1900,Rayleigh1900} (based on classical equipartition) predicted:
\begin{equation}
    u(\nu,T)\sim \nu^2 T
\end{equation}
which diverges as $\nu \to \infty$, this is the infamous "ultraviolet catastrophe". 
Max Planck tried to fit the observed spectrum by introducing a new assumption \cite{Planck:1901tja}:
\begin{equation}
    E_n=n\nu h , ~~~n=0,1,2...;
\end{equation}
where $h$ is a Planck's constant (new fundamental constant!) and $\nu$ is frequency of the mode. In other words oscillators can exchange energy with the radiation field only in discrete packets of size $h \nu$. This was radical: before this, energy was assumed to vary continuously. Planck then calculated the average energy of an oscillator using this quantization assumption and summation with the Boltzmann weights:
\begin{equation}
    \left< E\right>=\frac{\sum_{n}E_n e^{-\beta E_n}}{\sum_{n} e^{-\beta E_n}}=\frac{h \nu}{e^{\frac{h \nu}{kT}}-1}.
\end{equation}
From this, he derived the Planck blackbody formula for spectral energy density that is in agreement with the experiment:
\begin{equation}
    u(\nu,T)=\frac{8 \pi h \nu^3}{c^3}\frac{1}{e^{\frac{h \nu}{kT}}-1}.
\end{equation}
Later on the puzzle of atomic spectra occurred. Experimentalists in the late 19th century observed that:
atoms (especially hydrogen) emitted light at discrete wavelengths when excited, the hydrogen emission spectrum showed distinct lines that followed a precise pattern (Balmer series, etc.). This was a mystery: why would atoms emit light at only certain wavelengths? Classical electromagnetism predicted that electrons orbiting a nucleus should emit radiation continuously as they spiral inward — not as discrete lines!
Niels Bohr proposed a bold and radical model \cite{Bohr1913}. Electrons orbit the nucleus in specific allowed orbits with quantized angular momentum:
\begin{equation}
    L=n \hbar,~~ n=1,2....
\end{equation}
And radiation occurs only when an electron jumps between these quantized orbits  $n_i \to n_f$, with the energy of the emitted photon given by:
\begin{equation}
    \delta E=\; R_y\!\left(\frac{1}{n_f^{2}}-\frac{1}{n_i^{2}}\right)=h \nu,
\end{equation}
where \(R_y\approx 13.6\,\mathrm{eV}\) is the Rydberg energy.
This explained the stability of atoms (no spiraling into the nucleus) and the discrete lines in hydrogen’s emission spectrum.
By the mid-1920s, a deeper and more general framework emerged.
Heisenberg’s matrix mechanics (1925) \cite{Heisenberg1925}: treated observable quantities as matrices encoding transition amplitudes between states. Schrödinger’s wave mechanics (1926) \cite{Schrodinger1926}, described particles as wavefunctions governed by the Schrödinger equation:
\begin{equation}
    -\frac{\hbar^2}{2m} \nabla^2 \psi+V\psi=E \psi.
\end{equation}

Beyond blackbody radiation and atomic spectra, the discovery of exotic materials presented theorists with further surprises. In $1911$, Kamerlingh Onnes discovered superconductivity \cite{Onnes1911} — the complete disappearance of electrical resistance in mercury below $4.2$ K — a phenomenon that defied classical expectations and remained unexplained for decades. Likewise, ferromagnetism, long observed empirically, posed a deep theoretical challenge: why do certain materials spontaneously magnetize below a critical temperature? The resolution of these puzzles required the new tools of quantum mechanics and the development of quantum many-body theory, culminating in models such as the Ising and Heisenberg models \cite{Ising1925,Heisenberg1928} and, later, BCS theory of superconductivity \cite{Bardeen:1957mv}. These experimental discoveries revealed that collective behavior in condensed matter could exhibit fundamentally new physics, driving theory toward increasingly sophisticated frameworks.
Throughout much of the 20th century, this interplay between theory and experiment defined the natural rhythm of scientific progress. Experiments revealed surprising behaviors, and theorists responded by constructing new frameworks to explain them, frameworks that would in turn predict new phenomena to be tested.

In condensed matter physics, this dynamic led to enormous successes: the theory of metals, superconductivity \cite{Onnes1911}, semiconductors \cite{Bloch1929}, quantum Hall states \cite{vonKlitzing1980,Tsui1982,Laughlin1983}, and more. But as the field matured, the nature of inquiry began to shift. Many experimental frontiers became technically challenging, while the most celebrated puzzles — such as high-temperature superconductivity — resisted solution despite decades of effort.

Today, much of theoretical condensed matter physics is increasingly driven from within: by formalism, by classification schemes, by the desire to generalize and extend concepts \cite{Wen2004,Kitaev2009}. In many cases, the questions themselves are generated not by experiment but by the theoretical community’s internal dialogue. This mathematical turn has produced deep insights, but it also marks a departure from the experiment-led paradigm that characterized earlier generations. Today’s theorist often has to \emph{"invent"} what is interesting. 

Scientists, like all people, are guided by aesthetic sensibilities. Throughout the history of physics, concepts such as symmetry, simplicity, and unity \cite{Yang:1954ek} have often guided theoretical discovery. In the modern landscape of condensed matter physics—where experimental surprises have grown rarer—this aesthetic instinct has become increasingly influential. Elegant mathematical structures such as topology, geometry \cite{Polyakov:1975rr,Efetov:1983xg,Haldane:1983ru,Berry:1984jv,Nakahara2003}, and entanglement theory \cite{Kitaev:2005dm}  have emerged as organizing principles around which research communities cluster, shaping discourse and redefining what constitutes a \emph{meaningful question}.

Among the many such structures, my interests are especially drawn to two interconnected themes: \emph{quantum chaos} \cite{Srednicki1994,DAlessio2016} and \emph{field theory} \cite{Fradkin2013, Wen2004, Altland2010}. Quantum chaos lies at the heart of our understanding of thermalization and complexity in closed quantum systems. It offers a theoretical framework to track the spreading of information, whereby initially localized operators evolve into highly nonlocal ones under unitary dynamics—a process referred to as \emph{scrambling}. This phenomenon connects quantum chaos to questions of entanglement growth, operator spreading, and the emergence of statistical mechanics in isolated systems. Modern diagnostics such as out-of-time-order correlators (OTOCs)\cite{Maldacena2016, Brenes:2021bjr, Lerose:2020qlg}, spectral form factors, frame potentials \cite{Gross2007}, and the \emph{adiabatic gauge potential}(AGP) \cite{Kolodrubetz:2017ofs} offer diverse and complementary perspectives on chaotic dynamics. While OTOCs probe sensitivity to initial conditions, and frame potentials quantify pseudorandomness, the AGP captures the geometry of quantum states under slow parameter changes, providing insights into state distinguishability, and the structure of quantum adiabatic response. AGP, originally introduced in the context of quantum control, has more recently emerged as a diagnostic of chaos \cite{Pandey:2020tru,Kim:2024gmo,Orlov_2023,Pozsgay_2024,sharipov2024hilbertspacegeometryquantum}.

Field theory provides a language for emergent collective behavior in terms of coarse-grained, often universal, degrees of freedom \cite{Fradkin2013,Altland2010}. A canonical example is the quantum Ising chain near criticality, whose long-wavelength physics is captured by the Euclidean \(\mathbb{Z}_2\) Landau–Ginzburg theory
\begin{equation}
S[\phi]=\int \mathrm{d}\tau\,\mathrm{d}x\,
\Big[\tfrac{1}{2v^2}(\partial_\tau\phi)^2+\tfrac{1}{2}(\partial_x\phi)^2+\tfrac{r}{2}\phi^2+\tfrac{u}{4!}\phi^4\Big],
\end{equation}
where \(\phi\) is the coarse-grained order parameter, \(r\) tunes the transition, \(u>0\) ensures stability, and \(v\)  is the emergent velocity. This flows to the Ising CFT at criticality.

A complementary continuum description arises in Haldane’s mapping of antiferromagnetic spin chains to the \(O(3)\) nonlinear sigma model with a topological \(\theta\)-term,
\begin{equation}
S[\mathbf{n}]=\frac{1}{2g}\!\int\!\mathrm{d}\tau\,\mathrm{d}x\,(\partial_\mu\mathbf{n})^2
\;+\; i\,\frac{\theta}{4\pi}\!\int\!\mathrm{d}\tau\,\mathrm{d}x\;\mathbf{n}\!\cdot\!(\partial_\tau\mathbf{n}\times\partial_x\mathbf{n}),
\end{equation}
with unit vector \(\mathbf{n}\) and \(\theta=2\pi S\). Combined with Polyakov’s result that the \(O(3)\) model is gapped at \(\theta=0\), this explains the Haldane conjecture: integer-spin chains ( \(\theta\equiv 0\) ) are gapped, whereas half-integer chains ( \(\theta=\pi\) ) are gapless \cite{Haldane:1983ru,Polyakov:1975rr}.

Field-theoretic tools have also proven essential in the study of quantum chaos. Efetov’s nonlinear sigma model for disordered systems offered a path-integral approach to spectral correlations \cite{Efetov:1983xg}, while the Sachdev-Ye-Kitaev (SYK) model provided a solvable playground for strongly chaotic dynamics governed by an emergent low-energy Schwarzian action \cite{Sachdev1993,Kitaev2015,MaldacenaStanford2016}.

In this thesis, I draw on the perspectives of field theory and quantum chaos to study aspects of complex quantum dynamics. I focus on phenomena such as entanglement growth, scrambling, and measurement-induced phase transitions (MIPTs) in systems that combine unitary evolution with continuous measurement. To approach these problems, I use a range of analytical techniques, including nonlinear sigma models, replica methods, random matrix theory, and stochastic path integrals. MIPTs, in particular, offer a setting to explore dynamical phase transitions without equilibrium analogues, and provide a useful context for examining how information and entanglement evolve under non-unitary dynamics.

More broadly, this work contributes to the ongoing effort to identify universal features of far-from-equilibrium quantum systems \cite{DAlessioKafriPolkovnikovRigol2016,AbaninAltmanBlochSerbyn2019}. The goal is to build a conceptual and mathematical framework that can describe how complexity arises, how information propagates, and how quantum coherence and randomness coexist in many-body evolution \cite{LiebRobinson1972,vonKeyserlingkRakovszkyPollmannSondhi2018,NahumVijayHaah2018,GarciaMataEtAl2022,HarrowLow2009,BrandaoHarrowHorodecki2016,HunterJones2019}. While these questions are grounded in theory, they are increasingly relevant to quantum simulators \cite{GeorgescuAshhabNori2014,AltmanEtAl2021,GrossBloch2017}, which—though still in their technological infancy—have begun to probe regimes where these phenomena unfold in real time \cite{CheneauEtAl2012,NoelEtAl2022,KohEtAl2023,GoogleAI2023}. Despite rapid progress, the scalability, fidelity, and controllability required for systematically exploring such complex dynamics remain formidable challenges \cite{AcharyaEtAl2023,BluvsteinEtAl2024,GrahamEtAl2023,LisEtAl2023}. Nonetheless, the theoretical structures developed here may help guide and interpret future experimental breakthroughs as the capabilities of synthetic quantum platforms continue to evolve \cite{HalimehAidelsburgerYang2025,TarruellSanchezPalencia2018}.

\section{Unitary \texorpdfstring{$k$}{k}-Designs, Frame Potential, and Complexity}

In many quantum‐information and simulation tasks, one needs to average over truly random unitaries to guarantee unbiased performance or to erase unwanted correlations. For instance, randomized benchmarking makes use of unitary 2-designs to develop eﬃcient protocol for experimentally characterizing the fidelity of a quantum process \cite{Dankert2009}, and one-shot decoupling protocols—central to quantum channel coding and privacy amplification—rely on Haar averages to ensure complete information erasure \cite{Szehr2013}. Likewise, schemes such as quantum data hiding \cite{DiVincenzo2002,Hayden2005} or port-based teleportation \cite{Ishizaka2008} invoke higher moments of the Haar ensemble, and studies of quantum chaos benchmark operator spreading against Haar randomness via out-of-time-order correlators \cite{Maldacena2016,Roberts2017}. Yet exactly implementing a Haar–random \(D\times D\) unitary on \(n\) qubits (\(D=2^n\)) requires circuit size \(\Theta(D^2)=\Theta(4^n)\) in two–qubit gates \cite{Shende2006, Mottonen2004, Vartiainen2004}.

Unitary $k$-designs solve this bottleneck by reproducing the first $k$ statistical moments of the Haar measure. Concretely, an ensemble $\mathcal E$ of unitaries is a $k$-design if, for every polynomial $P$ of degree at most $k$ in the entries of $U$ and $U^\dagger$,
\[
  \mathbb{E}_{U\sim\mathcal E}\bigl[P(U,U^\dagger)\bigr]
  \;=\;
  \mathbb{E}_{U\sim\mathrm{Haar}}\bigl[P(U,U^\dagger)\bigr].
\]
Physically, this means no measurement involving up to $k$ copies of the evolution operator can distinguish $\mathcal E$ from true Haar randomness.  Exact $k$-designs are rare and typically require fine-tuned constructions. In practice, we settle for \emph{approximate} $k$-designs, where the agreement with Haar is not perfect but good enough. The degree of approximation can be measured in various ways, such as \emph{trace distance} or the \emph{diamond norm} between quantum channels. The larger $k$ is, the more fine-grained our probe of the ensemble becomes—and the harder it is for a structured process to pass as random.

A useful diagnostic for $k$-design behavior is \emph{frame potential}, defined as
\[
  \mathcal{F}^{(k)}_{\mathcal{E}}
  = \mathbb{E}_{U,V\sim\mathcal{E}}\bigl|\Tr(U^\dagger V)\bigr|^{2k}.
\]
It quantifies the amount of overlap or interference between different elements of the ensemble. A completely random ensemble (i.e., Haar-distributed unitaries) minimizes this quantity with $\mathcal{F}^{(k)}_{\rm Haar} = k!$ for large Hilbert space dimension \cite{Gross2007,Webb2016}. Deviations from this value indicate residual structure or correlations in the ensemble.  Writing the \(k\)-th moment operator
\(M_{\mathcal E}=\mathbb{E}_{U\sim\mathcal E}\!\big[U^{\otimes k}\otimes U^{\dagger\otimes k}\big]\),
one has the exact identity
\[
  \mathcal F^{(k)}_{\mathcal E}-\mathcal F^{(k)}_{\rm Haar}
  = D^{2k}\,\big\| M_{\mathcal E}-M_{\rm Haar}\big\|_2^2 \;\ge 0,
\]
so the decay of \(\mathcal F^{(k)}(t)\) directly tracks convergence to Haar in \(k\)-th moments, and (via norm inequalities) controls operational distances such as the diamond norm. In effect, the frame potential measures the “leftover” higher‐order interference when randomness is imperfect.

The frame potential was introduced in studies of unitary designs \cite{Gross2007,Webb2016} and—together with OTOC-based probes of scrambling (e.g., the decay of \(\langle W(t)V W(t)V\rangle\) or growth of \(\langle[W(t),V]^2\rangle\))—has become a standard tool for diagnosing quantum chaos and complexity growth \cite{Maldacena2016,Roberts2017,Cotler2017,Nahum2018}.
Since sampling the full Haar measure is generally infeasible, one turns to efficient constructions—most notably quantum circuits.  Early landmark works \cite{Dankert2009,Brandao2016} explored how random or pseudorandom quantum circuits can serve as approximate designs.  Given an $n$-qubit unitary $U$, its \emph{circuit complexity} $\mathcal{C}(U)$ is defined as the minimal number of elementary gates (from some fixed universal set) required to approximate $U$ to within operator‐norm error $\epsilon$.  Complexity may be measured by total gate count (“size”) or by circuit “depth” (the number of timesteps).

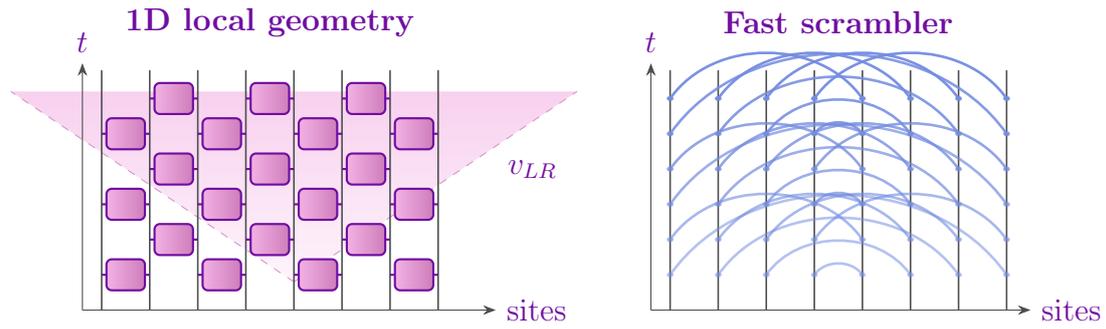
\begin{figure}[t]
\centering

\newdimen\Xunit    \Xunit=1.15cm   
\newdimen\Yunit    \Yunit=0.85cm   
\newdimen\PanelGap \PanelGap=2.1cm 

\def\NumQ{8}        
\def\Depth{6}       
\newcommand{\FigScale}{0.55}

\newcommand{\LabelSize}{\normalsize}  
\newcommand{\TitleSize}{\normalsize}       

\def\LRslope{0.95}  
\def\BrickFrac{0.80}
\def\BrickH{0.44}   

\newdimen\RightShift
\RightShift=\NumQ\Xunit \advance\RightShift by 2\Xunit \advance\RightShift by \PanelGap

\definecolor{DuneDeep}{RGB}{112,13,161}   
\definecolor{DuneMid}{RGB}{198,102,175}   
\definecolor{DuneLight}{RGB}{236,136,213} 
\definecolor{BlueVio}{RGB}{90,120,220}    

\begin{tikzpicture}[
  scale=\FigScale,
  x=\Xunit, y=\Yunit, >=Stealth,
  wire/.style   ={line width=0.6pt, draw=black!70},
  gate1D/.style ={rounded corners=2pt, draw=DuneDeep, line width=0.8pt,
                  top color=DuneLight!65, bottom color=DuneMid!85, shading angle=90},
  gateAA/.style ={draw=BlueVio!85, line width=1pt},
  txt/.style    ={font=\LabelSize, text=DuneDeep}
]

\pgfmathtruncatemacro{\ConeSite}{int(\NumQ/2 + 1)}

\begin{scope}
  \begin{scope}
  \clip (\ConeSite,0.8)
        -- ({\ConeSite-\LRslope*(\Depth+0.2)},\Depth+0.2)
        -- ({\ConeSite+\LRslope*(\Depth+0.2)},\Depth+0.2)
        -- cycle;
  \shade[shading angle=90, top color=DuneLight!55, bottom color=DuneLight!5]
    (-2,0) rectangle (\NumQ+3,\Depth+3); 
\end{scope}
  \draw[DuneMid!70, dashed] (\ConeSite,0.8) -- ({\ConeSite-\LRslope*(\Depth+0.2)},\Depth+0.2);
  \draw[DuneMid!70, dashed] (\ConeSite,0.8) -- ({\ConeSite+\LRslope*(\Depth+0.2)},\Depth+0.2);
  \node[txt, anchor=east]
    at (10.7, 4.0) {$v_{LR}$};

  \foreach \i in {1,...,\NumQ} {
    \draw[wire] (\i,0) -- (\i,\Depth+0.8);
  }

  \foreach \layer in {1,...,\Depth} {
    \pgfmathtruncatemacro{\phase}{mod(\layer,2)}
    \ifnum\phase=1
      \foreach \i in {1,3,...,\NumQ} {
        \ifnum\i<\NumQ
          \path[gate1D] (\i+0.5-0.5*\BrickFrac, \layer-\BrickH)
                        rectangle
                        (\i+0.5+0.5*\BrickFrac, \layer+\BrickH);
          \draw[DuneDeep, line width=0.8pt] (\i,\layer) -- (\i+0.5-0.5*\BrickFrac,\layer);
          \draw[DuneDeep, line width=0.8pt] (\i+1,\layer) -- (\i+0.5+0.5*\BrickFrac,\layer);
        \fi
      }
    \else
      \foreach \i in {2,4,...,\NumQ} {
        \ifnum\i<\NumQ
          \path[gate1D] (\i+0.5-0.5*\BrickFrac, \layer-\BrickH)
                        rectangle
                        (\i+0.5+0.5*\BrickFrac, \layer+\BrickH);
          \draw[DuneDeep, line width=0.8pt] (\i,\layer) -- (\i+0.5-0.5*\BrickFrac,\layer);
          \draw[DuneDeep, line width=0.8pt] (\i+1,\layer) -- (\i+0.5+0.5*\BrickFrac,\layer);
        \fi
      }
    \fi
  }

  \draw[->, black!70] (0.6,0) -- (0.6,\Depth+1.0) node[above,txt] {$t$};
  \draw[->, black!70] (0.6,0) -- (\NumQ+1.2,0) node[right,txt] {sites};
  \node[font=\bfseries\TitleSize, align=center, text=DuneDeep]
       at (\NumQ/2+0.5,\Depth+2.15) {1D local geometry};
\end{scope}

\begin{scope}[xshift=\RightShift]
  \foreach \i in {1,...,\NumQ} {
    \draw[wire] (\i,0) -- (\i,\Depth+0.8);
  }

  \pgfmathtruncatemacro{\Half}{\NumQ/2}
  \foreach \layer in {1,...,\Depth} {
    \pgfmathsetmacro{\arcopa}{0.45 + 0.5*\layer/(\Depth)} 
    \ifodd\layer
      \foreach \p in {1,...,\Half} {
        \pgfmathtruncatemacro{\q}{\NumQ+1-\p}
        \draw[gateAA, opacity=\arcopa] (\p,\layer) to[out=55,in=125] (\q,\layer);
        \fill[BlueVio!85, opacity=\arcopa] (\p,\layer) circle (0.07);
        \fill[BlueVio!85, opacity=\arcopa] (\q,\layer) circle (0.07);
      }
    \else
      \foreach \p in {1,...,\Half} {
        \pgfmathtruncatemacro{\q}{\p+\Half}
        \draw[gateAA, opacity=\arcopa] (\p,\layer) to[out=55,in=125] (\q,\layer);
        \fill[BlueVio!85, opacity=\arcopa] (\p,\layer) circle (0.07);
        \fill[BlueVio!85, opacity=\arcopa] (\q,\layer) circle (0.07);
      }
    \fi
  }

  \draw[->, black!70] (0.6,0) -- (0.6,\Depth+1.0) node[above,txt] {$t$};
  \draw[->, black!70] (0.6,0) -- (\NumQ+0.5,0) node[right,txt] {sites};
  \node[font=\bfseries\TitleSize, align=center, text=DuneDeep]
       at (\NumQ/2+0.5,\Depth+2.15) {Fast scrambler};
\end{scope}

\end{tikzpicture}

\caption{Geometry controls mixing times. Left: nearest-neighbor brickwork with a Lieb–Robinson light cone. Right: all-to-all pairings (fast scrambler).}\label{Fig:Lieb}
\end{figure}

Circuit geometry crucially affects mixing times.  In a one-dimensional chain with local interactions bounded by a Lieb–Robinson velocity $v_{LR}$ \cite{Lieb1972} see Fig.\ref{Fig:Lieb}, the depth $t$ needed to generate an $\epsilon$-approximate $k$-design scales as
\begin{equation}
  t_k \;=\; O\bigl(\,\mathrm{poly}(n,k)\,\ln\tfrac1\epsilon\bigr)
\end{equation}
The factor of $\mathrm{poly}(n,k)$ reflects the time to spread information across $n$ qubits and then assemble $k$-body correlations \cite{Harrow2009,Brandao2016}.  By contrast, in a fully connected or “fast‐scrambling” see Fig. \ref{Fig:Lieb} architecture—where every qubit can interact with every other in a single step—the $n$-dependence drops out, yielding
\[
  t_k \;\gtrsim\; k\,\ln\!\frac{1}{\epsilon}.
\]
Thus, generating an approximate $k$-design requires first spreading correlations throughout the entire $n$-qubit system—so that every qubit has “seen” the randomness—and then applying further random layers to equilibrate the $k-$th moments, exponentially suppressing their deviation from the Haar averages until they lie within $\epsilon$.

This narrative connects with broader themes in quantum complexity theory and even holographic duality, where the growth of circuit complexity has been conjectured to mirror the linear growth of the Einstein–Rosen bridge inside a black hole \cite{Susskind2014,Stanford2014}.  Just as a classical gas requires time to equilibrate, a quantum circuit demands depth to become sufficiently “random” up to the $k$th moment.

In the sequel, we will apply these general notions to concrete models of quantum many-body systems, identifying conditions under which dynamics intrinsically approximate k-designs. In doing so, we aim to clarify how locality, interactions, and noise conspire to produce effective randomness — a key diagnostic of complexity growth in modern quantum systems. To ground these abstract diagnostics—OTOCs, frame potentials, level statistics—in a fully solvable many-body setting, let us now turn to the Sachdev–Ye–Kitaev model.

\section{SYK as Paradigm of Quantum Chaos}

In close analogy to how random circuits offer a practical testbed for realizing unitary $k$-designs, the Brownian SYK model provides an analytically controlled arena for studying \emph{quantum scrambling} and \emph{operator complexity}. In both settings, the overarching question is how many simple, local ingredients—random gates or noisy $q$-body couplings—can collectively produce behavior that appears “random’’ at the level of many-body correlations.
\paragraph{Chaos}
Here we outline a link between correlation functions and classical chaos \cite{Maldacena2016,kitaev_hidden_correlations_2015}. Consider a system with configuration coordinates $q^i$ and conjugate momenta $p^i$, for $i=1,\dots,N$. Using Poisson brackets, sensitivity to initial conditions can be quantified by
\begin{equation} \label{eq:class_lyapunov}
    \{q^i(t),p^j(0)\}=\left| \frac{\partial q^i(t)}{\partial q^j(0)} \right|
    \sim e^{\lambda t},
\end{equation}
where $\lambda$ is the \emph{Lyapunov exponent} characterizing the exponential sensitivity to initial conditions. Turning to quantum mechanics, in the semiclassical limit the Poisson bracket maps to the commutator of the corresponding operators:
\begin{equation} \label{eq:quantize_mom}
    \{q^i(t),p^j(0)\}\sim -\frac{i}{\hbar}\,[q^i(t),p^j(0)], \qquad \hbar\to 0.
\end{equation}
Because $q^i$ and $p^j$ act at different times, the expression Eq.~\eqref{eq:quantize_mom} is nontrivial. This correspondence motivates extending the notions of classical chaos and a maximal Lyapunov exponent to general quantum systems \cite{kitaev_hidden_correlations_2015,de_Wijn_2012,Fine_2014,Aleiner_1996}. Heuristically, one seeks a quantity that faithfully captures the system’s sensitivity to initial conditions and reproduces the exponential growth \ref{eq:class_lyapunov} in the $\hbar \to 0$ limit for chaotic dynamics. 

Therefore, to diagnose the quantum analogue of classical sensitivity to initial conditions while avoiding depedence on a particular state, one works directly with the thermal average of the squared commutator
\begin{equation}
    C(t)=- \left<[q^i(t),p^j(0)]^2\right>_{\beta}=\frac{1}{Z}\sum_{n}e^{-\beta E_n} \left<n\right|[q^i(t),p^j(0)]^2 \left|n \right>.
\end{equation}
This choice avoids dependence on a particular pure state and remains meaningful in QFT (one often uses a regularized thermal OTOC with \(y=e^{-\beta H/4}/\sqrt{Z}\) to remove contact terms). In chaotic systems, \(C(t)\) shows early-time exponential growth before saturating near the scrambling time. Here \(\beta\) is the inverse temperature, \(E_n\) are energy eigenvalues, \(Z=\sum_n e^{-\beta E_n}\), and \(\langle\cdot\rangle_\beta\) denotes a thermal average. From Eq.~\eqref{eq:quantize_mom} one expects early-time growth \(C(t)\sim \hbar^2 e^{\lambda_L t}\) in the semiclassical regime. This construction generalizes naturally to large-\(N\) systems:
\begin{equation}\label{eq:corr}
  C(t)=-\big\langle [V(t),W(0)]^2 \big\rangle_{\beta},
\end{equation}
with \(V,W\) Hermitian, few-body (\(\mathcal O(1)\) support) and \(\langle V\rangle_\beta=\langle W\rangle_\beta=0\). We call the dynamics \emph{chaotic} if a broad class of such operator pairs exhibits an intermediate-time exponential growth window; in the semiclassical limit this reduces to classical Lyapunov growth (cf. Eq.~\eqref{eq:class_lyapunov}). In integrable models, \(C(t)\) may grow for some pairs but not generically (see, e.g., \cite{Roberts_2015}). The maximal growth rate defines the \emph{quantum Lyapunov exponent} \(\lambda_L\) (with the bound \(\lambda_L\le 2\pi/\beta\)). Two time scales then emerge: a \emph{dissipation time} \(t_d\) (when two-point functions have decayed and \(C(t)\) is still small) and a parametrically larger \emph{scrambling time} \(t_*\sim \lambda_L^{-1}\log N\). In many large-\(N\) models, for \(t_d\lesssim t\lesssim t_*\),
\begin{equation}\label{eq:Lyapunov}
  C(t)\sim \frac{1}{N}\,e^{\lambda_L t}.
\end{equation}
Two caveats are worth stressing. First, the reasoning above is intentionally heuristic; the asserted link between exponential growth of $C(t)$ and classical chaos is not ironclad. There is evidence both supporting \cite{Cotler_2018} and questioning \cite{Xu_2020,Hashimoto_2017} this connection. Thus it is prudent to distinguish “scrambling’’ (exponential $C(t)$ growth) from “chaos’’ (exponential separation of classical trajectories), even though they are often conflated. Second, correlators are not the sole probe of quantum chaos. 

A prominent alternative hinges on \emph{level statistics} at small energy spacings: agreement with Random Matrix Theory (RMT) is widely taken as a hallmark of chaos \cite{Gharibyan_2018,Haake2010,Ott2002,Stockmann1999}. This perspective is closely related to the Eigenstate Thermalization Hypothesis (ETH) \cite{Srednicki1994,Deutsch_2018,DAlessio2016}, which posits that, under appropriate conditions, matrix elements of a local operator take the thermal form
\begin{equation}
    O_{ij}=\langle i| O |j \rangle
    =\bar O\,\delta_{ij}
      + e^{-\frac{1}{2}S(E)}\, f(E,\omega)\, R_{ij},
\end{equation}
where $|i\rangle$ is an energy eigenstate, $S(E)=-{\rm tr}(\rho \log \rho)$, $\bar O(E)={\rm tr}(\rho O)$, $f(E,\omega)=f(E,-\omega)$ is smooth and real, and $R_{ij}$ is a Hermitian random matrix with zero mean and unit variance. Whether and how this framework is tied to correlators remains unsettled, though there is suggestive evidence \cite{Foini_2019,Murthy_2019,Parker_2019,Avdoshkin_2020,Huang_2019}. In particular, the SYK model and certain 2D CFTs at large central charge, under suitable assumptions, display RMT-like behavior \cite{Sonner_2017,Nayak_2019,Anous_2019}, and their correlators take the form Eq.~\eqref{eq:Lyapunov}.

\paragraph{SYK }

Among solvable many-body models, the Sachdev–Ye–Kitaev (SYK) model offers a paradigmatic setting for fast scrambling. It is defined in terms of $N$ Majorana fermions $\{\chi_i\}_{i=1}^{N}$ with $\{\chi_i,\chi_j\}=\delta_{ij}$ and Hilbert space dimension $\mathcal D = 2^{N/2}$. The standard (quenched) SYK Hamiltonian is
\begin{equation}
  H \;=\; i^{q/2}\!\!\!\sum_{1\le\mu_1<\cdots<\mu_q\le N}
     J_{\mu_1\cdots\mu_q}\,\chi_{\mu_1}\cdots\chi_{\mu_q},
  ~~
  \mathbb{E}\!\left[J_{\mu_1\cdots\mu_q}J_{\nu_1\cdots\nu_q}\right]
  = \sigma^2\,\delta_{\mu_1\nu_1}\cdots\delta_{\mu_q\nu_q}.
\end{equation}
In \emph{Brownian} SYK, the couplings fluctuate in time as independent Gaussian white noises,
\begin{equation}
  \mathbb{E}\!\left[J_{i_1\cdots i_q}(t)\,J_{j_1\cdots j_q}(t')\right]
  = \sigma^2\,\delta_{i_1 j_1}\!\cdots\delta_{i_q j_q}\,\delta(t{-}t'),
\end{equation}
with normalization $\sigma^2 \sim J$, where $J$ is the strength of the coupling (the precise coefficient will be fixed below). This time dependence implements a fresh random $q$-body interaction at each infinitesimal step, and the temporal independence enables a controlled large-$N$ analysis in real time.

In SYK, thermal averages such as Eq.~\eqref{eq:corr} can be computed explicitly in the large-$N$ limit~\cite{Sachdev1993,MaldacenaStanford2016,Saad2018ASR,Rosenhaus2019}. A convenient tool for organizing these computations is the thermofield double (TFD) state, which purifies the Gibbs ensemble into an entangled state on two copies of the Hilbert space~\cite{TakahashiUmezawa1975,Israel1976}. Besides streamlining manipulations of thermal traces, the TFD has a direct physical interpretation in holography as the two-sided black-hole geometry~\cite{Maldacena2003}. In the Sachdev–Ye–Kitaev context, the low-energy dynamics is captured by nearly anti–de Sitter space in two dimensions (AdS$_2$) / Jackiw–Teitelboim (JT) gravity, making the TFD the natural language to discuss scrambling and thermal correlators~\cite{MaldacenaStanfordYang2016,EngelsoyVerlindeVerlinde2016,KitaevSuh2018,Shenker2014,Maldacena2016}, and eventually the frame potential and unitary-design diagnostics~\cite{Gross2007,RoyScott2009}.

\begin{figure}[t]
\centering
\begin{tikzpicture}[x=1cm,y=1cm,line cap=round,line join=round]

\definecolor{DuneDeep}{RGB}{112,13,161}   
\definecolor{DuneMid}{RGB}{198,102,175}   
\definecolor{DuneLight}{RGB}{236,136,213} 
\definecolor{BlueVio}{RGB}{45,90,222}    
\definecolor{BlueSoft}{RGB}{120,165,245}  

\tikzset{
  edge/.style      ={draw=DuneDeep!60, line width=0.9pt},
  edgeThick/.style ={draw=DuneDeep!60, line width=1.0pt},
  rimFront/.style  ={draw=DuneDeep!60, line width=0.9pt, dashed},
  rimBack/.style   ={draw=DuneDeep!60, line width=0.9pt},
  txt/.style       ={font=\footnotesize, text=DuneDeep},
  title/.style     ={font=\bfseries\footnotesize, text=DuneDeep},
  eq/.style        ={font=\footnotesize, text=DuneDeep}, 
}

\begin{scope}[shift={(-5.4,0)}]
  \node[title] at (0,3.2) {(a) Thermal circle};
  \def\Rcircle{1.9}

  \shade[inner color=DuneLight!55, outer color=BlueSoft!0]
    (0,0) circle[radius=\Rcircle];

  \draw[edgeThick] (0,0) circle[radius=\Rcircle];
  \draw[edge] (\Rcircle,0) -- ++(0.35,0);

  \def\thO{120}
  \fill[DuneDeep!70] ({\Rcircle*cos(\thO)},{\Rcircle*sin(\thO)}) circle[radius=1.8pt];
  \node[txt,anchor=south west]
       at ({1.03*\Rcircle*cos(\thO)},{1.03*\Rcircle*sin(\thO)}) {$O$};

  \draw[->,edgeThick] (\Rcircle,0) arc[start angle=0, end angle=0, radius=\Rcircle];
  \node[txt,anchor=west]
       at ({\Rcircle*cos(8)},{\Rcircle*sin(8)}) {$\tau=0,\beta$};
    \node[eq, align=center] at (0.5,-3)
      {$\displaystyle \langle O\rangle_{\beta}
        = \frac{1}{Z(\beta)}\,\mathrm{Tr}\!\big(e^{-\beta H}\,O\big)$};
\end{scope}

\begin{scope}[shift={(3.4,0)}]
  \node[title] at (0,3.2) {(b) Two thermal circles coupled with TFD };

  \def\RadX{2.3}
  \def\RadY{0.60}
  \def\HalfH{1.9}
  \def\Curvature{0.45}
  \def\CtrlY{1.1}

  \begin{scope}
    \path[clip]
      (-\RadX,-\HalfH)
        .. controls (-\RadX+\Curvature,-\HalfH+\CtrlY)
                    and (-\RadX+\Curvature,\HalfH-\CtrlY) ..
      (-\RadX,\HalfH) -- (\RadX,\HalfH)
        .. controls (\RadX-\Curvature,\HalfH-\CtrlY)
                    and (\RadX-\Curvature,-\HalfH+\CtrlY) ..
      (\RadX,-\HalfH) -- cycle;
    \shade[shading angle=90, top color=DuneLight!0, bottom color=BlueSoft!0]
      (-\RadX-0.1,-\HalfH-0.1) rectangle (\RadX+0.1,\HalfH+0.1);
  \end{scope}

  \shade[inner color=BlueSoft!45, outer color=BlueSoft!0]
    (0,\HalfH) ellipse[x radius=\RadX, y radius=\RadY];
  \draw[rimFront] (0,\HalfH) ellipse[x radius=\RadX, y radius=\RadY];
  \begin{scope}
    \clip (-\RadX-1,\HalfH) rectangle (\RadX+1,\HalfH+1); 
    \draw[rimBack] (0,\HalfH) ellipse[x radius=\RadX, y radius=\RadY];
  \end{scope}

  \shade[inner color=DuneLight!45, outer color=DuneLight!0]
    (0,-\HalfH) ellipse[x radius=\RadX, y radius=\RadY];
  \draw[rimFront] (0,-\HalfH) ellipse[x radius=\RadX, y radius=\RadY];
  \begin{scope}
    \clip (-\RadX-1,-\HalfH) rectangle (\RadX+1,\HalfH+1); 
  \draw[rimBack] (0,-\HalfH) ellipse[x radius=\RadX, y radius=\RadY];
  \end{scope}

  \draw[edgeThick]
    (-\RadX,-\HalfH)
      .. controls (-\RadX+\Curvature,-\HalfH+\CtrlY)
                  and (-\RadX+\Curvature,\HalfH-\CtrlY) ..
    (-\RadX,\HalfH);
  \draw[edgeThick]
    ( \RadX,-\HalfH)
      .. controls ( \RadX-\Curvature,-\HalfH+\CtrlY)
                  and ( \RadX-\Curvature,\HalfH-\CtrlY) ..
    ( \RadX,\HalfH);

  \node[txt, anchor=east]
    at (-\RadX+\Curvature-0.20, 0.0) {$\left|\mathrm{TFD}\right>$};
  \node[txt, anchor=east]
    at (\RadX+\Curvature+0.50, 0.0) {$\left<\mathrm{TFD}\right|$};

  \def\theta{60}
  \fill[DuneDeep!70] ({\RadX*cos(\theta)},{\HalfH + \RadY*sin(\theta)}) circle[radius=1.7pt];
  \node[txt, anchor=west]
    at ({1.03*\RadX*cos(\theta)},{\HalfH + 1.4*\RadY*sin(\theta)}) {$O$};

  \draw[->,edgeThick] (\RadX,\HalfH) arc[start angle=0, end angle=0, x radius=\RadX, y radius=\RadY];
  \node[txt,anchor=west]
    at ({\RadX*cos(8)},{\HalfH+\RadY*sin(8)}) {$\tau=0$};
  \node[txt,anchor=east]
    at ({-\RadX*cos(8)},{\HalfH+\RadY*sin(8)}) {$\tau=\beta$};

  \draw[->,edgeThick] (\RadX,-\HalfH) arc[start angle=0, end angle=0, x radius=\RadX, y radius=\RadY];
  \node[txt,anchor=west]
    at ({\RadX*cos(8)},{-\HalfH + \RadY*sin(8)}) {$\tau=0$};
  \node[txt,anchor=east]
    at ({-\RadX*cos(8)},{-\HalfH - \RadY*sin(8)}) {$\tau=\beta$};
    \node[eq, align=center] at (0,-3.0)
      {$\displaystyle
        \big\langle \mathrm{TFD}\big|\, O(\tau)\otimes \mathbb{I}\, \big|\mathrm{TFD}\big\rangle
        $};
    %
  
\end{scope}

\end{tikzpicture}
\caption{Thermal average and its TFD representation. 
(a) Euclidean “thermal circle’’ of circumference $\beta$ with seam identifying $\tau{=}0\sim\beta$; an operator $O$ is inserted at Euclidean time $\tau$. 
(b) Two copies $L,R$ prepared in the thermofield–double state $\big|\mathrm{TFD}\big\rangle$; the top/bottom endcaps correspond to $\tau{=}0$ and $\tau{=}\beta$ (front rim dashed, back rim solid). 
The TFD matrix element reproduces the thermal trace,
$\langle O\rangle_\beta = Z(\beta)^{-1}\,\mathrm{Tr}\!\big(e^{-\beta H} O\big)
= \big\langle \mathrm{TFD}\big|\, O_L(\tau)\, \big|\mathrm{TFD}\big\rangle$ with $Z(\beta)=\mathrm{Tr}\!\big(e^{-\beta H}\big)$.}
\label{fig:TFD}
\end{figure}
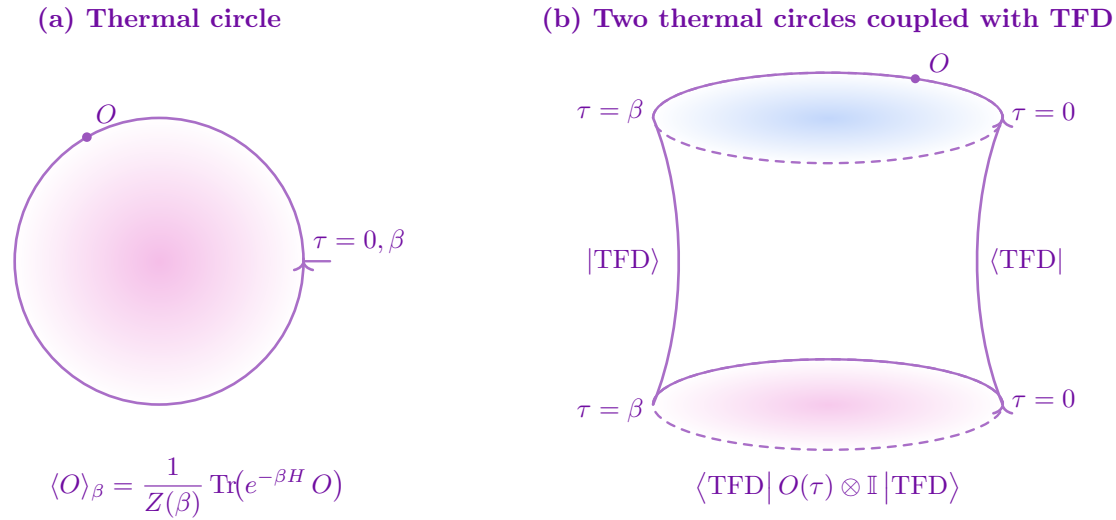

Concretely, the TFD state entangles two identical thermal systems $\mathcal{H}_L\otimes\mathcal{H}_R$~\cite{TakahashiUmezawa1975,Israel1976}:
\begin{equation}
  \label{eq:TFD-def-intro}
  \ket{\mathrm{TFD}_\beta}
  \;=\;
  \frac{1}{\sqrt{Z(\beta)}} \sum_{n} e^{-\beta E_n/2}\, \ket{n}_L \otimes \ket{n}_R,
\end{equation}
where the sum runs over a complete set of energy eigenstates 
\(\{|n\rangle\}\) of \(H\). TFD purifies the Gibbs state $\rho_\beta=e^{-\beta H}/Z$, so that thermal expectation values map to expectation values in a pure entangled state (see Fig.\ref{fig:TFD})~\cite{TakahashiUmezawa1975,Israel1976}:
\begin{equation}
\left<O\right>_\beta = \frac{1}{Z}\mathrm{Tr}(e^{-\beta H} O)
= \bra{\mathrm{TFD}_\beta} \, O_L \, \ket{\mathrm{TFD}_\beta},
\end{equation}
where $O_L=O\otimes \mathbb I$. This construction not only streamlines the computation of correlators but also naturally connects to entanglement measures such as mutual information between the left and right copies, with a geometric interpretation via the eternal AdS black hole~\cite{Maldacena2003,Shenker2014,Shenker2014-2,Hayden2007}. 

In SYK, these quantities display the characteristic signatures of fast scrambling: exponential growth of commutators with Lyapunov exponent $\lambda_L=2\pi/\beta$~\cite{Kitaev2015,Maldacena2016,MaldacenaStanford2016} and decay of mutual information on a time scale $t_0\sim \beta\log N$~\cite{Shenker2014,Shenker2014-2,Hayden2007}. In later chapters we show that, using the infinite-temperature TFD, the frame potential reduces to the $k$-th moment of the spectral form factor~\cite{RoyScott2009,Gross2007,Cotler2017,Saad2018ASR}. Thus, Brownian SYK links three standard diagnostics—OTOCs, spectral form factors, and frame potentials—in one solvable framework~\cite{Snderhauf2019,Jian2021,Zhang:2023vpm}.

\medskip

As noted in the quantum-chaos section, energy-level statistics provide another valuable diagnostic. 
In the SYK model, the spectral statistics depend sensitively on \(q \bmod 4\) and \(N \bmod 8\).
After unfolding and within a fixed fermion-parity sector, the nearest-neighbor level-spacing distribution agrees with the Wigner–Dyson form of the appropriate Altland–Zirnbauer (AZ) class. 
These congruences determine whether the relevant ensemble is GOE (\(\beta=1\)), GUE (\(\beta=2\)), GSE (\(\beta=4\)), or one of their chiral or superconducting/Bogoliubov–de Gennes (BdG) variants. 
By “chiral” we mean a sublattice symmetry \(S\) with \(S^2=1\) and \(\{S,H\}=0\), which renders \(H\) off-diagonal in the \(S\)-basis and enforces an \(E\!\leftrightarrow\!-E\) symmetry (AZ classes BDI/AIII/CII). 
“BdG” refers to particle–hole symmetric superconducting Hamiltonians with an antiunitary \(C\) obeying \(C H C^{-1}=-H\) (\(C^2=\pm1\)), leading to the AZ classes D, C, DIII, CI.
A complete AZ classification of SYK spectral statistics—including numerical confirmation of level spacings in each sector—was given in~\cite{Kanazawa2017}.

Baseline SYK offers large-$N$ solvability and maximal chaos (saturating $\lambda_L=2\pi/\beta$ at low temperature), making it a canonical testbed. The Brownian SYK variant retains the essential scrambling features while introducing temporally fluctuating couplings that render real-time dynamics analytically tractable—akin to a continuous random quantum circuit.

In the main part of the text we connect the discussion on correlators and quantum chaos with the calculation of the frame potential and quantum information in the Brownian SYK model. Starting the evolution from the infinite-temperature TFD state ($\beta=0$), we rewrite the time-dependent frame potential as the $k$-th moment of a spectral form factor, another well-known diagnostic of quantum dynamics closely related to the thermal correlator $C(t)$.

\medskip
\noindent\textbf{Our approach.} Rather than compute $C(t)$, we use the frame potential as a complementary, moment-based diagnostic of pseudorandomness and design growth. We develop a Keldysh path-integral formulation and a bilocal-field saddle to compute $\mathcal{F}^{(k)}(t)$ in Brownian SYK and expose distinct symmetry structures for $q=2$ versus $q>2$, connecting them to the corresponding random-matrix ensembles.

\section{Quantum measurements}

In quantum mechanics, a measurement extracts classical information from a quantum system \cite{nielsen2010quantum,Watrous2018,BuschLahtiBook2016}. Unlike unitary evolution—which is deterministic, reversible, and norm-preserving—measurements are stochastic and non-unitary: they yield information while perturbing the state, often in a highly nontrivial way \cite{nielsen2010quantum,BuschLahtiBook2016}.

Understanding how measurements reshape dynamics in many-body systems is central to nonequilibrium physics, quantum chaos, and decoherence. In this thesis, we focus on \emph{weak measurements (continuous monitoring)}. Continuous monitoring is the realistic description of how quantum systems are actually read out: detectors have finite bandwidth and efficiency, and signals are acquired in time rather than in instantaneous projective shots \cite{WisemanMilburn2010,Jacobs2014,Leonhardt1997,GardinerZoller2004}. This framework enables real-time feedback and control—stabilizing phases, cooling motion, tracking phases, and suppressing errors—by conditioning dynamics on the measurement record \cite{WisemanMilburn2010,GardinerZoller2004}. Conceptually and computationally, the trajectory picture (stochastic Schrödinger \emph{(diffusive)} and quantum-jump equations) provides a physically transparent, pure-state unraveling of open-system evolution and an efficient simulation tool \cite{GisinPercival1992,DalibardCastinMolmer1992,Carmichael1993}. From an information perspective, weak continuous measurements naturally connect to quantum filtering and metrology, clarifying how information and backaction accrue in time \cite{Jacobs2014,GardinerZoller2004}. Finally, when such monitoring competes with entangling unitary dynamics, it gives rise to qualitatively new nonequilibrium phenomena—measurement-induced phase transitions and critical regimes—in monitored circuits and many-body systems: unitary dynamics tends to \emph{scramble} information and generate volume-law entanglement, whereas continuous measurements continuously extract information, collapse local degrees of freedom, and suppress entanglement toward area-law behavior \cite{LiChenFisher2018,SkinnerRuhmanNahum2019,ChanNandkishorePretko2019,BaoChoiAltman2020,Zabalo2020,AlbertonBuchholdDiehl2021,ElseMachadoNayak2020}, with early continuous-monitoring evidence in free-fermion chains \cite{CaoTilloyDeLuca2019}. We begin by recalling the basic formalism.

\subsubsection*{General formalism of quantum measurements}

A simplest example of quantum measurements is \emph{projection-valued measure (PVMs).} A projective measurement with outcomes $m$ is a collection of orthogonal projectors $\{P_m\}$ satisfying
\[
P_m P_n = \delta_{mn} P_m, \qquad \sum_m P_m = \mathbb{I}.
\]
Given a pre-measurement state $\rho$, the outcome probabilities and (selective) post-measurement state are
\[
p(m)=\Tr(P_m \rho), \qquad \rho_m = \frac{P_m \rho P_m}{p(m)}.
\]
If the outcome is discarded (a nonselective measurement), the state updates to $\rho'=\sum_m P_m \rho P_m$ \cite{nielsen2010quantum,Watrous2018}. However PVMs are too restrictive on the system’s Hilbert space. Many physically relevant measurements (finite-resolution/noisy detectors, joint/approximate measurements of incompatible observables, heterodyne detection, weak/continuous monitoring) and many information-theoretic tasks (optimal state discrimination, informationally complete tomography) cannot be described as projectors on the system alone \cite{Helstrom1976,BuschLahtiWernerRMP2014,Leonhardt1997,Jacobs2014}. Therefore one needs a generalized notion of quantum measurements \cite{Holevo1982,Watrous2018,BuschLahtiBook2016}.

\emph{Generalized measurements:}
A general quantum measurement is specified by a positive-operator-valued measure (POVM), a set of positive operators $\{E_m\}$ with
\[
E_m \ge 0,\qquad \sum_m E_m=\mathbb{I},\qquad p(m)=\Tr(E_m\rho).
\]
Projective measurements (PVMs) are a special case: $E_m=P_m$ with $P_m^2=P_m$, $P_mP_n=\delta_{mn}P_m$. Thus every PVM is a POVM, but not conversely; POVMs need not be idempotent, orthogonal, or commuting, and one may have more than $d$ outcomes in $d$ dimensions \cite{Watrous2018,RenesSIC2004,BuschLahtiBook2016}.

\emph{Naĭmark dilation.}
However any POVM can be realized as a projective measurement on a larger space \cite{Watrous2018,BuschLahtiBook2016}: couple the system to an ancilla prepared in $|0\rangle^A$,
\begin{equation}
    \tilde\rho = \rho \otimes \ket{0}\bra{0}^A
\end{equation}
apply a joint unitary $U$, such that $\tilde \rho(t)=U [\rho \otimes \ket{0} \bra{0}^A]U^{\dagger}$ and measure the ancilla with orthogonal projectors $\{P_m\}$
\begin{equation}
    p(m)={\rm Tr}([\mathbb{I} \otimes P^A_m] U [\rho \otimes \ket{0} \bra{0}^A]U^{\dagger} ).
\end{equation}
Writing isometry $V:=U(\,\cdot\,\otimes|0\rangle)$, then the POVM operators become:
\[
E_m = V^\dagger(\mathbb{I}\otimes P^A_m)V.
\]
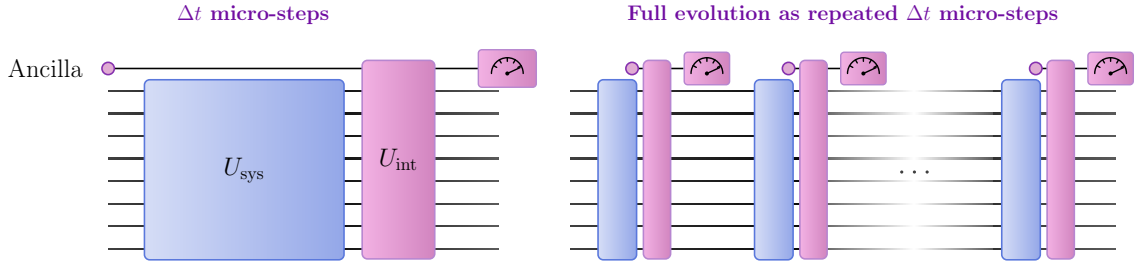
\begin{figure}[t]

\resizebox{\linewidth}{!}{
\begin{tikzpicture}[
  x=0.9cm, y=0.9cm, >=Stealth,
  txt/.style    ={font=\normalsize, text=DuneDeep},
  wire/.style   ={line width=0.9pt, draw=black},
  cwire/.style  ={line width=0.6pt, draw=black, densely dashed, -}, 
  gatebox/.style={rounded corners=3pt, draw=BlueVio, line width=0.9pt,
                  top color=BlueVio!25, bottom color=BlueVio!65, shading angle=90},
  couplebox/.style={rounded corners=3pt, draw=DuneDeep!50, line width=0.9pt,
                    top color=DuneLight!65, bottom color=DuneMid!80, shading angle=90},
  measbox/.style={rounded corners=2pt, draw=DuneDeep!50, line width=0.8pt,
                  top color=DuneLight!65, bottom color=DuneMid!80, shading angle=90,
                  minimum width=0.55cm, minimum height=0.46cm, align=center},
  ancprep/.style={circle, draw=DuneDeep!85, fill=DuneMid!55, line width=0.9pt, minimum size=7pt, inner sep=0pt},
  labelL/.style ={anchor=east, font=\large, text=black},
  labelR/.style ={anchor=west, font=\scriptsize, text=DuneDeep},
  fullbox/.style={rounded corners=3pt, draw=Teal!70, line width=0.9pt,
                  top color=Teal!25, bottom color=Teal!50, shading angle=90}
]
\tikzset{
  pics/meter/.style={
    code={
      \begin{scope}[line width=0.8pt, line cap=round, line join=round]
        \path[use as bounding box] (0,0) rectangle (28,18);

        \coordinate (C) at (14,9); 
        \def\R{7}                  
        \def\TickLen{1.6}          

        \draw (C) ++(180:\R) arc[start angle=180, end angle=0, radius=\R];

        \foreach \a in {150,120,90,60,30}{
          \draw (C) ++(\a:\R) -- ++(\a:-\TickLen);
        }

        \def\NeedleAngle{25}
        \draw (C) -- ++(\NeedleAngle:\R-0.8);
        \fill (C) circle (0.9);
      \end{scope}
    }
  }
}

\definecolor{DuneDeep}{RGB}{112,13,161}   
\definecolor{DuneMid}{RGB}{198,102,175}   
\definecolor{DuneLight}{RGB}{236,136,213} 
\definecolor{BlueVio}{RGB}{90,120,220}    
\definecolor{Teal}{RGB}{30,150,140}       

\def\Nanc{1}   
\def\Nsys{8}
\def\Hsep{1.0}
\def\wireh{0.035} 

\def\xL{0}
\def\xR{8.6}
\def\xGateL{2.0}
\def\xGateR{6.4}
\def\ypad{0.25} 

\pgfmathsetmacro{\xCoupleL}{\xGateR+0.20}
\pgfmathsetmacro{\xCoupleR}{\xGateR+1.20}
\def\ypadC{0.18}

\pgfmathsetmacro{\yTopAnc}{\Hsep*(\Nanc-1)} 
\pgfmathsetmacro{\ySysTop}{-\Hsep*1}        
\pgfmathsetmacro{\yBot}{-\Hsep*\Nsys}       

\foreach \j in {1,...,\Nsys} {
  \pgfmathsetmacro{\y}{-0.5*\Hsep*\j}
  \pgfmathsetmacro{\xMid}{0.5*(\xL+\xR)}
  \path[shading=axis, shading angle=0,
        left color=black!75, right color=black, draw=none]
       (\xL,\y-\wireh/3) rectangle (\xMid,\y+\wireh/3);
  \path[shading=axis, shading angle=0,
        left color=black, right color=black!65, draw=none]
       (\xMid,\y-\wireh/3) rectangle (\xR,\y+\wireh/3);
}

\draw[wire] (\xL,\yTopAnc) -- (\xR,\yTopAnc);
\node[labelL] at (\xL-0.40,\yTopAnc) {Ancilla};
\node[ancprep] at (\xL,\yTopAnc) {};

\path[gatebox] (\xGateL-1.2,\ySysTop+\ypad+0.5) rectangle (\xGateR-1.2,\yBot-\ypad+4);
\node[text=black,font=\large] at ({0.5*(\xGateL+\xGateR)-1.2},{0.5*(\ySysTop+\yBot+4.5)})
  {$U_{\mathrm{sys}}$};

\path[couplebox] (\xCoupleL-1,\yTopAnc+\ypadC) rectangle (\xCoupleR-0.4,\ySysTop-\ypadC-3.05);
\node[font=\large, text=black]
  at ({0.5*(\xCoupleL+\xCoupleR+0.6)-1},{0.5*(\yTopAnc+\ySysTop-3.05)}) {$U_{\mathrm{int}}$};

\node[measbox,
      minimum width=1.15cm,
      minimum height=0.75cm,
      inner sep=1pt,
      path picture={
        \node[inner sep=0pt] at ([yshift=-0.10cm]path picture bounding box.center)
          {\tikz[x=0.045cm, y=0.045cm]{\pic{meter};}};
      }] (ancMeas) at (8.8,\yTopAnc) {};

\def\panelgap{1.6}
\pgfmathsetmacro{\xLII}{\xR+\panelgap}
\pgfmathsetmacro{\xRII}{\xLII+12.0}

\def\StepW{1.40}
\def\StepGap{0.50}
\def\DropDepth{3.05}
\def\RightMargin{0.40}                 
\pgfmathsetmacro{\MeasOverhang}{0.70}   
\pgfmathsetmacro{\xStart}{\xLII+0.60}
\def\SecondShift{1.55}                  

\pgfdeclarelayer{bg}
\pgfsetlayers{bg,main}

\pgfmathsetmacro{\xsOne}{\xStart}
\pgfmathsetmacro{\xEOne}{\xsOne+\StepW}

\pgfmathsetmacro{\xsTwo}{\xStart + 1*(\StepW+\StepGap) + \SecondShift}
\pgfmathsetmacro{\xETwo}{\xsTwo + \StepW}

\pgfmathsetmacro{\xEThree}{\xRII - \RightMargin - \MeasOverhang}
\pgfmathsetmacro{\xsThree}{\xEThree - \StepW}

\pgfmathsetmacro{\xGapL}{\xETwo  }  
\pgfmathsetmacro{\xGapR}{\xsThree - 0.20}
\pgfmathsetmacro{\xGapM}{0.5*(\xGapL+\xGapR)+0.2}

\begin{pgfonlayer}{bg}
  \foreach \j in {1,...,\Nsys} {
    \pgfmathsetmacro{\y}{-0.5*\Hsep*\j}
    \pgfmathsetmacro{\xMid}{0.5*(\xLII+\xRII)}
    \path[shading=axis, shading angle=0,
          left color=black!75, right color=black, draw=none]
         (\xLII,\y-\wireh/3) rectangle (\xMid,\y+\wireh/3);
    \path[shading=axis, shading angle=0,
          left color=black, right color=black!65, draw=none]
         (\xMid,\y-\wireh/3) rectangle (\xRII,\y+\wireh/3);
  }
\end{pgfonlayer}
\foreach \j in {1,...,\Nsys} {
  \pgfmathsetmacro{\y}{-0.5*\Hsep*\j}
  \path[shading=axis, shading angle=0,
        left color=black!65, right color=white, draw=none]
       (\xGapL,\y-\wireh/3) rectangle (\xGapM,\y+\wireh/3);
  \path[shading=axis, shading angle=0,
        left color=white, right color=black!65, draw=none]
       (\xGapM,\y-\wireh/3) rectangle (\xGapR,\y+\wireh/3);
}

\foreach \s in {0,1,2}{
  \pgfmathsetmacro{\xsDefault}{\xStart + \s*(\StepW+\StepGap)}
  \pgfmathsetmacro{\xEDefault}{\xsDefault + \StepW}

  \ifnum\s=2
    \pgfmathsetmacro{\xE}{\xEThree}
    \pgfmathsetmacro{\xs}{\xsThree}
  \else\ifnum\s=1
    \pgfmathsetmacro{\xs}{\xsTwo}
    \pgfmathsetmacro{\xE}{\xETwo}
  \else
    \pgfmathsetmacro{\xs}{\xsOne}
    \pgfmathsetmacro{\xE}{\xEOne}
  \fi\fi

  \path[gatebox]
    (\xs, \ySysTop+\ypad+0.5)
    rectangle
    (\xE-0.55, \yBot-\ypad+4);

  \path[couplebox]
    (\xE-0.4, \yTopAnc+\ypadC)
    rectangle
    (\xE+0.2, \ySysTop-\ypadC-\DropDepth);

  \node[measbox,
        minimum width=0.90cm, minimum height=0.65cm, inner sep=1pt,
        path picture={
          \node[inner sep=0pt]
            at ([yshift=-0.10cm]path picture bounding box.center)
            {\tikz[x=0.045cm,y=0.045cm]{\pic{meter};}};
        }] (ancMeasR\s) at (\xE+1.0,\yTopAnc) {};

  \pgfmathsetmacro{\xIntL}{\xE - 0.40}      
  \pgfmathsetmacro{\xAstart}{\xIntL - 0.25} 
  \begin{pgfonlayer}{bg}
    \draw[wire] (\xAstart,\yTopAnc) -- (ancMeasR\s.west);
  \end{pgfonlayer}

  \node[ancprep] at (\xAstart,\yTopAnc) {};
}

\pgfmathsetmacro{\dsep}{0.28}
\fill[black!75] (\xGapM- \dsep,\yTopAnc-2.3) circle (0.035);
\fill[black!75] (\xGapM          ,\yTopAnc-2.3) circle (0.035);
\fill[black!75] (\xGapM+ \dsep,\yTopAnc-2.3) circle (0.035);

\node[txt, font=\bfseries] at ({0.5*(\xLII+\xRII)}, {\yTopAnc+1.2})
  {Full evolution as repeated $\Delta t$ micro-steps};

\node[txt, font=\bfseries] at ({0.5*(\xLII+\xRII)-13}, {\yTopAnc+1.2})
  {$\Delta t$ micro-steps};

\end{tikzpicture}}

\caption[Micro-step vs.\ full evolution]{\textbf{Left:} One monitored micro-step of duration $\Delta t$: the blue block $U_{\rm sys}$ acts on the system, then an ancilla–system interaction $U_{\rm int}$  is applied and the ancilla is measured. A dot on the ancilla line marks that a fresh ancilla is prepared each step (see App.~\ref{App:SSE}). \textbf{Right:} Full evolution shown as the repetition of blocks acting on system\(+\)ancilla.}\label{Fig:SSE_der}

\end{figure}

By Naĭmark, any measurement on the system can be realized by briefly coupling to a fresh ancilla and then projectively measuring the ancilla. A \emph{weak} measurement corresponds to making each system–ancilla interaction so short that the induced POVM on the system is close to the identity: each step reveals only a tiny amount of information and induces only a tiny backaction \cite{WisemanMilburn2010,Jacobs2014}. Repeating these steps with new ancillae turns the discrete readouts into a continuous noisy signal. Choosing the coupling so that the information per step scales like $\sqrt{\mathrm{d}t}$ see Fig.\ref{Fig:SSE_der} while the disturbance scales like $\mathrm{d}t$, a central-limit/Itô limit gives a stochastic, norm-preserving evolution for the conditioned state. For continuous monitoring of a Hermitian observable $M$ at rate $\Gamma$ (with Hamiltonian $H^{\rm uni}$) see Appendix(\ref{App:SSE}), the resulting diffusive stochastic Schrödinger equation is
which updates the state of the system $|\psi\rangle $ as:
\begin{equation} \label{eq:SSE}
    d \left| \psi \right>   =  -i dt  H^{\rm uni} \left|\psi \right>  
    +  \sqrt{\Gamma} \Bigg[ (M - \langle M \rangle ) dW  
     -  \frac{\sqrt{\Gamma} dt}{2} (M - \langle M \rangle )^2 \Bigg] | \psi \rangle , 
\end{equation}

where $\mathrm{d}W_t$ is a Wiener increment ($\mathbb{E}[\mathrm{d}W_t]=0$, $\mathrm{d}W_t^2=\mathrm{d}t$) \cite{GisinPercival1992,Carmichael1993,GardinerZoller2004}. Averaging over the noise (discarding the record) recovers the usual Lindblad dephasing toward the $M$ eigenbasis \cite{Lindblad1976,GKS1976}.

Continuous measurements thus generalize the notion of state collapse from discrete jumps to smooth, stochastic evolution \cite{DalibardCastinMolmer1992,WisemanMilburn2010}. While each trajectory remains pure, the ensemble-averaged state becomes mixed. Crucially, continuous monitoring retains the core property of reducing entanglement and coherence over time by gradually extracting information; purification provides a complementary operational diagnostic, with characteristic timescales and universality clarified for monitored fermions and broader nonunitary processes \cite{Loio:2023fxn,DeLucaUniversality2023}.

To build intuition for how even minimal measurements can encode nontrivial information, let us turn to a simple example involving Haar-random states—an example I first heard in Romain Vasseur’s lecture at the 2025 Les Houches Summer School \cite{VasseurRandomCircuitsNotes}. This highlights how a single measurement outcome can already carry a surprisingly large amount of information.

\subsubsection*{Two Haar-random states example}
Consider a simple scenario. Take two different known Haar-random pure states, $\ket{\psi}$ and $\ket{\phi}$, of an $L$-qubit system (with Hilbert-space dimension $D=2^L$, and $L$ large). Then choose one of these states with probability $p(\phi)=p(\psi)=\tfrac{1}{2}$ and measure it.
We perform a bitstring measurement (one projective measurement in the computational basis) and obtain a bitstring $n \in {0,1}^L$.
Now we try to guess which state was measured. A natural question to ask is: what is the strategy for guessing, and what is the probability of a successful guess?
If both states are equally likely and known in advance, the optimal strategy is simple: look at the probabilities $p(n|\psi)=|\braket{n}{\psi}|^2$ and $p(n|\phi)=|\braket{n}{\phi}|^2$ for the observed outcome $n$, and guess the state that assigns the larger probability. This is the so-called maximum-likelihood rule~\cite{VasseurRandomCircuitsNotes,nielsen2010quantum}.
The probability of success using this strategy is
\begin{equation}
\begin{split}
P_{\mathrm{succ}}
= \sum_n \big[p(n|\psi) \theta\big(p(n|\psi)-p(n|\phi)\big)p(\psi)-p(n|\phi)\theta\big(p(n|\phi)-p(n|\psi)\big)p(\phi)\big].
\end{split}
\end{equation}
Here $\theta$ is the Heaviside function. This expression has a simple interpretation: for each possible outcome $n$, the two candidate states assign probabilities $p(n|\psi)$ and $p(n|\phi)$. The optimal strategy is to guess the state with the larger probability, which succeeds with probability $\max{p(n|\psi),p(n|\phi)}$. Summing over all outcomes then gives the total average success probability: it is simply the fraction of measurement weight that lies under the larger of the two distributions, outcome by outcome.
For Haar-random states, the probabilities $|\braket{n}{\psi}|^2$ in a fixed basis fluctuate strongly. In fact, in the large-$D$ limit they follow the Porter–Thomas distribution~\cite{Haake2010,Mehta}:
\begin{equation}
p(x = |\braket{n}{\psi}|^2) = D e^{-Dx}.
\end{equation}
Using this distribution, one can compute the average success probability of the maximum-likelihood rule:
\begin{equation}
P_{\mathrm{succ}}
=\left. \frac{D}{2} \int_0^1 d x \int_0^1 d y  D^2 e^{-D(x+y)} \max(x,y)\right|_{D \to \infty}
= 0.75.
\end{equation}
The result is striking: even a single measurement in a fixed basis allows us to identify the correct state with probability $P{\mathrm{succ}} = 0.75$. 
At first sight, one might expect such a task to be impossible: Haar-random states look nearly featureless in any fixed basis, and the measurement probabilities are broadly spread. Nevertheless, by comparing the likelihoods of the two candidate states, one gains a constant, $\mathcal{O}(1)$ amount of information from just one shot.
This simple calculation illustrates a broader lesson: even minimal measurements can encode nontrivial information about quantum states. Understanding how such information competes with entanglement growth is a central theme in the study of monitored quantum systems. Building on this intuition, much recent work has focused on identifying universal features of measurement-driven dynamics in both random circuits and microscopic physical models.
\subsubsection{Measurements vs.\ Unitary Evolution}

\begin{figure}[t]
\centering
\resizebox{\linewidth}{!}{%
\begin{tikzpicture}[x=0.80cm,y=0.80cm]

\pgfdeclarelayer{bg}
\pgfsetlayers{bg,main}

\definecolor{DuneDeep}{RGB}{112,13,161}   
\definecolor{DuneMid}{RGB}{198,102,175}   
\definecolor{DuneLight}{RGB}{236,136,213} 
\definecolor{BlueVio}{RGB}{90,120,220}    
\definecolor{DustGray}{RGB}{184,189,198}  

\colorlet{DeepNavy}{DuneDeep}    
\colorlet{Aegean}{BlueVio}       
\colorlet{Terracotta}{DuneMid}   

\def\ArcScale{1.15}
\def\NodeSize{3.0mm}
\def\Xraise{0.60}
\def\Xarm{0.10}
\def\dx{0.60}

\tikzset{
  labelsmall/.style={font=\footnotesize, text=DeepNavy},
  labelmain/.style={font=\bfseries\footnotesize, text=DeepNavy},
  base/.style={draw=black!55, line width=0.40pt}, 
  singarc/.style={draw=Aegean!90, line width=0.75pt, line cap=round,
                  shorten >=0.4pt, shorten <=0.4pt},
  cutx/.style={draw=Terracotta, line width=0.50pt, line cap=round},
  spinU/.style={circle, draw=Aegean, line width=0.55pt,
                fill=Aegean!22, minimum size=\NodeSize, inner sep=0pt},
  spinM/.style={circle, draw=Terracotta, line width=0.55pt,
                fill=Terracotta!18, minimum size=\NodeSize, inner sep=0pt}
}

\newcommand{\entlink}[3]{%
  \begin{pgfonlayer}{bg}
    \draw[singarc]
      (#1) .. controls ($ (#1)!0.5!(#2) + (0,{#3*\ArcScale}) $)
            and      ($ (#1)!0.5!(#2) + (0,{#3*\ArcScale}) $)
      .. (#2);
  \end{pgfonlayer}
}

\node[labelmain, anchor=west] at (-0.2, 2.35)
  {Low rate $p<p_c$: \textcolor{Aegean}{Volume law}};
\node[labelmain, anchor=west] at (8.6, 2.35)
  {High rate $p>p_c$: \textcolor{Terracotta}{Area law}};

\def\xL{0}
\def\yL{0}

\foreach \i in {0,...,11} {
  \node[spinU] (L\i) at (\xL + \dx*\i, \yL) {};
}

\entlink{L0}{L7}{0.65}
\entlink{L1}{L11}{0.95}
\entlink{L2}{L6}{0.42}
\entlink{L3}{L9}{0.72}
\entlink{L4}{L7}{0.36}
\entlink{L5}{L8}{0.38}
\entlink{L6}{L9}{0.32}
\entlink{L7}{L10}{0.50}

\coordinate (Lleft)  at ($(L0)+(-0.40,0)$);
\coordinate (Lright) at ($(L11)+(0.40,0)$);
\begin{pgfonlayer}{bg}\draw[base] (Lleft) -- (Lright);\end{pgfonlayer}
\node[labelsmall, anchor=north] at ($(Lleft)!0.5!(Lright)$) {$L$};

\node[labelsmall, font=\normalsize, anchor=south]
  at ($(Lleft)!0.5!(Lright) + (0,1.30)$) {$S \sim L$};

\def\xR{9.1}
\def\yR{0}

\foreach \i in {0,2,5,7,10} {
  \node[spinU] (R\i) at (\xR + \dx*\i, \yR) {};
}
\foreach \j in {1,3,4,6,8,9,11} {
  \node[spinM] (R\j) at (\xR + \dx*\j, \yR) {};
}

\entlink{R0}{R2}{0.22}
\entlink{R5}{R7}{0.22}
\entlink{R7}{R10}{0.36}

\coordinate (Rleft)  at ($(R0)+(-0.40,0)$);
\coordinate (Rright) at ($(R11)+(0.40,0)$);
\begin{pgfonlayer}{bg}\draw[base] (Rleft) -- (Rright);\end{pgfonlayer}
\node[labelsmall, anchor=north] at ($(Rleft)!0.5!(Rright)$) {$L$};

\foreach \j in {1,3,4,6,8,9,11} {
  \draw[cutx] ($(R\j)+(0,\Xraise)$) +(-\Xarm,\Xarm) -- +(\Xarm,-\Xarm);
  \draw[cutx] ($(R\j)+(0,\Xraise)$) +(-\Xarm,-\Xarm) -- +(\Xarm,\Xarm);
}

\node[labelsmall, font=\normalsize, anchor=south]
  at ($(Rleft)!0.5!(Rright) + (-2.3,1.30)$) {$S \sim \mathrm{const}$};

\begin{scope}[shift={(12.4,1.85)}] 
  \draw[singarc] (0,0) -- (0.9,0);
  \node[anchor=west, font=\scriptsize, text=DeepNavy] at (1.05,0) {entanglement links};
\end{scope}
\begin{scope}[shift={(12.4,1.45)}]
  \draw[cutx] (0.50,0) +(-\Xarm,\Xarm) -- +(\Xarm,-\Xarm);
  \draw[cutx] (0.50,0) +(-\Xarm,-\Xarm) -- +(\Xarm,\Xarm);
  \node[anchor=west, font=\scriptsize, text=DeepNavy] at (1.05,0) {local measurements};
\end{scope}

\end{tikzpicture}
}
\caption{Entanglement structure at low ($p<p_c$) and high ($p>p_c$) measurement rates. Blue links dominate at low $p$ (volume law), while magenta measurements suppress long links at high $p$ (area law).}
\label{fig:vol-area-law-dune}
\end{figure}
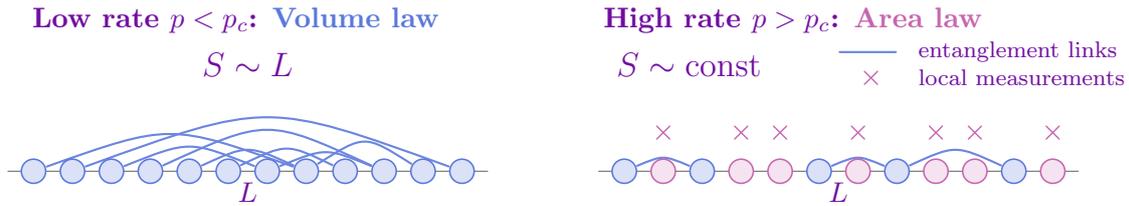

In isolated quantum systems, unitary dynamics generated by a Hamiltonian spread quantum information nonlocally, producing entanglement growth and (often) thermalization—i.e., scrambling. Local measurements, by contrast, suppress entanglement by projecting degrees of freedom onto more definite configurations. In monitored systems these tendencies compete, as illustrated in Fig.~\ref{fig:vol-area-law-dune}: in the left panel ($p<p_c$), unitary gates create long-range entanglement links (blue arches) across a chain of length $L$, yielding the volume-law scaling $S\sim L$; in the right panel ($p>p_c$), frequent local measurements (pink nodes marked $\times$) sever links so that only short-range correlations survive and the entropy saturates, $S\sim \mathrm{const}$. Tuning the measurement rate $p$ therefore drives a sharp dynamical transition—the \emph{measurement-induced phase transition} (MIPT)—between these volume-law and area-law regimes of entanglement entropy~\cite{LiChenFisher2018,SkinnerRuhmanNahum2019,ChanNandkishorePretko2019,BaoChoiAltman2020,Zabalo2020}.


\paragraph{Field-theoretic viewpoints.}Several complementary field-theory frameworks now organize our understanding of MIPTs.

(i) Replica/statistical-mechanics mappings. For hybrid Haar-random circuits, Rényi entropies map (via replicas) to a classical spin model of permutation degrees of freedom on the circuit’s spacetime lattice; the MIPT appears as an ordering transition in this model \cite{JianYouVasseurLudwig2020}. Closely related mappings are exact for stabilizer Clifford dynamics and random tensor networks, where the “spins’’ are permutations by Schur–Weyl duality \cite{LiVasseurFisherLudwig2024}. These approaches underpin the minimal-cut/percolation picture for $S_0$ and motivate domain-wall descriptions of entanglement fluctuations \cite{LiVijayFisher2023,NahumRuhmanAllToAll2021}.

(ii) Continuum and Landau–Ginzburg–Wilson (LGW) descriptions. By a LGW–type theory we mean a coarse-grained order-parameter field theory constrained by symmetries and locality. Such theories have been proposed for the measurement-induced transition \cite{BuchholdDiehl2021,NahumRuhmanAllToAll2021} and for forced-measurement transitions, where we study the ensemble of trajectories conditioned on a fixed spatiotemporal pattern of measurement outcomes (or enforce them adaptively) \cite{NahumRuhmanAllToAll2021,Turkeshi_2021,Feng_2023}. In this setting the key observable is the postselection cost—the exponential rate at which the success probability decays—and the corresponding order parameter differs from that of the MIPT \cite{Turkeshi_2021,Feng_2023}. At the critical point of $1{+}1$D circuits, evidence for emergent (nonunitary) conformal invariance has been found, with spacetime-conformal covariance of entanglement and mutual information \cite{LiChenLudwigFisherCFT2021}.

(iii) Keldysh/replica field theories for monitored matter. Continuous weak measurement admits stochastic-trajectory (unraveled) path-integral descriptions; in free-fermion and interacting cases one obtains replica Keldysh field theories whose symmetries determine whether volume- or area-law phases occur and how transitions emerge \cite{CaoTilloyDeLuca2019,FujiAshida2020,AlbertonBuchholdDiehl2021,GuoFosterJianLudwig2024}. Replica-limit subtleties specific to monitored systems are analyzed in \cite{GiachettiElusive2022}. These frameworks connect naturally to non-equilibrium techniques used elsewhere in many-body physics \cite{Kamenev2011}.

\paragraph{Beyond minimal settings.}
The universality class and even the phase structure can change with additional ingredients. Examples include long-range interactions \cite{BlockEtAl2022}, global or non-Abelian symmetries (yielding “spin-sharpening’’ or symmetry-enforced features), and SPT/topological structure and measurement-protected phases \cite{SangHsieh2021,LavasaniBarkeshli2020}. Disorder and dissipation offer further knobs: weak quenched disorder can stabilize extended critical regimes (often BKT-like) in monitored free fermions, while local dissipation can be incorporated in replica stat-mech models that deform the effective classical theory \cite{Szyniszewski2023,MonitoredDissipativePRB2023}. Related classical state-estimation problems also exhibit measurement-induced transitions via a directed-polymer mapping \cite{GerbinoDeLucaDirected2025,P_Kim_2025}.

\paragraph{Connections to the models studied in this thesis.}
The Brownian/SYK line of attack admits a controlled, large-$N$ field theory/replica formulation of monitored dynamics and their entanglement properties, enabling direct derivations of order parameters and saddle-point equations for the MIPT and related crossovers \cite{Jian2021Rep,Fava:2023tgg}. In monitored fermion chains (with or without conservation laws), the replica Keldysh approach clarifies when large replica symmetries enforce area-law phases and how interactions reduce symmetry to allow true transitions \cite{GuoFosterJianLudwig2024}. A recent work introduced such a field theory via the equation-of-motion approach \cite{Fava:2023tgg}. However, a derivation from first principles using the coherent-states formalism has not been attempted until now. In this thesis, we address this gap by deriving the emergent NLSM directly from the coherent-states path-integral representation \cite{ZhangFengGilmore1990,Nishiyama1981} of the replicated dynamics. We consider interacting SYK clusters each composed of $N_F$ Majorana fermions undergoing continuous monitoring of quadratic (two-fermion) idempotent operators, as modeled in the framework of the stochastic Schrödinger equation \cite{WisemanMilburn1993}. A complementary Dyson–Brownian-motion route to weak measurements in chaotic systems is provided by \cite{GerbinoDeLucaDyson2024}.

\chapter{Frame Potential of the Brownian SYK Model}
\label{chap:Frame potential}

\section{Motivation}

In quantum computing and quantum information, the ability to efficiently sample from Haar-random unitaries — or even approximate them — plays a crucial role. Random unitaries are essential for benchmarking quantum devices, modeling noise, and implementing cryptographic protocols. However, exact sampling from the Haar measure is exponentially hard in system size. This motivates the study of structured or dynamical processes that approximate Haar randomness in a physically realizable way.

One approach is to characterize how well a quantum system can emulate a Haar-random unitary over time. For example, if the system is allowed to evolve under a sufficiently complex Hamiltonian, does its time-evolution operator begin to mimic Haar randomness? Can this behavior be quantified? And at what point does the system's dynamics become indistinguishable — to some level of statistical scrutiny — from true randomness?

These questions lie at the intersection of quantum chaos, complexity theory, and statistical mechanics. In particular, they motivate the study of operator growth, scrambling, and the emergence of pseudorandomness in chaotic many-body systems.

Consider a simple operator, such as the Pauli matrix \( Z_1 \) acting on the first qubit of a many-body system. Under time evolution with a chaotic local Hamiltonian \( H \), the operator spreads and becomes increasingly complex:
\[
Z_1(t) = e^{iHt} Z_1 e^{-iHt}.
\]
At early times, \( Z_1(t) \) remains localized near site 1, but as time progresses, it evolves into a superposition of multi-qubit operators, supported on an ever-growing region of the system. Eventually, it becomes highly nonlocal, involving nearly all degrees of freedom — a signature of quantum chaos.

This process, known as \emph{operator growth}, is deeply connected to thermalization and to the scrambling of quantum information. It motivates a quantitative, moment-sensitive notion of how ``random’’ or ``complex’’ the dynamics have become in an information-theoretic sense.

A powerful and unifying framework for these ideas comes from \emph{unitary designs}. An ensemble $\mathcal{E}\subset\mathrm{U}(N)$ is a unitary $k$-design if, for every balanced polynomial $P$ of degree at most $k$ in $U$ and at most $k$ in $U^\dagger$,
\[
  \mathbb{E}_{U\sim\mathcal E}\!\bigl[P(U,U^\dagger)\bigr]
  \;=\;
  \mathbb{E}_{U\sim\mathrm{Haar}}\!\bigl[P(U,U^\dagger)\bigr].
\]
Equivalently, the $k$-copy twirl over $\mathcal{E}$ matches the Haar twirl. As chaotic dynamics $U(t)$ evolve, low moments typically converge first (small $k$), and higher moments later, yielding a structured hierarchy of pseudorandomness and complexity.

We study the design hierarchy in the \emph{Brownian SYK} model: $L$ Majorana (or Dirac) fermions with all-to-all $q$-body interactions (even $q$), whose couplings are Gaussian white noise. The time-dependent Hamiltonian is
\begin{equation}\label{eq:mot_sykh}
  H(t)=i^{\frac{q}{2}}
  \!\!\!\sum_{1 \le i_1<\cdots<i_q \le L}\!\!
  h_{i_1\cdots i_q}(t)\,\hat\chi_{i_1}\cdots \hat\chi_{i_q},
\end{equation}
where the Majoranas satisfy $\{\hat\chi_i,\hat\chi_j\}=\delta_{ij}$. The stochastic couplings obey
\begin{equation}\label{eq:noise}
  \big\langle h_{\boldsymbol{i}}(t_1)\,h_{\boldsymbol{j}}(t_2)\big\rangle
  = \frac{2^{\,q-1}q!}{q^2 L^{\,q-1}}\,
    \delta_{\boldsymbol{i},\boldsymbol{j}}\;\delta(t_1-t_2)
  \equiv \sigma^2\,\delta_{\boldsymbol{i},\boldsymbol{j}}\;\delta(t_1-t_2),
  \qquad
  \delta_{\boldsymbol{i},\boldsymbol{j}}=\delta_{i_1,j_1}\cdots\delta_{i_q,j_q}.
\end{equation}
 This Brownian form is closely related to the celebrated SYK model \cite{Sachdev1993,Gu:2019jub,Saad2018ASR} but with time-fluctuating couplings; it retains key features of scrambling, chaos, and thermalisation while allowing analytic control.

To quantify the approach to $k$-design behaviour \emph{along the orbit} of an initial state $\rho_0$, we compare two independent realisations of the stochastic evolution and define the state-dependent frame potential
\begin{equation}\label{eq:FP}
  F^{(k)}(T)
  = \Big\langle
      \big[\mathrm{Tr}\!\big(U_1(T)\rho_0 U_1^\dagger(T)\,U_2(T)\rho_0 U_2^\dagger(T)\big)\big]^k
    \Big\rangle.
\end{equation}
Here $U_{1,2}(T)$ are the evolution operators of two independent Brownian SYK Eq.~\eqref{eq:mot_sykh} systems (independent noise realisations), and $\langle\cdot\rangle$ averages over both; thus $F^{(k)}$ is the $k$-th moment of the two-copy overlap along the orbit of $\rho_0$.

\paragraph{Main results}

For sufficiently mixing Brownian SYK dynamics and a pure reference state $\rho_0=\ket{\Psi}\bra{\Psi}$, we find that $F^{(k)}(T)$ decays \emph{exponentially at early times} from its initial value $1$, reflecting the decreasing probability that two independently evolved copies remain close. At late times, the distribution of $U(T)$ approaches Haar \emph{within the symmetry sector} explored by $\rho_0$, so that
\begin{equation}\label{eq:latetimeconv}
  \rho(T)\stackrel{\mathrm{law}}{\longrightarrow} U^\dagger \rho_0 U,
  \qquad U \sim \mbox{Haar}(\text{sector}),
\end{equation}
and consequently $F^{(k)}(T)\to F_{\rm Haar}^{(k)}$, the corresponding sector-Haar value. As discussed in \cite{Ippoliti2022}, the Haar value minimises the frame potential among ensemble averages on the given Hilbert space (or sector), providing a sharp late-time benchmark for mixing.

We show that this convergence occurs for any $q>2$ (interactions coupling more than two fermions), generalising \cite{jian2022linear}. The Gaussian, integrable case $q=2$ mixes only within the manifold of Gaussian states \cite{10.21468/SciPostPhys.12.1.042,Bernard2021,swann2023spacetime}; there the late-time law is governed by a Gaussian–Haar (gHaar) measure \cite{PhysRevE.104.014146} and exhibits logarithmic-in-$L$ corrections from Goldstone fluctuations about the long-time saddle. In Dirac SYK, a global $U(1)$ charge splits the late-time plateaux by charge sector, leading to behaviour distinct from the Majorana case.

\begin{figure}
    \centering
    \includegraphics[width=1 \textwidth]{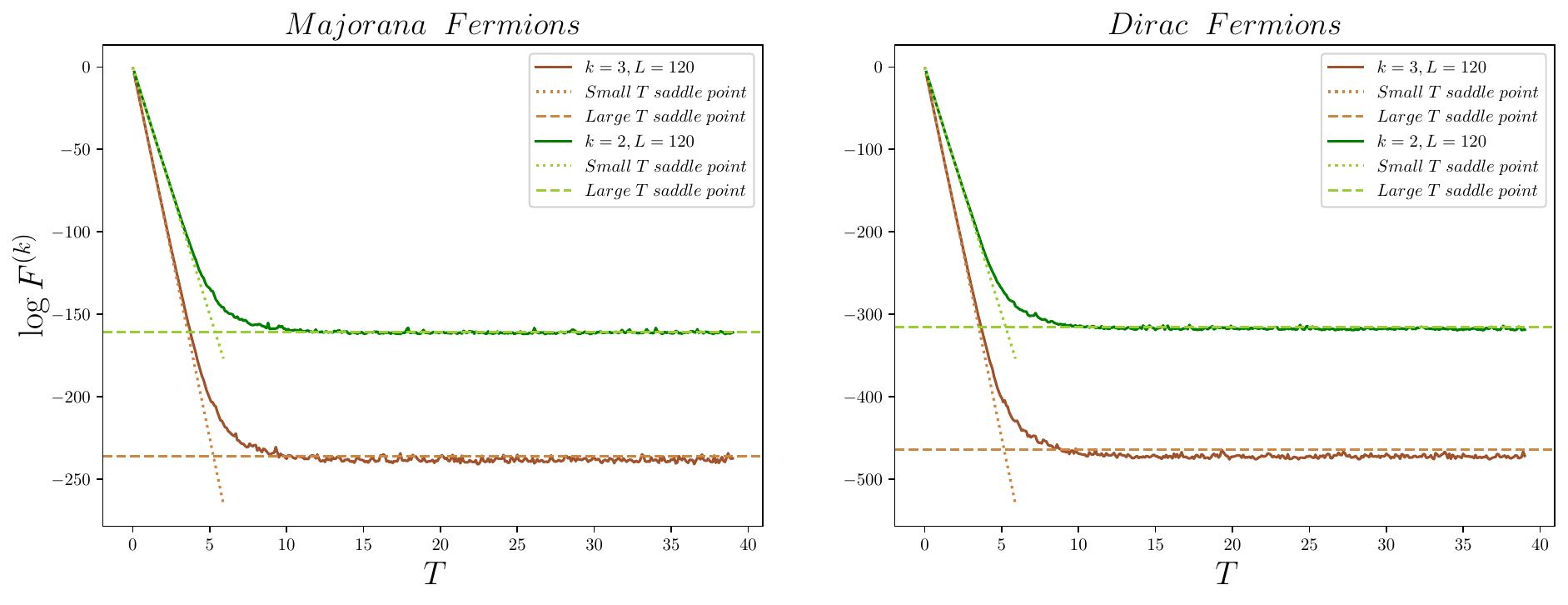}
    \caption{Rescaled state-dependent frame potential $F^{(k)}(T)$ for \emph{Brownian SYK} with $q=2$, plotted on a logarithmic (vertical) scale. Solid curves are exact numerics at $L=120$: green for $k=2$, brown for $k=3$. Dotted lines show the early-time Keldysh prediction; dashed lines show the late-time (sector-Haar) prediction. The saturation time $T_{\mathrm{sat}}$ is estimated by the intersection of the dotted and dashed analytics, up to finite-size corrections (larger for higher $k$). For $q=2$,
$T^{q=2}_{\rm Dirac}=4\!\left(2\log 2-\tfrac{k}{L}\log L\right)+O(L^{-1})$ and
$T^{q=2}_{\rm Majorana}=4\!\left(2\log 2-\tfrac{k-1}{L}\log L\right)+O(L^{-1})$;
for $q>2$, $T^{q>2}_{\rm Dirac}=T^{q>2}_{\rm Majorana}=q^{2}\cdot 2\log 2+O(L^{-1})$.}

    \label{fig:num}
\end{figure}
Summarizing our results here, we find the following: 
\paragraph{Short-time decay }
\[
F^{(k)}(T)\simeq
\begin{cases}
\exp\!\left(-\dfrac{LTk}{2q^2}\right), & \text{Majorana},\\[4pt]
\exp\!\left(-\dfrac{LTk}{q^2}\right), & \text{Dirac}.
\end{cases}
\]

\paragraph{Late-time saturation}
\[
F^{(k)}(T\!\to\!\infty)\simeq
\begin{cases}
\dfrac{k!}{\mathcal N^{2k}}, & \text{Majorana,\ } q>2,\\[6pt]
\dfrac{L^{k}k!}{\mathcal N^{2k}}, & \text{Dirac,\ } q>2,\\[8pt]
\dfrac{c_k\,L^{\frac{k(k-1)}{2}}}{\mathcal N^{2k}}+\dots, & \text{Majorana,\ } q=2,\\[8pt]
\dfrac{\tilde c_k\,L^{k^2}}{\mathcal N^{2k}}+\dots, & \text{Dirac,\ } q=2,
\end{cases}
\]
with $c_k=2^{1-\frac{(k-1)(k-2)}{2}}\prod_{i=0}^{k-1}\frac{1}{(2i-1)!!}$ and $\tilde c_k=\dfrac{{\rm sf}(k-1)^2}{{\rm sf}(2k-1)}$, ${\rm sf}(n)=\prod_{m=1}^{n} m!$.

The stochastic nature of the couplings enables us to use the \emph{Keldysh path integral formalism}, a powerful field-theoretic technique designed for averaging over non-equilibrium quantum evolutions. By reformulating the problem in terms of bi-local collective fields, we compute the frame potential analytically at both short and long times via a saddle-point approximation in the large-\( L \) limit.

We further show how global symmetries — such as $U(1)$ charge conservation in the Dirac case — constrain the dynamics and lead to sector-dependent randomness. These features result in different saturation values and finite-size corrections.

In summary, this chapter provides a detailed and pedagogical exploration of how the frame potential encodes the transition from structured to random dynamics in a solvable quantum system. By computing it explicitly in the Brownian SYK model, we make precise the idea that quantum chaos is not just randomness — it is the \emph{emergence of randomness by design}.

\section{Unitary \texorpdfstring{$k$}{k}-Designs, Complexity, and Chaos}

In many-body quantum systems, chaotic evolution can give rise to statistical properties that resemble those of completely random unitary operators. However, exact Haar-random unitaries are physically intractable due to their exponential complexity. Instead, one often considers ensembles of unitaries that approximate the Haar measure in a weaker sense. This leads to the concept of \emph{unitary $k$-designs}.

\textbf{Unitary $k$-Designs}

Let $\mathcal{E}$ be an ensemble of unitary operators on a Hilbert space $\mathcal{H}$ of dimension $d$, equipped with a probability measure. The $k$-fold twirling channel associated with $\mathcal{E}$ is defined as
\begin{equation}
    \Phi^{(k)}_{\mathcal{E}}(\mathcal{O}) = \int_{\mathcal{E}} dU\, U^{\otimes k} \mathcal{O} (U^\dagger)^{\otimes k},
\end{equation}
for any operator $\mathcal{O}$ acting on $\mathcal{H}^{\otimes k}$. We say that $\mathcal{E}$ is a \emph{unitary $k$-design} if
\begin{equation}
    \Phi^{(k)}_{\mathcal{E}}(\mathcal{O}) = \Phi^{(k)}_{\text{Haar}}(\mathcal{O})
\end{equation}
for all $\mathcal{O}$, where $\Phi^{(k)}_{\text{Haar}}$ denotes averaging over the full Haar measure on $U(d)$.

A $k$-design thus reproduces the $k$-th moment of the Haar distribution exactly. The case $k=1$ corresponds to thermalization of single-qubit observables, $k=2$ to scrambling and OTOCs, and higher $k$ to full pseudorandomness and complexity growth. For $k \geq 2$, exact constructions are rare; instead, one often works with approximate $k$-designs.

\textbf{Approximate designs.}
 The \emph{diamond norm} of a linear map $\Phi$ is
\[
  \|\Phi\|_\diamond \;:=\; \sup_{d\ge 1}\ \sup_{\|X\|_1=1}\ \big\| (\Phi \otimes \mathrm{id}_d)(X) \big\|_1
  \;,
\]
where $\|\cdot\|_1$ is the trace norm, $\mathrm{id}_d$ is the identity channel on a $d$-dimensional ancilla. \cite{Watrous2018}.
An ensemble $\mathcal{E}$ is an \emph{$\epsilon$-approximate $k$-design} if the diamond-norm distance between its $k$-fold twirl channel and the Haar twirl is bounded:
\begin{equation}
    \bigl\| \Phi^{(k)}_{\mathcal{E}} - \Phi^{(k)}_{\mathrm{Haar}} \bigr\|_\diamond \le \epsilon.
\end{equation}

In practice, the \emph{frame potential} provides a convenient, computable proxy for proximity to a $k$-design. Originally introduced in quantum information \cite{Gross2007}, it has become a standard diagnostic of chaos and complexity \cite{Roberts2017,Cotler2017,Haferkampha2022}.

\textbf{Frame Potential}
The $k$-th frame potential of an ensemble $\mathcal{E}$ is defined as
\begin{equation}
    F^{(k)}_{\mathcal{E}} = \int_{\mathcal{E}} dU \int_{\mathcal{E}} dV \left| \mathrm{Tr}(U^\dagger V) \right|^{2k}.
\end{equation}

Intuitively, the frame potential quantifies the average overlap between two unitaries drawn from the ensemble. Its decay signals the loss of structure and the emergence of random-like behavior.
This quantity is minimized by the Haar measure:
\[
F^{(k)}_{\mathcal{E}} \geq F^{(k)}_{\text{Haar}} ,
\]
with equality if and only if $\mathcal{E}$ is a $k$-design (for $k \leq d$). Moreover, the frame potential bounds the diamond norm distance as
\begin{equation}
    \left\| \Phi^{(k)}_{\mathcal{E}} - \Phi^{(k)}_{\text{Haar}} \right\|_\diamond \leq d^{2k} \left( F^{(k)}_{\mathcal{E}} - F^{(k)}_{\text{Haar}} \right),
\end{equation}
allowing a quantitative measure of Haar-randomness via moment statistics.

Beyond diagnosing proximity to Haar randomness, the frame potential also encodes information about the system’s \emph{complexity}. In quantum chaotic systems, the decay of the frame potential correlates with the growth of computational complexity — the minimal number of simple operations required to approximate the time-evolution operator \( U(t) \). While this connection was first explored in the context of random circuit models \cite{Roberts2017}, it extends to Hamiltonian systems as well: as the system evolves under chaotic dynamics, \( U(t) \) becomes increasingly complex, nonlocal, and harder to simulate.

In particular, ensembles that saturate the frame potential near its Haar value can no longer be efficiently distinguished from true randomness using \( k \)-moment statistics — implying a high degree of pseudorandomness and circuit complexity. In this sense, the decay of the frame potential serves not only as a diagnostic of scrambling, but also as a coarse-grained, dynamical probe of quantum complexity growth, even in continuous-time systems governed by stochastic Hamiltonians.

This perspective motivates the use of the frame potential as a physically accessible and analytically tractable measure of how complexity builds up in real time — a central theme in the Brownian SYK model analyzed in this chapter.

\textbf{Design Formation in Random Circuits and Chaotic Systems}

The dynamics of various physical systems have been studied to understand how and when approximate $k$-designs emerge. In random circuit models, forming a $k$-design typically requires circuit depth scaling with system size and moment order.

For example, Harrow and Low showed that local random circuits form approximate 2-designs after $O(n^2)$ steps \cite{Harrow2009}. Subsequent works analyzed the moment operator gap and extended these results to 3-designs and beyond \cite{Brandao2016}. In particular, Brownian circuits and parallelized all-to-all models form $k$-designs in depths scaling like $O(n \cdot \text{poly}(k))$ \cite{Cotler2017,Brown2015,Haferkampha2022}.

In the context of Hamiltonian dynamics, Jian and collaborators have studied the convergence to designs under time evolution with chaotic or Brownian Hamiltonians \cite{Jian2021,jian2022linear}. These studies reveal how the growth of operator size, complexity, and entanglement correlates with the decay of the frame potential, and how different physical platforms (e.g., SYK models, Brownian circuits, and holographic systems) exhibit varying time scales for design formation.

In this chapter, we will compute the frame potential analytically in the Brownian SYK model, providing a solvable and physically motivated setting in which to understand the emergence of pseudorandomness from chaotic dynamics.

\section{Keldysh Path Integral for Fermions}\label{sec:Keldysh}

To compute the frame potential in the Brownian SYK model, we employ the Keldysh path integral formalism.
We therefore begin with a brief review of the operator structure of the closed time contour and its representation in terms of fermionic coherent states.
Originally developed by Keldysh to handle non-equilibrium quantum systems, this formalism provides a systematic way to compute the expectation value of an observable $\hat{O}$ at some time $t$, given an initial density matrix $\hat{\rho}(-\infty)$ and a time-dependent Hamiltonian $\hat{H}(t)$. In the Schr\"odinger picture, the density matrix evolves according to the von Neumann equation
\begin{equation}
\partial_t \hat{\rho}(t) = -i \, [ \hat{H}(t), \hat{\rho}(t) ],
\end{equation}
with the formal solution
\begin{equation}
\hat{\rho}(t) = \hat{U}_{t,-\infty} \, \hat{\rho}(-\infty) \, \hat{U}_{-\infty,t},
\end{equation}
where $\hat{U}_{t,t'}$ is the unitary time-evolution operator,
\begin{equation}
\hat{U}_{t,t'} = T \exp\left[ -i \int_{t'}^t \! \hat{H}(\tau) \, d\tau \right].
\end{equation}

The expectation value is then
\begin{equation}
\langle \hat{O} \rangle(t) = 
\frac{\mathrm{Tr}\!\left\{ \hat{O} \, \hat{\rho}(t) \right\}}
{\mathrm{Tr}\{\hat{\rho}(t)\}}
= \frac{\mathrm{Tr}\!\left\{ \hat{U}_{-\infty,t} \, \hat{O} \, \hat{U}_{t,-\infty} \, \hat{\rho}(-\infty) \right\}}
{\mathrm{Tr}\{\hat{\rho}(-\infty)\}}.
\label{eq:exp_value}
\end{equation}
This expression already contains \emph{both forward and backward time evolution}: 
$\hat{U}_{t,-\infty}$ evolves the bra in the density matrix forward from $-\infty$ to $t$, while $\hat{U}_{-\infty,t}$ evolves the ket backward from $t$ to $-\infty$. 
In equilibrium, one can sometimes eliminate one branch using the adiabatic theorem and imaginary-time techniques; however, for genuinely non-equilibrium processes --- sudden quenches, periodic drives, coupling to external baths --- such simplifications fail, and both evolutions must be retained explicitly.

\begin{figure}[t]
\centering
\resizebox{\linewidth}{!}{%
\begin{tikzpicture}[x=1.00cm,y=1.00cm,>=Latex]

\pgfdeclarelayer{bg}
\pgfsetlayers{bg,main}

\definecolor{Saffron}{HTML}{FFB703}    
\definecolor{Terracotta}{HTML}{E76F51} 
\definecolor{Aegean}{HTML}{2A9D8F}     
\definecolor{DeepNavy}{HTML}{1D3557}   
\definecolor{DustGray}{HTML}{B8BDC6}   

\def\Xmin{-5.2}
\def\Xmax{ 5.2}
\def\Yfwd{ 1.20}
\def\Ybwd{-1.20}
\def\tPos{0.80}         
\def\EdgeThick{1.2pt}
\def\AxisThick{0.8pt}
\def\TickSize{0.10}
\def\Ytop{ 2.15}
\def\Ybot{-2.10}
\def\PlatePadX{0.55}
\def\PlatePadY{0.35}

\tikzset{
  lbl/.style={font=\scriptsize, text=DeepNavy},
  title/.style={font=\bfseries\footnotesize, text=DeepNavy},
  opnode/.style={draw=none, fill=Saffron, circle, minimum size=3.2mm, inner sep=0pt},
  rho/.style={draw=DeepNavy, rounded corners=2pt, line width=0.6pt, fill=DeepNavy!4, inner sep=2pt},
  branchFwd/.style={draw=Terracotta, line width=\EdgeThick},
  branchBwd/.style={draw=Aegean,    line width=\EdgeThick},
  axis/.style={draw=DustGray, line width=\AxisThick, dashed},
  tick/.style={draw=DustGray, line width=0.6pt},
  plate/.style={rounded corners=4pt, draw=DeepNavy!12, fill=DeepNavy!3, line width=0.5pt},
}

\begin{pgfonlayer}{bg}
  \draw[plate]
    (\Xmin-\PlatePadX-1.5, \Ybot-\PlatePadY) rectangle
    (\Xmax+\PlatePadX, \Ytop+\PlatePadY);
\end{pgfonlayer}

\node[title, anchor=west] at (\Xmin-0.2, \Ytop) {Closed-time contour with operator insertion};

\draw[branchFwd,->] (\Xmin,\Yfwd) -- (\Xmax,\Yfwd);

\draw[branchBwd,->] (\Xmax,\Ybwd) -- (\Xmin,\Ybwd);

\draw[branchFwd] (\Xmin,\Ybwd) -- (\Xmin,\Yfwd); 
\draw[branchBwd] (\Xmax,\Yfwd) -- (\Xmax,\Ybwd); 

\draw[axis] (\tPos, \Yfwd+0.22) -- (\tPos, \Ybwd-0.22);
\foreach \x/\txt in {\Xmin/$-\infty$, \tPos/$t$, \Xmax/$+\infty$} {
  \draw[tick] (\x,\Yfwd+\TickSize) -- (\x,\Yfwd-\TickSize);
  \draw[tick] (\x,\Ybwd+\TickSize) -- (\x,\Ybwd-\TickSize);
  \node[lbl, anchor=north] at (\x, \Ybwd-0.28) {\txt};
}

\node[lbl, anchor=south west] at (\Xmin+0.15, \Yfwd+0.15) {$U_{-\infty,+\infty}$};
\node[lbl, anchor=north west] at (\Xmin+0.35, \Ybwd-0.15) {$U_{+\infty,t}\,\hat O\,U_{t,-\infty}$};

\node[opnode] at (\tPos, \Ybwd) {};
\node[lbl, anchor=south] at (\tPos+0.3, \Ybwd-0.02) {$\hat{O}$};

\node[rho, anchor=east] at (\Xmin-0.38, 0.0) {$\hat\rho(-\infty)$};

\end{tikzpicture}%
}
\caption{Operator form of the Keldysh closed-time contour for Eq.~\eqref{eq:closed_contour}: the forward branch carries $U_{-\infty,+\infty}$, the backward branch carries $U_{+\infty,t}\,\hat O\,U_{t,-\infty}$; branches are glued at $+\infty$ by the trace, with initial weighting $\hat\rho(-\infty)$.}
\label{fig:keldysh-contour}
\end{figure}
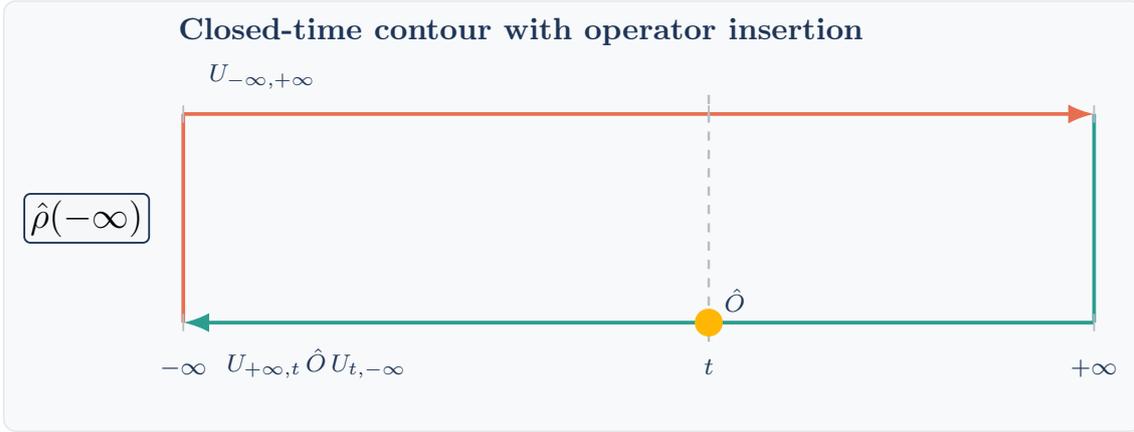

A convenient way to unify these two branches is to insert the identity
\begin{equation}
\hat{U}_{t,+\infty} \, \hat{U}_{+\infty,t} = \mathbb{I},
\end{equation}
into Eq.~\eqref{eq:exp_value}, extend the evolution to $+\infty$, and rearrange:
\begin{equation}
\langle \hat{O} \rangle(t) =
\frac{\mathrm{Tr}\!\left\{ 
\underbrace{\hat{U}_{-\infty,+\infty}}_{\text{forward branch}}
\underbrace{\hat{U}_{+\infty,t} \, \hat{O} \, \hat{U}_{t,-\infty}}_{\text{backward branch with insertion}}
\hat{\rho}(-\infty) \right\}}
{\mathrm{Tr}\{\hat{\rho}(-\infty)\}}.
\label{eq:closed_contour}
\end{equation}
This defines evolution along a \emph{closed time contour} $C$: first forward from $-\infty$ to $+\infty$, then backward to $-\infty$, with $\hat{O}$ inserted on one branch (or symmetrically as a half-sum over both) see Fig. \ref{fig:keldysh-contour}. 
The contour ordering guarantees that all operator insertions, regardless of branch, can be expressed through a single contour-ordered exponential.

When the initial state is thermal,
\begin{equation}
\hat{\rho}(-\infty) = \frac{e^{-\beta \hat{H}_{\mathrm{ini}}}}{Z(\beta)},
\end{equation}
the Boltzmann factor can be represented as \emph{imaginary-time evolution} over a duration $-i\beta$:
\begin{equation}
e^{-\beta \hat{H}} = e^{-i \hat{H} (-i\beta)}.
\end{equation}
This naturally extends the contour into the complex plane by adding a vertical segment of length $-i\beta$ at $t=-\infty$, implementing the thermal weighting as Euclidean time propagation. 
In the infinite-temperature limit ($\beta \to 0$) or for pure initial states, this vertical segment disappears, leaving a purely real-time contour.

Finally, by adding branch-dependent source terms,
\begin{equation}
\hat{H}^\pm_V(t) = \hat{H}(t) \pm \hat{O} \, V(t),
\end{equation}
the expectation value can be generated from the contour partition function
\begin{equation}
Z[V] = \frac{\mathrm{Tr}\{\hat{U}_C[V] \, \hat{\rho}(-\infty)\}}{\mathrm{Tr}\{\hat{\rho}(-\infty)\}},
\qquad 
\langle \hat{O} \rangle(t) = \frac{i}{2} \frac{\delta Z[V]}{\delta V(t)}\Big|_{V=0}.
\end{equation}
The closed-time-contour framework therefore packages the entire forward/backward evolution, initial-state preparation, and observable generation into a single, unified formalism. 
This makes it the natural starting point for the path-integral and diagrammatic techniques that will follow involving \emph{Grassmann-valued fields}.

\subsection*{Fermionic Path Integrals and Grassmann Variables}\label{subsec:Keldysh_Major}


We start with the simplest example: a single quantum level that can be either empty or occupied by a fermion. The Hilbert space is two-dimensional, spanned by
\[
|0\rangle, \qquad |1\rangle=\hat c^\dagger |0\rangle,
\]
so that \(|0\rangle\) denotes the empty level and \(|1\rangle\) the occupied one. The operators \(\hat c\) and \(\hat c^\dagger\) annihilate and create a fermion on this level and satisfy the canonical anticommutation relations
\[
\{\hat c,\hat c^\dagger\}=1,\qquad \{\hat c,\hat c\}=\{\hat c^\dagger,\hat c^\dagger\}=0.
\]
These relations encode Pauli exclusion: acting twice with \(\hat c^\dagger\) gives zero, so the occupation operator \(\hat n=\hat c^\dagger\hat c\) has eigenvalues \(0\) or \(1\).

To build a path integral that reproduces fermionic statistics using \(c\)-number fields, we introduce \emph{Grassmann variables} \(\psi(t)\) and \(\bar\psi(t)\), which anticommute:
\[
\psi\,\bar\psi=-\bar\psi\,\psi,\qquad \psi^2=\bar\psi^2=0,
\]
and also anticommute with the fermionic operators,
\[
\{\psi,\hat c\}=\{\psi,\hat c^\dagger\}=0.
\]
Because Grassmann numbers are nilpotent (\(\psi^2=0\)), any function of a single Grassmann variable truncates at linear order,
\[
f(\psi)=f_0+f_1\,\psi.
\]
Integration over Grassmann variables (Berezin integration) is defined so that it behaves like differentiation and correctly normalizes Gaussian integrals:
\[
\int d\psi\,1=0,\qquad \int d\psi\,\psi=1.
\]

Fermionic coherent states are defined to diagonalize the annihilation operator with a Grassmann eigenvalue. Specifically,
\begin{equation}
\label{eq:cohstate_fermion}
\hat c\,|\psi\rangle=\psi\,|\psi\rangle,
\end{equation}
which cannot be achieved with an ordinary complex superposition \(x|0\rangle+y|1\rangle\). Allowing a Grassmann coefficient fixes this:
\[
|\psi\rangle=|0\rangle-\psi\,|1\rangle=e^{-\psi\,\hat c^\dagger}|0\rangle,
\]
where the exponential truncates because \((\psi\,\hat c^\dagger)^2=0\). One checks directly that Eq.\eqref{eq:cohstate_fermion} holds. The corresponding left state is
\[
\langle\psi|=\langle 0|\,e^{-\hat c\,\bar\psi},
\]
with \(\bar\psi\) an independent from $\psi$ Grassmann variable. Coherent states are overcomplete; their overlap,
\[
\langle\psi|\psi'\rangle=1+\bar\psi\,\psi'=e^{\bar\psi\,\psi'},
\]
follows from nilpotency. Crucially, they resolve the identity,
\begin{equation}
\label{eq:id}
\int d\bar\psi\,d\psi\;e^{-\bar\psi\psi}\;|\psi\rangle\langle\psi|=\hat{\mathbb I},
\end{equation}
which we will insert between short-time evolution factors to convert operator traces into Grassmann path integrals. The Gaussian weight \(e^{-\bar\psi\psi}\) ensures correct normalization of the completeness relation.

On a closed real-time (Keldysh) contour \(C\)---forward and then backward in time---the normalized generating functional for an initial density matrix \(\hat\rho_0\) reads
\[
Z=\frac{{\rm Tr}\!\left(\hat U_C\,\hat\rho_0\right)}{{\rm Tr}(\hat\rho_0)}.
\]
Here \(\hat U_C\) is the contour-ordered evolution operator. The normalization by \({\rm Tr}(\hat\rho_0)\) guarantees \(Z=1\) in the absence of sources; it simply enforces that we compute expectation values with respect to \(\hat\rho_0\). To obtain a path integral, one Trotterizes \(\hat U_C\), inserts Eq.\eqref{eq:id} at each time slice, and evaluates short-time matrix elements of the form \(\langle\psi(t+\delta t)|e^{-i\delta t\,\hat H}|\psi(t)\rangle\). The overlap \(\langle\psi(t+\delta t)|\psi(t)\rangle\) produces the kinetic term, while the Hamiltonian matrix element yields the potential term.

For the single level with \(\hat H=\epsilon_0\,\hat c^\dagger\hat c\), the resulting contour action for the forward/backward branches \((+,-)\) is
\begin{equation}
\label{eq:action_dir}
S[\bar\psi_\pm,\psi_\pm]=\int_0^T\!dt\;\Big[\;\bar\psi_+(t)\,(i\partial_t-\epsilon_0)\,\psi_+(t)\;-\;\bar\psi_-(t)\,(i\partial_t-\epsilon_0)\,\psi_-(t)\;\Big].
\end{equation}
 The overall minus sign between the two contributions reflects the reversed time orientation on the backward branch. Boundary conditions at the turn-around of the contour glue the fields, while \(\hat\rho_0\) encodes the initial occupation (e.g., a thermal \(\hat\rho_0\) leads to Fermi factors in Keldysh Green’s functions). This quadratic action is the starting point for computing real-time propagators and for adding interactions, sources, or couplings to baths.

Since we are going to work with Majorana fermions we should adapt this approach to real fermions. 
For Dirac fermions, the Keldysh path integral follows directly from coherent states, as reviewed in Appendix~\ref{App:path_int_dir}. 
Majorana fermions, however, present a subtlety: a Majorana operator is Hermitian, 
\(\hat\chi = \hat\chi^\dagger\), and therefore simultaneously mixes creation and annihilation operators. 
As a consequence, no nontrivial Grassmann coherent states exist for a single Majorana mode.  

Two standard workarounds resolve this difficulty:
\begin{enumerate}
    \item \textbf{Auxiliary-field method.} One introduces an auxiliary Majorana \(\hat\xi\) to form a Dirac fermion 
    \[
    \hat c = \tfrac{1}{\sqrt{2}}(\hat\chi + i \hat\xi), \qquad \hat c^\dagger = \tfrac{1}{\sqrt{2}}(\hat\chi - i \hat\xi).
    \]
    The path integral can then be constructed for \(\hat c, \hat c^\dagger\) and re-expressed in terms of Grassmann fields \( \chi, \xi\). 
    The \(\xi\)-sector factorizes and contributes only an overall prefactor, which cancels once the partition function is properly normalized. This method we describe in the Appendix \ref{App:path_int_maj}.
    \item \textbf{Pairing method.} If the system contains an even number of Majorana operators, they can be paired into Dirac fermions. 
    The path integral is first constructed in the Dirac basis and subsequently rewritten in terms of Majorana variables.
\end{enumerate}

Both approaches yield the same effective Majorana path integral. The result is structurally identical to the Dirac case, 
but with two important differences: 
\begin{enumerate}
    \item the kinetic term acquires an overall factor of \(\tfrac{1}{2}\), and
    \item the integration is only over the real Majorana Grassmann fields \(\chi\), with no independent conjugates.
\end{enumerate}

Concretely, starting from a quadratic Hamiltonian $\hat H(\hat \chi)$ one obtains the contour action
\begin{equation}
    S(\chi,\xi) = \frac{i}{2} \int_C d t \sum_k \chi_k \, \partial_t \chi_k 
    - \int_C d t \, H(\chi) 
    + \frac{i}{2} \int_C d t \sum_k \xi_k \, \partial_t \xi_k ,
\end{equation}
where \(\int_C\) denotes integration along the closed Keldysh contour. 
The auxiliary $\xi$-sector completely decouples and contributes only a normalization factor, so the physical path integral 
is defined entirely in terms of the Majorana fields $\chi$. 
The detailed derivation is presented in Appendix~\ref{App:path_int_maj}.

\section{Brownian SYK model and Keldysh path integral representation of the Frame potential }
In this section we perform calculation on the Brownian SYK, which represents a useful tool for analytical calculations connected to study of quantum chaos.
We consider the time evolution generated by the time-dependent Hamiltonian
\begin{equation}\label{sykh}
    H(h,t)=i^{\frac{q}{2}} \sum_{1 \leq i_1<i_2<..<i_q \leq L} h_{i_1...i_q}(t) \hat \chi_{i_1}...\hat \chi_{i_q},
\end{equation}
where $\hat \chi_i$ are $L$ Majorana fermions  and anticommutation relations are applied $\{\hat \chi_i, \hat \chi_j\}=\delta_{i j}$, here $q$ is an even integer constant. This model has close analogies with the celebrated SYK~\cite{Sachdev1993,Gu:2019jub,Saad2018ASR}, but with the important difference that $h_{i_1\cdots i_q}$ are not constant in time here. On the contrary, we take them to follow a white noise distribution,
\begin{equation} \label{av}
    \left<h_{\ii}(t_1) h_{\jj}(t_2) \right>=\frac{2^{q-1}q!}{q^2 L^{q-1}  } \delta_{\ii, \jj}\delta(t_1-t_2)=\sigma^2 \delta_{\ii, \jj}\delta(t_1-t_2),
\end{equation}
where we denote the collective set of indices $\ii = (i_1,\ldots, i_q)$ (similarly for $\jj$) and set $\delta_{\ii, \jj} = \delta_{i_1, j_1} \ldots \delta_{i_q, j_q}$. For this reason, the current model is named Brownian SYK.
We assume $L$ is even, so that the Majorana operators $\{\hat\chi_i\}_{i=1}^L$ admit a faithful representation on a Hilbert space of dimension $\mathcal N = 2^{L/2}$.

We are interested in studying the scrambling dynamics induced by the time evolution Eq.\eqref{sykh}. We focus on a specific initial condition, known as \textit{thermofield double} $\left(\left|{\rm TFD}\right>\right)$ \cite{Gu2017SpreadOE,osti,Almheiri2020}, which has already been employed in the context of Brownian SYK in \cite{Jian2021}. In practice, we consider two copies of the system, prepared in a maximally entangled state. In the following, we shall address the two copies as left (L) and right (R) halves. The selection of the $\left|{\rm TFD}\right>$ state is not uniquely determined and relies on the choice of a basis. However, different definitions yield states that are connected through unitary transformations.  

To make the discussion concrete, consider two identical SYK systems, “left” (L) and “right” (R), each built from
\(2L\) Majorana operators \(\{\hat\chi_{a,\sigma}\}_{a=1}^{2L}\) with \(\sigma\in\{L,R\}\). A convenient per-site Dirac basis is
\begin{equation}
\hat c_{j,L} \;=\; \frac{1}{\sqrt{2}}\!\left(\hat\chi_{2j-1,L} + i\,\hat\chi_{2j,L}\right),
\qquad
\hat c_{j,R} \;=\; \frac{1}{\sqrt{2}}\!\left(\hat\chi_{2j,R} - i\,\hat\chi_{2j-1,R}\right),
\qquad j=1,\dots,L,
\end{equation}
which differs by a harmless phase convention on the right so as to implement the standard left–right “gluing”
used for the TFD.

At infinite temperature (\(\beta=0\)), the thermofield double \(|\mathrm{TFD}\rangle\) is the unique normalized state
annihilated by the pair of constraints, mode by mode,
\begin{equation}
\label{eq:tfd_dirac_constraints}
\bigl(\hat c_{j,L}-\hat c^\dagger_{j,R}\bigr)\,|\mathrm{TFD}\rangle=0,
\qquad
\bigl(\hat c^\dagger_{j,L}-\hat c_{j,R}\bigr)\,|\mathrm{TFD}\rangle=0,
\qquad j=1,\dots,L.
\end{equation}
Intuitively, Eq.\eqref{eq:tfd_dirac_constraints} enforces perfect left–right pairing of occupations:
whenever the left mode is empty/filled, so is the corresponding right mode.

In the number basis of the per-site Dirac modes
\(n_{j,\sigma}=\hat c^\dagger_{j,\sigma}\hat c_{j,\sigma}\) (\(\sigma=L,R\)),
Eq.~\eqref{eq:tfd_dirac_constraints} implies a product of Bell pairs:
\begin{equation}
\label{eq:tfd_bell_product}
|\mathrm{TFD}\rangle
\;=\;
2^{-L/2}\,\prod_{j=1}^L\Bigl(\,|0\rangle_{j,L}\,|0\rangle_{j,R}
\,+\,|1\rangle_{j,L}\,|1\rangle_{j,R}\Bigr),
\end{equation}
where \(|0\rangle_{j,\sigma}\) and \(|1\rangle_{j,\sigma}=\hat c^\dagger_{j,\sigma}|0\rangle_{j,\sigma}\) are the empty/filled
states of the \(j\)-th Dirac mode on side \(\sigma\). Equation \eqref{eq:tfd_bell_product} makes explicit that the
\(\beta=0\) TFD is maximally entangled between the two copies, mode by mode.

\medskip

\noindent\emph{Remark on finite temperature.} For \(\beta>0\) the (normalized) TFD is
\( |\mathrm{TFD}(\beta)\rangle = Z(\beta)^{-1/2}\sum_n e^{-\beta E_n/2}\,|n\rangle_L\otimes|n\rangle_R \),
or equivalently it is the state obtained by “imaginary-time evolving” the \(\beta=0\) TFD by \(\tfrac{\beta}{4}(H_L+H_R)\).
The simple Bell-pair form Eq.~\eqref{eq:tfd_bell_product} is specific to \(\beta=0\).
 More explicitly, we consider the initial state:

\begin{equation}\label{instate}
    \left|{\rm TFD} \right> = \frac{1}{\sqrt{\mathcal{N}}}\sum_{j=1}^{\mathcal{N}} \ket{j} \ket{j},~~~~~\rho_0 = \left|{\rm TFD} \right> \left<{\rm TFD} \right|,
\end{equation}where each $\left|j \right>$ represents a possible string of zeros and ones. Here $\mathcal{N}$ is the dimension of the Hilbert space.

As we already mentioned this state is particularly useful in SYK calculations and allows us to find the Frame potential in the form of $k-$th degree of spectral form factor. Here we provide the derivation of this statement.

This state has the following property, 
\begin{equation} \label{tfdprop}
    \Tr^{(2)}[A \otimes B \left|{\rm TFD} \right> \left<{\rm TFD} \right| C \otimes D] = 
    \frac{1}{\mathcal{N}} \Tr[A B^t D^t C],
\end{equation}
where the trace on the left is on the doubled Hilbert space, while the one on the right is on the single copy. This identity holds for arbitrary operators $A,B / C,D$ acting respectively on the first/second copy of the Hilbert space. 

Except for the initial entanglement, the two halves evolve according to two uncoupled unitary operators,
\begin{equation}
    \left|{\rm TFD} \right>(t) = U_{\boldsymbol{h}^L}(t) \otimes U_{\boldsymbol{h}^R}(t) \left|{\rm TFD} \right>, \qquad \rho(t) = \left|{\rm TFD}(t) \right>\left< {\rm TFD}(t)\right|,
\end{equation}
 where $U_{\boldsymbol{h}^L}$ and $U_{\boldsymbol{h}^R}$ are the unitary time evolution operators acting on the two halves. We assume that the time evolution in each half is generated by an independent realisation of the Brownian SYK
 \begin{equation}
     U_{\boldsymbol{h}^\sigma}(t + dt) = e^{-i H(\boldsymbol{h^\sigma},t) dt} U_{\boldsymbol{h}^\sigma}(t),
 \end{equation}
 where the subscript $\sigma = L,R$ labels the corresponding half 
 and $H_L(t), H_R(t)$ have the form of Eq.~\eqref{sykh} in terms of two sets of Majoranas $\hat\chi_{i,\sigma}$ and independently generated white noises $h^{\sigma}_{\ii}(t)$. 
Starting with this initial state, we calculate the $k$-th moments of the Frame potential averaged over Hamiltonian realisations with measure $d\eta(\boldsymbol{h})$:
\begin{equation}\label{fr}
 F^{(k)}(T)=\int  d \eta \left(\boldsymbol{h}_1 \right)d \eta \left(\boldsymbol{h}_2\right) {\rm Tr} \left(U_{\boldsymbol{h_1}}(T) \rho_0 U^{\dagger}_{\boldsymbol{h}_1}(T)  U_{\boldsymbol{h_2}}(T) \rho_0 U^{\dagger}_{\boldsymbol{h}_2}(T) \right)^k,
\end{equation}
here, $U_{\boldsymbol{h}_1}(T)=U_{\boldsymbol{h}^L_1}(T)\otimes U^t_{\boldsymbol{h}^R_1}(T)$, (and the same for $\boldsymbol{h}^{L(R)}_2$), where the couplings $\boldsymbol{h}^{L(R)}_{1(2)}$ are independent random variables with the distribution $d \eta\left(\boldsymbol{h}\right)=e^{-\frac{\boldsymbol{h}^2}{2\sigma^2}}d \boldsymbol{h}$ and variance defined in Eq.~(\ref{av}). The initial condition $\rho_0$ is the double thermofield state that we discussed earlier. Then, the Frame potential takes the form, for generic $k$, 
\begin{equation}\label{framep}
    F^{(k)}(T)=\frac{1}{\mathcal{N}^{2k}}\int d\eta(\boldsymbol{h}) \left|{\rm Tr}[U_{\boldsymbol{h}}(T)]\right|^{2k},
\end{equation}
where the evolution operator $U_{\boldsymbol{h}}(T)$ is now only acting on a single copy $R$ (or $L$) and $\mathcal{N}$ is the dimension of its Hilbert space, coming from the definition of the initial state Eq.~(\ref{instate}). To obtain this expression, we used the property Eq.~(\ref{tfdprop}) together with $U_{\boldsymbol{h}}=U_{\boldsymbol{h}^L_1}(t) U_{\boldsymbol{h}^R_1}(t) U_{\boldsymbol{h}^R_2}^{\dag}(t) U_{\boldsymbol{h}^L_2}^{\dag}(t)$.

 Further, in this chapter we consider path integral representation using Keldysh technique. Then, we shall average with respect to Gaussian random variables, and finally, calculate the path integral using the saddle-point approximation.

Using the formalism developed in Sec.~\ref{subsec:Keldysh_Major}, where we derived the Keldysh path–integral representation for Majorana fermions, we now switch to a Dirac–fermion description via the change of variables in Eq.~\eqref{eq:dir_major}. As explained there, this introduces an auxiliary Majorana mode $\hat\xi$. Inserting the resolution of the identity, Eq.~\eqref{eq:id}, at each discrete time slice along the Keldysh contour, we arrive at the following expression for the frame potential:

\begin{multline}\label{framecontour}
    F^{(k)}=\int \frac{ d \eta(h) d \boldsymbol{\overline{\psi}} d\boldsymbol{\psi } }{\mathcal{N}^{2k}}\prod_{l=1}^k  \left<\psi^{-,l}_1 \right| U_{1}(\epsilon) \left| \psi^{-,l}_2 \right>\left<\psi^{-,l}_2 \right|U_{2}(\epsilon) \left|\psi^{-,l}_3\right>...\left<\psi^{-,l}_{2N} | -\psi^{-,l}_{1} \right>\cdot\\
\cdot\left<\psi^{+,l}_{2N}\right|U^{\dag}_{2N}(\epsilon) \left| \psi^{+,l}_{2N-1} \right>\left<\psi^{+,l}_{2N-1}\right|U^{\dag}_{2N-1}(\epsilon) \left| \psi^{+,l}_{2N-2} \right>...\left<\psi^{+,l}_{1}| -\psi^{+,l}_{2N} \right>e^{-\overline{\psi}^{s,l}_i \psi^{s,l}_{i} },
\end{multline}where $s=\pm$ indicates one of the two Keldysh contours, $l$ counts replicas, and $i$ labels time. Each matrix element here has the form, 
\begin{equation}
    \left<\psi^{-,l}_1 \right| U_{1}(\epsilon) \left| \psi^{-,l}_2 \right>=\left<\psi^{-,l}_1 \right| e^{ -i^{\frac{q}{2}+1} \epsilon   h_{i_1 i_2..i_q}^1 (\hat c_{i_1}^{\dagger}+\hat c_{i_1})...(\hat c_{i_q}^{\dagger}+\hat c_{i_q})} \left| \psi^{-,l}_2 \right>,
\end{equation}and for matrix elements of monomials built from distinct modes,
\(
(\hat c^\dagger_{i_1}\!+\hat c_{i_1})\cdots(\hat c^\dagger_{i_q}\!+\hat c_{i_q}),
\)
the coherent-state symbol is obtained by the naive replacement
\(\hat c^\dagger\to\bar\psi,\ \hat c\to\psi\).
Normal ordering is unnecessary: any reordering of the fermion operators produces only a sign (no c-number, since \(i_a\) are all distinct), and the same sign arises when reordering the Grassmann factors. Leading to the standard Keldysh action
\begin{equation}\label{eq:keldyshFermions0}
S(\boldsymbol{\overline{\psi}}, \boldsymbol{\psi})=\sum_{l,s= \pm} s\int_0^T d t \left(  \sum_j \overline{\psi}_j^{s,l} \partial_t \psi_j^{s,l}+H( \overline{\boldsymbol{ \psi}}, \boldsymbol{ \psi})\right).
\end{equation}
 Here the continuous limit was taken $\epsilon \rightarrow 0$, $N \rightarrow \infty$, and $N\epsilon \rightarrow T$. We can now move back to Grassmann variables representing Majorana fermions(see \ref{subsec:Keldysh_Major}):
\begin{equation}
\chi^{s,l}_i=\frac{\overline{\psi}^{s,l}_i+\psi^{s,l}_{i}}{\sqrt{2}},~~~~\xi^{s,l}_i=i \frac{\overline{\psi}^{s,l}_i-\psi^{s,l}_{i}}{\sqrt{2}},
\end{equation}where $i$ is the index of the Fermionic degrees of freedom. In this limit, actions for $\chi$ and $\xi$ fields separate \cite{Shankar:2017zag}, and the partition function factorizes. Therefore, the part that depends on $\xi$ is a prefactor and can be integrated out Eq.~(\ref{framep}) giving a contribution to the overall normalisation.
This can be fixed using the fact that the frame potential must be equal to $1$ at time zero. Therefore, with correct normalisation, we write 
\begin{equation}
     F^{(k)} \equiv \frac{\tilde F^{(k)}(T )}{\tilde F^{(k)}(T = 0)}=\frac{Z(\xi)}{\tilde F^{(k)}(T = 0) \mathcal{N}^{2k}}\int d \eta(h) d \boldsymbol{\chi} e^{-S(\boldsymbol{\chi})},
\end{equation}
where we have
\begin{equation}
    S(\boldsymbol{\chi})=\sum_{l,s= \pm} s\int_0^T d t \left(\frac{1}{2}\sum_j  \chi_j^{s,l} \partial_t \chi_j^{s,l}+i^{\frac{q}{2}+1}\sum_{i_1<i_2<..<i_q} h_{i_1...i_q}(t) \chi^{s,l}_{i_1}... \chi^{s,l}_{i_q} \right),
\end{equation}
and $\tilde F^{(k)}(T)$ is un-normalised value of frame potential. This action is a part of the standard path integral formalism for the SYK model \cite{Saad2018ASR,Gu:2019jub,Lunkin:2020tbq}. Here, we are interested in averaging the Frame potential over the random noisy couplings $h$. To do so, we calculate the Gaussian integral over each of the variables and use the property given by equation (\ref{av}), which defines the variance $\sigma^2$:
\begin{equation}\label{gauss}
    \int d h_{i_1...i_q} e^{-\frac{1}{2 \sigma^2} h^2_{i_1..i_q}+i^{\frac{q}{2}+1} h_{i_1..i_q}M_{i_1..i_q}}= e^{-\frac{(-1)^{\frac{q}{2}}\sigma^2}{2}M^2_{i_1...i_q}},
\end{equation}
where $M_{i_1...i_q}=\sum_{l}(\chi^{-,l}_{i_1}...\chi^{-,l}_{i_q}-\chi^{+,l}_{i_1}...\chi^{+,l}_{i_q})$. Let us introduce the bi-local fields $\hat{G}(t_1,t_2)$, $\hat{\Sigma}(t_1,t_2)$ and perform Hubbard-Stratonovich transformation \footnote{Notice that here in the case of Brownian SYK one could also more simply introduce fields $\hat{G}(t)$ of only one time instead of two, as the interactions are local in time.   }:
\begin{equation}\label{normdelta}
\int d \hat{G} \prod_{s,s'} \delta \left(G^{s s'}_{ll'}(t_1,t_2)-\frac{1}{L} \sum_i \chi^{s,l}_i(t_1) \chi^{s',l'}_i(t_2)\right)= \int d \hat{\Sigma} d \hat{G} e^{- \frac{L}{2} \int d t_1 d t_2 \Sigma^{s s'}_{ll'}\left(G^{s s'}_{ll'}-\frac{1}{L} \chi^{s,l}_i \chi^{s',l'}_i\right)},
\end{equation}
here we suppose the summation over the repeating indexes. Inserting this delta function in the expression for the frame potential we obtain the following expression for an effective action:
\begin{multline}
    S =-\sum_{s,l} s\int_0^T \frac{1}{2} \chi_j^{s,l} \partial_t \chi_j^{s,l}+\frac{L}{4 q^2}\sum_{ l ,l',s,s'} \int_0^T dt_1 dt_2 c_{s s'} \left( 2 G_{l l'}^{s s'}\right)^{q}\delta(t_{12})-\\- \frac{L}{2} \sum_{l,l',s,s'}\int d t_1 d t_2 \Sigma^{s s'}_{ll'}(t_1,t_2)\left(G^{s s'}_{ll'}(t_1,t_2)-\frac{1}{L} \chi^{s,l}_i(t_1) \chi^{s',l'}_i(t_2)\right),
\end{multline}
where  $c_{ss'}= - s s'$. To form bi-local fields in the expression Eq.~(\ref{gauss}), we performed $\frac{q(q-1)}{2}$ transpositions of the Grassman variables, which gave us the prefactor in the interaction term $(-1)^{\frac{q(q-1)}{2}}$ together with $(-1)^{\frac{q}{2}}$ from the Gaussian integration, resulting in $(-1)^{\frac{q^2}{2}}$, which is unity due to taking $q$ even integer. Further, we integrate out the fields $\chi_i$, and the action becomes:

\begin{align}\label{act}
S=L \log({\rm Pf}(\hat \sigma_z \partial_t -\hat\Sigma))+\frac{L}{4 q^2}\sum\limits_{l,l's,s' } \int_0^T dt_1 dt_2 c_{s s'} \left( 2 G_{l l'}^{s s'}\right)^{q} \delta(t_{12})- \frac{L}{2} \text{Tr}(\hat \Sigma^T \hat G),
\end{align}
where the Pfaffian is taken with respect to indices in time, in Keldysh contours and replica spaces, and the same is applied to the trace.  
Here, the indexes $s$ and $s'$ take two different values, $+$ and $-$, which indicate one of the Keldysh contours. These contours are two loops, as can be seen from equation (\ref{framecontour}). Notice that after averaging over the noise, the fields from the different contours become mixed, and in the action, we have not only $G^{++}$ and $G^{--}$, but also $G^{+-}$ and $G^{-+}$. Therefore, we can think of these operators as matrices in the Keldysh contour $s,s'$ space. Also, let us take into account that the derivative term $\frac{1}{2} \chi \partial_t \chi$ has a positive sign for the $+$ contour and a negative sign for the $-$ contour. Therefore, this term can also be expressed as a $2 \times 2$ matrix with Keldysh contour indexes in the following way:

\begin{equation}
-\sum_{s,l} s\int_0^T \frac{1}{2} \sum_j \chi_j^{s,l} \partial_t \chi_j^{s,l}=-\frac{1}{2}\sum_{l} \sum_j \chi^{l}_j \left( \hat \sigma_z \partial_t \right) \chi^{l}_j.
\end{equation}Here, we introduced the Pauli matrix $\hat\sigma_z$ in the $ss'$ space, and we assume a Kronecker delta in the replica space for this term. In the next section, we are going to calculate this functional integral over the fields $\hat{G}$ and $\hat{\Sigma}$ using the saddle-point approximation at large $N$.

\section{Majorana SYK model and saddle points}\label{sec3}

We are now ready to derive the saddle-point (Schwinger–Dyson) equations by varying the action with respect to the bi-local fields. Denoting $t_{12}\equiv t_1-t_2$ and suppressing replica indices $(\ell,\ell')$ for readability, we obtain
\begin{equation}\label{eq:m}
\begin{split}
&\frac{\delta S}{\delta \hat{\Sigma}}=0 \;:\qquad 
\hat G^{-1}(t_1,t_2)=\hat\sigma_z\,\partial_t-\hat{\Sigma}(t_1,t_2),\\[2mm]
&\frac{\delta S}{\delta \hat{G}}=0 \;:\qquad 
\hat{\Sigma}(t_1,t_2)=\frac{\hat c}{q}\,\big(2\hat G\big)^{\,q-1}(t_1,t_2)\,\delta(t_{12}),
\end{split}
\end{equation}
where $\hat\sigma_z$ acts in Keldysh space and $\hat c_{ss'}\equiv c_{ss'}$ encodes the contour signs (with $c_{ss'}=-ss'$). The $\delta(t_{12})$ arises from the temporal locality of the Brownian couplings.

Setting the variations to zero and solving Eq.~\eqref{eq:m} yields two qualitatively distinct classes of saddles. 
(i) A \emph{diagonal-in-Keldysh} solution governs the \emph{short-time} regime and leads to an exponential decay of the frame potential with time. 
(ii) A \emph{non-diagonal} solution, mixing the Keldysh branches and structured by replica permutations (for $q>2$) or continuous replica rotations (for $q=2$), controls the \emph{long-time} regime and sets the late-time saturation values. 
We analyze these two branches in the following subsections.

\subsection{Short-time behaviour}\label{shorttime}
Let us now proceed naively by looking for a saddle point that is diagonal in the $+-$ space. Then the obvious solution is:
\begin{equation}
    \hat\Sigma=0,~~~~~\hat G^{-1}=\hat \sigma_z \partial_t.
\end{equation}
With this solution, the Frame potential can be expressed as:
\begin{equation}
      F^{(k)}(T=0)=\frac{1}{\mathcal{N}^{2k}} \det(\hat \sigma^z \partial_t)^{\frac{L}{2}} e^{-\frac{L}{4 q^2}\sum\limits_{l l'} \int_0^T d t  (2 G_{l l'}^{++})^q+(2 G_{l l'}^{--})^q}.
 \end{equation}This functional determinant can be calculated using diagonalisation of the operator $\hat \sigma^z \partial_t$. By establishing antiperiodic boundary conditions due to fermionic nature of this path integral (see the last time slice in Eq.~\eqref{framecontour}), we obtain eigenvalues  $\lambda_n=\pm \frac{\pi i (2n+1)}{T}$, where $n=0,\pm 1,\pm 2,..$. In the $+-$ space, we have two eigenvalues $\lambda_n$ and $-\lambda_n$. However, coherent states are over complete, therefore we consider $\lambda_n= \frac{\pi i}{T}, \frac{3 \pi i}{T}, \dots, \frac{(4N+1) \pi i}{4 T}$, which prevents us from treating matrices
 \begin{equation}
\begin{pmatrix}
-\lambda_n & 0\\
0 & \lambda_n 
\end{pmatrix} ~~~\text{and}~~~   \begin{pmatrix}
\lambda_n & 0\\
0 & -\lambda_n 
\end{pmatrix},
\end{equation}
as different matrices. Notice that this determinant should also be taken in the replica space, where we have $k$ copies of the system. Taking the limit $N \rightarrow \infty$ and using the $\zeta$-function regularization \cite{Nakahara2003}, we get
\begin{equation}
    \det(\hat \sigma_z \partial_t)^{\frac{L}{2}}=2^{\frac{2 k L}{2}},
\end{equation}where $2^{\frac{L}{2}}$ is the dimension of the Hilbert space for the initial operator $U$. Also the field $\hat{G}$ is  diagonal in the replica space. Therefore, $G^{++}_{l l'}=\frac{1}{2} \delta_{l l'}=-G_{l l'}^{--}$, and we have
\begin{equation}
    e^{-\frac{L}{4 q^2}\sum_{l l'} \int_0^T d t (2 G_{l l'}^{++})^q+(2 G_{l l'}^{--})^q}=e^{-\frac{LTk}{2 q^2}}.
\end{equation}The Frame potential on this solution takes the form
\begin{equation}
    F^{(k)}= e^{-\frac{LTk}{2 q^2}}.
\end{equation}
In this saddle the Frame potential decays exponentially in time and factorises over replicas, \(F^{(k)}(T)=[F^{(1)}(T)]^{k}\). Hence it governs the early-time regime but becomes exponentially subleading at late times.

\subsection{Long-time behaviour}\label{sec:long-time}

At late times two qualitatively different situations arise. For $q>2$, the disorder-averaged action is invariant only under discrete permutations of the replicas ($S_k$), whereas for $q=2$ it enjoys a continuous symmetry ($SO(k)\times SO(k)$ for Majorana, and $U(k)\times U(k)$ for Dirac at fixed charge). Consequently, the $q>2$ saddle manifold is discrete (no Goldstones), while for $q=2$ it is continuous and supports massless (Goldstone) modes; integrating over these zero modes produces the characteristic $L$–dependent prefactors in the Gaussian case.

The structure of the late-time saddle follows from two elementary facts. 

(i) The Brownian couplings are $\delta$–correlated in time, so after averaging the self-energy is local; at the stationary saddle we may write
\[
\hat\Sigma(t_1,t_2)=\hat M\,\delta(t_{12}),\qquad 
\hat G(t_1,t_2)=\hat G(t_{12}),
\]
where $\hat M$ is time independent.

(ii) In the time domain the Dyson equation reduces to a first-order ODE,
\[
\bigl(\hat\sigma_z\,\partial_t-\hat M\bigr)\,\hat G(t)=\delta(t)\,\hat{\mathbb{I}}.
\]
For $t\neq 0$ the solution is a matrix exponential, and imposing Keldysh causality (retarded/advanced support on the forward/backward branches) fixes the diagonal contour blocks to the universal sign structure,
\[
G^{++}(t)\ \propto\ +\tfrac{1}{2}\,\mathrm{sgn}(t)\,f(|t|),\qquad
G^{--}(t)\ \propto\ -\tfrac{1}{2}\,\mathrm{sgn}(t)\,f(|t|),
\]
with a common decaying envelope
\[
f(|t|)=\exp\!\big[-\,|t|\, M_{\!\rm eff}\big]
\]
(understood in the internal replica/copy space). The effective “mass” $ M_{\!\rm eff}$ is fixed self-consistently by the off-diagonal components generated by the interaction, and it sets the late-time relaxation rate.

The remaining freedom sits in a constant matrix that pairs forward and backward Keldysh branches across replicas. Minimizing the action selects off-diagonal Keldysh blocks proportional to a replica \emph{permutation} $\tau\in S_k$ for $q>2$, or to a \emph{rotation} $\theta\in SO(k)$ (or $U(k)$ in the Dirac case) for $q=2$. In compact form,
\begin{equation}\label{eq:ansatz}
\hat G(t)=\frac{f(|t|)}{2}\!
\begin{pmatrix}
\mathrm{sgn}(t)\,\mathbb{I}_k & -R^\top\\
R & -\mathrm{sgn}(t)\,\mathbb{I}_k
\end{pmatrix},
\qquad
R=
\begin{cases}
\tau\in S_k, & q>2,\\
\theta\in SO(k)\ \text{or}\ U(k), & q=2.
\end{cases}
\end{equation}
This “pairing across branches’’ mirrors Weingarten/Wick pairings. The discrete versus continuous nature of $R$ is precisely what distinguishes the non-Gaussian ($q>2$) late-time saturation from the Gaussian ($q=2$) case with Goldstone-induced finite-size corrections.

In what follows, we compute the frame potential by evaluating the on-shell action on these late-time saddle configurations and summing over all symmetry-related saddles.

\subsubsection{The non-Gaussian case $q > 2$}
In this subsection we consider the ansatz Eq.~\eqref{eq:ansatz} for $q>2$ and find the function $f(|t|)$. To do so we plug the ansatz in the equations of motion Eq.~\eqref{eq:m}, and solve them. Second saddle-point equation gives:
\begin{equation}
\hat\Sigma(t_1,t_2)=\frac{1}{q}\,f(0)^{\,q-1}\,\delta(t_{12})
\begin{pmatrix}
0 & -\hat\tau^{\!T}\\[2pt]
\hat\tau & 0
\end{pmatrix}.
\end{equation}
Fourier transforming and inserting into the first saddle equation yields
\begin{equation}
\hat G(\omega)=
\frac{1}{\omega^2+\left(\frac{f(0)^{\,q-1}}{q}\right)^2}
\begin{pmatrix}
 i\omega & -\tfrac{f(0)^{\,q-1}}{q}\,\hat\tau^{\!T}\\[2pt]
 \tfrac{f(0)^{\,q-1}}{q}\,\hat\tau & -i\omega
\end{pmatrix}.
\end{equation}
Inverting back to time variables and matching to the ansatz fixes $f(0)=1$ and the decay envelope
\begin{equation}\label{eq:sol_sp}
\begin{aligned}
\hat G(t_1,t_2)&=\frac{e^{-|t_{12}|/q}}{2}
\begin{pmatrix}
\Theta(t_{12})-\Theta(-t_{12}) & -\hat\tau^{\!T}\\
\hat\tau & -\Theta(t_{12})+\Theta(-t_{12})
\end{pmatrix},\\[4pt]
\hat\Sigma(t_1,t_2)&=\frac{1}{q}\,\delta(t_{12})
\begin{pmatrix}
0 & -\hat\tau^{\!T}\\
\hat\tau & 0
\end{pmatrix}=\hat\Sigma_0 \delta(t_{12}).
\end{aligned}
\end{equation}
As anticipated, the function $f(|t|)=e^{-|t|/q}$ produces an exponential time decay. We adopt the symmetric convention $\Theta(0)=\tfrac12$ so that $\mathrm{sgn}(t)=\Theta(t)-\Theta(-t)$ is never evaluated at $t=0$ in the action, and contact terms are regulated. 

We now evaluate the action on this saddle. 
The interaction term cancels due to the Keldysh sign structure $c_{ss'}$,
\begin{equation}
\frac{L}{4q^2}\sum_{l,l',s,s'}\!\int_0^T\! dt_1 dt_2\; c_{ss'}\big(2G_{ll'}^{\,ss'}\big)^q\delta(t_{12})=0,
\end{equation}
while the mixed term contributes
\begin{equation}
-\frac{L}{2}\sum_{l,l',s,s'}\!\int_0^T\! dt_1 dt_2\; \Sigma^{ss'}_{ll'}(t_1,t_2)\,
G^{ss'}_{ll'}(t_1,t_2)
=-\frac{L}{2}\,\mathrm{Tr}\,(\hat\Sigma^{\!T}\hat G)
=-\frac{LTk}{2q}.
\end{equation}
For the Pfaffian we use $\mathrm{Pf}(\hat\sigma_z\partial_t-\hat\Sigma)=\mathrm{Tr}\,e^{-T\hat H}$ with a quadratic Majorana Hamiltonian:
\begin{equation}
    \hat H= -\frac{1}{2}\begin{pmatrix}\hat \chi^+&i \hat \chi^-
        
    \end{pmatrix} \hat \Sigma_0 \begin{pmatrix}
        \hat \chi^+\\
        i \hat \chi^-
    \end{pmatrix},
\end{equation}
Here $\hat \Sigma_0$ is time independent part of $\hat \Sigma$ from Eq.~\eqref{eq:sol_sp}.
With the permutation ansatz, $\hat H$ block–diagonalizes into independent sectors labeled by the disjoint cycles of $\hat\tau$; each cycle of length $n_c$ maps to a Kitaev chain contributing $e^{n_cT/(2q)}$ at large $T$. 
Since $\sum_c n_c=k$, we obtain (see App.~\ref{App:pfaffian} for details)
\begin{equation}
e^{S_1}=\big(\mathrm{Pf}[\hat\sigma_z\partial_t-\hat\Sigma]\big)^L
= e^{\frac{LkT}{2q}}.
\end{equation}
Summing over all saddle points (permutations $\hat\tau$ of size $k$; we restrict to the parity compatible with the initial state, which only affects an overall $O(1)$ factor), we find for large Hilbert‐space dimension(for more details see Appendix \ref{App:pfaffian})
\begin{equation}
F^{(k)}=\sum_{\hat\tau}F^{(k)}(\hat\tau)
=\sum_{\hat\tau}\frac{1}{\mathcal{N}^{2k}}=\frac{k!}{\mathcal{N}^{2k}},
\end{equation}
i.e., the Haar value in the relevant sector. 
This shows that the late–time saddle reproduces the Haar frame potential.

\subsubsection{The Gaussian case $q=2$}\label{subsec:q=2}
In the case where $q=2$, the action has the continuous symmetry, $SO(k) \times SO(k)$ as discussed at the beginning of the section. Therefore, the ansatz that we examine here consists of orthogonal matrices instead of permutations. Using the same procedure as in the previous subsection, one can see that for $q=2$, the solution is:
\begin{align}
    \hat{G}(t_1,t_2)=\frac{e^{-\frac{|t_{12}|}{q}} }{2} \begin{pmatrix}
\text{{\rm sgn}}(t_{12})&-\hat{\theta}^T\\
\hat{\theta} & -\text{{\rm sgn}}(t_{12})
\end{pmatrix}, ~~~\hat{\Sigma}(t_1,t_2)= \frac{1}{q} \delta(t_{12}) \begin{pmatrix}
0&-\hat{\theta}^T\\
\hat{\theta} &0
\end{pmatrix},
\end{align}where $\hat{\theta}$ is now an orthogonal matrix in the replica space and not a permutation matrix as in the previous case. Again, we are interested in the calculation of the action on this solution. Notice that the second and third terms in the action are the same as in for permutation matrices. Therefore, our objective now is simply to extend the calculation to the first term, namely the Pfaffian. Let us then apply the same procedure to the orthogonal matrices. Any arbitrary orthogonal matrix can be expressed in block diagonal form:
\begin{align}
        \hat{\theta}_{2m}=
\left(\begin{array}{c|c|c|c|c}
  \hat R_1
  & 0 & 0&0&0 \\
\hline
  0 &
  \hat R_2& 0&0&0\\
  \hline 
  0&0&...&0&0\\
  \hline
  0 &
0 &  0&\hat R_{m-1}&0
\\
  \hline
  0 &
0 & 0& 0&\hat R_m
\end{array}\right), ~~~\text{and} ~~~    \hat{\theta}_{2m+1}=
\left(\begin{array}{c|c|c|c|c}
  \hat R_1
  & 0 & 0&0&0 \\
\hline
  0 &
  \hat R_2& 0&0&0\\
  \hline
  0 &
0 &  ...&0&0\\ \hline
  0 &
0 & 0& \hat R_m&0\\
\hline
  0 &
0 & 0& 0&\pm \hat 1\\
\end{array}\right),
\end{align}where, $m$ is a positive integer, and $\hat R_i$ is an orthogonal matrix in two-dimensional space. Then, ${\rm Tr}(e^{i \frac{T}{q} \hat\chi^- \hat{\theta} \hat\chi^+})={\rm Tr}(e^{i \frac{T}{q} \sum_i \hat\chi_i^- \hat R_i \hat\chi_i^+})$. Here, we should also take into account that not all orthogonal matrices will preserve the parity of the initial state. Therefore,  $\hat R_i$ should be rotation matrices, so $\det \hat R_i=1$.  Also, in the case of an odd size, the last element on the diagonal is positive unity as we want to preserve parity.
Here we map Hamiltonian into free fermion Hamiltonian that can be easily diagonalised(see Appendix \ref{App:pfaffian}):
\begin{multline}
        {\rm Tr}(e^{-T \hat H_R})=\det\left(1+\exp\left(\frac{T}{q}\begin{pmatrix}
1 & 0 \\
0 & -1 
\end{pmatrix}  \right) \right)=2+e^{\frac{T}{q}}+e^{-\frac{T}{q}}    \rightarrow e^{\frac{T}{q}}, ~~T\rightarrow \infty.
\end{multline}Therefore the Frame potential on the particular solution which depends on an orthogonal matrix $\hat{\theta}$ is
\begin{equation}
    F^{(k)}(\hat{\theta})=\frac{1}{\mathcal{N}^{2 k}}  e^{\frac{LkT}{2 q}-\frac{LkT}{2 q}}=\frac{1}{\mathcal{N}^{2 k}}. 
\end{equation}
As in the previous subsection, the saddle point solution with a given rotational matrix does not depend on the latter. Therefore, we should first consider the fluctuations around the saddle point, and sum over all possible solutions given by the space of orthogonal matrices. As we shall see in the coming section, the zero-mass fluctuations are responsible for different finite-size effects between the $q>2$ and the $q=2$ case. In the $q=2$ case, they carry  $\frac{\log(L)}{L}$ corrections, while with $q>2$ there are no zero-mass modes that can fluctuate around the saddle point solution.  
\subsubsection{Symmetries and Goldstone modes}

\begin{figure}[t]\centering
  \includegraphics[width=0.6\linewidth]{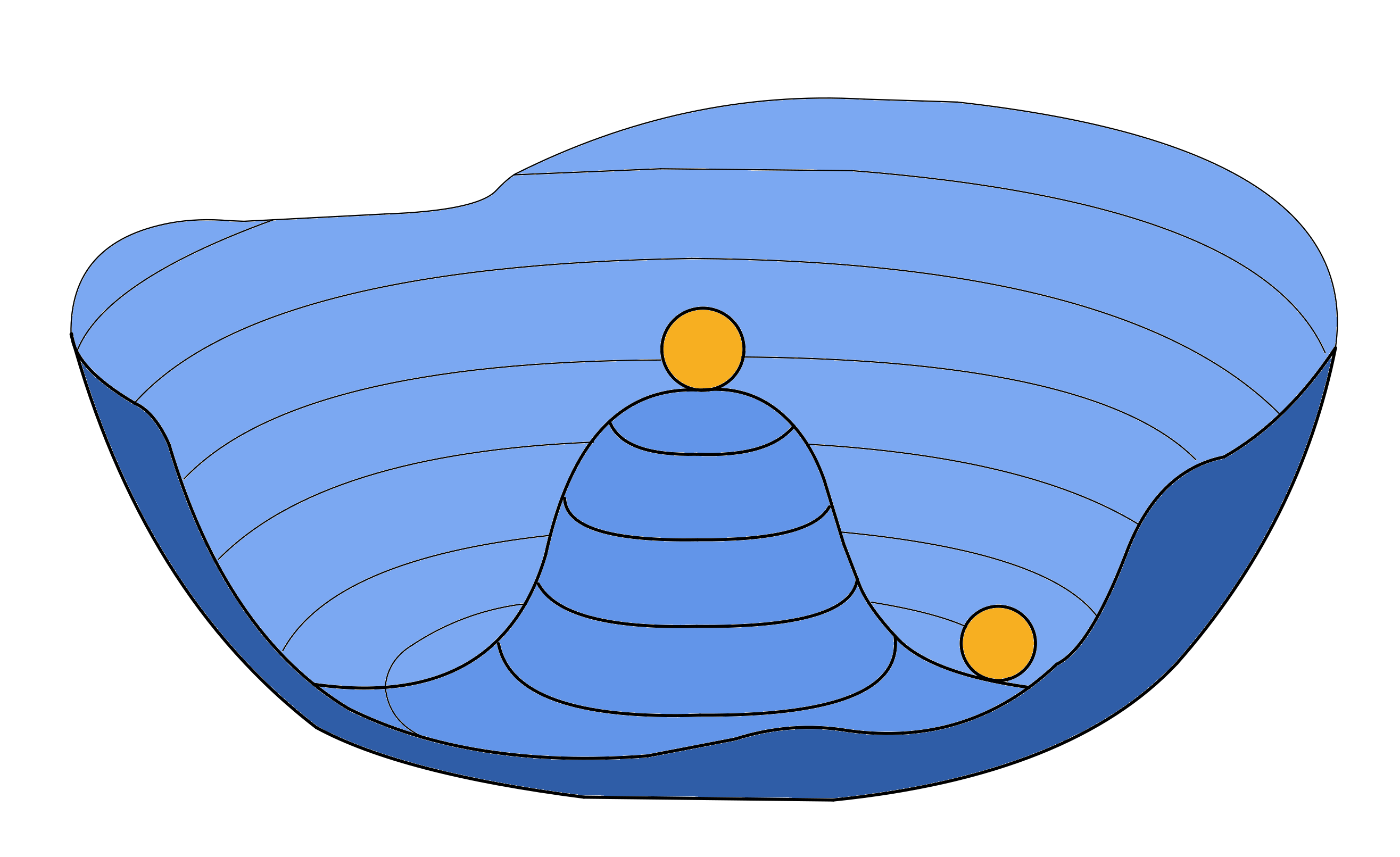}
  \caption{Mexican-hat potential \(V(\phi)=\lambda(|\phi|^2-v^2)^2\) (schematic).
The symmetric point at \(\phi=0\) is unstable, while the minima form the
degenerate manifold \(|\phi|=v\) (valley). Choosing one minimum (yellow dot)
spontaneously breaks the continuous symmetry; fluctuations split into a
massless Goldstone (tangential) mode and a massive radial (“Higgs”) mode.
\emph{We use this schematic in the present subsection to illustrate the symmetry
breaking and Goldstone counting in our \(q=2\) analysis; the valley represents
the vacuum manifold discussed in the text (the coset \(SO(2k)/U(k)\)).}}

  \label{fig:boat1}
\end{figure}

In the next subsection, we shall calculate fluctuations around the saddle point, which we discussed earlier. For intuition, the Mexican-hat landscape and the fluctuation directions are shown in Fig.~\ref{fig:boat1}. For the case $q=2$, we need to understand how many modes will be massive and carry non-trivial corrections to the Frame potential. To do so, we can use the Goldstone theorem. 

Let $B$ be a continuous group of a global symmetry of the action and $H$ is a subgroup of $B$ which leaves the solution of the equations of motion unchanged. Then the number of massless modes is equal to $R_{ml}={\rm dim}(B)-{\rm dim}(H)$. Therefore, the number of massive modes is $R_m=d-R_{ml}$ where $d$ is the number of degrees of freedom of our system. In the schematic, massless modes correspond to tangential (along-the-valley) motions, whereas massive modes are radial; see Fig.~\ref{fig:boat1}.

Let us start with the identification of the degrees of freedom. By the definition:
\begin{equation}
    G_{l l'}=\begin{pmatrix}
        G^{++}_{l l'} & G^{+-}_{l l'}\\
        G^{-+}_{l l'} & G^{--}_{l l'}
    \end{pmatrix},
\end{equation}
and $G^{++}_{l l'}=\frac{1}{L} \sum^L_{i=1} (\chi^+)^i_l (\chi^{+})^i_{l'}$, where $L$ is the number of fermionic modes, first we notice that the matrix $G_{l l'}$ is a skew-symmetric matrix due to the anticommutation relations of the field $\chi$, which contains $2k(2k-1)/2$ degrees of freedom. Let us check the commutation relations for this matrix
\begin{equation}
    [G_{l l'}, G_{m m'}]=\frac{1}{L^2} \sum_i \sum_j [\chi^i_l \chi^i_{l'},\chi^j_m \chi^j_{m'}],
\end{equation}expanding the commutator and using anticommutation relations $\{\chi^i_l,\chi^j_m \}=\frac{1}{2} \delta^{i j} \delta^{l m}$ we have, 
\begin{equation}
    [G_{l l'}, G_{m m'}]=\frac{1}{2 L}(\delta_{l' m}G_{l m'}-\delta_{l m'}G_{m l'}+\delta_{l m} G_{m' l'}-\delta_{l' m'}G_{l m}),
\end{equation}
which gives the commutation relations of rotation generators in $2k$ dimensions. Therefore, we see that by the definition our matrix $G_{l l'}$ is a generator of $so(2 k)$ algebra. In \cite{Fava:2023tgg}, it was shown that $S_{l l'}=i G_{l l'}$ satisfy orthogonality relation in the large $L$ limit:
\begin{equation}
    S S^T=\mathbb{I},
\end{equation}
which for the matrix $\hat{G}$ this implies
\begin{equation}
    \hat{G} \hat{G}^T=\mathbb{I}.
\end{equation}
Therefore this matrix should be not only skew-symmetric but also orthogonal. The space of these matrices isomorphic to the $SO(2k )/U(k)$ \cite{baker}, which gives us $d=\frac{2k(2k-1)}{2}-k^2=k(k-1)$ degrees of freedom. This coset is the vacuum manifold—the valley of degenerate minima in Fig.~\ref{fig:boat1}.

Now let us consider the group of the global symmetry of the action in more detail. Remind that at finite times, the action looks like Eq.~(\ref{act}). Therefore, after the integration over the fields $\hat{G}$ we have
\begin{equation}
    S = L \log({\rm Pf}(\hat \sigma_z \partial_t -\hat\Sigma))-\frac{L}{4 }\sum\limits_{l,l's,s' } \int_0^{T} dt_1 d t_2 c_{s s'}\left(\Sigma^{s s'}_{l l'} \right)^2 \delta(t_{12}).
\end{equation}
One can see that the problematic term here is $L \log({\rm Pf}(\hat \sigma_z \partial_t -\hat\Sigma))$, which as we already noticed has a rotation symmetry in replica space $O_1=e^{i \hat A_1 \hat \sigma_z }$, $O_2=e^{i \hat A_2 \hat 1 }$ with $\hat A_i$ skew-symmetric matrix of size $k$. Notice that matrices $O_0=e^{i \hat A}$ form a $SO(k)$ symmetry group, therefore the space of the symmetries is $SO(k) \times SO(k)$. Geometrically, these continuous rotations move us along the ring of minima (cf. Fig.~\ref{fig:boat1}). Therefore,  
\begin{equation}
    \det(\hat \sigma_z \partial_t -O_i\hat\Sigma O_i^T)=\det(O_i O_i^T)\det(O_i^{-1} \hat{\sigma_z} \partial_t (O_i^T)^{-1} -\hat\Sigma), 
\end{equation}
since matrices $O_i$ commute with $\hat \sigma_z$, we have
\begin{equation}
    \det(\hat \sigma_z \partial_t -O_i\hat\Sigma O_i^T)= \det(\hat \sigma_z \partial -\hat\Sigma).
\end{equation}
Other possible transformations are not symmetries of the action, therefore the group is $B=SO(k) \times SO(k)$ and ${\rm dim}(B)=k(k-1)$. 
Remind that the saddle point solution has the form
\begin{multline}
    \hat{G}(t_1,t_2)=\frac{e^{-\frac{|t_{12}|}{2}} }{2} \begin{pmatrix}
{\rm sgn}(t_{12})&-\hat{\theta}^T\\
\hat{\theta} & -{\rm sgn}(t_{12})
\end{pmatrix},~~~ \hat{\Sigma}= \frac{1}{2}  \delta(t_{12}) f(0) \begin{pmatrix}
0&-\hat{\theta}^T\\
\hat{\theta} &0
\end{pmatrix},\end{multline}
where $\hat{\theta}$ is an orthogonal matrix $k \times k$. Choosing $\hat{\theta}$ selects a specific vacuum on this manifold (yellow dot in Fig.~\ref{fig:boat1}). From the form of the solution, we see that the group that allows us to transform one solution into another one is $B/H=SO(k)$, which means that $H=SO(k)$ too. Therefore, $R_{ml}={\rm dim}(B)-{\rm dim}(H)=k(k-1)/2$. Therefore, we find
\begin{equation}
    R_{m}=d-R_{ml}=k(k-1)-k(k-1)/2=k(k-1)/2,
\end{equation}
giving the number of massive modes, which will allow us to calculate the integral over the relevant fluctuations.

 The saddle point approximation gives us the following expression for the Frame potential
 \begin{equation}\label{framepotsp}
     \tilde F^{(k)}(T) =\frac{1}{ \mathcal{N}^{2 k}}e^{S_0}\left<\int D \delta G D \delta \Sigma e^{\frac{1}{2}S^{(2)}(\delta G, \delta \Sigma)} \right>_{\hat{\theta}},
 \end{equation}where the on-shell action $S_0$ was found in the previous subsection \ref{subsec:q=2} and $\left< ...\right>_{\hat \theta}$ means the summation other all the solutions. Now let us  consider the quadratic fluctuation term $S^{(2)}(\delta G, \delta \Sigma)$:
 \begin{equation}\label{fluctact}
     S^{(2)}(\delta G, \delta \Sigma)=\frac{\delta^2 S}{\delta \hat G \delta \hat G} \delta \hat G^2+\frac{\delta^2 S}{\delta \hat \Sigma \delta \hat \Sigma} \delta \hat \Sigma^2+ \frac{\delta^2 S}{\delta \hat G \delta \hat \Sigma} \delta \hat \Sigma \delta \hat G,
 \end{equation}doing variation we can find 
 \begin{gather}
         \frac{\delta^2 S}{\delta \hat G \delta \hat G} \delta \hat G^2=\frac{L}{2} \sum\limits_{l,l's,s' } \int_0^T dt_1 dt_2  c^{s s'} (\delta G^{s s'}_{l l'})^2\delta(t_{12}), \\
             \frac{\delta^2 S}{\delta \hat G \delta \hat \Sigma} \delta \hat \Sigma \delta \hat G=-L\sum\limits_{l,l's,s' } \int d t_1 d t_2 \delta \Sigma^{s s'}_{ll'}(t_1,t_2) \delta G^{s s'}_{ll'}(t_1,t_2).
 \end{gather}
 \begin{equation}
    \frac{\delta^2 S}{\delta \hat \Sigma \delta \hat \Sigma} \delta \hat \Sigma^2=-\frac{L}{2}\int d t_1 d t_2 G^{s s'}_{l l'}(t_1,t_2) \delta  \Sigma^{s' \alpha}_{l' m}(t_1,t_2) G^{\alpha \alpha'}_{m m'}(t_1,t_2)  \delta \Sigma^{\alpha' s}_{m' l}(t_1,t_2), 
 \end{equation}where the field $\hat{G}$ is a solution of the saddle point equation, and we suppose the summation over the repeating indexes.  Integrating out the fluctuations, we obtain:
 \begin{equation}
     \int D \delta \hat{G} D \delta \hat{\Sigma} e^{\frac{1}{2}S^{(2)}(\delta \hat{G}, \delta \hat{\Sigma})}\sim L^{-R_m}.
 \end{equation}
Here $R_m$ is a number of massive modes. We also need to calculate the normalisation term, which is a standard Gaussian integral 
 \begin{multline}\label{eq:norm}
         \tilde F^{(k)}(0)=\int D \hat{G} D \hat{\Sigma} e^{\left[\frac{L}{4 }\sum\limits_{l,l's,s' } \int_0^{T} dt c_{s s'}\left(G^{s s'}_{l l'} \right)^2-\frac{L}{2} \int d t_1 d t_2 \Sigma^{s s'}_{ll'}(t_1,t_2)G^{s s'}_{ll'}(t_1,t_2)\right]}\sim\\    \sim L^{-\frac{d}{2}} \int D \delta \Sigma e^{\frac{L}{4} \sum\limits_{l,l's,s' } \int_0^T dt c^{s s'} ( \Sigma^{s s'}_{l l'})^2(t)}\sim L^{-d},
 \end{multline} and $d$ is a number of degrees of freedom. Putting it all together, we obtain for $q=2$:
 \begin{equation}
     F^{(k)}_{q=2}=\frac{\tilde F^{(k)}(T)}{\tilde F^{(k)}(0)}\sim \frac{ L^{d-R_m}}{\mathcal{N}^{2 k}} \sim \frac{ L^{R_{ml}}}{\mathcal{N}^{2 k}} ,
 \end{equation}
where $R_{ml}=\frac{k(k-1)}{2}$ is a number of massless modes, taking the logarithm of the Frame potential we can get:\begin{equation} 
     \lim_{L \rightarrow \infty}\left(\frac{1}{L} \log F^{(k)}_{q=2}\right)\sim -k \log(2)+\frac{k(k-1)}{2} \frac{1}{L} \log(L),
 \end{equation}which coincides with the result for Gaussian Haar calculation, see Appendix (\ref{Haarresof}).
 Notice that for the generic case of $q>2$ interaction, the symmetry of the action and the solution is discrete, therefore there are no Goldstone modes in this case. Due to this fact, the integration over the fluctuation must be carried over all the degrees of freedom which are ${\rm dim}(SO(2 k)/U(k))$, also, we have the same degrees of freedom at zero and at finite time. Therefore, the normalization will cancel the fluctuation part from the finite time action. Hence, for the generic $q>2$ case we have 
 \begin{equation}
     F^{(k)}_{q>2}\sim \frac{k!}{\mathcal{N}^{2 k}} ,
 \end{equation}
 together with,
 \begin{equation}\label{frp:qg2} 
     \lim_{L \rightarrow \infty}\left(\frac{1}{L} \log F^{(k)}_{q>2}\right)\sim -k \log(2),
 \end{equation}
 which indeed coincides with the first correction in system size for gHaar distributed Majorana fermions, see sec. \ref{sec:majo-ghaar}.

\section{Dirac SYK model and Kelsdysh saddle points}\label{sec4}
We shall now consider the Brownian evolution given by the complex SYK Hamiltonian
\begin{equation}\label{HamiltDirac}
    H(t)=\sum_{\substack{1\leq i_1\leq i_2...\leq i_{q/2}\leq L\\1\leq i_{q/2+1}\leq ...\leq i_{q}\leq L}} h_{i_1...i_q}\hat c^{\dagger}_{i_1}..\hat c^{\dagger}_{i_{q/2}}\hat c_{i_{q/2+1}}...\hat c_{i_{q}}.
\end{equation}
Here, $\hat c^{\dagger}$ and $\hat c$ are Dirac Fermions, and standard anti-commutation relations are applied $\{\hat c_{k}, \hat c^{\dagger}_{l}\}=\delta_{k l}$. As in the previous chapter, $h_{i_1...i_q}$ are normally distributed random variables, but this time, they are complex and have the variance:
\begin{equation}
    \left<h_{i_1...i_q }(t_1) h^{*}_{i_1...i_q}(t_2) \right>=\frac{2^{q}(q/2)!^2}{q^2 L^{q-1}  } \delta(t_1-t_2).
\end{equation}
We shall then repeat the analysis of the previous chapter.  The model now possesses an extra symmetry compared to the Majorana case, which is the global $U(1)$ charge conservation 
\begin{equation}
    \hat Q = \sum^L_{i=1} \hat c^\dagger_i \hat c_i . 
\end{equation}
We shall then show how this impacts the saddle point solutions at late time, both for $q>2$ and $q=2$. 

\subsection{Keldysh path integral and saddle points}
We again choose the {\rm TFD} state as the initial state, and the object of our interest has the form given in the Eq.~(\ref{framep}), which can be expressed as two Keldysh contours Eq.~(\ref{framecontour}). The evolution operator is given by $U_i(\Delta t)=e^{-i\Delta t H(t_i)}$, where the Hamiltonian is from the Eq.~(\ref{HamiltDirac}). 
Now we want to derive an effective action of the theory. We proceed similarly as in the Majorana case. Therefore,  the fermion density can be written as  $n=\frac{\left<Q \right>}{L}$.  The Frame potential can be expressed as 
\begin{equation}
    F^{(k)} =\frac{1}{\mathcal{N}^{2k}} \underbrace{\Tr\left[\Texp\left( - i \int_0^T dt H(t)\right)\right]
    \Tr\left[\Texp\left( i \int_0^T dt H(t)\right)\right]\ldots}_{k \mbox{ times}}~~.
\end{equation}Since each of the Hamiltonian conserves the total number of particles $\hat Q = \sum_i \hat c^\dag_i \hat c_i$, we can split each trace into smaller traces over each charge sector, namely
\begin{equation}
     \Tr\left[\Texp\left( \pm i \int_0^T dt H(t)\right)\right]=\sum_{Q^{\pm}} \Tr\left[\Texp\left( \pm i \int_0^T dt H(t)\right)P_{Q^{\pm}}\right],
   \end{equation}
with 
\begin{equation}
     P_{Q^\pm}=\int d \mu ^\pm e^{\pm  i L \mu^\pm\int_0^T dt  (\frac{1}{L}\sum_i \hat c^{\dagger}_i \hat c_i- n^\pm )}.
\end{equation}


Namely, for each Keldysh contour and each replica we introduce a chemical potential $\mu^s_\ell$, and we denote by $\boldsymbol{\mu}$ the whole set. 
Writing down the path integral representation for each trace and by averaging over Hamiltonian realisations, we obtain
\begin{equation}\label{eq:FPDirac-pathintegral}
    F^{(k)} =\frac{1}{\mathcal{N}^{2k}}  \int d\eta(\boldsymbol{h}) \sum_{\boldsymbol{Q} }  
    \int d \boldsymbol{\mu} \int \mathcal{D} \boldsymbol{\psi}    \mathcal{D} \bar{\boldsymbol{\psi}} e^{S(\boldsymbol{\overline{\psi}}, \boldsymbol{\psi})} \prod_{s,l} P_s(\mu^s_l),
\end{equation}
with the action $S(\boldsymbol{\overline{\psi}}, \boldsymbol{\psi})$ already introduced in Eq.\eqref{eq:keldyshFermions0} and 
with the projectors now written as 
\begin{equation}
    P_s(\mu^{s}_l) = e^{  i s  T  L \mu^{s}_l (\bar{\psi}^{s,l}(0^+) \psi^{s',l'}(0) - n^s_l + \delta_{s,+}\delta_{l l'})   }.
\end{equation}
Antiperiodic boundary conditions are enforced to result in proper traces. By performing the Hubbard-Stratonovich transformation as usual Eq.(\ref{normdelta}) and integrating over Grassmann fields, we obtain the action:
\begin{multline}\label{actferm}
        S = L \log(\det( \hat \sigma_z \partial_t -\hat\Sigma+i \hat \sigma_z \hat \mu  ))+\frac{L}{2 q^2}\sum\limits_{l,l's,s' } \int dt_1 dt_2 \delta(t_{12}) c_{s s'} (2 G^{s s'}_{l l'})^{q/2} (-2 G^{s' s}_{l' l})^{q/2}\\    -L \sum_{l ,l' s ,s'}\int d t_1 d t_2 \Sigma^{s s'}_{ll'}(t_1,t_2)G^{s s'}_{ll'}(t_1,t_2)+i L  T \mu^{+}_l (1-n^+_l)-i L  T \mu^{-}_l (-n^{-}_l).
\end{multline}Here $\hat \mu =\mu^{s}_{l} \delta_{ss'} \delta_{l l'} $ is a matrix diagonal in the Keldysh contours space $s,s'$ and in the replica spaces $l,l'$ .  And the fermionic two point function is given by $G^{s s'}_{l l'}(t)=\frac{1}{L}\left<\sum_i \psi^{s}_{i,l}(t) \overline{\psi}^{s'}_{i,l'}(0) \right>$. We are interested in the equations of motion. Repeating the procedure from the section with Majorana fermions, one can find:
\begin{gather}\label{eq:motionDirac}
        -(\hat G^{-1})^T(t_1,t_2)=  \hat\sigma_z \partial_t-\hat \Sigma(t_1,t_2)+i \sigma_z \hat{\mu} \delta(t_{12}), \\    \Sigma^{s s'}(t_1,t_2)=\frac{c_{ss'}}{q} (2 G^{s s'})^{\frac{q}{2}-1}(t_1,t_2)(-2G^{s' s})^{\frac{q}{2}}(t_1,t_2) \delta(t_{12}),\\
        G^{++}_{l,l'}(0^+)= \delta_{l,l'}(1-n^+_l),~~~G_{l,l'}^{--}(0^+)=- \delta_{l,l'} n^-_l. 
\end{gather}
In the first expression, we suppressed Keldysh and replica indexes, and in the second expression, we suppressed only replica indexes. Notice that for Majorana fermions, by definition, matrices $\hat G$ and $\hat \Sigma$ were real-valued. Now, they are allowed to be complex. 
In the case of single replica $k=1$, in the limit $t\rightarrow  0^{\pm}$ the Green function $G^{s s'}$ was found already in \cite{Zhang:2023vpm, Chen:2020bmq} but here we extend to generic replicas. 

The early-time behaviour can also be calculated similarly \ref{shorttime}, such that the corresponding Frame potential is given by 
\begin{equation}
    F^{(k)}(T\rightarrow 0)=e^{-\frac{LTk}{q^2}},
\end{equation}
which is different from Majorana calculation by the prefactor $\frac{1}{2}$ in the exponent.
\subsection{Non-Gaussian case q>2}
Now let us consider the late-time solution for $q>2$. Notice that the last two equations of motions, Eq. (\ref{eq:motionDirac}) define a boundary conditions for $\hat{G}$, fixed by the charge content. Also, notice that we have the sum over all possible values of  $n^{+(-)}$ in the Eq. (\ref{eq:FPDirac-pathintegral}). Therefore, let us first consider the terms with $n^+_l=n^-_l=n_l$. In this case, we have a replica diagonal solution, and it is given by charge-dependent matrix, see Appendix (\ref{A:SP})
\begin{equation}
\hat{G}_{\rm diag}=  \begin{pmatrix}
f(t_{12})(\Theta(t_{12})(\hat 1-\hat n))-\Theta(-t_{12})\hat n&-f(t_{12})\hat n\\
f(t_{12})(\hat 1-\hat n) & f(t_{12})(\Theta(-t_{12})(\hat 1-\hat n)-\Theta(t_{12})\hat n)
\end{pmatrix},
\end{equation}and 
\begin{equation}
    \hat \Sigma_{\rm diag}=\frac{2^{q-1}}{q}\begin{pmatrix}
        \frac{1}{2} \Gamma(1-2  \hat n)&-\Gamma(1-\hat n) \\
        \hat n\Gamma&\frac{1}{2} \Gamma(1-2 \hat n)
    \end{pmatrix}\delta(t_{12}),
\end{equation}with $f(t_{12})=e^{-\frac{\Gamma |t_{12}|}{2}-i \hat \mu t_{12}}$ and $\hat n={\rm diag}(n_1,n_2,..,n_k)$ is a diagonal matrix in replica space. This matrix consists of the fermion densities from different replicas and $\Gamma=(\hat n(1-\hat n))^{\frac{q}{2}-1}$ is also a diagonal matrix in replica space.
As in the case of Majorana Fermions, other possible solutions of Eq. (\ref{eq:motionDirac}) consists of permutation matrices, as these equations have permutation symmetry in the replica space. However, these solutions are valid only for the terms in the Eq. (\ref{eq:motionDirac}) which are given by permutation of the charge content $\hat n^{-}=\hat{\tau} \hat n^{+} \hat{\tau}^T$, due to the last two equations in Eq. (\ref{eq:motionDirac}). This gives the solution:
\begin{equation}
\hat{G}=  \begin{pmatrix}
f(t_{12})(\Theta(t_{12})(\hat 1-\hat n)-\Theta(-t_{12})\hat n)&- f(t_{12})\hat n\hat{\tau}^T\\
\hat{\tau} f(t_{12})(\hat 1-\hat n) & \hat{\tau} f(t_{12})(\Theta(-t_{12})(\hat 1-\hat n)-\Theta(t_{12})\hat n )\hat{\tau}^T
\end{pmatrix}.
\end{equation}Let us calculate the on-shell action using this solution.  We shall do our calculation for a replica diagonal solution, as the first three terms in the action are invariant under permutations. Therefore,  we can always  cast our solution to the replica diagonal one and get: 
\begin{equation} \label{det}
    e^{S_1}=(\det[\hat \sigma_z \partial_t-\hat \Sigma_d +i\sigma_z \hat\mu \delta(t_{12})])^L={\rm Tr}(e^{-T \hat H})^L,
\end{equation}where $H$ is a quafratic Hamiltonian in the fermions and diagonalisation gives(see Appendix \ref{App:pfaffian}) 
\begin{equation}
    {\rm det}(\sigma_z \partial_t-\Sigma_d+i\sigma_z \hat\mu \delta(t_{12}))^L\rightarrow e^{\frac{ 2^{q-1} LT\sum_{l=1}^k \Gamma_l}{2q}-iL T \sum_l \mu_l}.
\end{equation}The second term has prefactor $c_{ss'}$ that ensures its cancelling due to backward and forward evolution (see Appendix \ref{App:act_sp_dir})\begin{equation}
    S_2=\frac{L}{2q^2}\sum\limits_{l,l's,s' } \int dt_1 dt_2 \delta(t_{12}) c_{s s'} (2 G^{s s'}_{l l'})^{q/2} (-2 G^{s' s}_{l' l})^{q/2}=0,
\end{equation}
 The third term in the action gives (see Appendix \ref{App:act_sp_dir}),
\begin{equation}
       S_3=-L\sum_{l,l',s,s'} \int d t_1 d t_2 \Sigma^{s s'}_{ll'}(t_1,t_2)G^{s s'}_{ll'}(t_1,t_2)=-\frac{ 2^{q-1} LT\sum_{l=1}^k \Gamma_l}{2q}.
\end{equation}
We find therefore the final expression for the Frame potential
\begin{equation}
    F^{(k)}=\sum_{\hat{\tau}} \int  d \boldsymbol{\mu}^+ \sum_{\boldsymbol{Q}^{+}} \frac{e^{S_1+S_2+S_3}}{\mathcal{N}^{2 k}}=\frac{k! L^k}{\mathcal{N}^{2 k}},
\end{equation}
where we used  $\sum_{\boldsymbol{Q}^{+}} = L^k$. We have then recovered the expression obtained by integration over the Haar group in each charge sector, see Appendix \ref{sec:dirac-haar}. 

\subsection{The Gaussian case \texorpdfstring{$q=2$}{q=2}}
For $q=2$ the saddle equations read
\begin{gather}\label{eq:motionq2}
-(\hat G^{-1})^{\!T}(t_1,t_2)=\hat\sigma_z\partial_t-\hat\Sigma(t_1,t_2)+i\,\hat\sigma_z \hat\mu\,\delta(t_{12}),\\
\Sigma^{ss'}(t_1,t_2)=\frac{c_{ss'}}{2}\,\big(-2G^{s's}(t_1,t_2)\big)\,\delta(t_{12}),\\
G^{++}_l(0^+)=1-n_l^+,\qquad G^{--}_l(0^+)=-n_l^-.
\end{gather}
As in the $q>2$ case, begin with a replica–diagonal ansatz and impose $n_l^+=n_l^-$ (but allow them to differ across replicas). Writing
\[
\hat n=\mathrm{diag}(n_1,\ldots,n_k),\qquad
f(t)=\exp\!\left(-\frac{|t|}{2}-i\,\hat\mu\,t\right),
\]
the diagonal solution is
\begin{equation}
\hat G_{\mathrm{diag}}(t_1,t_2)=
\begin{pmatrix}
f(t_{12})\big(\Theta(t_{12})(\hat 1-\hat n)-\Theta(-t_{12})\hat n\big) & -f(t_{12})\,\hat n\\[3pt]
f(t_{12})(\hat 1-\hat n) & f(t_{12})\big(\Theta(-t_{12})(\hat 1-\hat n)-\Theta(t_{12})\hat n\big)
\end{pmatrix},
\end{equation}
with self–energy
\begin{equation}
\hat\Sigma_{\mathrm{diag}}(t_1,t_2)=
\begin{pmatrix}
\frac12(\hat 1-2\hat n) & -(\hat 1-\hat n)\\[3pt]
\hat n & \frac12(\hat 1-2\hat n)
\end{pmatrix}\delta(t_{12}).
\end{equation}

Because the $q=2$ action enjoys a continuous replica–rotation symmetry, we can generate a non–diagonal Keldysh solution by inserting a unitary matrix $\hat u$ in the off–diagonal blocks:
\begin{equation}
\hat G(t_1,t_2)=
\begin{pmatrix}
f(t_{12})\big(\Theta(t_{12})(\hat 1-\hat n)-\Theta(-t_{12})\hat n\big) & -f(t_{12})\,\hat n\,\hat u^\dagger\\[3pt]
\hat u\,f(t_{12})(\hat 1-\hat n) & \hat u\,f(t_{12})\big(\Theta(-t_{12})(\hat 1-\hat n)-\Theta(t_{12})\hat n\big)\hat u^\dagger
\end{pmatrix}.
\end{equation}
However, this form implies boundary conditions $\hat n^-=\hat u\,\hat n^+\,\hat u^\dagger$, which are not realized when we sum over charge sectors unless $\hat n$ is proportional to the identity. Thus we restrict to a uniform filling per replica,
\[
\hat n = n\,\hat 1_k,\qquad \hat\mu=\mu\,\hat 1_k,
\]
for which the boundary conditions are automatically satisfied for any unitary $\hat u$.
The saddle then simplifies to
\begin{equation}\label{eq:SPFerm}
\hat G(t_1,t_2)=
\begin{pmatrix}
f(t_{12})\big(\Theta(t_{12})(\hat 1-n\hat 1)-\Theta(-t_{12})\,n\hat 1\big)
& -\,f(t_{12})\,n\,\hat u^\dagger\\[3pt]
f(t_{12})(\hat 1-n\hat 1)\,\hat u
& f(t_{12})\big(\Theta(-t_{12})(\hat 1-n\hat 1)-\Theta(t_{12})\,n\hat 1\big)
\end{pmatrix},
\end{equation}
where $f(t)=e^{-\frac{|t|}{2}-i\mu t}$.
To preserve fermion parity of the initial state we further take $\det\hat u=1$, i.e. $\hat u\in SU(k)$. The evaluation of the action on this saddle proceeds exactly as in the Majorana case, with the replacement $\hat n\to n\hat 1$; we therefore omit repeated steps here.

\subsubsection{Symmetries and fluctuations}
We now characterize the fluctuation spectrum around the saddle~\eqref{eq:SPFerm}. The fields $(\psi,\bar\psi)$ live in a $2k$–dimensional replica space, so the kinematic group is $U(2k)$ with
\[
d\equiv \dim U(2k)=4k^2
\]
degrees of freedom available for fluctuations.

At $\hat n=n\hat 1$ and $\hat\mu=\mu\hat 1$ the action is invariant under independent replica rotations on the two Keldysh branches,
\[
U_1=e^{i\hat A_1\hat\sigma_z},\qquad U_2=e^{i\hat A_2\hat 1},\qquad \hat A_{1,2}^\dagger=\hat A_{1,2},
\]
i.e. the global symmetry group is
\[
B=U(k)\times U(k).
\]
The family of saddles~\eqref{eq:SPFerm} is parameterized by the unitary $\hat u$, and two pairs $(U_1,U_2)$ act as $\hat u\mapsto U_2\,\hat u\,U_1^\dagger$. The stabilizer $H$ leaves $\hat u$ unchanged, so the manifold of degenerate saddles is the coset $B/H\simeq SU(k)$ with
\[
R_{\mathrm{ml}}=\dim(B/H)=k^2-1
\]
massless (Goldstone) modes. The number of massive modes is then
\[
R_{\mathrm{m}}=d-R_{\mathrm{ml}}.
\]

Expanding the action to quadratic order,
\[
S=S_0(\hat G_0,\hat\Sigma_0)+S^{(2)}(\delta\hat G,\delta\hat\Sigma),
\]
the Gaussian integration over fluctuations produces, at leading order in $L$, a factor $L^{-R_{\mathrm{m}}}$ from the finite–time path integral, while the normalization (the $T=0$ Gaussian integral) contributes $L^{-d}$, cf.\ the Majorana analysis. In the Dirac case we also sum over charge sectors $Q=0,1,\dots,L$, which contributes an overall factor $\sim L$. Hence
\begin{equation}
\tilde F^{(k)}(T)\;\sim\;\frac{1}{\mathcal N^{2k}}\,L^{-R_{\mathrm{m}}}\times L,
\qquad
\tilde F^{(k)}(0)\;\sim\;L^{-d},
\end{equation}
and the frame potential scales as
\begin{equation}
F^{(k)}_{q=2}
=\frac{\tilde F^{(k)}(T)}{\tilde F^{(k)}(0)}
\;\sim\;\frac{1}{\mathcal N^{2k}}\,L^{\,1+d-R_{\mathrm{m}}}
=\frac{1}{\mathcal N^{2k}}\,L^{\,1+R_{\mathrm{ml}}}
=\frac{1}{\mathcal N^{2k}}\,L^{\,k^2}.
\end{equation}
Taking the logarithm and using $\mathcal N=2^L$ for $L$ Dirac modes,
\begin{equation}
\lim_{L\to\infty}\frac{1}{L}\log F^{(k)}_{q=2}
=-2k\log 2+\frac{k^2}{L}\log L+O\!\left(\frac{1}{L}\right),
\end{equation}
in agreement with the gHaar calculation, cf.\ Eq.~\eqref{Haarresferm}.

Finally, as in the Majorana case, for generic $q>2$ there are no continuous zero modes: the saddle manifold is discrete, the fluctuation determinants at finite and zero time cancel, and one finds
\begin{equation}
\lim_{L\to\infty}\frac{1}{L}\log F^{(k)}_{q>2}
=-2k\log 2+O\!\left(\frac{1}{L}\right),
\end{equation}
consistent with the gHaar–integrated result in Eq.~\eqref{Haaru}.

\section{Conclusion}\label{sec5}

In this chapter we quantified the emergence of pseudorandom dynamics in the Brownian SYK model by means of the \emph{frame potential} \(F^{(k)}(T)\), a moment-based diagnostic that certifies unitary \(k\)-design formation. Concretely, we evaluated
\[
F^{(k)}(T)=\Big\langle \mathrm{Tr}\!\left(U_1(T)\rho_0 U_1^\dagger(T)\,U_2(T)\rho_0 U_2^\dagger(T)\right)^k \Big\rangle,
\]
which measures the overlap of two independently evolved copies, and analyzed it using a Keldysh path-integral formulation with bi-local collective fields. A saddle-point treatment at large \(L\) exposes the full time evolution of \(F^{(k)}(T)\).

\medskip
\noindent\textbf{Early times.} From the short-time saddle we find an exponential decay of the frame potential, reflecting rapid operator growth and the onset of scrambling as probed by two-copy observables (e.g., purities and OTOCs). The decay rate is set by \(q\), the degree of the interaction, and is independent of microscopic details.

\medskip
\noindent\textbf{Late times: two universality classes.} The late-time behavior separates into two sharply distinct regimes:
\begin{itemize}
  \item \emph{Interacting, non-Gaussian (\(q>2\)).} The replica structure is governed by a \emph{discrete} saddle manifold; there are no Goldstone modes. Evaluating the action on the non-diagonal Keldysh saddle and summing over replica pairings yields saturation of \(F^{(k)}(T)\) to its Haar value (in the appropriate symmetry sector). Physically, the dynamics explore the full Hilbert space uniformly.
  \item \emph{Gaussian, integrable (\(q=2\)).} A \emph{continuous} symmetry of the Keldysh action generates massless (Goldstone) modes. The dynamics mix only within the manifold of Gaussian states, and \(F^{(k)}(T)\) converges to the \emph{Gaussian Haar} (gHaar) value. The late-time value agrees with the generic (\(q>2\)) Haar prediction in the \(L\!\to\!\infty\) limit, but acquires characteristic logarithmic finite-\(L\) corrections coming precisely from integrating over these zero modes.
\end{itemize}

\medskip
\noindent\textbf{Dirac versus Majorana.} In the complex (Dirac) SYK model, global \(U(1)\) charge conservation constrains the dynamics to become Haar-like \emph{within fixed-charge sectors}. Implementing charge projectors allows saddle points with different charge content across replicas. For \(q>2\) the sector-mismatched saddles dominate, whereas for \(q=2\) the saddle with all replicas in the same charge sector prevails, thanks to its larger number of massless modes. As a consequence, late-time values of \(F^{(k)}\) are larger in the Dirac case than in the Majorana case, as expected from the additional global constraint.

\medskip
Altogether, our analysis makes precise the hierarchy from thermalization to scrambling to pseudorandomness: \(F^{(k)}(T)\) decays exponentially at early times and, on a timescale controlled by \(q\), saturates to a value that is exponentially small in system size (via the Hilbert-space dimension) and coincides with the Haar (or gHaar) prediction once symmetries are accounted for. In this sense, the Brownian SYK model furnishes a clean example of \emph{randomness by design}: chaotic dynamics generate the statistics of Haar-random unitaries up to the relevant moment order.

\medskip
\noindent\textbf{Outlook.} Two natural extensions follow from our framework. First, one can study how continuous monitoring and measurement-induced transitions modify design formation and the frame potential in SYK-like settings \cite{Jian2021Rep,Fava:2023tgg,SkinnerRuhmanNahum2019,Antonini2022}. Second, it is compelling to investigate non-Brownian (e.g., time-correlated or deterministic chaotic) deformations of SYK to understand how temporal structure impacts the approach to Haar/gHaar saturation and the associated finite-size corrections.

    \chapter{Field theory for monitored Brownian SYK clusters}
\section{Motivation}
A central challenge in the study of non-equilibrium quantum matter is to understand how quantum information spreads, becomes scrambled, or is lost. Early progress came from analyzing isolated systems evolving unitarily. Transport, correlations, and entanglement growth have revealed universal structures even far from equilibrium, underpinning thermalization and also unconventional regimes where it fails~\cite{PhysRevA.43.2046,Srednicki1994,Rigol2008,PhysRevX.9.031048,Nahum2018,Maldacena2016,Deutsch_2018}.

Measurements provide a complementary perspective. They are discrete events producing stochastic outcomes, and thus a natural ingredient of non-equilibrium physics. Modern platforms now allow repeated measurements during dynamics, motivating the study of \emph{monitored systems}, where local projective measurements compete with coherent evolution~\cite{WisemanMilburn2010,Jacobs2014}. Far from passive probes, measurements act as dynamical channels of decoherence and feedback, giving rise to collective behavior absent in closed systems. Recent theory and experiments have shown that such dynamics can host measurement-induced phase transitions (MIPTs)~\cite{LiChenFisher2018,SkinnerRuhmanNahum2019,BaoChoiAltman2020}, which extend familiar mechanisms of equilibration. Continuous monitoring is naturally described by the stochastic Schrödinger equation (SSE), which encodes both coherent dynamics and stochastic backaction~\cite{WisemanMilburn2010,Jacobs2014}.

The advent of digital quantum simulators—superconducting qubits, trapped ions, and Rydberg arrays—has accelerated this direction. These platforms combine unitary gates with high-fidelity local measurements and feedback~\cite{Noel2022,Bernien2017,Preskill2018review}, opening a frontier where coherent evolution and observations compete to generate novel dynamical phases~\cite{VasseurPotter2019,JianYouVasseurLudwig2020}.

Early progress came from random circuits, where measurements are incorporated as mid-circuit operations. Numerical studies revealed MIPTs separating volume-law and area-law entanglement~\cite{LiChenFisher2018,SkinnerRuhmanNahum2019,BaoChoiAltman2020}, and mappings to classical statistical models~\cite{VasseurPotter2019,JianYouVasseurLudwig2020}. Analytic approaches are challenging because entanglement measures are nonlinear in the density matrix, necessitating replica-trick methods familiar from many-body localization~\cite{Vasseur2016,PotterVasseurParameswaran2015}. 
Notably, in hybrid random circuits the solvable large-\(q\) limit renders the replica mapping exact: R\'enyi entropies reduce to a classical percolation problem for domain walls, yielding an entanglement transition at \(p_c=\tfrac{1}{2}\) with percolation criticality~\cite{SkinnerRuhmanNahum2019,BaoChoiAltman2020,JianYouVasseurLudwig2020}.

Attention has since turned to monitored physical models of fermions and spins. Field-theoretical descriptions of free fermions under monitoring revealed nonlinear sigma models (NLSMs) as effective replicated theories~\cite{AlbertonBuchholdDiehl2021,Fava:2023tgg}. This raised debates: free Dirac fermions appear not to exhibit sharp MIPTs, while Majorana fermions do~\cite{CaoTilloyDeLuca2019,Lunt2020,PoboikoPRX2023,Eissler2024Unraveling,Tsitsishvili2024NonMarkovian}. Recent advances include equation-of-motion and metrological approaches~\cite{Fava:2023tgg,BuchholdDiehl2021,DiFresco2024Metrology,ChatterjeeModak2025Driven}, though a first-principles derivation from the coherent-state path integral is still lacking.

In this thesis, we contribute to this program by deriving the emergent NLSM directly from the coherent-state path integral representation~\cite{Kamenev2011,Altland2010} of replicated dynamics. We focus on Brownian SYK clusters of $N_F$ Majorana fermions under continuous monitoring of quadratic operators~\cite{Loio:2023fxn}. The SSE Eq.~\eqref{eq:SSE} provides a trajectory-level description of pure-state evolution under both coherent Hamiltonian dynamics and stochastic backaction~\cite{WisemanMilburn2010,Jacobs2014}. This framework lets us model weak continuous monitoring in a controlled way and connect to field-theoretic approaches for monitored fermions, obtaining a microscopic route to the effective NLSM of monitored SYK dynamics.

\section{Model and setup}
In this study, we consider a system composed of $L$ clusters of Majorana fermions $\{\hchi_{i \nu}\}$, where $i \in \{1,\dots ,L\}$ labels the cluster and $\nu \in \{1,...,N_F\}$ is a ``flavour'' index within each cluster. These operators satisfy the usual anticommutation relations $\{\hchi_{i \nu}, \hchi_{j \mu}\} = 2 \delta_{ij} \delta_{\nu \mu}$.  

\paragraph{Monitoring.}  
Recalling the stochastic Schrödinger equation (SSE) introduced in Introduction Eq.~\eqref{eq:SSE}, we now generalize it to the case of monitoring multiple operators, each associated with an independent Wiener process. Inside each cluster we monitor $q$-body operators of the form
\begin{equation}
    \hat{M}_{i, \boldsymbol{\tilde \nu}} = i^{q/2}\prod_{l=1}^{q}\hchi_{i\tilde \nu_l},
\end{equation}
with $\boldsymbol{\tilde \nu}=(\tilde \nu_1,\dots,\tilde \nu_q)$ an ordered set of flavours and $dW_{i,\boldsymbol{\tilde \nu}}$ the corresponding Wiener increments. In this chapter we focus primarily on the quadratic case $q=2$. Importantly, these bilinear operators are idempotent, $\hat{M}_{i,\boldsymbol{\tilde \nu}}^2=\hat{M}_{i,\boldsymbol{\tilde \nu}}$, so the non-Hermitian contribution to the SSE involves only quadratic combinations of Majoranas.  

\paragraph{Unitary interactions.}  
In addition to monitoring, fermions interact both within each cluster and between neighboring clusters via random couplings. For two clusters $j,\ell$, we introduce couplings of $q_J/2$ fermions in each cluster:
\begin{equation}
\label{eq:Hunijj}
  H^{\rm uni}_{j\ell}(t) = i^{\frac{q_J}{2}} \sum_{\boldsymbol{\mu},\boldsymbol{\nu}}  
  h^{j,\ell}_{ \boldsymbol{\mu}\boldsymbol{\nu}}(t)
  \prod_{k=1}^{q_J/2} \hchi_{j \mu_{k}}
  \prod_{m=1}^{q_J/2} \hchi_{\ell \nu_{m}} ,
\end{equation}
and within a single cluster $j$,
\begin{equation}
\label{eq:Hunij}
  H^{\rm uni}_{j}(t) = i^{\frac{q_J}{2}} \sum_{\boldsymbol{\tilde \mu}}  
  h^j_{ \boldsymbol{\tilde \mu}}(t)
  \prod_{k=1}^{q_J} \hchi_{j \tilde \mu_{k}} .
\end{equation}
Here sums over bold indices denote
\[
\sum_{\boldsymbol{\mu}} = \sum_{1 \leqslant \mu_1 < \cdots < \mu_{q_J/2} \leqslant N_F}, 
\qquad
\sum_{\boldsymbol{\tilde \mu}} = \sum_{1 \leqslant \tilde \mu_1 < \cdots < \tilde \mu_{q_J} \leqslant N_F}.
\]
The random couplings $h^{j,\ell}_{\boldsymbol{\mu}\boldsymbol{\nu}}(t)$ and $h^j_{\boldsymbol{\tilde \mu}}(t)$ are Gaussian white noises with unit variance for each choice of indices, and are taken in the Stratonovich convention.  

\begin{figure}[t]
\centering
\resizebox{\linewidth}{!}{\begin{tikzpicture}[x=1cm,y=1cm,>=Latex]

  \definecolor{PurpleDark}{HTML}{5928a8} 
  \definecolor{Ink}{HTML}{111111}        
  \definecolor{PinkIn}{HTML}{fab1df}     
  \definecolor{PinkOut}{HTML}{d6c0fc}    
  \definecolor{balls}{HTML}{8e57e6}

  \def\R{2.1}        
  \def\gap{4.6}      
  \def\lw{0.9pt}     
  \def\dotr{0.075}   
  \def\rsel{0.85}    

  \pgfdeclarelayer{bg}
  \pgfsetlayers{bg,main}

  \coordinate (C1) at (-\gap, 0);
  \coordinate (C2) at (0, 0);
  \coordinate (C3) at (\gap, 0);

  \newcommand{\ShadedCluster}[2]{
    \begin{pgfonlayer}{bg}
      \shadedraw[inner color=PinkIn, outer color=PinkOut, draw=PurpleDark, line width=\lw]
        #1 circle (#2);
    \end{pgfonlayer}
  }
  \pgfmathsetseed{20251004} 
  \newcommand{\ScatterBalls}[2]{
    \pgfmathtruncatemacro{\NN}{#2}%
    \foreach \i in {1,...,\NN} {%
      \pgfmathsetmacro{\ang}{360*rnd}%
      \pgfmathsetmacro{\rad}{\rsel*\R*sqrt(rnd)}%
      \path #1 ++(\ang:\rad) coordinate (p\i);
      \shade[shading=ball, ball color=balls] (p\i) circle (\dotr);
    }%
  }

  \ShadedCluster{(C1)}{\R}
  \ShadedCluster{(C2)}{\R}
  \ShadedCluster{(C3)}{\R}

  \ScatterBalls{(C1)}{26}
  \ScatterBalls{(C2)}{24}
  \ScatterBalls{(C3)}{22}

  \draw[PurpleDark, line width=\lw,dash pattern=on 6pt off 3pt]
    ($(C1)+(0.55,0.15)$) circle (0.75);
    
  \draw[PurpleDark, line width=\lw,dash pattern=on 6pt off 3pt]
    ($(C2)+(-0.30,0.30)$) circle (0.75);
    \draw[PurpleDark, line width=\lw,dash pattern=on 6pt off 3pt]
    ($(C2)+(0.70,-1.0)$) circle (0.75);
  \draw[PurpleDark, line width=\lw,dash pattern=on 6pt off 3pt]
    ($(C3)+(-0.30,-0.5)$) circle (0.75); 

\tikzset{
  snakeTop/.style={
    draw=PurpleDark, line width=\lw,
    dash pattern=on 6pt off 3pt,
    decorate, decoration={snake, amplitude=0.18cm, segment length=1.0cm},
    line cap=round
  },
  snakeBot/.style={
    draw=PurpleDark, line width=\lw,
    dash pattern=on 6pt off 3pt,
    decorate, decoration={snake, amplitude=0.18cm, segment length=1.0cm},
    line cap=round
  }
}

\draw[snakeTop]
  ($(C1)+(0.95,0.80)$)
    .. controls ($(C1)+(1.8,1.50)$) and ($(C2)+(-1.8,1.50)$) ..
  ($(C2)+(-0.90,0.80)$);

\draw[snakeBot]
  ($(C2)+(1.5,-0.85)$)
    .. controls ($(C2)+(1.9,-1.60)$) and ($(C3)+(-1.9,-1.60)$) ..
  ($(C3)+(-0.95,-0.85)$);
  
\newcommand{\PlaceBallsAt}[3]{
  \pgfmathtruncatemacro{\NN}{#3}%
  \foreach \i in {1,...,\NN} {%
    \pgfmathsetmacro{\ang}{360*rnd}%
    \pgfmathsetmacro{\rad}{0.82*#2*sqrt(rnd)}%
    \shade[shading=ball, ball color=balls]
      ($(#1)+(\ang:\rad)$) circle (\dotr);
  }%
}
\def\sr{0.75} 
\coordinate (S1) at ($(C1)+(0.55,0.15)$);
\coordinate (S2) at ($(C2)+(-0.30,0.30)$);
\coordinate (S3) at ($(C2)+(0.70,-1.0)$);
\coordinate (S4) at ($(C3)+(-0.30,-0.5)$);
\pgfmathsetseed{4241}\PlaceBallsAt{S1}{\sr}{0}
\pgfmathsetseed{4242}\PlaceBallsAt{S2}{\sr}{3}
\pgfmathsetseed{4243}\PlaceBallsAt{S3}{\sr}{5}
\pgfmathsetseed{4244}\PlaceBallsAt{S4}{\sr}{5}

\def\devS{0.65} 

\pgfmathsetmacro{\devW}{3.2*\devS}   
\pgfmathsetmacro{\devH}{2.3*\devS}   
\pgfmathsetmacro{\devRC}{4*\devS}    
\pgfmathsetmacro{\gR}{1.15*\devS}    
\pgfmathsetmacro{\tickIn}{0.14*\devS}
\pgfmathsetmacro{\tickLen}{0.18*\devS}
\pgfmathsetmacro{\tickLW}{0.6*\devS} 
\pgfmathsetmacro{\needleLW}{1.1*\devS}
\pgfmathsetmacro{\hubR}{0.055*\devS} 
\pgfmathsetmacro{\gaugeYOffset}{0.55*\devS} 

\coordinate (D) at (0,3.85); 
\node[
  draw=PurpleDark, line width=\lw, fill=white,
  rounded corners=\devRC pt,
  minimum width=\devW cm, minimum height=\devH cm
] (meter) at (D) {};

\draw[PurpleDark, line width=\lw]
  (meter.south) -- (0, 1.5);

\path (meter.south) ++(0,\gaugeYOffset) coordinate (gC);

\draw[PurpleDark, line width=\lw]
  ($(gC)+(-\gR,0)$) arc (180:0:\gR);

\foreach \a in {180,160,140,120,100,80,60,40,20,0} {
  \draw[PurpleDark, line width=\tickLW pt]
    ($(gC)+(\a:\gR-\tickIn)$) -- ++(\a:\tickLen);
}

\draw[PurpleDark, line width=\needleLW pt, line cap=round]
  (gC) -- ++(35:\gR*0.95);

\fill[PurpleDark] (gC) circle (\hubR cm);

\node[anchor=west, PurpleDark,font=\Large]
  at ($(meter.east)+(0.35,0)$) {$i\,\hat\chi_{i\nu_1}\,\hat\chi_{i\nu_2}$};

\node[anchor=north, PurpleDark,font=\Large] at ($(C1)+(0,-\R-0.45)$) {$i{-}1$};
\node[anchor=north, PurpleDark,font=\Large] at ($(C2)+(0,-\R-0.45)$) {$i$};
\node[anchor=north, PurpleDark,font=\Large] at ($(C3)+(0,-\R-0.45)$) {$i{+}1$};
\def\edot{0.055}   
\def\esp{0.32}     

\path ($(C1)+(-\R-0.70,0)$) coordinate (Lellip);
\foreach \i in {0,1,2} {
  \fill[PurpleDark] ($(Lellip)+(-\i*\esp,0)$) circle (\edot);
}

\path ($(C3)+(\R+0.70,0)$) coordinate (Rellip);
\foreach \i in {0,1,2} {
  \fill[PurpleDark] ($(Rellip)+(\i*\esp,0)$) circle (\edot);
}

\end{tikzpicture}}
\caption{1D chain of Majorana clusters under continuous monitoring of bilinears inside each cluster. 
Each pink cluster (sites \(i{-}1\), \(i\), \(i{+}1\)) contains $N_F$ Majorana fermions. 
Wavy dashed connectors suggest inter-cluster interaction of the unitary evolution. 
The top gauge symbolizes a readout of the Majorana bilinear. 
Three-dot ellipses on the sides indicate additional clusters beyond the frame.}
\label{fig:local-majorana-measurement}
\end{figure}
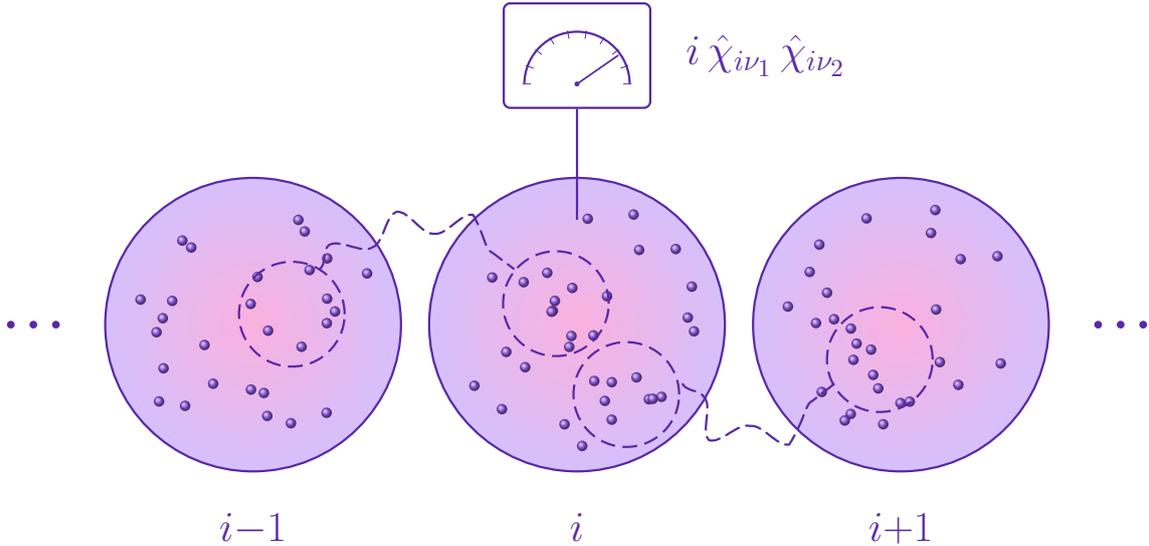

We place the clusters on a one-dimensional chain see Fig. \ref{fig:local-majorana-measurement}, restricting inter-cluster couplings to nearest neighbors. The full unitary Hamiltonian then reads
\begin{equation} \label{eq_H_uni}
  H^{\rm uni}(t) = \sqrt{J}\sum_{j} \Big(H^{\rm uni}_{j,j+1}(t) + H^{\rm uni}_{j}(t)\Big),
\end{equation}
where the overall prefactor $\sqrt{J}$ sets the interaction strength and is chosen for later convenience in replica averaging.

\paragraph{Trajectories.}  
A \emph{trajectory} is specified by a realization of all the Wiener processes (measurement outcomes) and of the stochastic unitary couplings. Given such a trajectory, the conditioned density matrix $\rho(t)$ evolves from its initial state $\rho(0)$ according to Eq.~\eqref{eq:SSE}. Our goal is to compute expectation values of scalar functionals $F[\rho(t)]$ at fixed time $t$, averaged over trajectories:
\[
    \overline{F[\rho]}_{\rm SSE}.
\]
While linear functionals of $\rho$ admit closed equations of motion, nonlinear quantities—most notably entanglement entropies—do not. The obstacle is the intrinsic nonlinearity of the SSE, which enforces normalization $\Tr \rho(t)=1$ along each trajectory but prevents straightforward averaging. In the next section we turn to an alternative formalism based on non-normalized density matrices and the replica trick, which circumvents this difficulty.

\section{Replica formalism for entanglement}\label{sec:replica_form}

As emphasized above, nonlinear observables such as entanglement entropies cannot be accessed directly from the SSE, since its nonlinear form prevents closed equations. A standard way forward is to drop the Born's rule constraint and work with an unnormalized state or density matrix. 
\subsection{From SSE to unnormalized density matrices}\label{sec:repl}
Concretely, one introduces a non-normalized operator $\check \rho(t)$~\cite{Fava:2023tgg, JianYouVasseurLudwig2020,GiachettiElusive2022}, related to the physical density matrix by
\[
\rho(t) = \frac{\check{\rho}(t)}{\Tr \check \rho (t)}.
\]
The operator $\check\rho(t)$ evolves linearly, governed by an effective non-Hermitian Hamiltonian (see Appendix \ref{App:linearSSE}). At the trajectory level, this corresponds to replacing the nonlinear, norm-preserving SSE by a linear stochastic equation for unnormalized states. The two descriptions are equivalent: normalizing $\check\rho(t)$ at each step recovers the SSE, while keeping it unnormalized makes analytic approaches (e.g. replicas, path integrals) tractable. In this sense the operator $K(t)$ below should be viewed as a non-unitary evolution operator generated by a non-Hermitian Hamiltonian.  

In particular, the unnormalized density matrix evolves as
\begin{equation}\label{eq:evol}
  \check{\rho}(t) = K(t)\,\rho(0)\,K^\dagger(t) , 
\end{equation}
where the (non-unitary) evolution operator $K(t)$ is defined by the time-ordered exponential 
\begin{equation}
  K(t) \equiv \mathcal{T} \exp\!\Big(-i \int_0^t H(s)\,ds\Big),
\end{equation}
with effective Hamiltonian
\begin{equation} \label{eq_H_uni_plus_mon}
  H(t)  = H^{\rm uni}(t) + i \sqrt{\Gamma} \sum_j H_{j}^{\mon}(t), ~~H^{\mon}_j  = i^{\frac{q}{2}} \sum_{\boldsymbol{\tilde \nu}} w^j_{\boldsymbol{\tilde \nu}}(t)\,
  \prod_{l=1}^{q}\hchi_{j\tilde \nu_l},
\end{equation}
and $w^j_{\boldsymbol{\tilde \nu}}=\frac{d W_{i,\boldsymbol{\tilde \nu}}}{dt}$ denotes Gaussian white noise—the formal time derivative of independent
Wiener processes—interpreted in the Stratonovich sense (see Appendix(\ref{App:linearSSE})). Now, we can circumvent the non-linear averaging discussed above by explicitly including the Born's weight $\Tr \check\rho(t)$ as a factor in the quantity to be averaged. As a consequence, the residual distribution for the couplings $w^j_{\boldsymbol{\tilde \nu}}(t)$ is unbiased, i.e. it amounts to uncorrelated white noises. We denote as $\mathbb{E}_{\rm G}$ the resulting average, with $\mathbb{E}_{\rm G}[w^j_{\boldsymbol{\tilde \nu}}(t) w^{j'}_{\boldsymbol{\tilde \nu'}}(t')] = \delta_{jj'} \delta_{\boldsymbol{\tilde \nu}\boldsymbol{\tilde \nu'}} \delta(t-t')$. Therefore, the final quantum trajectory-averaged expectation values  of  scalar functionals  $F[\rho]_{\rm SSE}$ are expressed in terms of the Gaussian averages as,
\begin{equation}
\label{eq:replicatrickF}
  \overline{F[\rho]}_{\rm SSE} \equiv 
  \frac{\mathbb{E}_{\rm G}\left[F[\rho] \Tr \check\rho(t) \right]}{\mathbb{E}_{\rm G}[\Tr \check\rho(t) ]} ,
\end{equation}
where the denominator in the right-hand side accounts for the proportionality constant in the Born probabilities.

A notable example is the trajectory-averaged purity of a subsystem $A$, corresponding to the choice $F[\rho] = \Tr[\rho_A^2]$, with $\rho_A = \Tr_{\bar{A}} \rho$, where $\bar{A}$ is the complementary of $A$. Applying Eq.~\eqref{eq:replicatrickF}, we can express it as
\begin{equation}
\label{eq:purityreplica}
\overline{\Tr\rho_A(t)^2}  = \frac{\mathbb{E}_{\rm G} \left[ \Tr  \check\rho_A(t)^2 (\Tr  \check\rho(t))^{-1} \right] }{\mathbb{E}_{\rm G}\left[\Tr  \check\rho(t) \right]}
  =  \lim_{n \rightarrow 1} \frac{\mathbb{E}_{\rm G} \left[ \Tr  \check\rho_A(t)^2 (\Tr  \check\rho(t))^{n-2} \right]}{\mathbb{E}_{\rm G}[\left(\Tr  \check\rho(t)\right)^n]} ,
\end{equation}
with $\check\rho_A = \Tr_{\bar{A}} \check\rho$. In the last step, we applied the replica trick to facilitate the Gaussian averaging of $\mathbb{E}_{\rm G} \left[ \Tr \check\rho_A(t)^2 \left(\Tr \check\rho(t)\right)^{-1} \right]$, by treating $n$ as a parameter.
In particular, {for integer $n \geqslant 2$}, we can replace the powers in Eq.~\eqref{eq:purityreplica} with a tensor product, arriving at
\begin{equation} \label{eq_purity_final}
  \overline{\Tr\rho_A(t)^2} =  
  \lim_{n \rightarrow 1} \frac{\Tr \left[ \rho^{(n)}(t) \left(\mathcal{C}_{A,2} \otimes \mathbb{I} \right)\right]}
  {\Tr \left[ \rho^{(n)}(t) \right]} ,
\end{equation}
where $\rho^{(n)}(t) = \mathbb{E}_{\rm G}\left[ \check\rho(t)^{\otimes n} \right]$
and $\mathcal{C}_{A,2}=\sum_{j,j'} \left| j \right>\left<j' \right|_A \otimes \left|j' \right>\left< j \right|_{A} \otimes \mathbb{I}$ is a swap operator that exchanges, within the region $A$, the first two replicas among the total $n$, while acting as the identity elsewhere. Here  $\left|j\right>$ is a basis vector on a subspace $A$.
More generally, we can introduce permutation operators $\mathcal{C}_{A,\sigma}$ for any permutation $\sigma$ in the symmetric group with $n$ elements. Denoting a permutation $P = (2^{k_2} 3^{k_3} \ldots )$ as its cycle decomposition, with $k_m$ the number of cycles of length $m$, we can express generic functionals of the density matrix. For instance,
considering the $\alpha$-Renyi entropy of the subsystem $A$, defined as,
\begin{equation}
S_A^{(\alpha)} = \frac{1}{1-\alpha} \log \Tr[\rho_A^\alpha]  ,  
\end{equation}
we can express its average over trajectory within the replica formalism as
\begin{equation} \label{eq_renyi_limit}
\overline{S^{(\alpha)}_A(t)} =     \frac{1}{1-\alpha} \lim_{k \rightarrow 0}  \lim_{n \rightarrow 1} 
    \frac{1}{k}\left(\frac{\Tr \left[ \rho^{(n)}(t) \left(\mathcal{C}_{A,\alpha^k}\otimes \mathbb{I} \right)\right]}{\Tr \left[ \rho^{(n)}(t) \right]} - 1\right) ,
\end{equation}
where $P = (\alpha^k)$ is any permutation with $k$ cycles of length $\alpha$.
While Eqs.~\eqref{eq_purity_final}–\eqref{eq_renyi_limit} provide a formal replica representation of entanglement measures, carrying out the Gaussian averaging of $\rho^{(n)}(t)$ remains nontrivial because the evolution involves both left and right multiplications by $K(t)$ Eq.\eqref{eq:evol}. To handle this structure systematically, we now turn to a vectorized representation of the density matrix, in which the replicated dynamics reduces to a Schrödinger-like evolution.
\subsection{Density matrix vectorization and replicated Hamiltonian} \label{sec:vect}

As discussed above, replica observables such as purities or Rényi entropies can be expressed in terms of nonlinear functionals of the unnormalized density matrix $\check\rho(t)$, weighted by powers of its trace [cf.~Eq.~\eqref{eq:purityreplica}].  
The difficulty is that the dynamics of $\check\rho(t)$ involves both left and right multiplications,
\begin{equation}
\check\rho(t) = K(t)\,\check\rho(0)\,K^\dagger(t),~~~K(t) = T \exp \left(- i \int_0^t \; dt' H(t')\right)
\end{equation}
so that after replication one faces a proliferation of $K$ and $K^\dagger$ factors acting on different sides of different copies.  
Directly averaging such expressions quickly becomes intractable.

To overcome this obstacle, we introduce a \emph{vectorized representation} of operators, sometimes called \emph{folding}.  
The idea is to embed the operator $\check\rho$ into a doubled Hilbert space where left and right multiplications act linearly,
\begin{equation}
  \ket{ \hat O} \equiv \sum_{kl} \bra{k} \hat O \ket{l}\, \ket{k}\otimes\ket{l}.
\end{equation}
In this representation, the two-sided action $K \hat O K^\dagger$ becomes a one-sided action of the tensor operator $K \otimes K^*$, so that the replicated object evolves according to a \emph{linear Schrödinger-like equation} generated by a (non-Hermitian) Hamiltonian. The corresponding generator is $H_L-H_R\equiv H\otimes\mathbb{I}-\mathbb{I}\otimes H^{\mathsf T}$ (see Appendix \ref{App:rep}), which is precisely the structure of a Keldysh contour with a forward ($+$) and backward ($-$) branch that we studied in the previous chapter \ref{sec:Keldysh}. In this language, left multiplication is the $+$ branch and right multiplication is the $-$ branch, and after a standard Keldysh rotation the dynamics naturally separates into ``classical'' (branch sum) and ``quantum'' (branch difference) combinations..

This change of perspective is more than a notational convenience.  
It provides two decisive advantages for the replica formalism:
\begin{enumerate}
  \item \textbf{Linear dynamics.} The replicated state evolves linearly under a single generator, allowing us to apply standard quantum-mechanical tools (path integrals, correlation functions, diagrammatics) without having to track simultaneous left/right actions.  
  \item \textbf{Overlap structure.} Nonlinear functionals of $\rho$, such as purities and Rényi entropies, become simple overlaps between the evolved replicated state and boundary operators implementing permutations (e.g. swap operators).  
  These overlaps play the role of transition amplitudes, which can be naturally represented and evaluated in a path-integral framework.
\end{enumerate}

In this way, vectorization translates the original problem of averaging nonlinear functionals of $\check\rho$ into the familiar problem of evaluating amplitudes in a replicated Hilbert space, paving the way for the constructions of the following sections.

For further consideration, let us introduce indices $\sigma = \pm$ to distinguish left and right actions. 
Specifically, we use the notation $+$ for operators multiplying on the left and $-$ for operators multiplying on the right, i.e.,
\[
  A \rho B \;\longmapsto\; A^+ (B^-)^t \ket{\rho},
\]
where $t$ denotes transposition. 
Therefore, the time evolution inside the average in Eq.~\eqref{eq_folded_replicated_rho} becomes
\begin{equation} 
  \left( K(t) \otimes K^*(t) \right)^{\otimes n} = \mathcal{T} \exp \left(-i \int_0^t H^{(n)}(s)\, ds \right),
\end{equation}
with the replicated Hamiltonian $H^{(n)}(t)$ derived in Appendix~\ref{App:rep}:

\begin{equation} \label{eq_H_rep_f}
  \begin{split}
    H^{(n)}(t) & = \sum_{\sigma,a,j} \left[ (i)^{\frac{q_J}{2}}f_\sigma^{\rm uni} \sqrt{J} \left[\sum_{{\boldsymbol{\tilde \mu}}}  h^j_{ \boldsymbol{\tilde \mu}}(t) \prod^{q_J}_{k=1} \hchi^{(\sigma, a)}_{j \tilde \mu_{k}} \right. \right. + \left. \sum_{{\boldsymbol{\nu}}} \sum_{{\boldsymbol{\mu}}}  h^j_{ \boldsymbol{\mu}\boldsymbol{\nu}}(t) \prod^{q_J/2}_{k=1} \hchi^{(\sigma, a)}_{j \mu_{k}} \prod_{l=1}^{q_J/2} \hchi^{(\sigma, a)}_{j+1 \nu_{l}} \right]\\
    &+\left. (i)^{\frac{q}{2} + 1}\sigma f_\sigma^{\rm mon} \sqrt{\Gamma} \sum_{\boldsymbol{\tilde \nu}} w^j_{\boldsymbol{\tilde \nu}} \prod_{l=1}^{q}\hchi^{(\sigma, a)}_{j\tilde \nu_l}  \right]  ,
\end{split}
\end{equation}
where $f^{\rm uni}_+ = i^{q_J+ 2}$ and $f_-^{\rm uni} = 1$,  and $f^{\rm mon}_+ = -i^{q+2}$, $f^{\rm mon}_- = -1$. And $(\sigma,a)$ are replica indexes $\sigma\in \{+,-\}$ and $a \in {1,...,n}$. Notice that in this form the averaging we need to calculate $\mathbb{E}_{\rm G}[\left( K(t) \otimes K^*(t) \right)^{\otimes n}]=\mathbb{E}_{\rm G}[\mathcal{T} \exp \left(-i \int_0^t H^{(n)}(s)\, ds \right)]$ is just calculating  gaussian integrals over random variables $h, \omega$.
We now perform averaging over the Gaussian distributed parameters $w$ and $h$ in order to find the evolution of the replicated state Eq. (\ref{eq_folded_replicated_rho}) (see Appendix \ref{App:aver}) :
\begin{equation}\label{eq:rep_ham}
    \mathbb{E}_G\left[\left( K(t) \otimes K^*(t) \right)^{\otimes n} \right]=e^{-t \mathcal{H}^{(n)}},
\end{equation}and obtain the effective Hamiltonian of the evolution 
\begin{equation} \label{repHam}
\begin{split}
  \mathcal{H}^{(n)}  = \frac{1}{2} & \left( J \sum_{j{\boldsymbol{\mu}} {\boldsymbol{\nu}}} \mathcal{H}^\uni_{j j+1, \boldsymbol{\mu}{\boldsymbol{\nu}}} \right. \left.+ J \sum_{j{\boldsymbol{\tilde \mu}}}\mathcal{H}^\uni_{j, \boldsymbol{\tilde \mu}} - \Gamma \sum_{j \boldsymbol{\tilde \nu} } \mathcal{H}^\mon_{j,\boldsymbol{\tilde \nu} }\right),
\end{split}
\end{equation}
with
\begin{equation}
\begin{split}
  & \mathcal{H}^\uni_{jj+1,{\boldsymbol{\mu}} {\boldsymbol{\nu}}} = \left(i^{\frac{q_J}{2}} \sum_{\sigma, a}  f_\sigma^{\rm uni} \prod^{q_J/2}_{k=1} \hchi^{(\sigma, a)}_{j \mu_{k}} \prod_{l=1}^{q_J/2} \hchi^{(\sigma ,a)}_{j+1 \nu_{l}} \right)^2,
 \mathcal{H}^\uni_{j,{\boldsymbol{\tilde \mu}}} = \left(i^{\frac{q_J}{2}} \sum_{\sigma, a}  f_\sigma^{\rm uni} \prod^{q_J}_{k=1} \hchi^{(\sigma ,a)}_{j{\tilde \mu}_{k}}\right)^2,
\end{split}
\end{equation}
and
\begin{equation}
    \mathcal{H}^\mon_{j,{\boldsymbol{\tilde \nu}}} = \left(i^{\frac{q}{2}} \sum_{\sigma, a} \sigma f_\sigma^{\rm mon} \prod_{l=1}^{q}\hchi^{(\sigma, a)}_{j{\tilde \nu}_l} \right)^2 ,
\end{equation}
where again $\sigma \in \{+,-\}$ and the summations are defined as  $\sum_{{\boldsymbol{\mu}}} = \sum_{1 \leqslant \mu_1 < \mu_2 < \dots < \mu_{q_J/2}\leqslant N_F}$ , $\sum_{{\boldsymbol{\tilde \mu}}} = \sum_{1 \leqslant {\tilde \mu}_1 < {\tilde \mu}_2 < \dots < {\tilde \mu}_{q_J}\leqslant N_F}$ and $\sum_{{\boldsymbol{\tilde \nu}}} = \sum_{1 \leqslant \nu_1 < \nu_2 < \dots < \nu_{q}\leqslant N_F}$. Notice that for replicated Majorana fermions, we have the commutation relations $\{ \hchi_{i \mu}^{(\sigma,a)}, 
    \hchi_{j \nu}^{(\sigma',a')}\} = \delta_{\sigma\sigma'} \delta_{a a'} \delta_{i j} \delta_{\mu \nu} $ that are distinct by the factor $2$ from the original ones. Finally, $n$-replica purity we are interested in can be expressed as a ratio of overlaps
\begin{equation} \label{eq:pur}
\frac{{\rm Tr}\left[\rho^{(n)}(t) \left(\mathcal{C}_{A,2} \otimes \mathbb{I}\right) \right]}
{{\rm Tr}\left[\rho^{(n)}(t) \right]}
=\frac{\left< \mathcal{C}_{A,2} |\rho^{(n)}(t) \right>}
{\left< \mathbb{I} |\rho^{(n)}(t) \right>}
.
\end{equation}
The folded replicated state can be expressed as an imaginary time Schr\"odinger evolution with the Hamiltonian Eq.~\eqref{repHam},
\begin{equation} \label{eq_folded_replicated_rho}
  \ket{\rho^{(n)}(t)} = e^{-t \mathcal{H}^{(n)}} \ket{\rho^{(n)}(0)} \
;.
\end{equation}
More general Renyi entropies can be represented similarly via Eq.~\eqref{eq_renyi_limit}, replacing $\bra{\mathcal{C}_{A,2}}$ with the appropriate boundary state $\ket{B}$.
According to this mapping, expectation values of operators are mapped onto overlaps:
\begin{equation} \label{eq_overlap_rho}
  \Tr \left[ \rho^{(n)} (t) \hat O \right] = \left.\left< \hat O \right|\rho^{(n)}(t)\right> .
\end{equation}
At late time, the imaginary time evolution will act as a projector onto the groundstate $\ket{\rm GS}$ of $\mathcal{H}^{(n)}$, thus we simplify the overlap as
\begin{equation}\label{eq:transitAmpl}
    \lim_{t \rightarrow \infty }
    \frac{\left< 
    B
    |\rho^{(n)}(t) \right>}
{\left< \mathbb{I} |\rho^{(n)}(t) \right>} = 
\frac{\braket{
B
}{\rm GS}}{\braket{\mathbb{I}}{\rm GS}}.
\end{equation}
In the next section, we discuss how to write this Hamiltonian in a compact way in terms of generators of the $SO(2n)$ algebra. Based on these generators, we will build a representation of the coherent states and write the path integral expression for the overlap Eq.~\eqref{eq:transitAmpl}.
\begin{figure}[t]

\centering
\begin{subfigure}[t]{\linewidth}
\centering
\def\FigScale{0.70}  
\begin{tikzpicture}[scale=\FigScale, transform shape, x=1cm,y=1cm,>=Latex]

  \definecolor{PurpleDark}{HTML}{6a32c7} 
  \definecolor{PinkIn}{HTML}{fab1df}     
  \definecolor{PinkOut}{HTML}{d6c0fc}    
  \definecolor{balls}{HTML}{8e57e6}

  \def\R{2.1}        
  \def\gap{4.6}      
  \def\lw{0.9pt}     
  \def\dotr{0.075}   
  \def\rsel{0.85}    
  \def\LayerSep{0.75} 

  \pgfdeclarelayer{bg}
  \pgfsetlayers{bg,main}

  \newcommand{\ShadedCluster}[2]{
    \begin{pgfonlayer}{bg}
      \shadedraw[inner color=PinkIn, outer color=PinkOut, draw=PurpleDark, line width=\lw]
        #1 circle (#2);
    \end{pgfonlayer}
  }
\renewcommand{\ShadedCluster}[2]{%
  \shadedraw[inner color=PinkIn, outer color=PinkOut,
             draw=PurpleDark, line width=\lw] #1 circle (#2);
}

  \pgfmathsetseed{20251004}
  \newcommand{\ScatterBalls}[2]{
    \pgfmathtruncatemacro{\NN}{#2}%
    \foreach \i in {1,...,\NN} {%
      \pgfmathsetmacro{\ang}{360*rnd}%
      \pgfmathsetmacro{\rad}{\rsel*\R*sqrt(rnd)}%
      \path #1 ++(\ang:\rad) coordinate (p\i);
      \shade[shading=ball, ball color=balls] (p\i) circle (\dotr);
    }%
  }

  \newcommand{\PlaceBallsAt}[3]{
    \pgfmathtruncatemacro{\NN}{#3}%
    \foreach \i in {1,...,\NN} {%
      \pgfmathsetmacro{\ang}{360*rnd}%
      \pgfmathsetmacro{\rad}{0.82*#2*sqrt(rnd)}%
      \shade[shading=ball, ball color=balls]
        ($(#1)+(\ang:\rad)$) circle (\dotr);
    }%
  }

  \tikzset{
    snakeTop/.style={
      draw=PurpleDark, line width=\lw,
      dash pattern=on 6pt off 3pt,
      decorate, decoration={snake, amplitude=0.18cm, segment length=1.0cm},
      line cap=round
    },
    snakeBot/.style={
      draw=PurpleDark, line width=\lw,
      dash pattern=on 6pt off 3pt,
      decorate, decoration={snake, amplitude=0.18cm, segment length=1.0cm},
      line cap=round
    }
  }

  \def\edot{0.055}   
  \def\esp{0.32}     

  \newcommand{\DrawOneLayer}[1]{
    \begin{scope}[shift={(0,#1)}]

      \coordinate (C1) at (-\gap, 0);
      \coordinate (C2) at (0, 0);
      \coordinate (C3) at (\gap, 0);

      \ShadedCluster{(C1)}{\R}
      \ShadedCluster{(C2)}{\R}
      \ShadedCluster{(C3)}{\R}

      \pgfmathsetseed{20251004} 
      \ScatterBalls{(C1)}{26}
      \ScatterBalls{(C2)}{24}
      \ScatterBalls{(C3)}{22}

      \def\sr{0.75}
      \draw[PurpleDark, line width=\lw, dash pattern=on 6pt off 3pt]
        ($(C1)+(0.55,0.15)$) circle (\sr);
      \draw[PurpleDark, line width=\lw, dash pattern=on 6pt off 3pt]
        ($(C2)+(-0.30,0.30)$) circle (\sr);
      \draw[PurpleDark, line width=\lw, dash pattern=on 6pt off 3pt]
        ($(C2)+(0.70,-1.0)$) circle (\sr);
      \draw[PurpleDark, line width=\lw, dash pattern=on 6pt off 3pt]
        ($(C3)+(-0.30,-0.5)$) circle (\sr);

      \coordinate (S1) at ($(C1)+(0.55,0.15)$);
      \coordinate (S2) at ($(C2)+(-0.30,0.30)$);
      \coordinate (S3) at ($(C2)+(0.70,-1.0)$);
      \coordinate (S4) at ($(C3)+(-0.30,-0.5)$);

      \pgfmathsetseed{4241}\PlaceBallsAt{S1}{\sr}{5}
      \pgfmathsetseed{4242}\PlaceBallsAt{S2}{\sr}{5}
      \pgfmathsetseed{4243}\PlaceBallsAt{S3}{\sr}{5}
      \pgfmathsetseed{4244}\PlaceBallsAt{S4}{\sr}{5}

      \draw[snakeTop]
        ($(C1)+(0.95,0.80)$)
          .. controls ($(C1)+(1.8,1.50)$) and ($(C2)+(-1.8,1.50)$) ..
        ($(C2)+(-0.90,0.80)$);

      \draw[snakeBot]
        ($(C2)+(1.5,-0.85)$)
          .. controls ($(C2)+(1.9,-1.60)$) and ($(C3)+(-1.9,-1.60)$) ..
        ($(C3)+(-0.95,-0.85)$);

    \end{scope}
  }

  \DrawOneLayer{-\LayerSep}
  \DrawOneLayer{0}
  \DrawOneLayer{\LayerSep}

\def\BracketScale{3.8} 
\def\EdgePad{0.9}      

\def\SubX{0.08}\def\SubY{0.10}

\def\BraceX{\gap+\R+0.22}                
\def\Ybot{-\LayerSep-\R}                 
\def\Ytop{\LayerSep+\R}                  
\draw[PurpleDark, decorate,
      decoration={brace, amplitude=6pt, raise=4pt, mirror}]
  (\BraceX,\Ybot) -- (\BraceX,\Ytop)
  node[midway, right=12pt, text=PurpleDark, font=\large] {$n$ };
\node (TopMid) at (current bounding box.south) {};

\pgfmathsetmacro{\BracketH}{2*(\LayerSep+\R)} 
\pgfmathsetmacro{\BracketW}{0.9}              

\coordinate (L) at (-\gap-\R-\EdgePad, 0);   
\coordinate (K) at (\gap+\R+\EdgePad+0.8, 0);    

\draw[black, line width=0.5pt, line join=round]
  ($ (L) + (\BracketW,  0.5*\BracketH) $) --
  (L) --
  ($ (L) + (\BracketW, -0.5*\BracketH) $);

\draw[black, line width=0.5pt, line join=round]
  ($ (K) + (-\BracketW,  0.5*\BracketH) $) --
  (K) --
  ($ (K) + (-\BracketW, -0.5*\BracketH) $);

\node[inner sep=0pt, outer sep=0pt, minimum size=0pt] (ket) at (K) {};

\node[anchor=north west, text=black, font=\Large]
  at ($ (ket) + (-0.3,-1.5) $) {$h_{\boldsymbol{\mu\nu}}$};

\end{tikzpicture}

\end{subfigure}

\vspace{0.8cm}

\begin{subfigure}[t]{\linewidth}
\centering
\begin{tikzpicture}[scale=1, transform shape, x=1cm,y=1cm,>=Latex]
\definecolor{DeepNavy}{HTML}{6a32c7}
  \definecolor{DustGray}{HTML}{fab1df}
  \definecolor{Aegean}{HTML}{6a32c7} 

  \tikzset{
    sphere/.style={draw=DeepNavy, line width=0.7pt},
    equatorFront/.style={draw=DustGray, line width=0.6pt, dashed},
    equatorBack/.style={draw=DeepNavy, line width=0.6pt},
    axisLine/.style={draw=DustGray, line width=0.5pt, ->}, 
    blochVec/.style={line width=0.9pt, ->},                
    conn/.style={draw=DeepNavy, line width=0.7pt},
    gridline/.style={draw=DeepNavy, line width=0.35pt},
    phiLabel/.style={font=\small, text=black},
  }

  \pgfmathtruncatemacro{\NS}{3}
  \def\R{1.5}
  \def\Gap{0.6}
  \def\Flat{0.36}
  \pgfmathsetseed{12345}
  \pgfmathsetmacro{\Step}{2*\R + \Gap}

  \pgfmathtruncatemacro{\NSm}{\NS-1}
  \foreach \i in {1,...,\NSm} {
    \draw[conn] ({(\i-1)*\Step+\R},0) -- ({\i*\Step-\R},0);
  }

  \foreach \i in {1,...,\NS} {
    \begin{scope}[shift={({(\i-1)*\Step},0)}]

      \begin{scope}
        \clip (0,0) circle (\R);
        \shade[inner color=DustGray, outer color=Aegean!14, shading=radial]
              (0,0) circle (\R);

        \foreach \phi in {-60,-30,30,60} {
          \pgfmathsetmacro{\a}{\R*cos(\phi)}       
          \pgfmathsetmacro{\b}{\Flat*\R*cos(\phi)} 
          \pgfmathsetmacro{\y}{\R*sin(\phi)}       
          \draw[gridline] (0,\y) ellipse ({\a} and {\b});
        }
        \foreach \t in {-60,-30,30,60} {
          \draw[gridline, rotate=\t] (0,0) ellipse ({\Flat*\R} and {\R});
        }
      \end{scope}

      \draw[sphere] (0,0) circle (\R);
      \draw[equatorBack] (0,0) ellipse ({\R} and {\Flat*\R});
      \begin{scope}
        \clip (-\R-0.05,0) rectangle (\R+0.05,\R+0.05);
        \draw[equatorFront] (0,0) ellipse ({\R} and {\Flat*\R});
      \end{scope}

      \pgfmathsetmacro{\ang}{360*rand}
      \pgfmathsetmacro{\len}{\R*0.75}
      \draw[blochVec] (0,0) -- ({\len*cos(\ang)},{-\len*sin(\ang)});

      \ifnum\i=1
        \node[phiLabel] at (0,-\R-0.55) {$\hat{\Phi}^{\alpha \alpha'}_{j-1}$};
      \fi
      \ifnum\i=2
        \node[phiLabel] at (0,-\R-0.55) {$\hat{\Phi}^{\alpha \alpha'}_{j}$};
      \fi
      \ifnum\i=3
        \node[phiLabel] at (0,-\R-0.55) {$\hat{\Phi}^{\alpha \alpha'}_{j+1}$};
      \fi

    \end{scope}
  }

  \node[phiLabel] at ({(\NS-1)*\Step/2}, -\R-1.55)
    {$\hat{\Phi}^{\alpha \alpha'}_k \in \mathrm{so}(2n)$ \; for $k\in\{1,n(n-1)\}$};
\node (BotMid) at (current bounding box.north) {};

\end{tikzpicture}

\end{subfigure}
\begin{tikzpicture}[remember picture, overlay, >={Implies[length=12pt,width=16pt]}]
  \draw[black,
        line width=2.6pt,
        -{Implies[length=12pt,width=16pt]},
        shorten >=8pt, shorten <=8pt]
    (0,6.5) -- (0,5.3); 
\end{tikzpicture}

\caption{(a) Top: an $n$-replica chain of Majorana clusters with disordered couplings $h_{\mu\nu}$, brace indicates the $n$ copies (Illustration of  Eqs.~\eqref{eq:rep_ham} \eqref{repHam}). (b) Bottom: the resulting chain of coset fields $\hat{\Phi}^{\alpha\alpha'}_{j}\in \mathrm{so}(2n)$ obtained from the replicated Majorana system after disorder averaging. The double arrow schematically denotes the averaging/decoupling map from (a) to (b). Here $j$ indexes chain sites, $\mu,\nu$ label Majorana modes within a cluster, and $\alpha,\alpha'$ are replica indices (Illustration of  Eqs.~\eqref{eq:Htotal}).}\label{Fig:dual}

\end{figure}
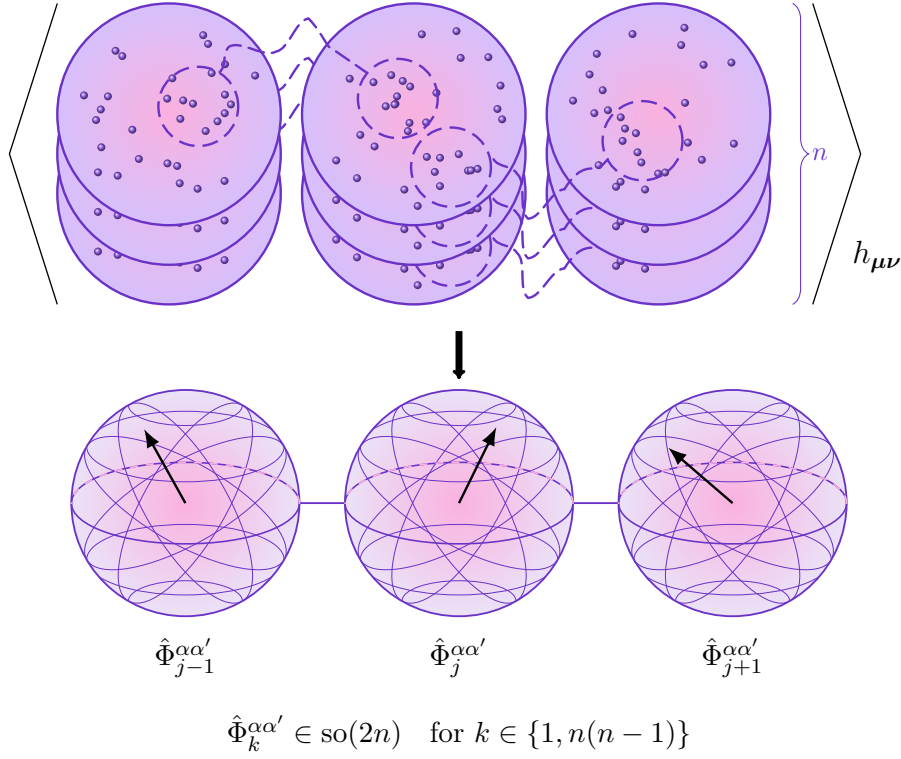

\subsection{The fermionic \texorpdfstring{$SO(2n)$}{SO(2n} fields}\label{sec:so(2n)}

The replicated Hamiltonian can be expressed in terms of the following field operators
\begin{equation}\label{eq:fieldoperators}
  \hat{\Phi}_j^{(\sigma,a),(\sigma',a')} = \frac{i}{2 N_F} \sum_{\nu} [\hchi_{j\nu}^{(\sigma,a)}, \hchi_{j\nu}^{{(\sigma',a')}}] . 
\end{equation}
It is simple to show that they satisfy the same commutation relations as the elements of the $SO(2n)$ algebra~\cite{Fava:2023tgg}. 
Representing the replica indices with a single flattened index, the commutation relations of the new operators can be simplified as    
\begin{equation}\label{eq:commPhi}
   \left[\hat{\Phi}_i^{\alpha \beta}, \hat{\Phi}_j^{\lambda \xi}\right]  =   \delta_{ij} \frac{i}{N_F} \times  (\delta_{\beta, \lambda} \hat\Phi_{i}^{\alpha \xi}
   -\delta_{\beta, \xi} \hat\Phi_{i}^{\alpha \lambda}
   +\delta_{\alpha, \xi} \hat\Phi_{i}^{\beta \lambda}
   -\delta_{\alpha, \lambda} \hat\Phi_{i}^{\beta \xi}
   ).
\end{equation}
To help clarify this algebra, we
recall that defining for $1\leq \alpha < \beta \leq 2n$, the $2n\times 2n$ matrices 
\begin{equation}\label{eq:Erepmajo}
[E^{\alpha\beta}]_{\alpha' \beta'} = \delta_{\alpha'}^\alpha \delta_{\beta'}^\beta   
\end{equation}
we see that this is the same algebra obtained replacing at each site $i$: $\hat{\Phi}^{\alpha\beta}_i \to W^{\alpha \beta} = \frac{i}{N_F} (E^{\alpha\beta} - E^{\beta \alpha})$, i.e. the antisymmetric matrices that generate $so(2n)$.

The replicated Hamiltonian expressed in terms of these fields therefore takes the form
\begin{equation}\label{eq:Htotal}
    \mathcal{H} =   \mathcal{H}_{\rm uni} -   \mathcal{H}_{\rm mon} ,
\end{equation}
with
\begin{equation}\label{eq:Huni}
\begin{split}
      \mathcal{H}_{\rm uni} =  
      \frac{J N_F^{q_J }}{2 (q_J/2)!^2} \sum_{j\alpha \alpha'} \Bigg[ f^{\rm uni}_{\sigma'} f^{\rm uni}_\sigma (\Phi^{\alpha \alpha'}_j)^{\frac{q_J}{2}}(\Phi^{\alpha \alpha'}_{j+1})^{\frac{q_J}{2}} 
      + \frac{(q_J/2)!^2}{ q_J !}f^{\rm uni}_{\sigma'} f^{\rm uni}_\sigma (\Phi^{\alpha \alpha'}_j)^{q_J}\Bigg],
\end{split}
\end{equation}
and
\begin{equation}
    \mathcal{H}_{\rm mon} = \Gamma \frac{N_F^q}{ 2 q!} \sum_{j\alpha \alpha'}\sigma \sigma' f^{\rm mon}_\sigma    f^{\rm mon}_{\sigma'} (\hat \Phi^{\alpha \alpha'}_j)^q .
\end{equation}From now on, we will focus on the case where the monitoring is quadratic $q=2$, which allows for the simplification
\begin{equation}\label{eq:Hmeas}
\begin{split}
   \mathcal{H}_{\rm mon}\Big|_{q=2}=  & \Gamma  \frac{N_F^2}{4} \sum_{j \alpha \alpha'} \sigma \sigma' \Big(\hat \Phi^{\alpha \alpha'}_j\Big)^2,
    \end{split}
\end{equation}
where $\{\alpha, \alpha'\}=\{(\sigma, a), (\sigma',a')\}$ are multi-indices.
The operators defined in Eq.~\eqref{eq:fieldoperators} are the basic building blocks of our construction, enabling a coherent-state representation for an arbitrary number of replicas. In brief, after replicating the Majorana chain (to $n$ copies) and averaging over the stochastic couplings $h_{\boldsymbol\mu\boldsymbol\nu}$, we arrive at an effective theory expressed in terms of the Lie-algebra generators $\hat{\Phi}$ of $\mathfrak{so}(2n)$ see Fig.~\ref{Fig:dual}.

In the next two sections, we develop a field-theoretic framework for computing entanglement measures directly in terms of these replica generators $\hat{\Phi}$ from Eq.~\eqref{eq:fieldoperators}. We begin with the illustrative case $n=2$, where $\mathfrak{so}(4)\simeq\mathfrak{su}(2)\oplus\mathfrak{su}(2)$ and the dynamics can be mapped to an $SU(2)$ spin system; this warm-up provides intuition and a concrete derivation of the associated nonlinear sigma model. We then turn to generic $n$, where coherent states on the coset $SO(2n)/U(n)$ furnish the full path-integral representation.

\section{The case with \texorpdfstring{$n=2$}{n=2} }\label{sec:n2q2}As a warm-up to the general replica construction, we focus here on the case $n=2$. 
For two replicas the algebra $so(4)$ reduces to $su(2)\oplus su(2)$, and—upon restricting to the symmetric subspace—one obtains an effective $SU(2)$ spin representation. 
This mapping provides an intuitive and tractable setting: coherent states of $SU(2)$ spins furnish a natural basis for the path integral, allowing us to translate the replicated dynamics into semiclassical trajectories on the Bloch sphere.

Within this framework we proceed in two steps. 
First, we derive the effective nonlinear sigma model (NLSM) emerging in the quadratic case ($q_J=2$), which already captures the structure of the continuum limit. 
Second, we use the spin–coherent path integral to compute the late-time purity for a system of two clusters, both for quadratic ($q_J=2$) and quartic ($q_J=4$) interactions. 
These calculations illustrate, in the simplest setting, how coherent states connect the replicated operator formalism with field-theoretic descriptions of monitored dynamics.

For $n=2$, the replica algebra admits a direct mapping to an $SU(2)$ spin representation~\cite{BAO2021168618}. We now construct the mapping explicitly using Dirac fermions \begin{equation}
\begin{split}
   & c_{a j \nu}^{\dagger} = \frac{\chi_{j\nu}^{(+,a)} - i \chi_{j \nu}^{(-,a)}}{\sqrt{2}} ,~~  c_{a j \nu} =  \frac{\chi_{i\nu}^{(+,a)} + i \chi_{j \nu}^{(-,a)}}{\sqrt{2}}\;.
\end{split}
\end{equation}To do so, we firstly introduce two sets of spin $1/2$ operators, one for each replica $a = 1,2$, us the Jordan-Wigner transformation. To simplify the notation, we omit cluster index $j$. Then we combine $\sigma = \pm1$ replicas introducing $c^{\dag}_{a \nu} c_{\nu a}=\frac{1}{2}(S^z_{a\nu}+\mathbb{I})$. And define raising/lowering spin operators as
\begin{equation}
\begin{split}
    c_{a\nu}^{\dag}=e^{i \pi \sum_{k=1}^{a-1} S_{k\nu}^+S_{k\nu}^-} S_{a\nu}^+,~~~c_{a\nu}=e^{-i \pi \sum_{k=1}^{a-1} S_{k\nu}^+S_{k\nu}^-} S_{a\nu}^-\;.
    \end{split}
\end{equation}
 We thus obtain six generators in total for each flavour $\nu$,  namely $\{S_{1, \nu}^z, S_{1, \nu}^{\pm}, S_{2 }^z, S_{2, \nu}^{\pm} \}$, consistent with the decomposition $so(4) \simeq su(2) \oplus su(2)$ with a representation of dimension $6$ (two spins $1/2$). We can further reduce the relevant Hilbert space using the conservation of parity. Labelling the states as the eigenvectors of $S^z_1, S^z_2$, we have that the two states $\{\ket{\uparrow \uparrow}, \ket{\downarrow \downarrow}\}$ are decoupled in the dynamics from $\{\ket{\uparrow \downarrow}, \ket{\downarrow \uparrow}\}$.

Since the identity matrix we choose as initial condition always lies in the former subspace, we can introduce a 
pseudospin basis with only two states: $\ket{\uparrow \uparrow}, \ket{\downarrow \downarrow}$. 
So, we can further reduce the operators for $a=1,2$ to a single spin $1/2$ in the space $\ket{\Uparrow} := \ket{\uparrow\uparrow}, \ket{\Downarrow} := \ket{\downarrow \downarrow}$. Let us denote simply as $S^x_{\nu},S^y_{\nu},S^z_{\nu}$ the spin operators acting on this $2$-dimensional space. Then,
\begin{equation}
\begin{split}
&S^x_{1\nu} S^x_{2\nu} = 
- S^y_{1\nu} S^y_{2\nu} 
= \frac{S^y_{\nu}}{2} , ~
S^x_{1\nu} S^y_{2\nu} = 
 S^y_{1\nu} S^x_{2\nu} 
= \frac{S^x_{\nu}}{2} , ~S^{z}_{1\nu}  = 
S^{z}_{2\nu} =   
 \frac{S^z_{\nu}}{2}  .
\end{split}
\end{equation}
It is useful to write explicitly the matrix $\Phi$ in Eq.~\eqref{eq:fieldoperators} in terms of this $su(2)$ algebra. Introducing the total spin generators $ \sum_{\nu} \vec{S}_{\nu}=\vec{S}$ and ordering the basis $(\sigma, a)$ as $((+,1), (+,2), (-,1),(-,2))$. Then the Eq.~\eqref{eq:fieldoperators} reads
\begin{equation}\label{eq:paramPhi} 
  \hat{\Phi} = \frac{1}{N_F} \begin{pmatrix}
    0 &  -S^y & S^z & S^x \\ S^y & 0 & -S^x & S^z  \\ -S^z & S^x & 0 & S^y \\ -S^x & - S^z & -S^y & 0 
  \end{pmatrix}
\end{equation}where $S=\frac{N_F}{2}$. And the cluster index can be recovered simply by replacement $\hat \Phi \rightarrow \hat \Phi_j$.
We stress how the original number of parameters expected from the $so(2n = 4)$ algebra, which simply required $\hat{\Phi}$ to be an antisymmetric matrix of generators, has been reduced from 6 to just $3$. We will discuss the nature of this reduction for generic replicas in section \ref{sec:coherent}.

\subsubsection{Non-interacting case $q_J=2$}\label{sec:q_J=2,n=2}
\begin{figure}[t]
  \centering
  \includegraphics[width=0.5\linewidth]{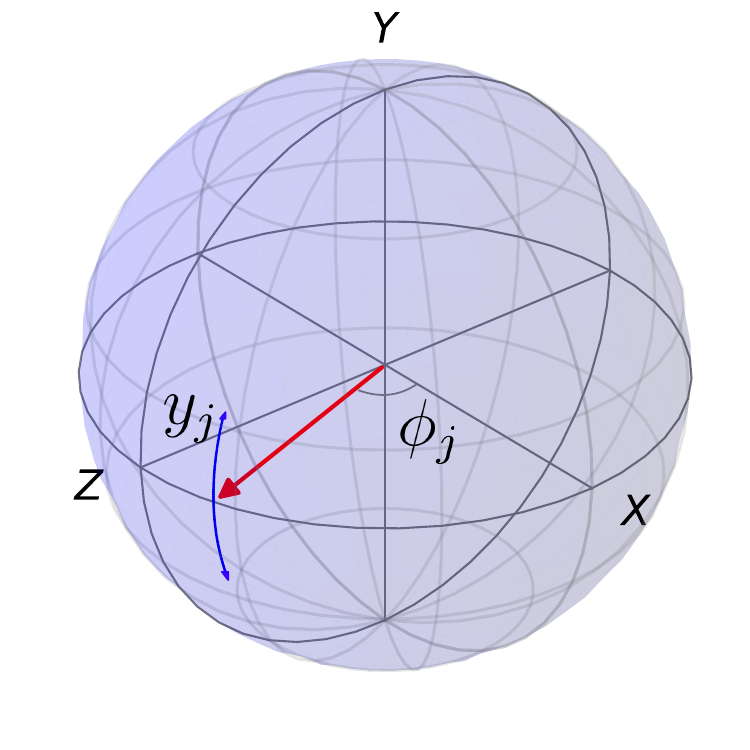} 
  \caption{Representation of $j$-th spin vector on the Bloch sphere (the red line). Monitoring part of the evolution forces the vector to be in $XZ$ plane, however coupling of $y_j$ with $\phi_j$ through the kinetic term allows fluctuations in $Y$ direction (the blue line) see Eq.~\eqref{eq:act2}.}
  \label{fig:Bloch}
\end{figure}

By mapping the $n=2$ Hamiltonian Eq.~\eqref{eq:Htotal} to spin operators Eq.~\eqref{eq:paramPhi}, one obtains
\begin{equation}\label{eq:spinHam}
  \mathcal{H} = - 2 J  \sum_j \vec{S}_j \cdot \vec{S}_{j+1} + 2 \Gamma \sum_j  (S_j^y)^2 .
\end{equation}
Each flavour contributes a $2$-dimensional representation of $su(2)$, so a cluster with $N_F$ flavours spans a $2^{N_F}$-dimensional Hilbert space. This representation is not irreducible, but the observables of interest---in particular, the late-time purity of a cluster $A$ (with complement $B=\bar A$)---always involve states symmetric under exchange of spins within a cluster. For instance,
\begin{equation}\label{eq:puroverlSpin}
 \lim_{t \rightarrow \infty} \frac{\Tr \!\left[ \rho^{(2)}(t) (\mathcal{C}_{2,A} \otimes \mathbb{I})\right]}{\Tr \rho^{(2)}(t)}
   = \frac{1}{Z}\langle X_A \otimes_{i \in B} Z_i | GS\rangle,
\end{equation}
where $Z=\langle \mathbb{I} | GS\rangle$ normalizes the overlap, and the boundary states are fully polarized product states in the $x$- or $z$-direction, e.g. $\ket{Z_i} = \otimes_{\mu=1}^{N_F}\ket{\Uparrow_{i,\mu}}$. All states entering Eq.~\eqref{eq:puroverlSpin} belong to the symmetric representation with maximal spin $S=N_F/2$ and dimension $N_F+1$. The ground state $\ket{GS}$ of Eq.~\eqref{eq:spinHam} is known to exhibit a Kosterlitz–Thouless transition~\cite{BAO2021168618}, though its behavior is modified in the replica limit $n\to 1$~\cite{Fava:2023tgg}.  

In the case of large $N_F$, the basis of spin coherent states offer a useful framework to deal with this computation. We will see that for $L=2$ clusters, this formalism leads to an accurate prediction for the behavior of the purity as a function of $N_F$ (see Appendix \ref{App:SpinCoherent}). Instead, in the continuum space limit, this leads to the NLSM action.  Introducing coherent states polarized along the $y$-axis,
\begin{equation}
\label{eq:su2cohe}
\ket{\Omega_{\hat{n}}} := e^{-i \frac{{\rm arccos}\, y}{ \sqrt{1-y^2}} (\hat{\mathbf{n}} \times \hat{\mathbf{y}})\cdot \mathbf{S}} \ket{\Uparrow^y},
\end{equation}
with $\hat{n}=(\sqrt{1-y^2}\sin\phi,\,y,\,\sqrt{1-y^2}\cos\phi)$, one arrives at the path integral
\begin{equation}
\label{eq:overlapn2}
    \langle GS | X_A, Z_B \rangle \sim \lim_{T \rightarrow \infty} 
    \prod_{j=1}^N \int_{-1}^{1} dy_j \int_{0}^{2 \pi} d \phi_j \; e^{-\mathcal{S}},
\end{equation}
with action
\begin{equation}\label{eq:act}
    \mathcal{S} = \int_0^t dt \left( \sum_{j=1}^N K_j + \sum_{j=1}^N H_{j,j+1} \right),
    \qquad K_j = -i S \dot \phi_j (1-y_j),
\end{equation}
and interaction
\begin{equation}\label{eq:Huni2}
\begin{split}
    H_{j,j+1} =  - 2 J S^2 \big[ \sqrt{(1-y_j^2)(1-y_{j+1}^2)} \cos (\phi_j - \phi_{j+1}) +  y_j y_{j+1} \big] 
    + 2\Gamma S^2 (y_j^2 + y_{j+1}^2).
\end{split}
\end{equation}

{We now take the limit of large $S$ (namely $N_F \gg 1 $ in each cluster) and redefine $y \rightarrow y/S$, so that $y \in [-S, S]$.}
We can then expand with respect to $y$ to obtain a quadratic action in these variables,
\begin{equation}\label{eq:act2}
\begin{split}
    \mathcal{S} & = \int^T_0 dt \Bigg[   -2 J S^2 \sum_{j=1}^L \cos (\phi_j- \phi_{j+1})    + 2 \Gamma \sum_{j=1}^L y_j^2 - i \sum_{j=1}^L  y_j \dot{\phi}_j + O(y/S) \Bigg],
\end{split}
\end{equation}
where we neglected the quadratic order in $y$ from the unitary part, assuming the scaling of the coupling $J\sim S^{\alpha}$, with $\alpha \in (-2,0)$. And we integrate over $y$ to find the effective action
\begin{equation}\label{eq:act3}
    \mathcal{S} = \int^T_0 dt \left[ \sum_{j=1}^L \frac{\dot{\phi}_j ^2}{8 \Gamma} -2 JS^2\sum_{j=1}^L \cos (\phi_j - \phi_{j+1}) \right].
\end{equation}
To take the continuous limit and derive the NLSM action, we assume $\phi_{j+1} - \phi_{j} = \Delta x \partial_x \phi_j$, where $\Delta x$ is the lattice distance, and expand the action with respect to $\Delta x$. Taking the limit $\Delta x \rightarrow 0$, $L \rightarrow \infty$ one obtains
\begin{equation}
    \mathcal{S} = \int d^2x \left[\frac{(\dot{\phi}(x)) ^2}{8 \Delta x \Gamma} + JS^2 \Delta x (\partial_x \phi(x))^2 \right].
\end{equation}

\paragraph{Two clusters.}  
As a complementary check, it is instructive to consider the smallest nontrivial geometry, $L=2$ clusters Eq.~\eqref{eq:act3}. In this case the effective action simplifies to
\begin{equation}
     \mathcal{S} = \int^T_0 dt \left[ \frac{\dot{\tilde \theta}^2}{4 \Gamma} + \frac{\dot{\tilde \phi}^2}{4 \Gamma} -2 JS^2 \cos (2 \tilde \phi) \right],
\end{equation}
with relative and center-of-mass coordinates $\tilde \phi=(\phi_2-\phi_1)/2$ and $\tilde \theta=(\phi_1+\phi_2)/2$.  
The late-time purity reduces to the overlap between the boundary state $\ket{X_A,Z_B}$ that comes from the swap operator action in Eq.~\eqref{eq:puroverlSpin} and the ground state of a particle in a pendulum potential,
\begin{equation}
    \mathcal{H} =  \frac{\dot{\tilde{\phi}}^2}{4 \Gamma} - 2 J S^2 \cos 2 \tilde \phi .
\end{equation}

which in the large-$S$ limit is well approximated by a harmonic oscillator. This yields
\begin{equation}
    \lim_{t \to \infty} 
    \frac{\Tr [ \rho_A^{(2)}(t) (\mathcal{C}_2 \otimes \mathbb{I})]}{\Tr \rho^{(2)}(t)}
    \sim \exp\!\left[-\,2 S \sqrt{\tfrac{J}{\Gamma}} \,\tfrac{\pi^2}{8}\right],
\end{equation}
with $J\sim S^\alpha$, $\alpha\in(-2,0)$.  
Details of the overlap calculation are provided in Appendix~\ref{App:SpinCoherent}.  

\subsubsection{Interacting case $q_J=4$}

The same formalism extends naturally to quartic interactions. Mapping the replicated Hamiltonian to spin operators via Eq.~\eqref{eq:paramPhi}, one finds
\begin{equation}
\begin{split}
    \mathcal{H}  &= \frac{J}{2} \sum_j  \Big[(S^y_j)^{2} (S^y_{j+1})^{2} - (S^x_j)^{2} (S^x_{j+1})^{2} +(S^z_j)^{2} (S^z_{j+1})^{2} \Big]  \\
    &\quad + \frac{J}{12}\sum_j \Big[(S^y_j)^{4} -(S^x_j)^4 +(S^z_j)^4 \Big] 
    + 2 \Gamma \sum_j (S^y_j)^2 .
\end{split}
\end{equation}
We can then repeat the procedure of the previous sections, calculating the effective action for the path integral and integrating out the massive degrees of freedom as we did in the $q_J=2$ case, we find the following effective action 
\begin{equation}
\begin{split}
    \mathcal{S} &  = \int^t_0 d \tau \ \left[ \frac{\dot{\tilde \theta}_j^2}{4 \Gamma}  + \frac{\dot{\tilde \phi}_j^2}{4 \Gamma} \right.  \left.- \frac{J S^4}{24}(3+\cos  4 \tilde\phi_j)(3+ \cos 4 \tilde  \theta_j) \right],
\end{split}
\end{equation}
with $\tilde \phi_j = (\phi_{j+1}-\phi_j)/2$ and $\tilde \theta_j =(\phi_{j+1}+\phi_j)/2$.Unlike the quadratic case ($q_J=2$), the quartic interaction pins the angles to a discrete set ($\tilde\phi,\tilde\theta\in \tfrac{\pi}{2}\mathbb{Z}$) through the lattice anisotropy term. As a consequence, the long-wavelength expansion does not yield a nontrivial continuum sigma model: gradients are short-ranged and all fluctuations are gapped. For this reason, we focus directly on the minimal nontrivial geometry $L=2$, where the path integral reduces to an explicit overlap problem that cleanly captures the late-time purity scaling.
\paragraph{Two clusters.}
In the case of two clusters, for the large times, we can again compute the overlap explicitly, mapping the partition function into the Hamiltonian evolution.
This problem reduces to finding the ground state of the following Hamiltonian
\begin{equation}
\begin{split}
   H  = \frac{p_{\tilde \phi}^2}{2m} + \frac{p_{\tilde \theta}^2}{2m} - \frac{J S^4}{24}(3+\cos 4\tilde \phi)(3+ \cos  4\tilde \theta),
\end{split}
\end{equation}
where $m=\frac{1}{2 \Gamma}$ and $p_{\sigma}=\dot\sigma/(2\Gamma)$.
Expanding around the minima $\tilde\phi=n\pi/2$, $\tilde\theta=m\pi/2$, the problem reduces again to two uncoupled oscillators with ground state
\[
   \Psi_{\overline{GS}}(\tilde\phi,\tilde\theta) \sim e^{-m\omega(\tilde\phi^2+\tilde\theta^2)/2}, 
   \qquad m\omega^2=\tfrac{8 J S^4}{3}.
\]

The resulting purity decays as
\begin{equation}
    \lim_{t \to \infty} 
    \frac{\Tr [ \rho_A^{(2)}(t) (\mathcal{C}_2 \otimes \mathbb{I})]}{\Tr \rho^{(2)}(t)}
    \sim \exp\!\left[-\,S^2 \sqrt{\tfrac{J}{3\Gamma}}\left(\tfrac{\pi}{2}\right)^2\right],
\end{equation}
valid for $1/S^4<J \leq 1/S^2$.  
At stronger coupling the overlap is dominated by a Gaussian integral, giving instead a volume law $\sim e^{-4 S \log 2}$. The full derivation is given in Appendix~\ref{App:SpinCoherent}.  

The two-cluster problem will reappear in the section \ref{sec:purity2} as a benchmark for generic $n$ problem: it is analytically tractable, yet rich enough to provide nontrivial checks against numerical simulations.

\section{Coherent states representation for \texorpdfstring{$n$}{n} replicas and the NLSM}\label{sec:coherent}
\subsection{Coherent states: intuition from oscillators and spins}

Before diving into the general framework, it is useful to recall again two familiar families of coherent states that illustrate their main features.

\paragraph{Oscillator coherent states.}  
For the harmonic oscillator, coherent states are defined as eigenstates of the annihilation operator,
\[
  a \ket{\alpha} = \alpha \ket{\alpha}, \qquad \alpha \in \mathbb C.
\]
They form displaced versions of the ground state, $\ket{\alpha} = e^{\alpha a^\dagger - \alpha^\ast a}\ket{0}$, and minimize the Heisenberg uncertainty relation.  
As a result they follow classical trajectories in phase space with only small quantum fluctuations.  
Their overcompleteness,
\[
  \int \frac{d^2\alpha}{\pi} \ket{\alpha}\bra{\alpha} = \mathbb I,
\]
is the key ingredient for building path integrals: by inserting this resolution of identity at each time step, one obtains the well–known phase–space path integral whose action contains both the classical Hamiltonian and the Berry phase $i\alpha^\ast \dot\alpha$.

\paragraph{Spin coherent states.}  
For a spin $S$, coherent states are obtained by rotating the fully polarized state $\ket{\Uparrow}$:
\[
  \ket{n} = e^{-i\theta \hat u\cdot \vec S}\ket{\Uparrow}, \qquad  n \in S^2,
\]
so that each state is labeled by a point on the Bloch sphere (see Appendix \ref{App:SpinCoherent}).  
The overlap of two nearby states encodes a Berry phase proportional to the solid angle swept by $ n(t)$.  
This geometric term gives the symplectic structure of the sphere, and the corresponding path integral describes semiclassical spin dynamics.

In both cases, coherent states provide a dictionary between operator dynamics and trajectories on a classical phase space (complex plane for oscillators, Bloch sphere for spins).  
They yield path integrals of the form
\[
  S[n] = \int dt \,\big( i\langle n|\dot n\rangle - \langle n|H|n\rangle \big),
\]
where the first term encodes geometry (the Berry phase) and the second the classical Hamiltonian.  
Stationary trajectories reproduce classical equations of motion, while fluctuations encode quantum corrections.

These examples will guide our construction for replicated dynamics.  
In the case of $n$ replicas, the relevant algebra is $so(2n)$, and the coherent states live on the coset $SO(2n)/U(n)$.  
Just as oscillator and spin coherent states provided natural bases for their respective phase spaces, generalised coherent states on the group will provide the natural language for describing the replicated Hamiltonian evolution in a path–integral representation. 
\subsection{General procedure for coherent states}
Before discussing the details of the fermionic coherent states for the $SO(2n)$ Lie group, it is useful to briefly remark about the general construction and how it applies to the familiar example of $SU(2)$ that we discussed above. 
In order to construct the coherent state representation of a given Lie group $\mathcal{G}$, we assume that $\{\hhc_a, \eec_{\vec\rho}, \eec_{-\vec\rho}\}$ denote the generators of its Lie algebra in the standard Cartan-Weyl basis and $\vec\rho \in \mathfrak{R}^+$, the set of positive roots. For a specific description of this basis in the case of $so(2n)$ using fermionic bilinears, see Appendix~\ref{sec:cartan}. 
Then, the coherent states can be written as
\begin{equation}
\label{eq:naivecoh}
\ket{g} = \Omega(g) \ket{\mathbb{I}}   , 
\end{equation}
where $g \in \mathcal{G}$ and $\Omega$ is an irreducible representation $\mathcal{G}$. The state $\ket{\mathbb{I}}$ is a reference state within the chosen representation that we take as normalised $\braket{\mathbb{I}}{\mathbb{I}} = 1$. Since we are dealing with unitary representations, this definition ensures that the coherent state $\ket{g}$ is normalised 
\begin{equation}
\label{eq:cohenorm}
    \braket{g}{g} = 1\;.
\end{equation}
There are multiple choices for the state $\ket{\mathbb{I}}$, but a essential requirement is that it is a highest-weight of the representation~\cite{STONE1989399}.
As it is well known, the coherent states are a largely overcomplete basis. However, 
one can use Haar measure over $\mathcal{G}$ to define a resolution of the identity in the form
\begin{equation}
\label{eq:resolid}
    \int_{\operatorname{Haar}} dg \ket{g} \bra{g} = \mathbb{I},
\end{equation}
which simply follows from an application of Schur's lemma. By repeatedly inserting the resolution of the identity Eq.~\eqref{eq:resolid} in between small time steps of the evolution Eq.~\eqref{eq_folded_replicated_rho}, one obtains the prototypical path-integral representation
\begin{equation}
\label{eq:pathintgen}
    \left<B \right|e^{-t \mathcal{H}^{(n)}}\left|\rho^{(n)}(0)\right> = 
    \lim\limits_{\substack{N_t \to \infty \\ \delta t \to 0}} \int \prod_{j=1}^{N_t} 
    d g(t_j) 
    \braket{g(t_j) |e^{-\delta t \mathcal{H}}}{g(t_{j+1})} \\:=
    \int \mathcal{D}[g] e^{-S} ,
\end{equation}
where the boundary states determine the boundary conditions for $g$ within the path integral. The action $S := \int_0^t dt L[g]$ and the Lagrangian is obtaining by expanding $\braket{g(t_j) |e^{-\delta t \mathcal{H}}}{g(t_j)}$ at first order in $\delta t$ and reads
\begin{equation}
   L[g] := \braket{\dot g}{g} + \braket{g|\mathcal{H}^{(n)}}{g}  \;. \label{eq:Lgeneral}
\end{equation}
Note that $\bra{\dot g} = \lim_{\delta t \to 0} (\bra{g(t_{j+1})} - \bra{g(t_{j})})/\delta t$ is nothing more than a formal expression at this stage. In fact, beyond being overcomplete, the definition Eq.~\eqref{eq:naivecoh} provides multiple representatives of the \textit{same} quantum state and this leads to a problem in defining two states that are close one to the other. 
In particular, this redundancy precludes a proper definition of the Berry phase, the first term of Eq.~\eqref{eq:Lgeneral}. To overcome this difficulty, we denote as $\mathcal{H} \subset \mathcal{G}$ the stability subgroup of the reference state, defined by
\begin{equation}
    \mathcal{H}=\left\{h \in \mathcal{G},~~\Omega(h) \ket{\mathbb{I}} = e^{i \phi_h} \ket{\mathbb{I}}\right\},
\end{equation}
i.e. the elements of $\mathcal{G}$ which leave the reference state unchanged, up to a phase with $\phi_h \in [0,2\pi)$. An appropriate evaluation of the overlap $\braket{\dot g}{g}$ requires removing this redundancy. Formally, this is done by reducing the volume of integration from the whole group to the coset $\mathcal{G} \to \mathcal{G}/\mathcal{H}$, so that each coherent state is associated with a unique element $\tilde{g} \in \mathcal{G}/\mathcal{H}$.

This construction is familiar in the case of the standard $SU(2)$ that we used explicitly in Eq.~\eqref{eq:su2cohe}. In such a case, the stabilizer group $\mathcal{H} \simeq U(1)$, corresponding to the transformations generated by $S_y$. It follows that the coherent states are in one-to-one correspondence with $SU(2)/U(1) \simeq S^2$, i.e. the 2-dimensional sphere, parameterised in Eq.~\eqref{eq:su2cohe} with cylindrical coordinates $(y, \phi)$.

The Cartan-Weyl basis provides an effective way to express the elements of the coset. Indeed, the requirement that $\ket{\mathbb{I}}$ is a highest-weight ensures that it is a common eigenvector for all elementns of the Cartan subalgebra $\hhc_a$ annihilated by $\eec_{\vec\rho}$ for all positive roots
\begin{align}
    &\mathfrak{h}_a \ket{\mathbb{I}} = \lambda_a \ket{\mathbb{I}} , \quad \lambda_a \in \mathbb{C} 
    ,
    \label{eq:cartaneigen}
    \\ &\mathfrak{e}_{\vec\rho} \ket{\mathbb{I}} = 0 , \quad \forall \vec\rho\in\mathfrak{R}^+ .\label{eq:annihilateplus}
\end{align}
Beyond this general properties, in many relevant cases, the reference state $\ket{\mathbb{I}}$ might also be annihilated by a subset of the negative roots. We therefore are decomposing the negative roots $\mathfrak{R}^- = \mathfrak{R}^-_0 \cup \mathfrak{R}^-_{\mathcal{H}}$, such that
\begin{equation}
\eec_{\vec{\rho}} \ket{\mathbb{I}} = 0 , \quad \forall \vec{\rho} \in \mathfrak{R}^{-}_0 .
\label{eq:annihilateminus}
\end{equation}
By means of the exponential map, one has that the generators of the stability group $\mathcal{H}$ are precisely those in eqs.~(\ref{eq:cartaneigen}, \ref{eq:annihilateplus}, \ref{eq:annihilateminus}). Thus, this provides an effective representation for the elements of the coset in terms of the residual elements that do not annihilate the reference state, i.e.
\begin{equation}
\label{eq:cohstEtaGen}
    \Omega(\eta) \ket{\mathbb{I}} := \exp\left[{\sum\limits_{\vec \rho \in \mathfrak{R}^-_{\mathcal{H}}}
    \eta_{\vec\rho} \eec_{\vec\rho} - \text{c.c.}
    }\right] \ket{\mathbb{I}},
\end{equation}
with $\eta_{\vec\rho} \in \mathbb{C}$ and the minus sign ensures unitarity of the representation. Note that the normalisation condition Eq.~\eqref{eq:cohenorm} remains true, so we will be dealing with normalised coherent states.
In the following, we will see how this general construction applies to the case of $SO(2n)$ coherent states built in Sec.~\ref{sec:cartan}.
\subsection{Generalised coherent states for \texorpdfstring{$SO(2n)$}{SO(2n)}}

We momentarily focus on a single cluster. As we have shown in Sec.~\ref{sec:so(2n)}, the fields $\hat \Phi^{\alpha \beta}$ span the $so(2n)$ Lie algebra and represent a complete set of generators. 
As we have already seen, the Hamiltonian can be expressed solely in terms of these operators. As clarified in the previous section, another ingredient in the definition of the coherent states is the reference state, which we will act on with the group elements generated by the operators $\hat \Phi$. 
Following \cite{Fava:2023tgg}, we define the reference state $\ket{\mathbb{I}}$ to be Gaussian and satisfy the condition
\begin{equation}\label{eq:ref}
  \bra{\mathbb{I}} \hat \Phi \ket{\mathbb{I}} = \frac{\Sigma}{2},
\end{equation}
with the usual symplectic matrix given by
\begin{equation}
\Sigma = \begin{pmatrix}
0 & \mathbb{I}_n \\ - \mathbb{I}_n & 0 
\end{pmatrix}
.
\end{equation}
Such a state is also a highest-weight for the $so(2n)$ algebra. We
provide a summary of the Cartan-Weyl basis for $so(2n)$ in Appendix~\ref{sec:cartan}. More directly, we set $\ket{\mathbb{I}}$ as the fully occupied state
\begin{equation}
\label{eq:cdacI}
c^\dag_{a \nu} c_{a \nu} \ket{\mathbb{I}} = \ket{\mathbb{I}}  , 
\end{equation}
where $a = 1,\ldots, n$ and $\mu = 1,\ldots, N_F$. Note that for the case $n = 2$, this state reduces to $\ket{\mathbb{I}} = \ket{\Uparrow}$, so there is a discrepancy with the convention we took in Eq.~\eqref{eq:su2cohe}, where the $y$ direction was used.
The state defined by Eq.~\eqref{eq:cdacI} is also consistent with  Eq.~\eqref{eq:ref}.
As explained in Appendix~\ref{sec:cartan}, the generators $\hhc_a = c^\dag_a c_a - 1/2$ (see Eq.~\eqref{eq:cartangen}), so that Eq.~\eqref{eq:cartaneigen} holds with $\lambda_a = 1/2$. Instead, $\eec_{\vec\rho}$ associated to the positive roots $\vec\rho \in \mathfrak{R}^+$ split into two kinds: $\{c^\dag_a c^\dag_{a'}\}_{a<a'}$ and $\{c^\dag_a c_{a'}\}_{a<a'}$ and both annihilate the reference state in agreement with Eq.~\eqref{eq:annihilateplus}.

Taking the complex conjugate, we can write the generators associated with negative roots as two families: $\{c_a c_a'\}_{a<a'}$ and $\{c_a^\dag c_{a'}\}_{a>a'}$. Consistently with Eq.~\eqref{eq:annihilateminus}, we see that the latter also annihilate the reference state.
We can thus simply characterize the stability group as generated by the set of $n^2$ operators $\{\frac{1}{N_F} \sum_\nu c^{\dagger}_{a \nu} c_{a'\nu} - \frac{1}{2} \delta_{aa'}\}_{a,a'=1}^n $. As these are the generators of the unitary group $U(n) \simeq \mathcal{H}$, we deduce that the coherent states are isomorphic to $SO(2n)/U(n)$. Following the general prescription in Eq.~\eqref{eq:cohstEtaGen}, we can represent the coherent states using the generators that do not annihilate the references state
\begin{equation}\label{eq:cohstEta}
\begin{split}
   & \left| O_\eta\right> = \exp\left[{\sum\limits_{\nu,1 \leq a < a' \leq n} \eta_{a a'} c^{\dagger}_{a\nu} c^{\dagger}_{a'\nu} - \text{h.c.}} \right] \left| \mathbb{I} \right> 
    \\& \equiv T_{\eta}  \left| \mathbb{I} \right> .
\end{split}
\end{equation}
As we can see, the presence of the stability group explains the reduction in the effective dimensionality of the coherent states: While the group $SO(2n)$ is a manifold of size $n(2n-1)$, $SO(2n)/U(n)$ has dimension $n(n-1)$. In the case $n=2$, this reduces to $2$, consistent with the coordinates $(y, \phi)$ parameterizing the sphere in Eq.~\eqref{eq:overlapn2}.

Notice that we easily recover the cluster index and write the generalisation as a tensor product over multiple clusters
\begin{equation}
\begin{split}
    \bigotimes_{j=1}^L & \left| O_{\eta^j}\right>  = \exp\left[{\sum\limits_{j,\nu,1 \leq a< a' \leq n} \eta^j_{a a'} c^{\dagger}_{a\nu j} c^{\dagger}_{a'\nu j} - \text{h.c.}}\right] \left| \mathbb{I} \right> .
\end{split}
\end{equation}

In the following, we will discuss how this formalism can be used to compute the relevant quantities. For instance, following the derivation in Eq.~\eqref{eq:pathintgen}, but with the replacement $\ket{g} \to \ket{O_\eta}$, we arrive at the expression of the purity as a path integral
\begin{equation}
    I:=\left< \mathcal{C}_{A,2} \right|e^{-t \mathcal{H}^{(n)}}\left|\rho^{(n)}(0)\right>=
    \int \mathcal{D}[\eta] e^{-S}\;.
\end{equation}
As explained in Eq.~\eqref{eq:Lgeneral}, 
this formal expression involves three ingredients: 
a resolution of the identity in terms of the Haar measure on the coset space, the overlap between two close coherent states at two consequent times $\braket{O_\eta(t_j)}{ O_\eta(t_{j+1})}$, leading to the Berry phase and the expectation value of the Hamiltonian on the coherent states $\braket{O_\eta | \mathcal{H}^{(n)}}{ O_\eta }$. We will start deriving an expression for the latter. Then, we will discuss the Berry phase and the geometric structure of the coset manifold, thus leading to the Haar measure and the resolution of the identity.
\subsection{The Hamiltonian in the coherent states basis}
As we saw, one of the ingredients of the path integral is the calculation of the expectation value of the Hamiltonian on the coherent states (see Appendix \ref{App:matel}). One can notice that our Hamiltonian depends on $\hat \Phi_i$, so we calculate the expectation values of the powers of $\hat \Phi_i$.  The operator $\hat\Phi$ in Eq.~\eqref{eq:fieldoperators} is expressed as an average over a large number of independent operators. Thus, quantum fluctuations in higher-order moments are subleading and one has
\begin{equation}
\left<O_{\eta}\right| (\hat\Phi^{\alpha \alpha'}_i)^{m} \left|O_{\eta} \right> 
   =  \left<O_{\eta}\right| \hat\Phi^{\alpha \alpha'}_i \left|O_{\eta} \right>^m \left(1+O\left(\frac{1}{N_F}\right)\right).\end{equation}
Therefore we need to determine the expectation values of the operators $\hat \Phi$. Moreover, we can express the operators $\hat \Phi$ in terms of the Dirac fermions, as was discussed in the previous subsection. 
As a result, we reduce the problem to finding the expectation values of the quadratic operators consisting of $c^{\dagger}$ and $c$. Let us consider one of these quadratic operators,
\begin{equation}
\begin{split}
    &\left<O_\eta \right| c^{\dagger}_{a\nu}c_{a' \nu}\left| O_\eta \right> = \left<\mathbb{I} \right| T^{-1}_{\eta} c^{\dagger}_{a\nu} T_{\eta }T^{-1}_{\eta} c_{a' \nu} T_{\eta}\left| \mathbb{I} \right> ,
\end{split}
\end{equation}
where $T_{\eta}$ is defined in Eq.~\eqref{eq:cohstEta}.
So we are interested in finding the transformation of each fermion under the operator $T_{\eta}$. As the transformation has to preserve the fermion algebra, it must take the form of a Bogoliubov transformation, that we parameterise as
\begin{equation}\label{Ttr}
\begin{bmatrix}
    T_{\eta} c_{a\nu} T^{-1}_{\eta} \\
    T_{\eta} c^{\dagger}_{a\nu} T^{-1}_{\eta}
\end{bmatrix} = \mathbb{T}^{\dagger} \begin{bmatrix}
    c_{a \nu} \\
    c^{\dagger}_{a \nu}
\end{bmatrix}
= \begin{bmatrix}
    U^{\dagger}_{ab} c_{b \nu} + V^{\dagger}_{ab} c^{\dagger}_{b\nu} \\
    V^T_{ab} c_{b \nu} + U^T_{ab} c^{\dagger}_{b \nu}
\end{bmatrix},
\end{equation}
with
\begin{equation}
    \mathbb{T}^{\dagger} = \begin{pmatrix}
        U^{\dagger} & V^{\dagger} \\
        V^t & U^t
    \end{pmatrix}.
\end{equation}
Here the summation over repeating replica index $b$ is implicit. 
We can explictly connect these matrices $U$ and $V$ with the matrix $\eta$ in Eq.\eqref{eq:cohstEta} (see Appendix \ref{App:matel}) as 
\begin{equation}
\label{eq:UVeta}
    U^{\dagger} = U = \cos(\sqrt{ \eta \eta^{\dagger}})\ , \ V^{\dagger} =- \frac{\sin(\sqrt{ \eta \eta^{\dagger}})}{\sqrt{ \eta \eta^{\dagger}}} \eta .
\end{equation}
Finally, we express the expectation value $\left<O_\eta\right|\hat \Phi^{\alpha \alpha'} \left|O_\eta \right> = \Phi^{\alpha \alpha'}$ in terms of these variables. 
We define
\begin{equation}
\hat\Psi^{a,b} = \frac{1}{N_F}\sum_\mu\left( \begin{array}{c|c}
       c_{a,\mu}^\dag c_{b,\mu}  & c_{a,\mu}^\dag c_{b,\mu}^\dag  \\ \hline
       c_{a,\mu} c_{b,\mu}  & 
      c_{a,\mu} c_{b,\mu}^\dag  
     \end{array} \right) \;.
\end{equation}
Then, on the reference state we simply get
\begin{equation}
\label{eq:Psi0def}
    \Psi_0 := \braket{\mathbb{I} | \hat\Psi}{\mathbb{I}} = \left(\begin{array}{c|c}
        \mathbb{I} & 0 \\ \hline
        0 & 0
    \end{array} \right)
    ,
\end{equation}
and on any coherent state
\begin{equation}
\label{eq:PsiTT}
\begin{split}
    \Psi := 
    \left<O_\eta \right| \hat\Psi \left| O_\eta \right>= \mathbb{T}^* \Psi_0 \mathbb{T}^{t} = \begin{pmatrix}
        U^t U^t & U^t V^t \\
        -V^{\dagger} U^t & -V^{\dagger} V^t
    \end{pmatrix} \;.
    \end{split}
\end{equation}
Finally, we can go back to expressing the expectation value of the Majorana operators. This is easily done by introducing the transformation to go from Dirac to Majorana fermions. We can write compactly
\begin{equation}
    \begin{pmatrix}
        c^\dag\\
        c
    \end{pmatrix} = R^\dag 
        \begin{pmatrix}
        \chi^+ \\
        \chi^-
    \end{pmatrix}\;,
\end{equation}
where $(c^\dag, c)$ is a shortcut for $(c^\dag_1,\ldots, c_n^\dag, c_1,\ldots, c_n)$ and similarly for $(\chi^+,\chi^-)$. The matrix $R$ takes the form of a $2n\times 2n$ matrix
\begin{equation}
    R = \frac{1}{\sqrt{2}} \begin{pmatrix}
        1 & 1 \\
        i & -i
    \end{pmatrix} \otimes \mathds{1}_n
    ,
\end{equation}
where the first matrix acts in the $(+,-)$ index space and the identity acts in the $a=1,\ldots,n$ replica space. We finally arrive at
\begin{equation}
    \Phi = \begin{pmatrix}
        \Phi^{++} & \Phi^{+-} \\
        \Phi^{-+} & \Phi^{--}
    \end{pmatrix} 
    = i R \Psi R^{\dagger} - i \frac{\mathds{1}}{2}
    \;.
    \label{eq:PhiFromPsi}
\end{equation} 
Combining Eq.~(\ref{eq:UVeta},\ref{eq:PsiTT},\ref{eq:PhiFromPsi}), each of the components of the matrix $\Phi$ is now expressed in terms of the parameters $\eta$. In the next subsection, we proceed similarly for the kinetic term of the path integral.

\subsection{The kinetic term of the action}\label{sec:kin}
We now focus on expressing the kinetic term (or Berry phase) in the parameterization Eq.~\eqref{eq:cohstEta}. The fundamental ingredient is the overlap 
$\braket{O_\eta}{O_{\eta'}}
$ and we now show how it can be obtained by a particular form of Baker-Campbell-Hausdorff (BCH)~\cite{RevModPhys.62.867}. 
To simplify the notation, we momentarily work at $N_F=1$ avoiding the sum over the greek flavour index. General formulas can be recovered at the end by tensor product over the flavour degrees of freedom. 
We have the identity
\begin{equation}\label{eq:BCH1}
\exp\left[{\sum_{a< a'} \eta_{a a'} c^{\dagger}_{a} c^{\dagger}_{a'} - \text{h.c.}} \right] = W_{-\tau}^\dag 
    \exp[\sum_{a,b = 1}^n \xi_{a,b} (c^\dag_a c_{b}- \delta_{ab}/2)]
    W_\tau    \;,
\end{equation}
with $W_\tau = \exp[\sum_{a<a'} \tau_{a,a'} c_a^\dag c_{a'}^\dag]$ and where $\tau$ and $\gamma$ are two complex $n\times n$ matrices. Clearly, one can choose $\tau^t = - \tau$. As this relation only depends on the commutation relations within the algebra, its proof and the explicit relation between the matrices $\gamma$ and $\tau$ with $\eta$ can be derived by using the fundamental representation (see Appendix \ref{App:matel}). We obtain
\begin{equation}
\label{eq:etatau}
    z = \eta \frac{\sin(\sqrt{\eta^{\dagger} \eta})}{\sqrt{\eta^{\dagger} \eta}}, ~e^\xi = (1 - z z^\dag )^{1/2} ,~ \tau = z(1 - z^{\dagger} z)^{-1/2}. 
\end{equation}
Eq.~\eqref{eq:BCH1} is particularly useful when acting on the reference state $\ket{\mathbb{I}}$. Indeed, as this state is completely filled
$W_{\tau} \ket{\mathbb{I}} = \ket{\mathbb{I}}$. Additionally, expanding the exponential, one clearly has
\begin{equation}
\label{eq:Agammasimpl}
     \exp[\sum_{a,b} \xi_{a,b} (c^\dag_a c_{b}-1/2 \delta_{ab})] \ket{\mathbb{I}}  =
     \exp[\sum_{a} \xi_{a,a} (c^\dag_a c_{a}-1/2)] \ket{\mathbb{I}} = A_\gamma \ket{\mathbb{I}}\;,
\end{equation}
with $A_\gamma = \prod_a e^{\gamma_{a,a}/2}$.
Thus, we simply have the representation
\begin{align}
    \ket{O_\eta} = A_\gamma  W^\dag_{-\tau}\ket{\mathbb{I}} \;.
\end{align}
Note that $A_\gamma$ is implicitly a function of the matrix $\eta$, so we write $A_\gamma = A[\eta]$. Let us use this expression to find the overlap between two arbitrary coherent states,
\begin{equation}
\label{eq:OAAN}
    \braket{O_\eta}{O_{\eta'}} = A[\eta] A[\eta'] \mathcal{N}[\tau, \tau^{\prime}] ,
\end{equation}
where we have set
\begin{equation}
\mathcal{N}[\tau, \tau^{\prime}] = \bra{\mathbb{I}}  W_{-\tau'} W_{-\tau}^\dag \ket{\mathbb{I}} .
\end{equation}
The explicit form of this factor can be found using direct manipulations of fermion operators~(see for instance~\cite{PhysRevB.87.245107}). One has
\begin{equation}
\label{eq:dettau}
    \mathcal{N}[\tau, \tau'] = \det(\mathbb{I} + \tau^{\dagger} \tau')^{N_F/2} \;,
\end{equation}
where in this expression we reintroduced an arbitrary number of flavors, simply noting that the overlap is factorised over them so $N_F$ just appears in the exponent. Finally, the expression of $A[\eta]$ could be obtained by direct manipulations. However,  it is more convenient to use the fact that 
coherent states $\ket{O_\eta}$ are normalised (see Eq.~\eqref{eq:cohenorm}), so that setting $\eta' = \eta$ (and thus $\tau' = \tau$), we must have
\begin{equation}
    A[\eta]^{-2} = \mathcal{N}[\tau, \tau] \;.
\end{equation}
Using the mapping Eq.~\eqref{eq:etatau}, we can always relate the matrix $\tau$ to $\eta$ parameterising the coherent state. Also, we observe that $\mathcal{N}[\tau, \tau']$ depends on the complex variables $\tau, \tau'$ but it is a real function when $\tau' = \tau$. So, it is convenient to use the standard notation for complex functions $\mathcal{N}[\tau, \tau'] \to \mathcal{N}[\tau, \tau^\ast]$. Thus, 
\begin{equation}
\label{eq:overlapN}
    \braket{O_\eta}{O_{\eta'}} = 
    \frac{\mathcal{N}[\tau, \tau^{\ast \prime}]}{\sqrt{\mathcal{N}[\tau, \tau^\ast]\mathcal{N}[\tau', \tau^{\ast \prime}]}} \;.
\end{equation}
As a side note, we observe that the square modulus of the overlap can be obtained more directly by noting that $\rho_\tau := \ket{O_\tau}\bra{O_\tau}$ defines a density matrix and the overlap $\Tr[\rho_\tau \rho_{\tau'}]$ between two density matrices has a simple form in terms of their correlation matrices~\cite{PhysRevB.87.245107}, thus
\begin{equation}
\label{eq:overlapPhi}
|\braket{O_\tau}{O_{\tau'}}|^2  = \det\left(\frac{\mathbb{I} - 4 \Phi_\tau \Phi_{\tau'}}{2}\right)^{N_F/2} \;.
\end{equation}
and using the explicit relation between $\Phi_\tau$ and $\tau$, which follows by Eqs.~(\ref{eq:PhiFromPsi}, \ref{eq:etatau}), one can verify that Eqs.~(\ref{eq:overlapN},\ref{eq:overlapPhi}) are consistent.

Finally, setting $\eta' = \eta + d \eta$, we can  expand with respect to $d \eta$ to find
\begin{equation}
\begin{split}
    &\ln \braket{O_\eta}{O_{\eta + d\eta}} = -\frac{1}{2} \sum_{a<b}   \left[\partial_{\eta_{ab}}(\ln \mathcal{N}) d\eta_{ab} - \partial_{\eta^\ast_{ab}}(\ln \mathcal{N}) d\eta^\ast_{ab} \right] .
\end{split}
\end{equation}
We see that this expression is invariant under reparameterizations of the coherent states. So, it is also convenient to use Eq.~\eqref{eq:etatau} to express the coherent states directly in terms of $\tau$, i.e. $\ket{O_\eta} \to \ket{O_\tau}$. Then, we can express the kinetic term as
\begin{equation}\label{eq:kin}
\begin{split}
    &K = -\lim_{dt \to 0}\frac{\ln \braket{O_\tau}{O_{\tau + \dot{\tau} dt}}}{dt} =  \frac{1}{2} \sum_{a<b}   \left[\partial_{\tau_{ab}} (\ln \mathcal{N}) \dot{\tau}_{ab} - \partial_{\tau_{ab}^\ast} (\ln \mathcal{N}) \dot{\tau}^\ast_{ab} \right],
\end{split}
\end{equation}
and using the explicit expression Eq.~\eqref{eq:dettau}
\begin{equation}
\label{eq:kintau}
    K = \frac{N_F}{4} \Tr \Big[( \mathbb{I} + \tau^{\dagger} \tau)^{-1} (\tau^{\dagger} \dot \tau - \dot \tau^\dagger \tau)\Big] .
\end{equation}
Here let us make the notation more intuitive and compare the calculation with the case $n=2$. First, we can parameterized $\tau$ in terms of the coordinates on the Bloch sphere:
\begin{equation}
    \tau=e^{-i \phi}\begin{pmatrix} 0& \tan \frac{\theta}{2}\\
    -\tan \frac{\theta}{2}&0
 \end{pmatrix}.
\end{equation}
Then the kinetic term takes the form
\begin{equation}\label{eq:kin_tau_spin}
    K = -i \frac{N_F}{2} \int dt \, (1 - \cos \theta) \, \dot{\phi}\;,
\end{equation}
which corresponds to the kinetic term we derived in Section~\ref{sec:n2q2} for the spin chain. Note that this parameterization uses the $z$-axis as a reference axis. Therefore, to obtain a complete mapping for $n = 2$, one must perform a coordinate transformation, $(x, y, z) \rightarrow (z, x, y)$. In this case, the kinetic term obtained will differ from Eq.~\eqref{eq:kin_tau_spin} by a total derivative.
\subsection{Geometric considerations}
A few observations are in order. The manifold defined by the coset $SO(2n)/U(n)$ (and more generally by the Lie group over the stability group $\mathcal{G}/\mathcal{H}$) naturally has the structure of a K\"ahler manifold~\cite{WIEGMANN1989311}. It has real dimensions $2n(n-1)$, which we can parameterise in terms of the  $n(n-1)$  complex components of the antisymmetric matrix $\tau$. Then, the K\"ahler potential and the metric $\mathtt{g}$ are provided by
\begin{equation}
\label{eq:kahler}
    F[\tau, \tau^\ast] = \log \mathcal{N}[\tau, \tau^\ast] , \; \mathtt{g}_{ab, cd} = \frac{\partial^2 F}{\partial \tau_{ab} \partial \tau^\ast_{cd}}\;,
\end{equation}
where $g_{ab, cd}$ is a $n(n-1)/2 \times n(n-1)/2$ matrix. The metric allows us to express explicitly the Haar measure in terms of the current parametrization in the usual form
\begin{equation}
\label{eq:haarmeas}
    \int_{\rm Haar} dO_\tau := \int \left[\prod_{a<b} d\tau_{ab} d\tau_{ab}^\ast\right] |\det \mathtt{g}|\;.
\end{equation}
From Eq.~\eqref{eq:dettau} and Eq.~\eqref{eq:kahler}, we obtain the expression
\begin{equation}
    \det \mathtt{g} \propto \det(\mathbb{I} + \tau^\dag \tau)^{-(n-1)}\;,
\end{equation}
where we neglect normalisation factors that are irrelevant in the ratio Eq.~\eqref{eq:transitAmpl}.

Also, when considering closed paths $\tau(t) = \tau(0)$, it is well known that the total contribution of the kinetic term
has a geometric interpretation as the 2-dimensional integral of a 2-form inside the coset manifold $SO(2n)/U(n)$. This is seen easily adding a fictitious extra coordinate $s \in [0,1]$ which connects the orbit to the identity element: choosing any smooth $\tau(s,t)$ such that $\tau(s = 0, t) = 0$ and $\tau(s = 1, t) = \tau(t)$, we can write
\begin{equation}
    \int_0^t dt' K = \int_{\substack{s \in [0,1] \\ t' \in [0,t]}} ds dt' \sum_{\substack{a<b\\ c<d}} g_{ab,cd} 
        (\partial_t \tau_{a,b} \partial_s \tau_{c,d}^\ast
        -\partial_s \tau_{a,b} \partial_t \tau_{c,d}^\ast
        )\;,
\end{equation}
which can be recognised as the Wess-Zumino-Witten term in the current parametrization~\cite{Witten:1983ar, Felder1988}.

We now have all the ingredients to write the path integral for our system.
In the next sections, we derive the NLSM action for $q_J=2$ using this formalism. 
To do so, we perform the integration over the massive modes coupled to the measurement rate, as we did in the case $n=2$. 
\subsection{Integration over the massive modes}\label{ssec:integ_tau}
In this subsection, we write the path integral and perform the integration over the heavy modes obtaining an effective description of the residual light ones. Parameterization in terms of the complex antisymmetric matrix $\tau$ provides a convenient set of variables for this analysis. In fact, writing $\tau = \tau_R + i \tau_I$ where $\tau_R, \tau_I$ are real antisymmetric matrices, we will see that the heavy modes are spanned by $\tau_I$. In terms of this parameterization, the measure in Eq.~\eqref{eq:haarmeas} takes the form
\begin{equation}
\label{eq:haarmeasri}
    \int_{\rm Haar} dO_\tau := \int \frac{d\tau_{R} d\tau_{I}}{\det(\mathbb{I} + \tau^\dag \tau)^{n-1}}\;,
\end{equation}
where will  collectively indicate $d\tau_R = \prod_{a<b} d\tau_{R,ab}$ and analogously for the imaginary part.
We momentarily drop the unitary part of the dynamics focusing on monitoring since the modes it consists of do not participate in the Gaussian integration to the leading order (see Appendix \ref{App:tauexp}), similarly to the case $n=2$. 
Let us first consider the path integral for a single cluster undergoing only local measurements,
\begin{equation}\label{action}
I  = \int \frac{d \tau_R d \tau_I}{[\det(\mathbb{I} + \tau^{\dagger} \tau)]^{(n-1)}}  \times    \exp\left(-\int_0^T dt [K(\tau_R, \tau_I) + \mathcal{H}_{\rm mon}(\tau_R, \tau_I)] \right) ,
\end{equation}
where the $\mathcal{H}_{\rm mon}$ is written in Eq.~\eqref{eq:Hmeas} for $q_J=2$ and for a single cluster and the kinetic term is given in Eq.~\eqref{eq:kintau}.
Proceeding analogously to what was done for the $n=2$ case, we wish to integrate the quantum fluctuations, represented by $\tau_I$, which become small due to the measurement part of the Hamiltonian. At the zero-th order in $\tau_I$, the field operator is given by 
\begin{equation}
\label{eq:PhitauR}
    \Phi = \frac{1}{2}\begin{pmatrix}
        O(\tau_I) & \frac{1 - \tau_R}{1 + \tau_R} +  O(\tau_I^2) \\
        -\frac{1 + \tau_R}{1 - \tau_R} +  O(\tau_I^2) &  O(\tau_I)
    \end{pmatrix}
    .
\end{equation}
By expanding to quadratic order for the measurement in a single cluster, we obtain
\begin{equation}
    \mathcal{H}_{\rm mon} = \Gamma N_F^2 \text{tr}\left( \frac{1}{1 - \tau_R^2} \tau_I \frac{1}{1 - \tau_R^2} \tau_I \right) \; .
\end{equation}
We can also expand the kinetic term to first order in $\tau_I$ (see Appendix \ref{App:tauexp}),
\begin{equation}
    K = i N_F \text{tr}\left(  \dot{\tau}_R \frac{1}{1 - \tau_R^2} \tau_I \frac{1}{1 - \tau_R^2}\right) + O(\tau_I^3).
\end{equation}

Finally, we arrive at the action written in terms of $\tau_R$ and $\tau_I$,

\begin{equation}\label{eq:fullNLSM}
    I = \int \frac{d\tau_R}{[\det(1 - \tau_R^2)]^{(n-1)/2}} \int d\tau_I \ e^{-i N_F \int dt \ \text{tr}\left( \dot{\tau}_R \frac{1}{1 - \tau_R^2} \tau_I \frac{1}{1 - \tau_R^2}\right) +  \Gamma N_F^2 \int dt \ \text{tr}\left( \frac{1}{1 - \tau_R^2} \tau_I \frac{1}{1 - \tau_R^2} \tau_I \right)} .
\end{equation}

To perform the computation, we diagonalize the matrix $\tau^2_R = O D O^{T}$ (the matrix $\tau^2_R$ is symmetric so it can be diagonalized with an orthogonal transformation), where $D = \text{diag}(\lambda_{R,1}^2, ..., \lambda_{R,1}^2)$ and $\lambda_{R,i}$ are the eigenvalues of $\tau_R$. 
By applying the change of variables
\begin{equation}
    \tau_I' = O \tau_I O^t, \quad \tau_R' = \tau_R ,
\end{equation}
which has unit Jacobian, and by rescaling the variables as $\tau_I' \rightarrow \tau_I' / N_F$, the Gaussian integration over $(\tau_I)_{ij}$ gives
\begin{equation}\label{eq:smtau}
    \int \frac{d\tau_R}{[\det(1 - \tau_R^2)]^{\frac{(n-1)}{2}}} e^{-\frac{1}{4 \Gamma} \int dt \ \text{tr}\left( \frac{1}{1 - \tau_R^2} \dot{\tau}_R \frac{1}{1 - \tau_R^2} \dot{\tau}^T_R \right)} .
\end{equation}
One can easily verify that the matrix $Q = ({1 - \tau_R})/({1 + \tau_R})$ appearing in Eq.~\eqref{eq:PhitauR} is an orthogonal matrix in $SO(n)$; such a representation goes under the name of Cayley transform, see Appendix \ref{App:Caley}. The Jacobian of the change of variable is precisely the determinant appearing in Eq.~\eqref{eq:smtau}, namely 
\begin{equation}
     \int \frac{d\tau_R}{[\det(1 - \tau_R^2)]^{\frac{(n-1)}{2}}} = \int_{\rm Haar} dQ\;,
\end{equation}
where the integral in the right-hand side is over the Haar measure in $SO(n)$. Therefore by changing variables, the integral in Eq.~\eqref{eq:smtau} can be expressed in terms of $Q$, by noticing the following relation
\begin{equation}
    \text{tr} \left( \partial_t Q \partial_t Q^t \right) = 4 \text{tr} \left( \dot{\tau}_R \frac{1}{1 - \tau_R^2} \dot{\tau}^T_R \frac{1}{1 - \tau_R^2} \right),
\end{equation}
from which we can finally conclude that the integral of Eq.~\eqref{eq:fullNLSM} is the same as a free NLSM path integral, with the usual kinetic term 
\begin{equation}
    I = \int dQ e^{-\frac{1}{16 \Gamma} \int dt \ \text{tr} \left( \partial_t Q \partial_t Q^t \right)} .
\end{equation}

\subsection{Many-clusters continuous limit and NLSM }\label{ssec:NLSM}
Now we can recover the cluster index and express the action in the continuous limit. The generalization of the kinetic part can be achieved by adding an extra index,
\begin{equation}
    K=\frac{1}{16 \Gamma} \int dt \sum_{i}~~{\rm tr} \Big(\partial_t Q_i \partial_t Q_i^t \Big) .
\end{equation}
For the Hamiltonian part, we use the definition given in Eq.~\eqref{eq:Huni}. Note that we have already assumed that $\tau_I$ is small and we have integrated it out in the path integral. Therefore, in the unitary part, we only need the leading order, which we obtain by setting $\tau_I=0$,
\begin{equation}
\begin{aligned}
    \mathcal{H}_{\rm uni}\Big|_{q_J=2} & = \frac{J N_F^2}{4} \sum_{i} {\rm tr}(Q_i Q_{i+1}^t)+O(\tau^2_I)\\ 
    \mathcal{H}_{\rm uni}\Big|_{q_J=4} & = -\frac{J N_F^4}{64}\sum_{i,aa'}   \left( (Q^{aa'}_i)^2 (Q^{aa'}_{i+1})^2 \right.\left.
        +\frac{1}{6}  (Q^{aa'}_i)^4\right)+O(\tau^2_I).
    \end{aligned}
\end{equation}
Expanding the unitary part to zeroth order in $\tau_I$ results in the same procedure as in the example with two replicas considered earlier. 
To take the continuum limit we associate the coarse-grained field to bonds:
let $x_i \equiv (i+\tfrac12)a$ be the midpoint between sites $i$ and $i{+}1$
and set $Q_i \equiv Q(x_i)$.Writing $Q_{i+1}=Q(x_i+\Delta x)$ and
$Q_i=Q(x_i-\Delta x)$ with $\Delta x=\tfrac{a}{2}$, a centered Taylor expansion gives

\begin{equation}
\begin{split}
  &  Q_{i+1}+Q_{i}\approx 2  Q(x_i)\\&
  Q_{i+1}-Q_i\approx 2 \Delta x \partial_x Q(x_i),
     \end{split}
\end{equation}
The kinetic term then becomes
\begin{equation}
\begin{split}
    &\lim_{L \rightarrow \infty} K=\frac{1}{16 \Delta x \Gamma} \int d^2x~~{\rm tr} \Big(\partial_t  Q  \partial_t  Q^t  \Big).
    \end{split}
\end{equation}
In the $q_J=2$ case up to an additive constant ${\rm tr}(\Phi^t_i \Phi_i)=\frac{n}{2}$, we get the spatial part of the NLSM
\begin{equation}
\lim_{L \rightarrow \infty} \mathcal{H}_{\rm uni}\Big|_{q_J=2}\sim- \frac{JN_F^2}{2}\Delta x^2 {\rm tr}(\partial_x  Q^t \partial_x  Q)+O(\Delta x^3) .
\end{equation}
Performing the same type of expansion in the case $q=4$ and keeping the second order term in $\Delta x $ we get

\begin{equation}
\begin{split}
     \lim_{L \rightarrow \infty} \mathcal{H}_{\rm uni}\Big|_{q_J=4}\sim 
     -\frac{JN_F^4}{64}\left( \frac{7 }{6}  \sum_{a a',i}  [Q^{aa'}(x_i)^4\right.\left.-Q^{aa'}(x_i)^2 (\Delta x)^2 (\partial_x Q^{aa'}(x_i))^2\right.\\ \left.-\frac{2}{3} Q^{aa'}(x_i)^3 \Delta x \partial_x Q^{aa'}(x_i) +O(\Delta x)]\right )
  .
  \end{split}
\end{equation}
Finally, the actions for both models in the continuous limit are expressed as

\begin{subequations}
\begin{align}
    \mathcal{S}\Big|_{q_J=2} & =\int d^2 x~~\left(\frac{1}{16 \Gamma \Delta x}{\rm tr} \Big(\partial_t  Q \partial_t  Q^t \Big)+ \frac{J N_F^2}{2} \Delta x {\rm tr}(\partial_x  Q^t \partial_x  Q)\right)\;, \\
    \mathcal{S}\Big|_{q_J=4} & =\int d^2 x~~\left(\frac{1}{16 \Gamma \Delta x}{\rm tr} \Big(\partial_t  Q \partial_t  Q^t \Big)+\frac{J N_F^4}{64}  \sum_{aa'}\left( \frac{1}{\Delta x }(Q^{aa'})^4-\right.\right.\\  &\left.\left. (Q^{aa'})^2 \Delta x (\partial_x Q^{aa'})^2-\frac{2}{3} (Q^{aa'})^3 \Delta x \partial_x Q^{aa'}  \right) \right)\; .\notag
\end{align}
\end{subequations}

In the case \(q=2\), we obtain the celebrated nonlinear sigma model for the \(SO(n)\) group manifold, and the explicit coefficients coincide with those derived in \cite{Fava:2023tgg} after the change of variables
\[
  J \;\longrightarrow\;\frac{2J}{\Delta x\,N_F}\,, 
  \qquad
  \Gamma \;\longrightarrow\;\frac{\Gamma}{\Delta x\,N_F}\,,
\]
which accounts for different conventions and the fact that here measurements act on a single cluster rather than coupling adjacent ones.

A renormalisation‐group analysis then reveals distinct phases \cite{Fava:2023tgg}. In the disentangling phase, the steady‐state entanglement entropy behaves as area-law.
In the stable non‐trivial phase, the entanglement entropy exhibits log-squared scaling. 
At \(t\gg\tau_P\), where $\tau_P$ is a purification time-scale, the system reaches a pure‐state trajectory whose bipartite entanglement across an interval of length \(L/2\) obeys
\[
  S_n \;\sim\;\,(\ln L)^2 \;.
\]At \(J=0\) and $N_F=1$ as we discuss in Sec.\ref{sec:measurementdynamics} the effective target manifold becomes $\frac{SO(2N)}{U(N)}$, and the $\theta$-term drives the theory to an unstable fixed point at \(\theta=\pi\).  This gives a genuine critical point on the measurement-only axis, at which scale invariance is restored and the entanglement law becomes purely logarithmic.  In particular,

\[
  S \sim \,\ln L \,.
\]
Instead, in the non-gaussian case $q > 2$,  we do not have a continuous $SO(2n)$ symmetry, but the field theory we derived has the expected permutation symmetry $Q \to \Pi_1 Q \Pi_2$, with $\Pi_{1,2} \in S_n$, the $n$-elements symmetric group. 
In the next sections, we show other applications of the path integral approach. \footnote{Similar results were found in the recent works \cite{Guo_2025,Poboiko_2025} where the interacting case was as well considered.}
In particular, we calculate the replicated purity in the systems of two-clusters.
\section{Two–cluster stationary purity}\label{sec:purity2}
We now revisit the two–cluster geometry ($A$ and $B$) at generic replica number $n$, using the $SO(n)$ sigma model. For $q_J=2$ the action reads
\begin{equation}
    \mathcal{S}
    =\int dt\;\mathrm{tr}\!\left[\frac{1}{16\Gamma}\sum_{i=A,B} (\partial_t Q_i^t)(\partial_t Q_i)
    -\frac{J N_F^2}{4}\, Q_A Q_B^t\right], \qquad Q_i\in SO(n).
\end{equation}
Introducing center–of–mass and relative fields
\begin{equation}
    Q_A=Q_+ Q_-^t,\qquad Q_B=Q_+ Q_-,
\end{equation}
and expanding around weak inter–cluster fluctuations $Q_-=e^{\delta Q}$(with $\delta Q$ skew–symmetric) (see Appendix \ref{App:purity2}), the action reduces at large $N_F$ to a set of decoupled harmonic oscillators for the $n(n-1)/2$ independent entries of $\delta Q$:
\begin{equation}
    \mathcal{S}\simeq \frac{1}{4\Gamma}\int dt\sum_{i>k}\left[\delta \dot{Q}_{ik}^2+4\Gamma J N_F^2\,\delta Q_{ik}^2\right].
\end{equation}
At late times the path integral is dominated by the oscillator ground state. The stationary purity is the overlap of this ground state with the boundary state implementing the swap on $A$ (see App.~\ref{App:purity2},~\ref{App:SpinCoherent}):
\begin{equation}
\left.\overline{\Tr\rho_A(t)^2}\right|_{t\to\infty}
\sim \exp\!\left[-\,\frac{N_F}{2}\sqrt{\frac{J}{\Gamma}}\left(\frac{\pi}{2}\right)^2\right],
\qquad (q_J=2,\;1/N_F^2<J\lesssim 1).
\label{eq:purity-2cluster-q2-main}
\end{equation}
This reproduces the $n=2$ result and shows an interpolation from area law ($J\lesssim N_F^{-2}$) to volume law ($J\gtrsim 1$).

For quartic unitary dynamics ($q_J=4$), the same procedure leads to an effective oscillator with frequency enhanced by $S\sim N_F$:
\begin{equation}
\left.\overline{\Tr\rho_A(t)^2}\right|_{t\to\infty}
\sim \exp\!\left[-\,\frac{N_F^2}{4}\sqrt{\frac{J}{3\Gamma}}\left(\frac{\pi}{2}\right)^2\right],
\qquad \Big(\tfrac{1}{N_F^4}<J\le \tfrac{1}{N_F^2}\Big),
\label{eq:purity-2cluster-q4-main}
\end{equation}
while for stronger coupling $1/N_F^2<J<1$ the overlap reduces to a Gaussian integral, giving a volume law
\begin{equation}
\left.\overline{\Tr\rho_A(t)^2}\right|_{t\to\infty}\sim e^{-\,N_F\log 2}.
\end{equation}

\paragraph{Remarks.}
Eqs.~\eqref{eq:purity-2cluster-q2-main}–\eqref{eq:purity-2cluster-q4-main} follow from a saddle that locks $Q_+$ and yields Gaussian fluctuations of $Q_-$; all intermediate steps (choice of $Q_0$, overlap determinants, rescalings, and minima structure for $q_J=4$) are given in App.~\ref{App:purity2}. Exact-diagonalization benchmarks and the scaling exponent $\alpha$ for $J=N_F^\beta$ are presented in App.~\ref{App:purity2}, showing good agreement within finite-size effects.
\section{Field theory for the measurement-only dynamics}\label{sec:measurementdynamics}
In this section, we change the protocol of our model slightly, showing how the method based on replicated fields can be flexible and used in different setups. In particular, we consider a model in which there is no unit dynamics but quadratic operator measurements are performed on all possible choices of operators between adjacent clusters. More explicitly, the model is described by the non-hermitian Hamiltonian Eq.~\eqref{eq_H_uni_plus_mon} where the unitary part is set to zero and where now the monitoring part acts on two consequential sites   
\begin{equation}
     H_{j,j+1}^{\rm mon} = i \sum_{\mu,\nu} M^{j}_{\mu \nu} \hat \chi_{j, \mu} \hat \chi_{j+1,\nu} ,
\end{equation}
where, as usual, the monitoring coefficients are white noises
\begin{equation}
    \mathbb{E}_{\rm G}[M^j_{\mu\nu}(t_1)M^i_{\alpha \beta}(t_1)]=\delta_{i j} \delta_{\mu \alpha} \delta_{\nu \beta} \delta(t_1-t_2) .
\end{equation}
When looking at this model on a $1D$ chain of $N$ clusters with $i,j = 1,\ldots, N$, the non-trivial dynamics results from the competition between measurements trying to project onto entangled dimers between 
even/odd sites ($2i, 2i +1$) and odd/even ones ($2i-1, 2i$). To lift the degeneracy between these two processes, we can introduce the staggering of the measurement strength $\Gamma$ as $\Gamma_j=[1+(-1)^j \Delta]\Gamma$. 
In  \cite{Fava:2023tgg}, it was argued, based on symmetry considerations, that the measurement-only phase can also be described by a NLSM.
It was conjectured that the sigma model for this system exhibits a larger symmetry, namely $SO(2n)/U(n)$ and have a topological term, which changes the phase diagram. 
In this context, we proceed at explicitly deriving this model conferming the conjectured form of the NLSM including the topological term. 
Performing the average on the replicated system as usual, we find the replicated Hamiltonian
\begin{equation}
\begin{split}
   \mathcal{H}^{\rm mon}=  &  \frac{N_F^2}{4}\sum_{\alpha,\alpha'} \Gamma_i \sigma \sigma'   \Phi^{\alpha \alpha'}_i \Phi^{\alpha \alpha'}_{i+1} .
    \end{split}
\end{equation}
First, we perform a canonical change of variables on odd sites that does not change commutation relations $\Phi^{+-}_{2i+1} \rightarrow -\Phi^{+-}_{2i+1}$ and $\Phi^{-+}_{2i+1} \rightarrow -\Phi^{-+}_{2i+1}$.
 Then for $n=2$, via the mapping to spins Eq.~\eqref{eq:paramPhi}, this model describes a $SU(2)$ antiferromagnet and was extensively studied in the literature, see for example \cite{Affleck:1985jy}. Note that antiferromagnetic chains can be divided into two interpenetrating subchains, which exhibit a staggered spin arrangement
 \begin{equation}
     S_{i} \rightarrow (-1)^{i+1} S_{i} .
 \end{equation}
 Then, by constructing the coherent states path integral, one can arrive at the celebrated $O(3)$ non-linear sigma model action, which here we shall generalise to the generic replica,  $SO(2n)/U(n)$ group.
Before presenting the explicit procedure, it is useful to describe the general outline. As before, we will obtain the NLSM first through expansion in coherent states and then by integrating fast modes, leading to an invariant field theory in the geometry of the manifold. In this case, we anticipate that the manifold on which the coherent states live is as before $SO(2n)/U(n)$ but it coincides with the manifold of the NLSM itself: the origin of this fact lies in the fact that we will integrate out half of the modes corresponding to fluctuations within a dimer, going from a system length $L$ to one length $L/2$. This differs from the procedure presented in Sect.~\ref{ssec:integ_tau} in which the system kept the original length, but part of the manifold (parametrized by the imaginary part of $\tau$) was integrated out.
We proceed therefore in an analogous way, and we perform the change of variable on even sites 
 \begin{equation}\label{eq:staggeringphi}
     \Phi_{2 i} \rightarrow - \Phi_{2 i}.
 \end{equation}
We then arbitrarily arrange the spins into different dimers containing even and odd sites $\{(12),(34),..,(N-1,N)\}$, as done also in \cite{Affleck:1985jy}. 
As explained in Sec.~\ref{sec:coherent}, the matrix $\Phi_i$ is parametrized in terms of the complex antisymmetric matrix $\tau_i$.
Aiming at a continuous space limit, we rewrite the space dependence of the $\tau_i$'s as follows
\begin{equation}
\begin{split}
   &\Delta x \delta \tau_{2 i-1}= \tau_{2i-1}-\tau_{2i}, \\&
   \Delta x  \partial_x \tau_{2 i}= \tau_{2i+1}-\tau_{2i}.
   \end{split}
\end{equation}
Note that at this stage this division is somewhat arbitrary, as the first field $\delta \tau$ gives the increment of the field $\tau$ inside each dimer, and the second field $\partial_x \tau $ the increment between two neigbouring dimers. 
Using the method we introduced in the previous sections, we compute the kinetic term and the matrix element of the Hamiltonian on the coherent states, keeping only the leading order in $N_F$. We therefore obtain the following action.  
\begin{equation}
\begin{split}
   - \mathcal{S}=\int d t~\sum_{i} & \bigg(K_{2i-1}-K_{2i}  
   +  H_{2i-1, 2i}+H_{2i,2i+1}\bigg),
   \end{split}
\end{equation}
where the kinetic terms $K_i$ are given by 
\begin{equation}
\begin{split}
     &K_i=\ln \braket{O_{\tau_i}}{O_{\tau  + \dot \tau_i d t}} =  \frac{1}{2} \sum_{\alpha \beta} \left[\partial_{\tau_{\alpha \beta}}(\ln \mathcal{N}_i) \dot \tau_{i,\alpha \beta} - \partial_{\tau_{i,\alpha \beta}^\ast}(\ln \mathcal{N}_i) \dot \tau^\ast_{i,\alpha \beta} \right] d t.
    \end{split}
\end{equation}
We note that the kinetic terms of even sites have different signs inside the sum due to the change of variables in Eq.~\eqref{eq:staggeringphi}. 
We can now expand the difference of the kinetic terms within each dimer with respect to $\delta \tau_{2 i-1}$,
\begin{equation}
    K_{2i-1}-K_{2i}= 2 N_F \mathtt{g}_{\alpha \beta}(\tau^{2i-1}) \dot \tau^{2i-1}_{\alpha} (\delta\tau^{2i-1}_{\beta})^*-c.c. ,
\end{equation}
where $\mathtt{g}_{\alpha \beta}= {\partial^2 \ln \mathcal{N}}/{\partial \tau_{\alpha } \partial \tau^*_{\beta}}$ is the metric tensor introduced in Eq.~\eqref{eq:kahler}. Here, to simplify the notation we are using the index $\alpha$ indicating the $n(n-1)/2$ ordered pairs $(a,b)$. We also expand the Hamiltonian in these new variables see Appendix \ref{App:mon}. 
Finally, taking the continuous limit $N \rightarrow \infty$, $\Delta x \rightarrow 0$  results in the action

\begin{equation}
\begin{split}
    -S & =  2 N_F \int d^2 x \mathtt{g}_{\alpha \beta}(\tau)  \dot \tau_{\alpha} \delta  \tau_{\beta}^*-c.c.-8 N_F^2\Delta x \Gamma \int d^2x  \mathtt{g}_{\alpha\beta}(\tau) \delta \tau_\alpha \delta  \tau_\beta^\ast\\
    & -4N_F^2 \Gamma(1+\Delta) \Delta x \int d^2x  \mathtt{g}_{\alpha \beta }(\tau) \delta  \tau_{\alpha } \partial_x \tau^*_\beta-4 N_F^2 \Gamma(1+\Delta) \Delta x \int d^2x  \mathtt{g}_{\alpha \beta }(\tau) \partial_x  \tau_{\alpha}\delta  \tau^*_{\beta } \\
    & -4 N_F^2\Gamma(1+\Delta) \Delta x \int d^2x \mathtt{g}_{\alpha \beta }(\tau) \partial_x  \tau_{\alpha} \partial_x \tau^*_{\beta } .
\end{split}
\end{equation}

Now we can perform the Gaussian field integral of $\delta \tau$ (see Appendix \ref{App:mon}) and finally obtain the NLSM action

\begin{equation}
\begin{aligned}
    -S_{\rm NLSM } & =2 N_F^2 \Gamma(1-\Delta^2) \Delta x \int d^2x~ \partial_x \tau_{\alpha} \mathtt{g}_{\alpha \beta} \partial_x \tau^*_{\beta} \\
    & -\frac{1}{2 \Gamma \Delta x} \int d^2 x~ \dot \tau_{\alpha} \mathtt{g}_{\alpha \beta} \dot \tau^*_{\beta}-
    \frac{\Theta}{ \pi}  \int d^2x~ (\dot \tau_{\alpha} \mathtt{g}_{\alpha\beta} \partial_x \tau^*_{\beta}-\partial_x \tau_{\alpha} \mathtt{g}_{\alpha \beta} \dot \tau^*_{\beta}) ,
    \end{aligned}
\end{equation}

where we obtained the angle $\Theta=\pi N_F (1+\Delta)$, associated to the celebrated topological term of the action. 
We leave the study of the phase diagram via the renormalisation group analysis of this field theory to future work (see also the discussion in~\cite{Fava:2023tgg}).

\section{Conclusions}

In this work, we have shown how to extend the formalism of spin coherent states to generic $SO(2n)$ symmetry for any number of replicas $n$. 
Borrowing from some previous works on coherent states  \cite{RevModPhys.62.867} we have extended and clarified many aspects of the formalism and used it to integrate out the heavy degrees of freedom and derive the effective field theories for monitored SYK clusters. 

Our work provides a systematic, first-principles route to deriving effective field theories for generic monitored fermionic systems, both Gaussian and interacting. 
As the first application of the method, we have derived the stationary purity of two SYK clusters, both interacting and not, in terms of the quantum fluctuations of the field theory. 
We have also demonstrated that when clusters interact only through a nearest-neighbor, quadratic,  monitoring Hamiltonian, the effective field theory is the $SO(2n)/U(n)$ NLSM, which extends the $O(3)$ field theory for antiferromagnetic spin chains \cite{Affleck:1985jy}. 
The case of intra-cluster monitoring and nearest-neighbor unitary interactions instead gives the $O(n)$ NLSM for orthogonal matrices, as previously found in \cite{Fava:2023tgg,2302.09094} with a different method.

Our approach is sufficiently general to allow for different applications and generalisations.
The entanglement structure of the monitored fermions could be investigated \cite{PhysRevE.104.014146,Carollo_2022,Szyniszewski2023,Piccitto2024,Russomanno_2023}, the role of the $U(1)$ or $SU(2)$ symmetry in interacting or Gaussian fermions \cite{Fava:2024tfu,PoboikoPRX2023,Barratt_2022,Agrawal_2022,Barratt_2022}, and in cases with boundary driving \cite{Turkeshi_2022} or structured measurements. 
We leave these exciting questions for near future works.

    \chapter{Infinite Randomness in Monitored Fermions}

A central theme of this thesis is to understand how \emph{measurements reshape quantum many-body states}.  
In clean systems, continuous monitoring competes with unitary evolution, driving  
\emph{measurement-induced phase transitions} (MIPTs) between volume-law and area-law entangled phases  
\cite{LiChenFisher2018,SkinnerRuhmanNahum2019,ChoiBaoQiAltman2020,NahumRuhmanAllToAll2021,BaoChoiAltman2020}.  
Most studies have focused on homogeneous settings, where all degrees of freedom are monitored at comparable rates.  

Here we explore \emph{measurement-only dynamics with strong disorder}. We consider a chain of $L$ clusters of Majorana fermions $\{\hchi_{i\nu}\}$, where $i\in\{1,\dots,L\}$ indexes clusters and $\nu\in\{1,\dots,N_F\}$ labels flavours. The operators satisfy
$\{\hchi_{i\nu},\hchi_{j\mu}\}=2\,\delta_{ij}\delta_{\nu\mu}$. We monitor nearest-neighbour bilinears
\begin{equation}
  \hat M_{i,i+1;\,\nu_1\nu_2} \;=\; i\,\hchi_{i\nu_1}\hchi_{i+1\,\nu_2}.
\end{equation}

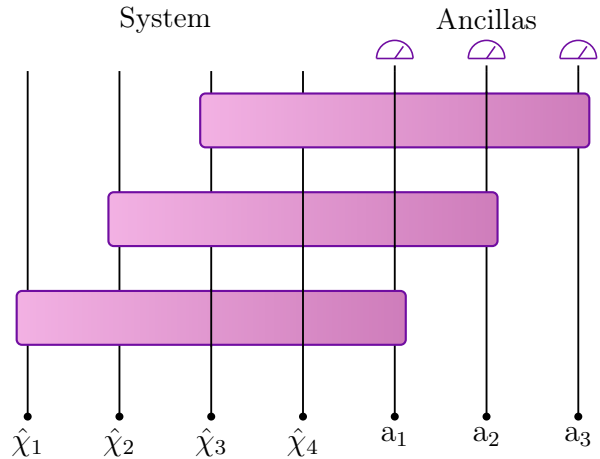
\begin{figure}[t]
\centering

\newdimen\Xunit    \Xunit=1.15cm   
\newdimen\Yunit    \Yunit=0.85cm   

\def\Nsys{4}       
\def\Nanc{3}   
\newcommand{\FigScale}{0.7}

\def\Hsep{1.50}    
\def\Gap{6.00}     
\def\Tsep{2.20}    
\def\GateH{0.60}   
\def\Pad{0.18}     
\def\HiW{0.22}     
\def\HiOp{0.55}    

\pgfmathsetmacro{\Tmax}{(\Nsys-1)*\Tsep + 1.1} 

\definecolor{DuneDeep}{RGB}{112,13,161}   
\definecolor{DuneMid}{RGB}{198,102,175}   
\definecolor{DuneLight}{RGB}{236,136,213} 

\begin{tikzpicture}[
  scale=\FigScale,
  x=\Xunit, y=\Yunit, >=Stealth,
  wire/.style   ={line width=0.7pt, draw=black},
  dot/.style    ={circle, fill=black, inner sep=0pt, minimum size=3pt},
  txt/.style    ={font=\normalsize, text=black},
  gateTrip/.style={rounded corners=2pt, draw=DuneDeep, line width=0.8pt,
                   top color=DuneLight!65, bottom color=DuneMid!85, shading angle=90},
  hiBar/.style  ={draw=none, fill=DuneDeep!50, opacity=\HiOp}
]
\pgfdeclarelayer{bg}
\pgfdeclarelayer{fg}
\pgfsetlayers{bg,main,fg}
\foreach \i in {1,...,\Nsys} {
  \pgfmathsetmacro{\xs}{(\i-1)*\Hsep}
  \draw[wire] (\xs,0) -- (\xs,\Tmax);
  \node[dot] at (\xs,0) {};
  \node[txt,anchor=north] at (\xs,0) {$\hchi_{\i}$};
}
\foreach \i in {1,...,\Nanc} {
  \pgfmathsetmacro{\xa}{\Gap + (\i-1)*\Hsep}
  \draw[wire] (\xa,0) -- (\xa,\Tmax);
  \node[dot] at (\xa,0) {};
  \node[txt,anchor=north] at (\xa,0) {$\mathrm{a}_{\i}$};
}

\pgfmathsetmacro{\SysCenter}{((\Nsys-1)*\Hsep)/2}
\pgfmathsetmacro{\AncCenter}{\Gap + ((\Nanc-1)*\Hsep)/2}
\node[font=\small, text=black] at (\SysCenter,\Tmax+1.15) {System};
\node[font=\small, text=black] at (\AncCenter,\Tmax+1.15) {Ancillas};

\foreach \i in {1,...,\numexpr\Nsys-1\relax} {
  \pgfmathsetmacro{\xsone}{(\i-1)*\Hsep}             
  \pgfmathsetmacro{\xstwo}{(\i)*\Hsep}               
  \pgfmathsetmacro{\xa}{\Gap + (\i-1)*\Hsep}         
  \pgfmathsetmacro{\y}{\i*\Tsep}                     

  \path[gateTrip] (\xsone-\Pad,\y-\GateH) rectangle (\xa+\Pad,\y+\GateH);

}
\def\Eps{0.03} 
\pgfmathsetmacro{\xsThree}{(3-1)*\Hsep} 
\pgfmathsetmacro{\xsFour }{(4-1)*\Hsep} 
\pgfmathsetmacro{\yOne}{1*\Tsep}        
\pgfmathsetmacro{\yTwo}{2*\Tsep}        
\pgfmathsetmacro{\yTwo}{2*\Tsep}
\pgfmathsetmacro{\yThree}{3*\Tsep}

\pgfmathsetmacro{\xaOne}{\Gap + 0*\Hsep} 
\pgfmathsetmacro{\xaTwo}{\Gap + 1*\Hsep} 

\def\Eps{0.04}

\tikzset{
  meter/.pic={
    \begin{scope}[x=1cm,y=1cm]
      \def\r{0.24}
      \draw[DuneDeep, line width=0.6pt, fill=white]
        (-\r,0) arc[start angle=180,end angle=0,radius=\r];
      \draw[DuneDeep, line width=0.6pt] (-\r,0)--(\r,0);
      \draw[DuneDeep, line width=0.6pt] (0,0) -- ++(55:{0.82*\r});
    \end{scope}
  }
}
\def\MeterLift{0.28}

\begin{pgfonlayer}{fg}
  \draw[wire] (\xsThree, \yOne-\GateH-\Eps) -- (\xsThree, \yOne+\GateH+\Eps);

  \draw[wire] (\xsFour , \yOne-\GateH-\Eps) -- (\xsFour , \yOne+\GateH+\Eps);
  \draw[wire] (\xsFour , \yTwo-\GateH-\Eps) -- (\xsFour , \yTwo+\GateH+\Eps);
\end{pgfonlayer}
\begin{pgfonlayer}{fg}
  \draw[wire] (\xaOne,\yTwo-\GateH-\Eps) -- (\xaOne,\yTwo+\GateH+\Eps);
  \draw[wire] (\xaOne,\yThree-\GateH-\Eps) -- (\xaOne,\yThree+\GateH+\Eps);

  \draw[wire] (\xaTwo,\yThree-\GateH-\Eps) -- (\xaTwo,\yThree+\GateH+\Eps);

  \foreach \i in {1,...,\Nanc} {
    \pgfmathsetmacro{\xa}{\Gap + (\i-1)*\Hsep}
    \draw[wire] (\xa,\Tmax) -- (\xa,\Tmax+0.45*\MeterLift);
    \pic at (\xa,\Tmax+\MeterLift) {meter};
  }
\end{pgfonlayer}

\end{tikzpicture}

\caption{Measurement-only circuit with tripartite couplings. At each time layer~$i$, a weak entangling ``brick'' couples the Majorana modes $\chi_i$ and $\chi_{i+1}$ to a fresh ancilla $a_i$. The ancilla is then weakly measured (meter icon) and discarded, implementing continuous monitoring of the parity $\hat M_{ii+1} = i\,\chi_i\chi_{i+1}$ with bond-dependent strength $\lambda^2_i$. Wires drawn on top of a brick indicate non-participating lines at that layer. Repetition along the chain realizes measurement-only dynamics with strong disorder in the monitoring rates.}
\label{fig:meas_sse_anc}
\end{figure}

The figure \ref{fig:meas_sse_anc} depicts the \emph{single-flavour} case $N_F=1$, for which we drop flavour indices and write
\[
\hchi_i\equiv \hchi_{i,1},\qquad \hat M_{i,i+1}\equiv i\,\hchi_i\hchi_{i+1}.
\]
As already mentioned in \cite{attal2003repeatedcontinuousquantuminteractions,Ciccarello2017,WisemanMilburn2010,Jacobs2014}, a convenient microscopic implementation is: for each monitored pair $\hat M_{i,i+1}$, couple the system weakly to a fresh ancilla qubit for a short time $\Delta t$ via
\[
H_{\mathrm{int}}^{(i)}=\lambda_i\,\hat M_{i,i+1}\otimes \hat\sigma_y^{(i)},
\]
then perform a weak measurement of $\hat\sigma_x^{(i)}$ on the ancilla and reset/discard it. In the continuum limit $\Delta t\to0$ with $\lambda_i^2\Delta t$ held fixed, one recovers Eq.~\eqref{eq:SSE} for measurement-only dynamics (see App.~\ref{App:SSE}). In the diagram, each “brick’’ denotes the weak entangling pulse, and each meter icon denotes the subsequent weak readout.

The same setup extends to multiple flavours by introducing \emph{one ancilla per monitored observable}. Concretely, if at bond $i$ we choose a set $\mathcal{O}_i\subseteq\{(\nu_1,\nu_2)\}$ of bilinears to monitor, we attach $|\mathcal{O}_i|$ fresh ancillas and implement
\[
H_{\mathrm{int}}^{(i)}=\sum_{(\nu_1,\nu_2)\in\mathcal{O}_i}
\lambda_{i;\nu_1\nu_2}\,\hat M_{i,i+1;\nu_1\nu_2}\otimes \hat\sigma_{y}^{(i,\nu_1\nu_2)},
\]
followed by weak readout of each corresponding $\hat\sigma_x^{(i,\nu_1\nu_2)}$ and ancilla reset. Taking $\Delta t\to0$ with $\lambda_{i;\nu_1\nu_2}^2\Delta t\to \Gamma_{i}$ fixed yields measurement-only dynamics with independent local monitoring rates $\Gamma_{i}$ for every measured bilinear.

The novelty lies in their statistics:  $\Gamma_i$ are drawn from a heavy-tailed distribution. Then the experiment is performed to obtain the statistics over $\Gamma_i$. 
This produces a regime where a few measurements are anomalously strong while most are weak.  
The central question is whether such dynamics flow to an \emph{infinite-randomness fixed point} (IRFP), and how entanglement scaling is affected.  

Here the measurement rates $\{\Gamma_i\}$ are quenched (fixed during a trajectory). We first analyze the trajectory statistics at fixed $\{\Gamma_i\}$, and only afterward average—or more appropriately, consider the typical behavior—over the disorder distribution of $\Gamma_i$

\subsection*{Ordinary renormalization group and scale invariance}

The renormalization group (RG) analyzes how physical systems evolve under changes of length scale.  
The procedure is to integrate out short-distance degrees of freedom, rescale, and track how effective couplings transform.  
Under a rescaling by a factor $b>1$,
\[
x \to x' = x/b, 
\qquad 
E \to E' = b^z E,
\]
where $z$ is the dynamical exponent relating length and energy (or time).  
At an \emph{ordinary} critical point, correlation functions decay algebraically,
\[
C(r) \sim r^{-\Delta},
\]
with $\Delta$ a scaling dimension, while the characteristic energy scale decreases as a power law with system size,
\[
\Omega \sim L^{-z}.
\]
This power-law scaling embodies ordinary scale invariance: under $L \to \lambda L$, observables transform with well-defined exponents.  
In disordered systems, if disorder is irrelevant in the Harris sense or flows
to a finite-disorder fixed point (the relative width of the coupling
distribution remains finite under the RG), critical scaling remains power-law
with a finite dynamical exponent $z$
\cite{Harris1974,Chayes1986,Cardy1996,Goldenfeld1992,Sachdev2011}.

\subsection*{From ordinary RG to infinite randomness}
By contrast, when the distribution broadens without bound the flow reaches an
\emph{infinite-randomness} fixed point with activated dynamics
($\ln \tau \sim L^{\psi}$) and effectively divergent $z$
\cite{Fisher1992,Fisher1995,IGLOI_2005,Vojta2006};
in the adjacent Griffiths regime $z$ varies continuously and can become very
large.
For example in the random antiferromagnetic Heisenberg chain
\begin{equation}\label{eq:Hisenb}
H = \sum_j \Gamma_j \, \mathbf{S}_j \cdot \mathbf{S}_{j+1}, \qquad \Gamma_j > 0,
\end{equation}
the strong-disorder RG (Ma--Dasgupta decimation) drives the system to an IRFP.  
\begin{figure}[t]
\centering
\begin{tikzpicture}[x=1cm,y=1cm]

\definecolor{PurpA}{RGB}{250,165,188}   
\definecolor{PurpB}{RGB}{214,165,250}    
\definecolor{PurpC}{RGB}{112,13,161} 

\tikzset{
  spin/.style   ={circle, draw=none, minimum size=11mm, inner sep=0pt, fill=white},
  spinP/.style  ={circle, draw=none, minimum size=11mm, inner sep=0pt,
                  shading=radial, inner color=PurpA, outer color=PurpB},
  lab/.style    ={font=\footnotesize, inner sep=1pt, outer sep=2pt, draw=none, fill=white},
  link/.style      ={draw=PurpB, line width=0.9pt, line cap=round},
  stronglink/.style={draw=PurpB, line width=1.6pt, line cap=round},
  eff/.style={draw=PurpB, line width=2pt, dashed}
}

\begin{scope}[shift={(0,0)}]
  \node[lab, font=\bfseries\small] at (3.1,1.55) {Before decimation};

  \node[spinP] (s1) at (0,0) {};   \node[text=PurpC] at (s1) {$\mathbf{S}_{j-1}$};
  \node[spinP] (s2) at (2.1,0) {}; \node[text=PurpC] at (s2) {$\mathbf{S}_{j}$};
  \node[spinP] (s3) at (4.2,0) {}; \node[text=PurpC] at (s3) {$\mathbf{S}_{j+1}$};
  \node[spinP] (s4) at (6.3,0) {}; \node[text=PurpC] at (s4) {$\mathbf{S}_{j+2}$};

  \draw[link]       (s1.east) -- (s2.west) node[midway, above, lab] {$\Gamma_L$};
  \draw[stronglink] (s2.east) -- (s3.west) node[midway, above, lab] {$\Omega$};
  \draw[link]       (s3.east) -- (s4.west) node[midway, above, lab] {$\Gamma_R$};
\end{scope}

\begin{scope}[shift={(0,-3.2)}]
  \node[lab, font=\bfseries\small] at (3.1,1.55) {After decimation};
  \node[lab, font=\bfseries\small] at (3.1,-1.75) {$\displaystyle \Gamma'=\frac{\Gamma_L \Gamma_R}{2\Omega}$};

  \node[spinP] (t1) at (0,0) {};   \node[text=PurpC] at (t1) {$\mathbf{S}_{j-1}$};
  \node[spinP] (t2) at (2.1,0) {}; \node[text=PurpC] at (t2) {$\mathbf{S}_{j}$};
  \node[spinP] (t3) at (4.2,0) {}; \node[text=PurpC] at (t3) {$\mathbf{S}_{j+1}$};
  \node[spinP] (t4) at (6.3,0) {}; \node[text=PurpC] at (t4) {$\mathbf{S}_{j+2}$};

  \begin{scope}[on background layer]
    \draw[line width=1pt, shorten >=2mm, shorten <=2mm,color=PurpB]
      (t1.east) to[out=-45, in=-135, looseness=1.15] (t4.west);
  \end{scope}

  \draw[eff, shorten >=1.2mm, shorten <=1.2mm,color=PurpB]
    (t2.east) to[out=22, in=158, looseness=1.12] (t3.west);

\end{scope}

\end{tikzpicture}
\caption{Strong-disorder RG decimation in a random Heisenberg chain.
Top: strongest bond $\Omega$ between $\mathbf{S}_{j}$ and $\mathbf{S}_{j+1}$; neighbors are coupled by $\Gamma_L$ and $\Gamma_R$ .
Bottom: after decimation, a singlet forms (dashed line) and an effective coupling $\Gamma'$ connects $\mathbf{S}_{j-1}$ and $\mathbf{S}_{j+2}$.}
\label{fig:SDRG_decimation}
\end{figure}
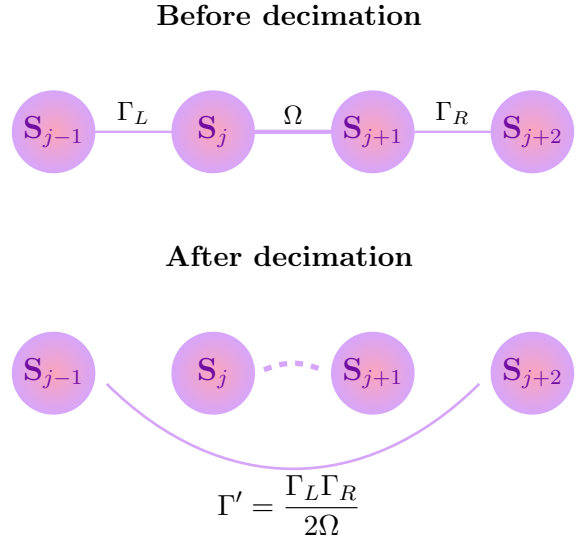
At each step the largest bond $\Omega=\max_j \Gamma_j$ is frozen and generates
an effective coupling $\Gamma'\simeq \Gamma_L\Gamma_R/(2\Omega)$ between its
neighbors (Fig.~\ref{fig:SDRG_decimation}) \cite{MaDasguptaHu1980,Fisher1995}. Introducing $\beta\equiv\ln(\Omega/\Gamma)\ge 0$, this rule reads
$\beta'=\beta_L+\beta_R+\ln 2$, so the logarithms \emph{add}. Consequently the
distribution $P(\beta)$ broadens under the RG with a
width parameter $\ell=\ln(\Omega_0/\Omega)$ that grows without bound
(infinite-randomness). The fraction of active spins obeys
$dn/d\ell=-2P(0)n=-2n/\ell$, giving $n\sim \ell^{-2}$ and a typical length
$L\sim \ell^2$. Hence $\ln(\Omega_0/\Omega)\sim \sqrt{L}$ and the
dynamics is \emph{activated}, $\ln\tau\sim L^{\psi}$ with $\psi=1/2$ (so
$z=\infty$).
 \cite{IGLOI_2005}.  
Correlations are dominated by rare events: the \emph{average} spin–spin correlation decays as $r^{-2}$ due to rare long-distance singlets even though the \emph{typical} correlation is exponentially small.   Under the Ma–Dasgupta RG the largest bond is frozen into a singlet, so the ground state of Eq.~\eqref{eq:Hisenb} flows to a \emph{random-singlet}
state.  Entanglement of an interval $A$ is then obtained by \emph{counting
singlets that cross the $A|\bar A$ boundary}: each crossing contributes
$\ln 2$.  The RG then gives 
\[
S_A \sim\; \frac{\ln 2}{3}\,\ln L_A,
\]
as in Refael–Moore \cite{Refael:2004zz}.

\subsection*{Motivation for monitored systems}

The key question of this chapter is whether such infinite-randomness physics for the measurements rates can persist in \emph{monitored} systems, where dynamics are stochastic and (in effective descriptions) non-Hermitian.  
Do measurements destabilize infinite randomness, or do they provide a new arena where IRFP behavior survives?  
We derive Dasgupta–Ma SDRG rules within the \(SO(2n)\) replica algebra. Starting from heavy-tailed monitoring rates \(\Gamma_j\). We will show that, for any $n>1$ the SDRG flow again broadens without bound, converging to an IRFP with logarithmic entanglement scaling and broad, non-self-averaging fluctuations. However we encounter an order of limits ambiguity $n \to 1, \Gamma \to \infty$, furher analytical studies of the flow equations are needed to conclude this case.

This analysis is of interest for three reasons.  
\emph{Conceptually}, it extends the random singlet paradigm into measurement-driven settings and connects to the broader MIPT literature \cite{LiChenFisher2018,SkinnerRuhmanNahum2019,BaoChoiAltman2020,NahumRuhmanAllToAll2021,AlbertonBuchholdDiehl2021}.  
\emph{From the perspective of entanglement}, it identifies a universality class distinct from both clean MIPTs and conventional disordered phases.  
\emph{Experimentally}, our results yield concrete predictions—logarithmic entanglement growth and broad entanglement fluctuations—testable with modern randomized-measurement protocols \cite{Brydges_2019,Elben_2018}.

\section{Replica construction with \texorpdfstring{$SO(2n)$}{SO(2n)} generators \texorpdfstring{\ensuremath{\Phi}}{Phi}}
\label{sec:replica-Phi}

\subsection{Setup and replica trick}

 We consider a chain of $L$ clusters of Majorana fermions $\{\hchi_{i\nu}\}$, where $i\in\{1,\dots,L\}$ indexes clusters and $\nu\in\{1,\dots,N_F\}$ labels flavours. The operators satisfy
$\{\hchi_{i\nu},\hchi_{j\mu}\}=2\,\delta_{ij}\delta_{\nu\mu}$.
Following previous chapter \ref{sec:repl} in particular Eq.\eqref{eq_H_rep_f} we can write the stochastic evolution in Fig.\ref{fig:meas_sse_anc} as non-Hermitian Hamiltonian: 
\begin{equation}\label{eq:mon_only}
H = - \sum_j \sum_{\mu,\nu} w_{\mu\nu,jj+1}\,\hat \chi_{\mu j} \hat \chi_{\nu j+1},
\end{equation}
where $j$ labels clusters, $\mu,\nu$ enumerate fermions within each cluster, and $w_{\mu\nu,jj+1}$ are random monitoring-induced couplings.  
Concretely, we place Majorana fermions $\{\hchi_j^\mu\}$ on a one-dimensional chain of clusters labeled by $j$, with flavor index $\mu=1,\dots,N_F$ inside each cluster. 

We assume that $w_{\mu\nu,jj+1}$ are zero-mean Gaussian white noises with variance set by a cluster-dependent monitoring rate $\Gamma_j$,  
\begin{equation}
  \mathbb{E}_{\rm G}[w_{\mu\nu,jj+1}(t) w_{\mu'\nu',j'j'+1}(t')] 
  = \delta_{jj'}\,\delta_{\mu\mu'}\,\delta_{\nu\nu'}\,\Gamma_j \,\delta(t-t').
\end{equation}
Importantly, the disorder distribution of $\Gamma_j$ can be broad and even heavy-tailed, such that rare bonds dominate the long-time dynamics.  
This places the model in close analogy with strongly disordered spin chains studied by Fisher’s strong-disorder renormalization group (SDRG).

Our main observable of interest is the entanglement entropy across a bipartition of the chain.  
Specifically, we study the statistics of the second Rényi entropy of a subsystem $A$,
\begin{equation}
  S_A^{(2)}(t) = -\log \Tr[\rho_A^2(t)] ,
  \qquad \rho_A(t) = \Tr_{\bar A} \rho(t),
\end{equation}
where $\rho(t)$ is the physical density matrix of the monitored system and $\bar A$ its complement.  
Because the dynamics is stochastic, $S_A^{(2)}(t)$ fluctuates from trajectory to trajectory; we are interested in its typical scaling, fluctuations, and distribution at long times.

To compute such nonlinear observables, it is advantageous to reformulate the dynamics in terms of unnormalized density matrices and replicas, as we now review.

Here we remind the reader what we discussed in Sect. \ref{sec:repl}. For nonlinear observables such as entanglement entropies, it is convenient to work with an
\emph{unnormalized} density matrix~\cite{Fava:2023tgg,JianYouVasseurLudwig2020,GiachettiElusive2022},  
\begin{equation}
  \rho(t) \;=\; \frac{\check\rho(t)}{\Tr \check\rho(t)} ,
\end{equation}
which evolves linearly under a non-Hermitian Hamiltonian.  
This replaces the nonlinear stochastic Schrödinger equation (SSE) with a linear stochastic equation for $\check\rho(t)$, simplifying analytic approaches.

The evolution reads
\begin{equation}
  \check\rho(t) = K(t)\,\rho(0)\,K^\dagger(t), 
  \qquad 
  K(t)=\mathcal{T}\exp\!\Big(-i \int_0^t H(s)\,ds\Big),
\end{equation}
with $H(t)$ containing both unitary and monitoring contributions.  
Averages over trajectories are obtained by weighting with $\Tr \check\rho(t)$: for any scalar functional $F[\rho]$,
\begin{equation}
  \overline{F[\rho]}_{\rm SSE}
  \;=\;
  \frac{\mathbb{E}_{\rm G}\!\left[ F[\rho]\;\Tr \check\rho(t)\right]}
  {\mathbb{E}_{\rm G}[\Tr \check\rho(t)]},
\end{equation}
where $\mathbb{E}_{\rm G}$ denotes Gaussian averaging over the noise variables.

A key example is the trajectory-averaged purity of a subsystem $A$.  
Introducing replicas, one obtains
\begin{equation}
\label{eq:purity-recap}
  \overline{\Tr \rho_A^2}
  \;=\;
  \lim_{n\to 1}\;
  \frac{\Tr\!\left[\rho^{(n)}(t)\,(\mathcal{C}_{A,2}\otimes\mathbb{I})\right]}
       {\Tr \rho^{(n)}(t)} ,
  \qquad 
  \rho^{(n)}(t)=\mathbb{E}_{\rm G}[\check\rho(t)^{\otimes n}],
\end{equation}
where $\mathcal{C}_{A,2}$ swaps the first two replicas inside region $A$.  
More generally, Rényi entropies can be expressed in terms of permutation operators $\mathcal{C}_{A,\sigma}$ acting on replicas.
\begin{equation} 
\overline{S^{(\alpha)}_A(t)} =     \frac{1}{1-\alpha} \lim_{k \rightarrow 0}  \lim_{n \rightarrow 1} 
    \frac{1}{k}\left(\frac{\Tr \left[ \rho^{(n)}(t) \left(\mathcal{C}_{A,\alpha^k}\otimes \mathbb{I} \right)\right]}{\Tr \left[ \rho^{(n)}(t) \right]} - 1\right) ,
\end{equation}
  As discussed in Sec.~\ref{sec:vect} of the previous chapter, vectorization then maps the replicated dynamics to a Schrödinger-like evolution with generator $\mathcal{H}^{(n)}$
\begin{equation}\label{eq:evol_rho}
  \ket{\rho^{(n)}(t)} = \mathbb{E}_{\rm G}\!\left[(K\otimes K^*)^{\otimes n}\right]\ket{\rho^{(n)}(0)}
  \;=\; e^{-t\,\mathcal{H}^{(n)}}\ket{\rho^{(n)}(0)},
\end{equation}
providing a natural entry point for path-integral and field-theoretic techniques.
The replicated, averaged generator can be written in terms of the $SO(2n)$ site generators
\(
  \hat{\Phi}_j^{\alpha\beta}=\frac{i}{2N_F}\!\sum_\nu[\chi_{j\nu}^{\alpha},\chi_{j\nu}^{\beta}]
\)
(see Sec.~\ref{sec:so(2n)}), which satisfy the $so(2n)$ algebra.This formalism translates the evaluation of nonlinear functionals into the calculation of
overlaps in a replicated Hilbert space see Eq.\eqref{eq:transitAmpl}.

\subsection{ Fermionic \texorpdfstring{$SO(2n)$}{SO(2n)} fields}

As introduced in Sec.~\ref{sec:vect}, the dynamics of the replicated density matrix can be recast in terms of a Schrödinger-like evolution.  
A natural basis for this formulation is provided by the bilinear field operators
\begin{equation} \label{eq:Phi_def}
  \hat{\Phi}_j^{(\sigma,a),(\sigma',a')} 
  = \frac{i}{2N_F}\sum_\nu 
  [\chi_{j\nu}^{(\sigma,a)},\chi_{j\nu}^{(\sigma',a')}],
\end{equation}
which we first defined in Sec.~\ref{sec:so(2n)} of the previous chapter.  
These operators satisfy the commutation relations of the $so(2n)$ Lie algebra, and equivalently realize the same algebra as antisymmetric $2n\times 2n$ matrices.  
They therefore provide a natural set of generators for the replica theory.

For quadratic monitoring, the Hamiltonian takes the form
\begin{equation}
  \mathcal{H}_{\rm mon}
  =  \,\frac{N_F^2}{4}
    \sum_{j,\alpha,\alpha'} \Gamma_j\sigma\sigma'\,
    \hat\Phi_j^{\alpha\alpha'} \hat\Phi_{j+1}^{\alpha\alpha'},
\end{equation}
where $\{\alpha, \alpha'\}=\{(\sigma, a), (\sigma',a')\}$ are multi-indices and $(\sigma,a)$ are replica indexes $\sigma\in \{+,-\}$ and $a \in {1,...,n}$.
The operators $\hat\Phi_j$ thus serve as the main building blocks of the theory.

\section{Warm--up with \texorpdfstring{$n=2$}{n=2} replicas}

As discussed in Sec.~\ref{sec:n2q2} of the previous chapter, the case of two replicas admits a particularly simple description. The $so(4)$ algebra reduces to $su(2)\oplus su(2)$, and—after exploiting parity conservation—the relevant Hilbert space maps onto a single pseudospin-$\tfrac{1}{2}$ sector. In this basis, the field operators $\hat\Phi$ Eq.~\eqref{eq:paramPhi} collapse to three independent generators, so that the replicated dynamics reduces to an effective $SU(2)$ spin system.

This warm--up provides both intuition and a concrete example of how replica fermions can be reformulated in terms of spin degrees of freedom. It also sets the stage for the general $n$ case.

Specializing to the monitoring--only dynamics, the mapping between $\hat\Phi$ and spins (Eq.~\eqref{eq:paramPhi}) yields the lattice Hamiltonian with spin value $S=\frac{N_F}{2}$
\begin{equation}
  \mathcal{H}_{\rm mon}
  = \sum_j \Gamma_j \big(S^y_j S^y_{j+1} - S^x_j S^x_{j+1} - S^z_j S^z_{j+1}\big),
  \qquad \Gamma_j>0.
\end{equation}
A staggered $\pi$ rotation around the $y$-axis on the odd sites,
$U_y=\prod_{j\in{\rm odd}} e^{-i\pi S^y_j}$, flips $S^{x,z}_{2j+1}\!\to\!-S^{x,z}_{2j+1}$ while leaving $S^y$ invariant, and thus
\begin{equation}\label{eq:Mon_Ham}
  \mathcal{H}_{\rm mon} \;\longrightarrow\; U_y \mathcal{H}_{\rm mon} U_y^\dagger
  = \sum_j \Gamma_j\, \vec S_j \!\cdot\! \vec S_{j+1},
\end{equation}
i.e., a random antiferromagnetic Heisenberg chain with positive, site-dependent couplings $\{\Gamma_j\}$.

Our objective in this section is to characterize the trajectory--averaged purity. As shown in our previous work, its statistics can be recast in terms of ground--state properties of the replicated Hamiltonian (see Eqs.~\eqref{eq_folded_replicated_rho}, \eqref{eq_overlap_rho}, and \eqref{eq:transitAmpl}). For $n=2$ the replicated Hamiltonian is precisely the random Heisenberg antiferromagnet, whose ground--state physics is by now well-known; in particular, the Dasgupta--Ma strong--disorder real--space RG (SDRG) provides an asymptotically exact description at strong disorder. In our setting, monitoring produces broad, short--range, $SU(2)$--symmetric couplings $\{\Gamma_j\}$, placing the model squarely within the SDRG regime. In the next subsection we consider standard Dagupta-Ma perturbation theory procedure that will be useful for calculation of generic $n$ pertubation theory.

\subsection{Dasgupta--Ma perturbation theory}

The strong-disorder renormalization group (SDRG) proceeds by iteratively
\emph{decimating} the strongest bond in the chain Eq.~\eqref{eq:Mon_Ham}. To illustrate the logic,
consider four consecutive spins with Hamiltonian
\begin{equation}
H = \Gamma_1\,\vec S_1\cdot \vec S_2
  + \Gamma'\,\vec S_2\cdot \vec S_3
  + \Gamma_2\,\vec S_3\cdot \vec S_4,
\end{equation}
where the central coupling is much stronger than its neighbors,
\(\Gamma' \gg \Gamma_1,\Gamma_2\).
Physically, spins 2 and 3 are much more tightly bound to each other than to the rest of the chain,
so in the low-energy subspace they form an almost rigid object. The SDRG step makes this intuition precise:

\paragraph{Step 1: unperturbed versus perturbing parts.}
We treat the strong bond as the unperturbed Hamiltonian,
\begin{equation}
H_0 = \Gamma'\,\vec S_2\cdot \vec S_3, 
\end{equation} and the weaker neighboring couplings as the perturbation,
\begin{equation}\label{eq:pot}
    V = \Gamma_1 \vec S_1\cdot \vec S_2 + \Gamma_2 \vec S_3\cdot \vec S_4.
\end{equation}

\paragraph{Step 2: spectrum of the central pair.}
The eigenstates of \(H_0\) are classified by the total spin of sites \(2\) and \(3\).
A useful identity for two spins is
\begin{equation}
\label{eq:S2S3-identity}
\vec S_2\cdot \vec S_3 = \tfrac12\big[J(J{+}1)-2S(S{+}1)\big],
\end{equation}
where $S$ is the length of each individual spin and $J$ is the total spin of the pair.
This relation follows from the operator identity
\[
\vec S_2\cdot \vec S_3
= \tfrac12\left(\vec J^{\,2}-\vec S_2^{\,2}-\vec S_3^{\,2}\right),
\qquad \vec J=\vec S_2+\vec S_3,
\]
together with $\vec S_2^{\,2}=\vec S_3^{\,2}=S(S{+}1)$.

As a check, for spin-$\tfrac12$ one has $S(S{+}1)=3/4$.
In this case the total spin can be either a singlet $J=0$ or a triplet $J=1$,
and Eq.~\eqref{eq:S2S3-identity} gives the familiar eigenvalue equations
\begin{equation}\label{eq:singl_ch4}
\vec S_2\cdot \vec S_3\,|s\rangle = -\tfrac34\,|s\rangle, 
\qquad
\vec S_2\cdot \vec S_3\,|t,m\rangle = +\tfrac14\,|t,m\rangle,
\end{equation}
where $|s\rangle$ is the singlet state and $|t,m\rangle$ the triplet manifold.

\paragraph{Step 3: effective Hamiltonian from second order.}

To obtain the low-energy Hamiltonian we integrate out the high-energy
triplet manifold using second-order degenerate perturbation theory
(Schrieffer--Wolff transformation). The Hilbert space of the central pair
$(2,3)$ naturally decomposes into two sectors:

The \emph{singlet subspace} $J=0$, spanned by the unique singlet
  state $|s\rangle$ with energy $E_s$. The projector onto this sector is $ P_s = |s\rangle\langle s| $.

 The \emph{triplet subspace} $J=1$, spanned by the three states
  $|t,m\rangle$ with $m=-1,0,+1$, each at energy $E_t=E_s+\Delta$.
  The projector is $Q_t = \sum_{m=-1}^1 |t,m\rangle\langle t,m| $.
  
For spin-$\tfrac12$, these two sectors exhaust the Hilbert space of the
two-spin system, so that $P_s+Q_t=\mathbb{I}$.

Perturbation theory constructs an operator acting \emph{within} the singlet
sector that reproduces the spectrum of $H$ up to order $V^2$. Explicitly,
the effective Hamiltonian is
\begin{equation}
  H_{\rm eff} = P_s H_0 P_s + P_s V P_s
  - P_s V Q_t \frac{1}{H_0-E_s}\,Q_t V P_s + \mathcal{O}(V^3),
\end{equation}
with $V$ given by Eq.~\eqref{eq:pot}.
The first two terms are trivial constants in the singlet sector:
$P_s H_0 P_s = E_s P_s$ and $P_s V P_s = \langle s|V|s\rangle P_s$. The
nontrivial contribution comes from the third term, which describes virtual
excitations out of the singlet into the triplet manifold and back.

Since all three triplet states have the same energy separation
$\Delta=E_t-E_s=\Gamma'$, the resolvent simplifies to
\[
\frac{1}{H_0-E_s}\,Q_t = \frac{1}{E_t-E_s}\,Q_t
= \frac{1}{\Delta}\,Q_t.
\]
Substituting this into the effective Hamiltonian, and dropping constants
which only shift the overall ground-state energy, we arrive at the compact
form
\begin{equation}
  H_{\rm eff} = -\,P_s V \frac{1}{\Delta} Q_t V P_s,
\end{equation}
which is the expression used in the following steps. This term encodes the
virtual process $|s\rangle \to |t,m\rangle \to |s\rangle$, i.e.~a virtual
excitation of the $(2,3)$ bond into the triplet followed by relaxation back
to the singlet, mediated by the weaker neighboring couplings
$\Gamma_1,\Gamma_2$.

In more details the Hamiltonian can be written as
\[
H_{\rm eff} = -\,P_s V \frac{1}{\Delta} Q_t V P_s=\sum_{m} \frac{\left|\left<s \left|\Gamma_1 \vec{S}_1 \cdot \vec{S}_2+\Gamma_2 \vec{S}_3 \cdot\vec{ S}_4 \right|t,m\right>\right|^2}{-\Gamma}.
\]
Therefore
the relevant building blocks are overlaps where spin operators $S_2^\alpha$
or $S_3^\alpha$ act on the singlet $|s\rangle$ and connect it to the triplet
sector. Thus, the core task in this step is to compute $\sum_{m} \langle s|S_i^\alpha|t,m\rangle \langle t,m|S_j^\beta|s\rangle$, with $i,j\in\{2,3\}$, which encodes the propagation from the singlet to a
triplet and back.

Two key identities make this evaluation tractable. In the $J=0$ singlet state of spins $(2,3)$ the two spins cancel, so that their operators act as negatives of each other:
\begin{equation}\label{eq:singl}
    \vec S_2|s\rangle = -\,\vec S_3|s\rangle.
  \end{equation}
  This relation allows us to replace $S_3$ by $-S_2$ whenever the state
  $|s\rangle$ is present. 
  Summing over the intermediate triplet states that appear in $Q_t$,
  \[
  Q_t = \sum_{m=-1}^{+1}|t,m\rangle\langle t,m|,
  \]
  we encounter matrix elements of the form $\sum_{m}\langle s|S_2^\alpha|t,m\rangle\langle t,m|S_2^\beta|s\rangle $. Notice that $S^{\alpha}_2 \left|s\right>$ lives in a triplet space. Therefore the sum over the basis in triplet space $\sum_{m} \left|t,m\right>\left<t,m\right|$ in this expression acts as identity. Thus
  \begin{equation}\label{eq:rot_symm}
  \sum_{m}\langle s|S_2^\alpha|t,m\rangle\langle t,m|S_2^\beta|s\rangle
  =\left<s \right|S_2^{\alpha}S_2^{\beta} \left|s\right> = \frac{S(S{+}1)}{3}\,\delta_{\alpha\beta}.
  \end{equation}

The first identity Eq.~\eqref{eq:singl} expresses the antisymmetry of the singlet wavefunction,
while the second Eq.~\eqref{eq:rot_symm} encapsulates the rotational symmetry of the spin algebra:
all three spin components couple equally strongly to the triplet manifold.
Together, these relations drastically simplify the evaluation of
$H_{\rm eff}$ in the next step.

Using these relations one finds that the diagonal terms (\(\propto \Gamma_1^2,\Gamma_2^2\)) shift the ground-state energy, while the cross term generates a new coupling between spins 1 and 4:
\begin{equation}
  H_{\rm eff}
  = \frac{2S(S+1)}{3}\,\frac{\Gamma_1\Gamma_2}{\Gamma'}\,\vec S_1\cdot \vec S_4
  \;+\; \text{const}.
\end{equation}
Thus the effective antiferromagnetic interaction is $\Gamma_{\rm eff} = \frac{2S(S+1)}{3}\,\frac{\Gamma_1\Gamma_2}{\Gamma'}.$

\paragraph{Special case \(S=\tfrac12\).}
For spin-\(\tfrac12\), \(S(S+1)=3/4\), which gives
\begin{equation}
  \Gamma_{\rm eff} = \frac{\Gamma_1\Gamma_2}{2\Gamma'}.
\end{equation}
This is the classic Dasgupta--Ma rule for the random Heisenberg chain (see Fig.~\ref{fig:SDRG_decimation}).

\paragraph{Physical interpretation.}
Each decimation step removes a strong bond by forming a singlet, while generating a weaker effective coupling between its neighbors.
Since \(\Gamma_{\rm eff}\propto (\Gamma_1\Gamma_2)/\Gamma'\), the new bond is always smaller than the eliminated one.
Iterating the step across the chain broadens the distribution of \(\ln \Gamma\), driving the flow to an infinite-randomness fixed point where entanglement and correlations are governed by rare long-distance singlets. 

In the next subsection following \cite{Refael:2004zz, Refael_2009} we illustrate how entanglement is connected to singlet counts over the boundary of the region.
\subsection{Second R\'enyi entropy from singlet counting}
\label{subsec:S2_example}

For $n=2$ replicas the purity of a region $A$ can be written as an overlap in the replicated spin language (cf.\ Sec.~\ref{sec:n2q2}):
\begin{equation}
\label{eq:puroverlSpin_repeat}
 \mathcal{P}_2(A)\;=\;\lim_{t\to\infty}
 \frac{\Tr \!\left[ \rho^{(2)}(t)\,(\mathcal{C}_{2,A}\!\otimes\!\mathbb{I})\right]}{\Tr \rho^{(2)}(t)}
 \;=\;\frac{1}{Z}\,
 \big\langle \otimes_{j\in A}\!\rightarrow_j \,\otimes_{i\in B}\!\uparrow_i \,\big|\, \mathrm{GS}\big\rangle,
 \qquad 
 Z=\langle \mathbb{I}\,|\,\mathrm{GS}\rangle,
\end{equation}
where $|\mathrm{GS}\rangle$ is the ground state of the (random) antiferromagnetic Heisenberg Hamiltonian obtained above,
$\ket{\rightarrow_j}$ and $\ket{\uparrow_i}$ are fully polarized product states in the $+x$ and $+z$ directions, and
$\mathcal{C}_{2,A}$ is the replica swap on $A$. 

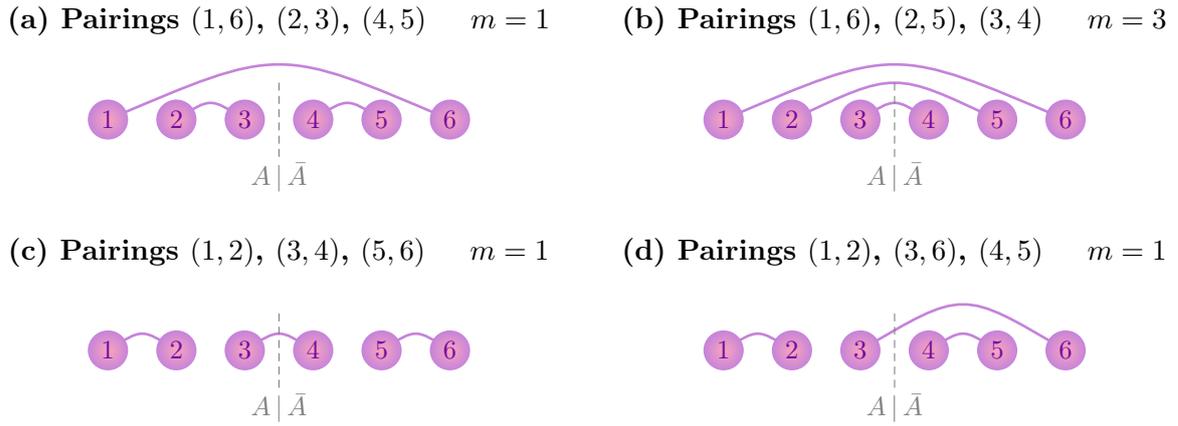
\begin{figure}[t]
\centering
\begin{tikzpicture}[x=0.9cm,y=0.9cm]

\pgfdeclarelayer{bg}
\pgfsetlayers{bg,main}

\definecolor{PurpA}{RGB}{250,165,188}   
\definecolor{PurpB}{RGB}{197,131,218}   
\definecolor{PurpC}{RGB}{112,13,161}    

\tikzset{
  spinP/.style  ={circle, draw=none, minimum size=5.3mm, inner sep=0pt,
                  shading=radial, inner color=PurpA, outer color=PurpB},
  nlab/.style   ={font=\footnotesize, inner sep=0pt, outer sep=0pt, text=PurpC},
  lab/.style    ={font=\footnotesize, inner sep=1pt, outer sep=1pt},
  title/.style  ={font=\bfseries\small},
  cutline/.style={densely dashed, color=gray},
  singarc/.style={draw=PurpB, line width=1.0pt, line cap=round,
                  shorten >=0.6pt, shorten <=0.6pt}, 
}

\def\ArcScale{1.8} 

\newcommand{\singlet}[3]{%
  \begin{pgfonlayer}{bg}
    \draw[singarc]
      (#1,0) .. controls ($ (#1,0)!0.5!(#2,0) + (0,{#3*\ArcScale}) $)
              and      ($ (#1,0)!0.5!(#2,0) + (0,{#3*\ArcScale}) $)
      .. (#2,0);
  \end{pgfonlayer}
}

\newcommand{\panel}[4]{%
  \begin{scope}[shift={({#1},{#2})}]
    \node[title] at (3.5,1.45) {#4};

    #3

    \foreach \i in {1,...,6} {
      \node[spinP] (p\i) at (\i,0) {};
      \node[nlab] at (p\i) {\i};
    }

    \draw[cutline] (3.5,-0.55)--(3.5,0.55);
    \node[lab, gray] at (3.5,-0.85) {$A\,|\,\bar A$};
  \end{scope}
}


\panel{0}{0}{%
  \singlet{1}{6}{0.60}
  \singlet{2}{3}{0.18}
  \singlet{4}{5}{0.18}
}{(a) Pairings $(1,6)$, $(2,3)$, $(4,5)$ \quad $m=1$}

\panel{9}{0}{%
  \singlet{1}{6}{0.60}
  \singlet{2}{5}{0.40}
  \singlet{3}{4}{0.18}
}{(b) Pairings $(1,6)$, $(2,5)$, $(3,4)$ \quad $m=3$}

\panel{0}{-3.4}{%
  \singlet{1}{2}{0.18}
  \singlet{3}{4}{0.18}
  \singlet{5}{6}{0.18}
}{(c) Pairings $(1,2)$, $(3,4)$, $(5,6)$ \quad $m=1$}

\panel{9}{-3.4}{%
  \singlet{1}{2}{0.18}
  \singlet{3}{6}{0.50}
  \singlet{4}{5}{0.18}
}{(d) Pairings $(1,2)$, $(3,6)$, $(4,5)$ \quad $m=1$}

\end{tikzpicture}
\caption{Six-spin singlet coverings across an $A|\bar A$ cut between sites $3$ and $4$. A singlet crossing the dashed cut contributes one crossing; the number of crossings $m$ determines the purity
$\mathcal{P}_2=2^{-m/2}$ and the second R\'enyi entropy $S_2=\tfrac{m}{2}\log 2$.}
\label{fig:6spins}
\end{figure}

Recall that the staggered $\pi$ rotation $U_y$
flips $S^{x,z}$ on odd sites, so the boundary product state must be chosen consistently with flipped spins:
$\ket{\rightarrow_{2j+1}}\!\mapsto\!\ket{\leftarrow_{2j+1}}$ and
$\ket{\uparrow_{2j+1}}\!\mapsto\!\ket{\downarrow_{2j+1}}$.
Further we consider example of $6$ spins to show the connection betwenn purity and singlet counting.
\paragraph{Six--spin example.}
Consider $L{=}6$ with $A=\{1,2,3\}$ and $\overline{A}=\{4,5,6\}$. For a given realization, the SDRG ground state is a product of singlets,
\(
|\mathrm{GS}\rangle = \bigotimes_{\langle i,j\rangle} s_{ij},
\)
with $s_{ij}=(|\uparrow_i\downarrow_j\rangle-|\downarrow_i\uparrow_j\rangle)/\sqrt{2}$.
Figure~\ref{fig:6spins} shows four representative singlet coverings; evaluating the overlap in
Eq.~\eqref{eq:puroverlSpin_repeat} one finds for each realisation of the singlet covering
\begin{equation}
\label{eq:6spin_overlaps}
\begin{aligned}
     &\text{(a)}\;\; \frac{1}{Z}\,\big\langle \leftarrow_1\,\rightarrow_2\,\leftarrow_3\,\uparrow_4\,\downarrow_5\,\uparrow_6 \big|\, s_{16}\!\otimes\! s_{23}\!\otimes\! s_{45}\big\rangle
     \;=\;\frac{1}{\sqrt{2}},\\[2pt]
     &\text{(b)}\;\; \frac{1}{Z}\,\big\langle \leftarrow_1\,\rightarrow_2\,\leftarrow_3\,\uparrow_4\,\downarrow_5\,\uparrow_6 \big|\, s_{16}\!\otimes\! s_{25}\!\otimes\! s_{34}\big\rangle
     \;=\;\frac{1}{2\sqrt{2}},\\[2pt]
     &\text{(c)}\;\; \frac{1}{Z}\,\big\langle \leftarrow_1\,\rightarrow_2\,\leftarrow_3\,\uparrow_4\,\downarrow_5\,\uparrow_6 \big|\, s_{12}\!\otimes\! s_{34}\!\otimes\! s_{56}\big\rangle
     \;=\;\frac{1}{\sqrt{2}},\\[2pt]
     &\text{(d)}\;\; \frac{1}{Z}\,\big\langle \leftarrow_1\,\rightarrow_2\,\leftarrow_3\,\uparrow_4\,\downarrow_5\,\uparrow_6 \big|\, s_{12}\!\otimes\! s_{36}\!\otimes\! s_{45}\big\rangle
     \;=\;\frac{1}{\sqrt{2}}.
\end{aligned}
\end{equation}
In all cases the normalization is $Z=\langle\downarrow_1\uparrow_2\downarrow_3\uparrow_4\downarrow_5\uparrow_6|\mathrm{GS}\rangle$.The pattern in Eq.~\eqref{eq:6spin_overlaps} is generic and admits a simple rule:

\begin{quote}
\textit{Each singlet whose two endpoints lie on opposite sides of the $A|\bar A$ cut contributes a factor $2^{-1/2}$ to the overlap in Eq.~\eqref{eq:puroverlSpin_repeat}; singlets contained entirely within $A$ or entirely within $\bar A$ contribute a factor $1$.}
\end{quote}

This rule follows from (i) factorization of the GS overlap across independent singlets and
(ii) the elementary projections of a two--spin singlet onto product states on opposite sides of the cut
(with the staggered basis choice inherited from $U_y$).
If we denote by $m$ the number of singlet links crossing the $A|\bar A$ boundary in the singlet covering,
the purity is therefore
\begin{equation}
\label{eq:purity_rule}
  \mathcal{P}_2(A)\;=\;2^{-\,m/2},
  \qquad\Rightarrow\qquad
  \boxed{\; S_2(A)\;=\;-\log \mathcal{P}_2(A)\;=\;\frac{m}{2}\,\log 2.\;}
\end{equation}

In the six--spin examples of Fig.~\ref{fig:6spins}, panels (a), (c), and (d) have a single crossing
($m{=}1$), while panel (b) has three crossings ($m{=}3$), precisely reproducing the values in Eq.~
\eqref{eq:6spin_overlaps}. Equation~\eqref{eq:purity_rule} is the microscopic origin of the
singlet--counting interpretation of $S_2$ used throughout this section. Since $S_2=\frac{m}{2}\ln 2$, the law of $S_2$ is the pushforward of the crossing–number distribution.
Let $\Pr\big(m=M\big)$ be the probability that a block of large length $L$ has $M$ crossings with the outside area under the SDRG. Then
\begin{equation}\label{eq:links_ent}
\begin{split}
P_{S_2}(s)=\sum_{M\ge0} \Pr\big(m=M\big)\,\delta\!\left(s-\frac{M}{2}\ln 2\right),\\\qquad
\langle S_2\rangle=\frac{\ln 2}{2}\,\langle m\rangle,\quad
\mathrm{Var}(S_2)=\left(\frac{\ln 2}{2}\right)^2\mathrm{Var}(m).
\end{split}
\end{equation}
Thus the entanglement problem reduces to determining $\Pr\big(m=M\big)$  from the RG flow of boundary–crossing singlets. Or in simpler words, it reduces to finding a probability distribution of the number of links formed over the boundary of the region.

\subsection{RG flow of the strengths of couplings}
At each step we decimate the largest bond $\Omega$ (forming a singlet across it) and replace its neighbors
$\Gamma_L,\Gamma_R$ by an effective coupling
$\Gamma'=\Gamma_L\Gamma_R/(2\Omega)$, see Fig. \ref{fig:SDRG_decimation}. The RG leads to an integrodifferential flow equation
for the bond coupling distribution (see Appendix \ref{app:RG-bulk-P}). 
\begin{equation}
\label{eq:evolPOmega}
    \partial_\Omega P(\Gamma, \Omega) =
    - P(\Omega, \Omega) \int d\Gamma_1 d\Gamma_2 P(\Gamma_L, \Omega) P(\Gamma_R, \Omega) 
    \delta\left(\Gamma - \frac{\Gamma_L \Gamma_R}{2\Omega}\right).
\end{equation}

And in logarithmic variables $\beta=\ln(\Omega/\Gamma)$ with RG time
$\ell=\ln(\Omega_0/\Omega)$ the deccimation rule is
\[
\beta'=\beta_1+\beta_2+\ln 2,
\] where $\Omega_0$ is the
Hamiltonian’s initial energy scale, and $\Omega$ is its reduced
energy scale at a given RG step.  And in terms of these variables we have the flow equation for the couplings distribution:
\begin{equation}
    \partial_\ell P_\ell(\beta) - \partial_\beta P_\ell(\beta) =  P_\ell(0) \int d\beta_{1}d\beta_2 P_{\ell}(\beta_1) P_{\ell}(\beta_2) \delta(\beta - (\beta_1 + \beta_2 + \ln 2)),
\end{equation}
where $P_{\ell} (\beta) = P(\Gamma,\Omega) |\frac{d\Gamma}{d\beta}| = P(\Gamma, \Omega) e^{-\beta}\Omega$.
Notice that these probabilities are not the probabilities of forming singlets across the boundary $\Pr\big(m=M\big)$, this we will discover in the future sections. $P_{\ell}(\beta)$ is describing the probability distribution of  strenghts of couplings over all the chain.
In the limit of large $\ell$ (so after many RG iterations), we look for a fixed point of this equation in the form
\begin{equation}
    P_{\ell}(\beta) = \frac{1}{\ell} g(\beta/\ell)
\end{equation}
which says that the $\beta$ are growing with the scale $\ell$. Replacing this ansatz in the equation
we arrive at
\begin{equation}
    g(x) + (1+x)g'(x) = - g(0) \int dx_1 dx_2 g(x_1) g(x_2) \delta(x - x_1 - x_2)
\end{equation}
which is easily solved by $g(x) = e^{-x}$. So we obtain the asymptotic solution at large $\Gamma$
\begin{equation}
\label{eq:fixedpoint}
    P_\ell(\beta) = \frac{1}{\ell} e^{-\beta/\ell}.
\end{equation}
In the next subsection we find the connection between $P_{\ell}(\beta)$ and probability distribution of the number of singlets $\Pr(m=M)$.
\subsection{Distribution of singlets across a marked bond }
\label{sec:singlet-dist-summary}
Here we fix a bond $B$ and let $\rho(t)$ be the survival probability that no new singlet forms across $B$ up to RG time $t:=\ln(\ell/\ell_0)$ after the last decimation at $\ell_0$.
At the infinite-randomness fixed point (IRFP) Eq.~\eqref{eq:fixedpoint}, the post-decimation state of $B$ is scale-invariant, so inter-decimation intervals are i.i.d.; singlet events form a classical renewal process. All derivations are deferred to App.~\ref{app:renewal-moments}; here we summarize the formulas used later.

\paragraph{Survival and waiting time.}
The survival probability found in Appendix \ref{app:renewal-moments} is
\begin{equation}
\label{eq:rho-main}
\rho(t)=\frac{e^{-3t/2}}{5}\!\left[\,3\sqrt{5}\,\sinh\!\Big(\tfrac{\sqrt{5}\,t}{2}\Big)
+5\,\cosh\!\Big(\tfrac{\sqrt{5}\,t}{2}\Big)\right],
\end{equation}
and the waiting-time density is $f(t)=-\rho'(t)$.
Equivalently, in Laplace space,
\begin{equation}\label{eq:ff}
\tilde f(u)=\mathcal{L}\{f\}(u)=\frac{1}{u^2+3u+1}\,.
\end{equation}
\paragraph{Counts via renewal theory.}
Let $m(t)$ be the number of singlets across $B$ up to time $t$.
“Exactly $M$ events by time $t$” is the convolution of $M$ waiting times with a final survival:
\[
\Pr\big(m(t)=M\big) = (f^{*M} * \rho)(t).
\]
In Laplace space this becomes a product,
\begin{equation}
\label{eq:exactM-laplace-main}
\mathcal{L}\{\Pr(m(t)=M)\}(u)
=\tilde f(u)^M\,\tilde\rho(u)
=\frac{u+3}{\big(u^2+3u+1\big)^{M+1}}.
\end{equation}
Closed-form inversion is unwieldy; for large $t$ (or large $M$) one may evaluate the Bromwich integral by steepest descent.  
Rather than approximate the full distribution, in the next subsection we compute the moments $\langle m(t)^n\rangle$ exactly via the generating function.  
Using, $\widehat Z(q,u)$, derivatives at $q=1$ yield the Laplace transforms of all moments; explicit formulas and inversions are collected in Appendix~\ref{app:renewal-moments}.

\subsection{Average entropy and higher moments}
We consider the open Heisenberg chain in Eq.~\eqref{eq:Mon_Ham} with area $A$ attached to the edge, so only a single boundary contributes to the count of singlets. Since, one can connect the moments of the entropy to the moments of number of singlet across the boundary through Eq.~\eqref{eq:links_ent}, we compute moments of the number of boundary–crossing singlets $m(t)$ via the joint Laplace–$t$ transform of the probability–generating function $Z(q,t)=\sum_{M\ge0}\Pr(m(t)=M) q^M$. Using convolution in Laplace space, one obtains (see App.~\ref{app:renewal-moments})
\begin{equation}
\label{eq:ZLap}
\widehat Z(q,u):=\mathcal{L}\{Z(q,t)\}(u)=\frac{1-\tilde f(u)}{u[1-q \tilde f(u)]},
\end{equation}
where $\tilde f(u)$ is given by Eq.~\eqref{eq:ff}.
Derivatives at $q=1$ generate the Laplace transforms of the moments, e.g.\ $\mathcal{L}\{\langle m(t)\rangle\}(u)=\partial_q \widehat Z_{q=1}$ and similarly for higher cumulants. Therefore we find
\begin{equation}
\label{eq:momnets}
\begin{split}
\langle m(t)\rangle = \frac{t}{3} + \mathcal{O}(1) + \mathcal{O}(e^{-3t}),~~\mathrm{Var}\,m(t) = \frac{7}{27}\,t + \mathcal{O}(1) + \mathcal{O}(e^{-3t}),\\\mathrm{Skew}~m(t)= \frac{13}{7}\sqrt{\frac{3}{7}}\;\frac{1}{\sqrt{t}} \;+\; \mathcal{O}\!\left(t^{-3/2}\right),~~\mathrm{Kur}~m(t)= 3+\frac{41}{49}\frac{1}{t}+\; \mathcal{O}\!\left(\frac{1}{t^2}\right).
\end{split}
\end{equation}

the mean count grows linearly. The variance is diffusive, so fluctuations grow $\propto t$ with the universal coefficient $7/27$, up to subleading constants and exponentially small terms. Standardized higher moments relax to their Gaussian limits. The skewness decay is
indicating that the distribution becomes increasingly symmetric at late times, and the kurtosis approaches the Gaussian value.

Combining Eq.~\eqref{eq:links_ent} with Eqs.~\eqref{eq:momnets} we find behaviour of the moments of the second Renyi entropy:
\begin{equation}
\begin{split}
&\langle S_2(A;t)\rangle=\frac{\ln 2}{2} \left<m\right>,
\qquad
\mathrm{Var}\,S_2(A;t)=\left(\frac{\ln 2}{2}\right)^2\,\mathrm{Var}\,m(t),\\
&\mathrm{Skew}~S_2(A;t)= \mathrm{Skew}~m(t),\qquad\mathrm{Kur}~S_2(A;t)= \mathrm{Kur}~m(t),
\end{split}
\end{equation}
Full derivations are collected in App.~\ref{app:renewal-moments}. 
In the next subsection we repeat the SDRG derivations developed for SO($N$) random chains 
\cite{Quito2019_HighlySymmetric,QuIto2020_EmergentSUNinSON} 
(see also the SU($N$) generalization \cite{HoyosMiranda2004_SU_N} 
and the spin-1 case with emergent SU(3) \cite{Quito2015_SU3}), 
we obtain the moments of the second R\'enyi entropy for measurement-only dynamics.

\section{SRGD for $n$ replicas in the \texorpdfstring{$\mathfrak{so}(2n)$}{so(2n)} generator language}
\label{sec:DM-Phi}
In this section we consider strong disorder renomalisation group approach for generic $n$. We adapt the Dasgupta–Ma strong-disorder renormalization procedure to our replica generators $\hat \Phi$, then take the replica limit $n\to 1$ to extract results for the physical system. At each site $i$ we use adjoint-index generators $\hat\Phi_i^{\alpha\beta}\in\mathfrak{so}(2n)$ with commutation relations from the previous chapter  
\begin{equation}\label{eq:comm_ch4}
   \left[\hat{\Phi}_i^{\alpha \beta}, \hat{\Phi}_j^{\lambda \xi}\right]  =   \delta_{ij} \frac{i}{N_F} \times  (\delta_{\beta, \lambda} \hat\Phi_{i}^{\alpha \xi}
   -\delta_{\beta, \xi} \hat\Phi_{i}^{\alpha \lambda}
   +\delta_{\alpha, \xi} \hat\Phi_{i}^{\beta \lambda}
   -\delta_{\alpha, \lambda} \hat\Phi_{i}^{\beta \xi}
   ).
\end{equation}

The nearest-neighbour Hamiltonian Eq.\eqref{eq:Hmeas} is
\begin{equation}\label{eq:Ham_mon_ch4}
  \mathcal{H}_{\rm mon}
  =  \,\frac{N_F^2}{4}
    \sum_{j,\alpha,\alpha'} \Gamma_j\sigma\sigma'\,
    \hat\Phi_j^{\alpha\alpha'} \hat\Phi_{j+1}^{\alpha\alpha'},
\end{equation}
where $\Gamma_j>0$. As in the previous section we can perform canonical transformation and cast the Hamiltonian in form of "antiferromagnet". So we set on the odd sites $\hat \Phi^{(+a,-,a')}_{2j+1}\rightarrow  -\hat\Phi^{(+a,-,a')}_{2j+1}$ and $\hat \Phi^{(-a,+,a')}_{2j+1}\rightarrow  -\hat\Phi^{(-a,+,a')}_{2j+1}$. With complete analogy to Eq.~\eqref{eq:Mon_Ham} we get the Hamiltonian
\begin{equation}\label{eq:Ham_mon_ch4_b}
  \mathcal{H}_{\rm mon}
  =  \,\frac{N_F^2}{4}
    \sum_{j,\alpha,\alpha'} \Gamma_j
    \hat\Phi_j^{\alpha\alpha'} \hat\Phi_{j+1}^{\alpha\alpha'}.
\end{equation}
So now the problem is about generalisation of the Dasgupta-Ma procedure to the group $SO(2n)$. Similar porblems were already considered in \cite{Quito2019_HighlySymmetric,QuIto2020_EmergentSUNinSON}  following these papers we derive the procedure for our case.
\subsection{Dasgupta-Ma procedure generic n} Here we again consider one step of RG procedure . We take the chain of $4$ particles where the middle bond has much stronger coupling than two others $\Gamma' \gg \Gamma_1,\Gamma_2$. 
\paragraph{Step 1: unperturbed versus perturbing parts.}
We again treat the strong bond as the unperturbed Hamiltonian,
$H_0 =\frac{N_F^2}{4} \Gamma'\, \sum_{\alpha, \alpha'}\hat \Phi^{\alpha \alpha'}_2 \hat \Phi^{\alpha \alpha'}_3 $, and the weaker neighboring couplings as the perturbation,
$V =\frac{N_F^2}{4}  \Gamma_1 \sum_{\alpha, \alpha'}\hat \Phi^{\alpha \alpha'}_1 \hat \Phi^{\alpha \alpha'}_2 +\frac{N_F^2}{4}  \Gamma_2 \sum_{\alpha, \alpha'}\hat \Phi^{\alpha \alpha'}_3 \hat \Phi^{\alpha \alpha'}_4$.
\paragraph{Step 2: spectrum of the central pair.}
Let the on–site algebra be \(\mathfrak g=\mathfrak{so}(2n)\) with generators
\(\{\Phi^{\alpha\alpha'}\}_{\alpha<\alpha'}\) acting in a fixed irrep \(R\) at each site.
We write the strong bond as
\begin{equation}
\label{eq:H0-Phi}
H_0 \;=\; \frac{N_F^2}{4} \Gamma'\, \sum_{A} \Phi^{A}_{2}\,\Phi^{A}_{3},
\qquad
A\equiv(\alpha\alpha'),\ \alpha<\alpha',
\end{equation}
where the sum is contracted with the invariant metric (we suppress it for brevity).
Introduce the ``total'' generators on the pair \((2,3)\),
\begin{equation}
J^{A}\ :=\ \Phi^{A}_{2}+\Phi^{A}_{3}\,,
\qquad
J^2\ :=\ \sum_A J^{A}J^{A},
\end{equation}
which realize the diagonal action of \(\mathfrak g\) on \(R\otimes R\) \cite{FultonHarris,Yamatsu2015}.
Using \((\Phi_2+\Phi_3)^2=\Phi_2^2+\Phi_3^2+2\sum_A \Phi^A_2\Phi^A_3\) we obtain the operator identity
\begin{equation}
\label{eq:Phi-identity}
\sum_A \Phi^{A}_{2}\,\Phi^{A}_{3}
\;=\;
\frac12\Big(J^2 - \Phi_2^2 - \Phi_3^2\Big)
\;=\;
\frac12\Big( C_2^{(23)} - C_2^{(2)} - C_2^{(3)}\Big),
\end{equation}
where \(C_2^{(i)}:=\sum_A \Phi^A_i\Phi^A_i\) is the quadratic Casimir on site \(i\), and
\(C_2^{(23)}:=\sum_A J^A J^A\) is the Casimir for the \emph{pair}.

The tensor product \(R\otimes R\) decomposes into irreps
\(
R\otimes R = \bigoplus_{\Lambda} \Lambda
\),
and on the subspace where the pair transforms in a particular irrep \(\Lambda\),
\begin{equation}
\label{eq:eigenvalue-Phi}
\sum_A \Phi^{A}_{2}\,\Phi^{A}_{3}
\ \longrightarrow\
\frac12\Big(C_2(\Lambda) - 2\,C_2(R)\Big)\,\mathbb{I}_{\Lambda}.
\end{equation}
Equivalently,
\begin{equation}
\label{eq:projector-decomp}
\sum_A \Phi^{A}_{2}\,\Phi^{A}_{3}
\;=\;
\sum_{\Lambda\subset R\otimes R}
\frac12\Big(C_2(\Lambda) - 2\,C_2(R)\Big)\,\mathcal P^{(23)}_{\Lambda},
\end{equation}
with \(\mathcal P^{(23)}_{\Lambda}\) the projector onto the \(\Lambda\)–isotypic component. For the \(n=2\) warm–up, the two–site space decomposes as \(R\otimes R=\mathbf{1}\oplus\mathbf{3}\), i.e. \(\Lambda\in\{s,t\}\) (singlet and triplet).

Therefore the eigenstates of the strong bond \(H_0\) are classified by the
\emph{total} irrep \(\Lambda\) of the pair \((2,3)\), and their energies are
\begin{equation}
\label{eq:E-Lambda}
E(\Lambda)
\;=\;
\frac{N_F^2}{8}\Gamma'\,\Big(C_2(\Lambda) - 2\,C_2(R)\Big).
\end{equation}
For antiferromagnetic coupling \(\Gamma'>0\), the ground multiplet is the \(\Lambda\) with the
\emph{smallest} Casimir \(C_2(\Lambda)\) among those appearing in \(R\otimes R\) (e.g.\ the singlet if present). 

\paragraph{On-site irrep fixed by flavor symmetry.}
Because each site carries $N_F$ identical flavors in the vector of $SO(2n)$ and both the initial state and the couplings involve only flavor–summed generators (hence are invariant under flavor permutations), the dynamics remains in the fully symmetric sector. Consequently, the on–site irrep is \emph{fixed} to be the completely symmetric rank-$N_F$ representation,
\[
R=\mathrm{Sym}^{N_F}(2n),
\]
i.e., the highest–weight irrep with weight $N_F\,\omega_1$. This is the $SO(2n)$ analogue of taking spin $S=N_F/2$ for $SU(2)$: $N_F$ fundamentals projected onto the ferromagnetically aligned, flavor–symmetric sector (one–row Young diagram of length $N_F$)
 \cite{FultonHarris,Slansky1981,Yamatsu2015,CornwellVol1}. 
Operationally we use flavor-summed generators with an \(N_F\)-independent normalization,
\begin{equation}
\label{eq:Phi-hat-def}
\hat\Phi^{A}\;=\;\frac{1}{N_F}\sum_{\nu=1}^{N_F} t^{A}_{(\nu)} \qquad
\big[t^A,t^B\big]=f^{AB}{}_C\,t^C,
\end{equation}
so that the single-site quadratic Casimir in this normalization stays \(O(1)\) at large \(N_F\)
\cite{CvitanovicBook,Georgi:1999wka,vanRitbergen1999}.
In order to find this representation we perform a Jordan–Wigner transformation \cite{JordanWigner1928}  to express \(\hat \Phi^A\) in terms of spins-1/2, which fixes the \(t^A_{(\nu)}\) (see Appendix~\ref{app:JW-Phi}). 
Notice that in our case the generators’ normalization is not conventional because of the \(\frac{1}{N_F}\) factor in Eq.~\eqref{eq:comm_ch4}. Therefore the difference between our Casimir \(C_2\) and the usual \(\mathfrak{so}(2n)\) Casimir \(\widehat C_2\) is
\begin{equation}
\label{eq:Casimir-Phi}
\sum_A \hat\Phi^A \hat\Phi^A =  C_2(R)\,\mathbb{I},\qquad
 C_2(R):=\frac{\widehat C_2(R)}{N_F^2}.
\end{equation}
For the symmetric traceless tensors (highest weight \(N_F\,\omega_1\)) the standard eigenvalue (long roots of length-squared \(2\)) is \(\widehat C_2({\rm Sym}^{N_F})=N_F(2n+N_F-2)\) \cite{OkuboPatera1983,Yamatsu2015,Slansky1981}, hence with our normalization Eq.~\eqref{eq:Phi-hat-def},
\begin{equation}\label{eq:Casimir}
C_2({\rm Sym}^{N_F})\;=\;\frac{1}{N_F^2}\,N_F\big(N_F+2n-2\big).
\end{equation}
With this \(R\), the tensor product \(R\otimes R\) contains a singlet (full contraction of the two symmetric tensors) \cite{FultonHarris,CornwellVol1}; therefore for antiferromagnetic \(\Gamma'>0\) the ground multiplet of the strong bond is the singlet \(\Lambda=\mathbf{1}\), and Eq.~\eqref{eq:E-Lambda} gives
\[
E_s=E(\mathbf{1})=-\frac{N_F^2}{4}\,\Gamma'\,C_2(R)
=-\Gamma'\,\frac{N_F\,(N_F+2n-2)}{4}\,.
\]

\paragraph{Step 3: effective Hamiltonian from second order.}

Finally, in the \emph{second–order} decimation 
\begin{equation}\label{eq:PertPhi}
\begin{split}
   & H_{\rm eff}=E_s+\left<s \left|V \right|s \right>+\sum_{n} \frac{|\left<s \left|V \right|n\right>|^2}{E_s-E_n}\\
   &\qquad=\frac{N_F^4}{16} \sum_n \frac{2 \Gamma_1 \Gamma_2 \hat \Phi^{\alpha \beta}_1\left<s\right|\hat \Phi^{\alpha \beta}_2 \left|n\right>\left<n\right|\hat \Phi^{\gamma \delta}_3  \left|s\right>\hat \Phi^{\gamma \delta}_4}{E_s-E_n}+{\rm const},
\end{split}
\end{equation}
which is the standard second–order (Schrieffer–Wolff/Kato) effective Hamiltonian \cite{SchriefferWolff1966,KatoPT,Bravyi2011}. 
Here let us show that $\hat \Phi$ transforms in the adjoint representation. 
Let $J^B=\Phi_2^B+\Phi_3^B$ and let $\ket{s}$ be the $(2,3)$ singlet, so $J^B\ket{s}=0$. 
Using the Lie brackets $[J^B,\Phi_2^A]=i f^{BA}{}_{C}\,\Phi_2^{C}$ (same on site~3), we have
\begin{equation}
\label{eq:adjoint-multiplet}
J^B\big(\Phi_2^A\ket{s}\big)
=\big[\,J^B,\Phi_2^A\,\big]\ket{s}
= i f^{BA}{}_{C}\,\Phi_2^{C}\ket{s},
\end{equation}
so the $\dim\mathfrak g$ states $\{\Phi_2^A\ket{s}\}$ transform with matrices $(i f^{B})^{A}{}_{C}$, i.e.\ as the \emph{adjoint} irrep of the pair \cite{Georgi:1999wka,CvitanovicBook}. 
In particular, only the adjoint intermediate manifold can be reached by a single action of $V$ on $\ket{s}$; thus the second–order sum can be restricted to the adjoint subspace (projector/completeness in the adjoint) \cite{CvitanovicBook}. 
Therefore for the effective Hamiltonian we have:
\begin{equation}
\begin{split}
       & H_{\rm eff}=\frac{N_F^4}{16} \sum_{\rm adj} \frac{2 \Gamma_1 \Gamma_2 \hat \Phi^{\alpha \beta}_1\left<s\right|\hat \Phi^{\alpha \beta}_2 \left|{\rm adj}\right>\left<{\rm adj}\right|\hat \Phi^{\gamma \delta}_3  \left|s\right>\hat \Phi^{\gamma \delta}_4}{E_s-E_{\rm adj}}.
    \end{split}
\end{equation}

For $R\otimes R=\mathbf 1\oplus \mathrm{adj}\oplus(\text{traceless symm})$, the gap to the adjoint sector from Eq.~\eqref{eq:E-Lambda} is
\begin{equation}
\label{eq:Delta-generic}
\Delta \equiv E_{\mathrm{adj}}-E_s=\frac{\Gamma' N_F^2}{8}\,C_2(\mathrm{adj})=
\frac{\Gamma'}{4}\,(n{-}1)\qquad(n\ge 3),
\end{equation}
since $C_2(\mathrm{adj})=\frac{2(n{-}1)}{N_F^2}$ for $SO(2n)$ in the long–root normalization \cite{OkuboPatera1983,Yamatsu2015,Slansky1981}. 
Since $\hat \Phi_2\ket{s}$ is in the adjoint, the sum $\sum_{\rm adj} \ket{\rm adj}\bra{\rm adj}$ acts as the identity on this subspace.

On the singlet of $(2,3)$ one also has $(\hat\Phi_2^A+\hat\Phi_3^A)|s\rangle=0$. Hence
\begin{equation}\label{eq:Heff_Phi}
\begin{split}
       & H_{\rm eff}=\frac{N_F^4}{16}\; \frac{2 \Gamma_1 \Gamma_2 \; \hat \Phi^{\alpha \beta}_1\;\left<s\right|\hat \Phi^{\alpha \beta}_2 \hat \Phi^{\gamma \delta}_2  \left|s\right>\;\hat \Phi^{\gamma \delta}_4}{E_{\rm adj}-E_{s}}.
    \end{split}
\end{equation}
For the overlap, invariance fixes
\begin{equation}
\label{eq:singlet-relations}
\langle s|\,\hat\Phi_2^A \hat\Phi_2^B\,|s\rangle
=\frac{C_2(R)}{\,d_G}\,\delta^{AB}.
\end{equation}
Inserting Eq.~\eqref{eq:Casimir} into Eq.~\eqref{eq:Heff_Phi} with $d_G=n(2n{-}1)$ and $C_2(\mathrm{adj})=2(n{-}1)$ yields, for $n>2$,
\begin{equation}
\label{eq:Heff-final}
\boxed{\;
\mathcal H_{\rm eff}
=\frac{N_F^2}{4}\;
\frac{\Gamma_1\Gamma_2}{\Gamma'\,(n-1)(2n-1)}\;
N_F\big[N_F+2(n-1)\big]\;
\mathrm{tr}\!\big(\hat\Phi_1\hat\Phi_4^{\mathsf T}\big).
\;}
\end{equation}

\paragraph{Special case $n=2$.}
$\mathfrak{so}(4)$ is not simple:
\(
\mathfrak{so}(4)\cong \mathfrak{su}(2)_L\oplus\mathfrak{su}(2)_R
\),
so the adjoint splits as
\(
\mathrm{Adj}=(\mathbf 3,\mathbf 1)\oplus(\mathbf 1,\mathbf 3)
\)
with equal gaps; both blocks contribute equally in the second–order sum, doubling the matrix–element sum \cite{CornwellVol2,Georgi:1999wka}. 
In the parity-reduced $n{=}2$ pseudospin mapping used in Sec.~\ref{sec:n2q2}, this appears directly as
\(
\mathcal H_{23}=\Gamma'\,\vec S_2\!\cdot\!\vec S_3
\Rightarrow \Delta=\Gamma'
\)
for the $j{=}0\to1$ excitation of two spins.

\paragraph{Adjoint gap and the $n\to1$ limit.}
From Eq.~\eqref{eq:E-Lambda} with $\Lambda=\mathrm{Adj}$ we obtain Eq.~\eqref{eq:Delta-generic}. 
Thus the Schrieffer–Wolff energy denominator vanishes linearly as $n\to1$ (since $SO(2)\simeq U(1)$ is abelian), and the induced coupling develops a simple pole:
\[
\Gamma_{\rm eff}
\;=\;
\underbrace{\frac{N_F^{3}}{4}\,\frac{N_F+2(n-1)}{(n-1)(2n-1)}}_{=:K(n,N_F)}
\frac{\Gamma_1\Gamma_2}{\Gamma'}\,.
\]

The result depends on the order of limits. 

If we first take $\Gamma'\to\infty$ at fixed $n>1$ and only then send $n\to1$, the prefactor $K(n,N_F)$ does not alter the fixed point: the flow converges to the same Infinite Randomness fixed point as in the Heisenberg chain. 
However, if we set $n\to1$ \emph{before} the decimation, the pole does not allow one to discard this prefactor within the RG derivation (see Appendix~\ref{app:RG-bulk-P}). 
A full comparison of these procedures and their subleading consequences is left for future work.

\subsection{Numerical check at \texorpdfstring{$n{=}1$}{n=1}}
\label{subsec:numerics-n1}

\begin{figure}[t]
  \centering
  \includegraphics[width=\linewidth]{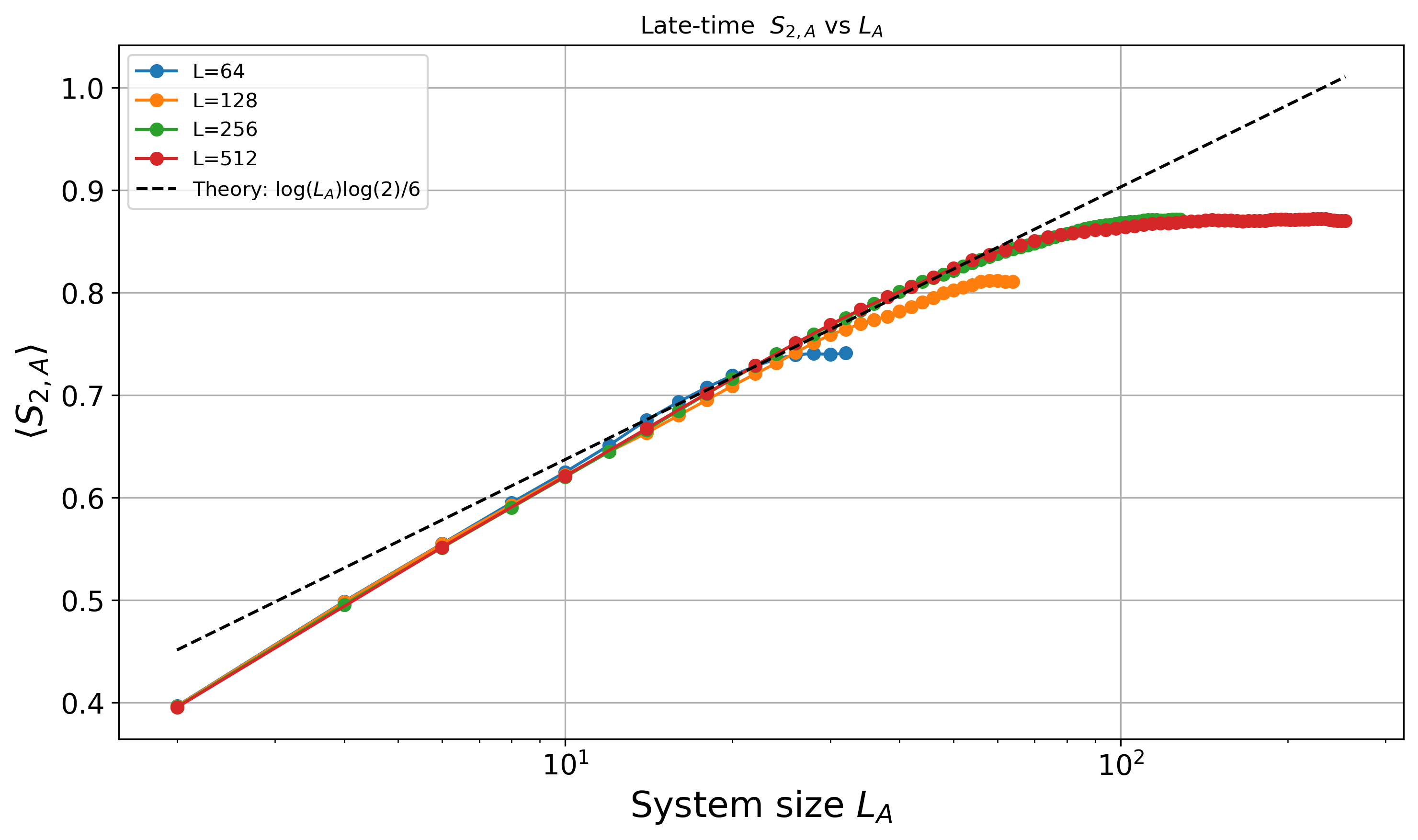}
  \caption{%
  Numerical check at \(n{=}1\).
  Late-time Rényi-2 entropy \(S_2(A)\) versus subregion size \(L_A\),
  obtained from Gaussian-Majorana simulations of a monitored chain
  with bond-resolved measurement rates.
  Each point is averaged over disorder realizations and over the last
  fraction of the time window. 
  The qualitative sublinear growth is consistent with the SDRG rule
  \eqref{eq:Heff-final} in the \(n\!\to\!1\) limit.
  }
  \label{fig:SDRG-n1-check}
\end{figure}

To test Eq.~\eqref{eq:Heff-final} at \(n{=}1\) we simulate the dynamics using the
\emph{stochastic Schrödinger equation (SSE)} for a monitored free-fermion
(Gaussian Majorana) chain, implemented in the covariance-matrix formalism.
Concretely:  monitoring is local on bonds with
heterogeneous, bond-resolved rates drawn from Cauchy distribution.
The state is represented by an orthogonal matrix \(O\) acting on the vacuum
covariance (see App.~\ref{app:gaussian-n1-code} ;
the Rényi-2 entropy of a contiguous block \(A\) of size \(L_A\) is computed from
the restricted covariance as \(S_2(A)=L_A\ln 2 - \sum_k \ln(1+\lambda_k^2)\),
with \(\{\lambda_k\}\) eigenvalues of \(iM_A\).
For each disorder realization we sample even/odd bond rates, evolve for with step \texttt{dt}, and average \(S_2(A)\) over the last
fraction of the time window (implementation details match the script used for
Fig.~\ref{fig:SDRG-n1-check}).

Even thought the limit $n\to1 $ of SDRG Eq.~
\eqref{eq:Heff-final} is ambiguous numerical analysis suggests the infinite-randomness (random-singlet) scenario.
And the entanglement is dominated by boundary-crossing singlets,
so \(S_2(A)\) grows sublinearly with \(L_A\) and, for broad disorder, is compatible
with the random-singlet prediction \(S_2(A)\sim \frac{\ln 2}{6}\ln L_A + \text{const}\) up to non-universal offsets.%
\footnote{See e.g.\ Fisher (1994, 1995) and Refael–Moore (2004).}
Figure~\ref{fig:SDRG-n1-check} shows the late-time \(S_2(A)\) versus \(L_A\) at \(n{=}1\),
averaged over disorder realizations. The data display the expected sublinear,
concave growth consistent with the infinite-randomness picture encoded by Eq.~
\eqref{eq:Heff-final}. A log fit \(S_2(A)\approx a+b\ln L_A\)
provides a slope compatible with the random-singlet coefficient within uncertainties.

\section{Conclusion}

In this chapter we asked whether the physics of infinite randomness—well established in random quantum magnets—can persist in \emph{monitored} fermionic systems with strongly inhomogeneous rates. Starting from the measurement-only model in Eq.~\eqref{eq:mon_only} with heavy-tailed $\Gamma_j$, we developed a strong-disorder renormalization procedure adapted to stochastic, effectively non-Hermitian dynamics. For any replica number $n>1$, the decimation step freezes the strongest monitored bond and generates a \emph{multiplicative} effective coupling between its neighbors. Iterating this rule broadens the distribution of logarithmic couplings without bound, and the flow exhibits an infinite-randomness fixed point (IRFP) with activated scaling,
\[
\ln\!\frac{\Omega_0}{\Omega(L)} \sim L^{\psi}\qquad (z=\infty).
\]
However, in the physical limit $n\to1$ an order-of-limits subtlety in the replica treatment ($n\!\to\!1$ versus the decimation $\Gamma'\!\to\!\infty$) introduces a simple pole and renders the leading RG recursions non-transparent. Consequently, our analytic control at $n \to 1$ is incomplete; establishing (or ruling out) an IRFP in the monitored setting requires further refinement of the IRFP derivation(see Appendix~\ref{app:RG-bulk-P}) in $n\to 1$ setting .

On the numerical side, stochastic Schrödinger–equation (SSE) trajectories at $n=1$ display entanglement features consistent with rare-event dominance in a putative IR regime. As in random-singlet phases, long-distance singlets crossing a cut control $S_A$. Within a renewal-process description of crossings we obtain that the \emph{average} second Rényi entropy scales logarithmically with subsystem size,
\[
S_{2,A}(L) \sim \frac{c_{\rm eff}}{6}\,\log L,
\]
with $c_{\rm eff}$ set by rare-bond statistics (recovering $c_{\rm eff}=\ln 2$ for the canonical spin-$1/2$ random-singlet case \cite{Refael:2004zz}). Our numerical data for the \emph{average} $S_{2}$ are consistent with this prediction. However, due to extremely broad distributions and lack of self-averaging, our disorder sampling was insufficient to obtain converged estimates for the \emph{variance} and higher moments; we therefore refrain from making firm statements about full distributional properties.

Conceptually, the picture is compatible with extending the random-singlet paradigm to measurement-driven dynamics, offering a disorder-controlled alternative to clean measurement-induced criticality \cite{LiChenFisher2018,SkinnerRuhmanNahum2019,BaoChoiAltman2020,NahumRuhmanAllToAll2021}. Practically, this motivates concrete, testable signatures—logarithmic growth of the \emph{average} entanglement with subsystem size and strong run-to-run fluctuations—amenable to randomized-measurement protocols \cite{Brydges_2019,Elben_2018}. Still, given the present order-of-limits ambiguity and the limited convergence of higher moments, these signatures should be interpreted with caution until a replica-stable (or replica-free) analysis and larger-scale averaging clarify the fixed-point structure.

\medskip
\noindent\textit{Outlook.} Promising directions include: (i) a controlled resolution of the $n\!\to\!1$ versus $\Gamma'\!\to\!\infty$ limit (e.g., analytic continuation from $n>1$ with explicit regulators, or a replica-free Keldysh/SUSY approach); (ii) rare-event–aware numerics (importance sampling, splitting methods) to converge higher moments; and (iii) experimental probes designed to discriminate true IR scaling from long preasymptotic transients.

  \appendix
  \chapter{Path integral derivation Dirac fermions}\label{App:path_int_dir}
In this appendix we derive Keldysh path integral formalism for Dirac fermions. We start with the example of single fermionic mode.

Consider the Hamiltonian
\[
\hat{H} = \epsilon_0 \, \hat{c}^\dagger \hat{c}.
\]
We take the partition function in the closed-time-contour (CTC) formalism as
\begin{equation}\label{eq:part_func_exp}
Z = \frac{\mathrm{Tr}\left( \hat{U}_C \, \hat{\rho}_0 \right)}{\mathrm{Tr}(\hat{\rho}_0)},
\end{equation}
where $\hat{U}_C$ is the contour-ordered evolution operator and $\hat{\rho}_0$ is the initial density matrix.
We choose $\hat \rho_0=e^{-\beta(\hat H-\mu \hat N)}=e^{-\beta(\epsilon_0-\mu)\hat c^{\dagger}\hat c}$, where $\beta$ is inverse temperature and $\mu$ is a chemical potential. For this density matrix we have ${\rm Tr}(\hat \rho_0)=\sum_{n=0}^{1} e^{-\beta(\epsilon_0-\mu)n}=1+e^{-\beta(\epsilon_0-\mu)}$.

To represent $Z$ as a functional integral, we discretize the contour $C$ into $2N-2$ time slices of length $\delta t$ and insert the identity in coherent-state form at each point $j= 1,2,...,2N$ along the contour :
\[
\hat{\mathbb{I}} = \int d\bar{\psi}_j \, d\psi_j \; e^{-\bar{\psi}_j \psi_j} \; |\psi_j\rangle \langle \psi_j|.
\]
The trace over states at the end of the contour enforces \(\psi_{N+1} = -\psi_1\) for fermions (antiperiodic boundary conditions).

Between two adjacent time slices $t_j$ and $t_{j+1}$, we have
\[
\langle \psi_{j+1} | e^{-i \hat{H}(\hat c^{\dagger}, \hat c) \delta t} | \psi_j \rangle
= e^{-i  H(\bar \psi, \psi) \delta t} \langle \psi_{j+1} | \psi_j \rangle \
.
\]
Then the path integral takes the form
\begin{equation}
    Z=\frac{1}{{\rm Tr}\hat \rho_0} \int \int \prod_{j=1}^{2N} d \overline\psi_j d \psi_j e^{i \sum_{j,j'=1}^{2N} \bar\psi_j G_{jj}^{-1}\psi_j},
\end{equation}
where for example $N=3$:
\begin{equation}
i G^{-1}_{j j'}=\begin{bmatrix}
-1&&&&&-e^{-\beta(\epsilon_0-\mu)}\\
1-i \epsilon_0 \delta t&-1&&&&&\\
&1-i \epsilon_0\delta t&-1&&&&\\
&&1&-1&&&\\
&&&1+i \epsilon_0\delta t&-1&\\
&&&&1+i \epsilon_0\delta t&-1&
\end{bmatrix},
\end{equation}
where the top-right corner encodes the thermal boundary condition $\hat \rho_0$ in the nominator of the partition function \ref{eq:part_func_exp}.
Multiplying over all slices and taking the continuum limit \(\delta t \to 0\) yields
\[\label{eq:part_func_dir}
Z = \int \mathcal{D}[\bar{\psi},\psi] \;
\exp\left\{
i \int_C dt \; \bar{\psi}(t) \left[ i \partial_t - \epsilon_0 \right] \psi(t)
\right\},
\]
where \(\int_C\) indicates integration along the closed time contour.

This procedure generalizes directly to multiple fermionic modes and interacting systems, and is the standart starting point for the Keldysh functional integral for fermions.

Instead of considering evolution over the contour we split it into forward and backward evolutions and introduce Grassmann fields \( \psi_+(t), \bar{\psi}_+(t) \) on the forward branch, and \( \psi_-(t), \bar{\psi}_-(t) \) on the backward branch. Then the action in Keldysh formalism becomes:
\begin{equation}\label{eq:action_dir_app}
S[\bar{\psi}_\pm, \psi_\pm] = \int_0^T dt \left[ \bar{\psi}_+(t) (i \partial_t - \epsilon_0) \psi_+(t) - \bar{\psi}_-(t) (i \partial_t - \epsilon_0) \psi_-(t) \right].
\end{equation}
This structure automatically accounts for the operator orderings required to compute observables like the frame potential, OTOCs, or time-evolved correlation functions.

  \chapter{Path integral derivation Majorana fermions}\label{App:path_int_maj}
In this appendix we extend the derivation of the path integral into Majorana fermions.
Whereas the Dirac fermion $\hat c$ has a distinct adjoint  $\hat c^{\dagger}$, the Majorana fermion operator is
Hermitian:
\begin{equation}
    \hat \chi=\hat \chi^{\dagger}, ~~\text{and}~~\{\hat \chi_i, \hat \chi_j\}=\delta_{i j}.
\end{equation}
Here we can try to define the coherent state in standart fermionic way using Grassmann variables  \ref{eq:cohstate_fermion}:
\begin{equation}
    \hat\chi \left| \chi \right>=\chi \left| \chi \right>,
\end{equation}
however we directly see that applying $\hat \chi$ once more we get the square of the eigenvalue $ \chi^2=1$ due to commutation relations, which disagrees with usual requirement for the Grassmann variables $\chi^2=0$. In more details coherent states are built as eigenstates of annihilation operators with Grassmann eigenvalues; a Majorana operator is Hermitian and mixes creation/annihilation, so there is no nontrivial Grassmann-eigenstate construction for a single Majorana operator.
How do we find the corresponding path integral, given that coherent states of $\left| \chi \right>$ do not exist?

One option is to introduce a second Majorana operator $\hat \xi$ with no term in the Hamiltonian:
\begin{equation}
    \hat H(\hat \chi, \hat \xi)=\frac{i}{2} \sum_{k,j} \hat \chi_k  h_{k j} \hat \chi_j.
\end{equation}
Now we can form Dirac fermion and its adjoint:
\begin{equation}\label{eq:dir_major}
    \hat c_k=\frac{1}{\sqrt{2}}(\hat \chi_k+i \hat \xi_k),~~~\hat c^{\dagger}_k=\frac{1}{\sqrt{2}}(\hat \chi_k-i \hat \xi_k),
\end{equation}
which obey 
\begin{equation}
    \{\hat c^{\dagger}_k,\hat c_l\}=\delta_{k l},
\end{equation}
then the Hamiltonian becomes 
\begin{equation}
    H=\frac{i}{4} \sum_{kl }(\hat c_k+\hat c^{\dagger}_k)h_{k l} (\hat c_l+\hat c^{\dagger}_l),
\end{equation}
this approach can be easily generalised to non quadratic Hamiltonians. After one can write the path integral using Grassmann variables formalism for dirac fermions. The resulting action will have similar form to the one we got for partition function of Dirac fermions \ref{eq:action_dir}.
And introducing new Grassmann variables $\chi$ and $\xi$
\begin{equation}
    \psi_i=\frac{1}{\sqrt{2}}(\chi_i+i\xi_i),~~ \overline{\psi}_i=\frac{1}{\sqrt{2}}(\chi_i-i\xi_i),
\end{equation}
one obtains factorised action:
\begin{equation}
    S(\chi,\xi)=\frac{i}{2} \int_C d t \sum_k \chi_k \partial_t \chi_k-\int_C d t H(\chi, \xi)+\frac{i}{2} \int_C d t  \sum_k\xi_k \partial_t \xi_k
\end{equation}
There are no cross terms between the $\chi$ and $\xi$ fields in the kinetic term. Any such terms can be shown to vanish after integrating by parts and using the anticommutativity of Grassmann variables. As a result, the action factorizes: the $\chi$ and $\xi$ sectors completely decouple.

For the physical degrees of freedom, this means that the path integral can be written as if we were dealing with Dirac fermions, but with two key differences:  the kinetic term carries an extra factor of $\tfrac{1}{2}$ in front, and the path integral involves only the real Majorana field $\chi$, without any independent field $\bar\chi$. The additional $\xi$ sector contributes only an overall prefactor, which cancels once the partition function is properly normalized. In this sense, the path integral behaves as though coherent states for Majorana fermions actually existed.

An alternative route—when the system contains an even number of Majorana operators—is to pair them directly into Dirac fermions and their adjoints, construct the path integral using Dirac coherent states, and then rewrite the result back in terms of Majorana Grassmann variables. This procedure avoids introducing the auxiliary $\xi$ fields, but is slightly more cumbersome algebraically. Either way, the final result is the same: the Majorana path integral is well defined and equivalent to working with an effective set of coherent states.

With this machinery in hand—frame potential, Keldysh formalism, and the Majorana path integral—we are ready to proceed to the analysis of the Brownian SYK model.
  
\chapter{Haar-averaged Frame potential}\label{App:Frame_potntial_RM}In this appendix, we calculate the Haar values for the Frame potential.
First, we consider the case when the evolution matrix $U$ is selected from the corresponding group using the Haar measure. The groups we consider here are the orthogonal group for Majorana fermions and the unitary group for Dirac fermions. These Haar values correspond to the evolution with an arbitrary number of interacting fermions ($q>2$).
Second, we consider the case where the evolution matrix is quadratic in fermions ($q=2$), i.e., $U=e^{c^{\dagger}h c}$. In this case, the matrix values are again chosen with respect to the Haar measure, but the space of integration is smaller compared to the full Haar average.



\section{Gaussian-Haar averages for Majorana fermions}\label{sec:majo-ghaar}
If we consider Gaussian evolution, then its Haar distribution is the unitary evolution is a Gaussian $U$, i.e. $U=e^{\frac{1}{2} h_{i j} \chi_i \chi_j}$, where $\chi_j$ are Majorana fields, then we can express  ${\rm Tr}(U)={\rm Pf}(1+e^{h_{i j}})$ (as for example shown in \cite{Klich2014ANO})  where $e^{h_{i j}}=\hat{u}$ is an orthogonal matrix with the size $L\times L$  ($L$ is even). We are interested in the case when matrices $\hat{u}$ are random with respect to the Haar measure. Notice that we are averaging over matrices with $\det(\hat{u})=1$, as $\hat{u}=e^{\frac{1}{2}h}$ where $h$ is an antisymmetric matrix (all the eigenvalues of $h$ have conjugated pairs and lie on the imaginary axis, it means that eigenvalues of $\hat{u}$ are $\lambda_1=e^{i b_1}, \lambda^*_1=e^{-i b_1},...$ and their product is 1). This gives us the following averaging
\begin{multline}
         F^k _{\rm gHaar}=\frac{1}{\mathcal{N}^{2k}} \int \left|\det(1+\hat{u}) \right|^{ k} d \mu(\hat{u})=\\    =\frac{1}{\mathcal{N}^{2k}} \int \left|\prod^L_{j=1} (1+e^{i \theta_j}) \right|^{k} \prod_{i< j} \left| 2(\cos \theta_i-\cos \theta_j ) \right|^2 \frac{d \theta_1..d \theta_{L/2}}{\left(\frac{L}{2} \right)! (2 \pi)^{\frac{L}{2}}}=\\ =\frac{1}{\mathcal{N}^{2k}} \int \left|\prod^{L/2}_{j=1} (2+2 \cos(\theta_j)) \right|^{k} \prod_{i< j} \left| 2(\cos \theta_i-\cos \theta_j ) \right|^2 \frac{d \theta_1..d \theta_{L/2}}{\left(\frac{L}{2} \right)! (2 \pi)^{\frac{L}{2}}}. 
\end{multline}In the third line, we used the fact that all eigenvalues have their conjugated pair. The measure in this expression comes from the fact that we integrate other orthogonal matrices with the determinant equal to 1 and even size(see chapter 2.6 in \cite{Forrester}).
Doing change of variables $  \cos(\theta)=t$ and calculating the Jacobian of this transformation we can get 
\begin{multline}\label{framehaar}
         F^k _{\rm gHaar}=\frac{1}{\mathcal{N}^{2k}}\frac{1}{F^0}\int_{-1}^1 d t_1 ..\int_{-1}^1dt_{L/2} \prod^{L/2}_{j=1} |2(1+t_j)|^{k}\prod_{i<j} |t_i-t_j|^2 \prod_{k=1}^{L/2} (1-t_k^2)^{-\frac{1}{2}}.
\end{multline}
Now we need to calculate these integrals. For simplicity, we can introduce the function to name these integrals :
\begin{equation}
    F=2^{\frac{Lk}{2}} \int_{-1}^1 d t_1 ..\int_{-1}^1dt_{L/2} \prod^{L/2}_{j=1} |1+t_j|^{k}\prod_{i<j} |t_i-t_j|^2 \prod_{k=1}^{L/2} (1-t_k^2)^{-\frac{1}{2}}.
\end{equation}Here we can notice that it is just a special case of Selberg integral (see chapters $3.6, 3.7 , 4.1, 4.7$ in \cite{Forrester})
\begin{equation}
    F=2^{\frac{Lk}{2}}2^{\frac{L}{2}+\frac{L(k-1)}{2}+\frac{L}{2}(\frac{L}{2}-1)}S_{\frac{L}{2}}\left(k-\frac{1}{2},-\frac{1}{2},1\right)=2^{Lk}2^{\frac{L}{2}(\frac{L}{2}-1)}S_{\frac{L}{2}}\left(k-\frac{1}{2},-\frac{1}{2},1\right).
\end{equation}
The Selberg integral has a well-known expression in terms of Gamma functions, which is:
\begin{multline}
        S_L(\lambda_1,\lambda_2,\lambda)=\int_0^1 d t_1...\int_0^1 d t_L \prod_{l=1}^L t_l^{\lambda_1}(1-t_l)^{\lambda_2} \prod_{k<j}|t_k-t_j|^{2 \lambda}=\\=\prod_{j=0}^{L-1}\frac{\Gamma(\lambda_1+1+j \lambda) \Gamma(\lambda_2+1+j \lambda)\Gamma(1+(j+1)\lambda)}{\Gamma(\lambda_1+\lambda_2+2+(L+j-1)\lambda) \Gamma(\lambda+1)}.
\end{multline}Inserting this solution into the eq.~(\ref{framehaar}) we can have the final expression
\begin{multline}
         F^k _{\rm gHaar}= \frac{1}{\mathcal{N}^{2k}}\frac{2^{Lk}2^{\frac{L}{2}(\frac{L}{2}-1)}S_{\frac{L}{2}}\left(k-\frac{1}{2},-\frac{1}{2},1\right)}{2^{\frac{L}{2}(\frac{L}{2}-1)}S_{\frac{L}{2}}\left(-\frac{1}{2},-\frac{1}{2},1\right)}=\frac{S_{\frac{L}{2}}\left(k-\frac{1}{2},-\frac{1}{2},1\right)}{S_{\frac{L}{2}}\left(-\frac{1}{2},-\frac{1}{2},1\right)}=\\    =\prod_{j=0}^{\frac{L}{2}-1}\frac{\Gamma(k+\frac{1}{2}+j)}{\Gamma(\frac{1}{2}+j)} \frac{\Gamma(\frac{L}{2}+j)}{\Gamma(k+\frac{N}{2}+j)}\sim \frac{c_k L^{k(k-1)/2}}{\mathcal{N}^{2k}}+...,
\end{multline}where $c_k=\frac{2}{2^{\frac{(k-1)(k-2)}{2}}} \prod_{i=0}^{k-1}\frac{1}{(2i-1)!!} $ and we define $(-1)!!=1$. Therefore, for $k=1$ we have
\begin{equation}
     F^1 _{\rm gHaar}=\frac{2}{2^L}= F^1 _{\rm Haar},
\end{equation}
which is equal to the Frame potential in the case when the evolution matrix was taken from the Haar ensemble of Orthogonal matrices. 
We are interested in the large $L$ limit expansion of the Frame potential for arbitrary k, therefore we can use a "Stirling-like"  formula for the Barnes functions, which gives: 
\begin{equation}\label{Haarresof}
    \lim_{L \rightarrow \infty} \left(\frac{1}{L}\log\left( F^k _{\rm gHaar}\right)\right)\sim-k \log(2)+\frac{k(k-1)}{2}\frac{1}{L} \log(L).
\end{equation}

\section{Haar average of frame potential for Dirac fermions}\label{sec:dirac-haar}
Now we focus on the Dirac fermions case. In this case, the large time Haar distribution for the evolution operator is split into different charge blocks, where each block is a unitary random matrix. The Frame potential then takes the form  
\begin{equation}
    F^{(k)}_{\rm Haar}=\frac{1}{\mathcal{N}^{2k}} \Big[\prod_Q \int_{\rm Haar}  d U_Q \Big] \left| \sum_Q {\rm Tr}\left[ U_Q\right] \right|^{2 k}.
\end{equation}
Given that only paired traces give a finite result, we obtain $k!$ pairs whose average is always equal to $1$, as
\begin{equation}
      \int_{\rm Haar}   d U \left|   {\rm Tr}\left[ U\right] \right|^{2 }= 1,
\end{equation}
therefore using $\sum_Q = L$, we obtain 
\begin{equation}\label{Haaru}
    \frac{1}{\mathcal{N}^{2k}} \int_{\rm Haar} d U \left|{\rm Tr}\left[ U\right] \right|^{2 k}=\frac{L^k}{\mathcal{N}^{2k}}(k)!.
\end{equation} 

\section{Gaussian-Haar averages for Dirac fermions}\label{sec:dirac-ghaar}
In the case of Gaussian evolution, we consider a unitary Gaussian operator $U$, i.e. $U=e^{i h_{i j} c^{\dagger}_i c_j}$, where $c_j$ are Dirac fermions. Then we can express  ${\rm Tr}(U)={\rm det}(1+\hat{u})$ where $e^{i h_{i j}}=\hat{u}$ is a unitary matrix with the size $L\times L$  (with $L$ even). The matrices $\hat{u}$ are again random with respect to the Haar measure. This gives us the following average,
\begin{multline}
      F^{(k)}_{\rm gHaar}=\frac{1}{\mathcal{N}^{2k}} \int \left|\det(1+\hat{u}) \right|^{2 k} d \mu(\hat{u})= \\     =\frac{1}{\mathcal{N}^{2k}} \int \prod^L_{j=1} \left|(1+e^{i \theta_j}) \right|^{ 2k} \prod_{i< j} \left| e^{i \theta_i}-e^{i \theta_j} \right|^2 \frac{d \theta_1..d \theta_N}{N! (2 \pi)^N}.
\end{multline}
The measure in this expression comes from the fact that we integrate other unitary matrices (see chapter 2.6 in \cite{Forrester}).  This integral is a Morris integral which can be expressed in terms of Gamma functions:
\begin{multline}
        M_L(a,b,\lambda)=\int_{-1/2}^{1/2} \prod^L_{j=1} e^{i \theta_i (a-b)/2} \left|(1+e^{i \theta_j}) \right|^{ a+b} \prod_{i< j} \left| e^{i \theta_i}-e^{i \theta_j} \right|^{2\lambda}d \theta_1..d \theta_N=\\=\prod_{j=0}^{L-1}\frac{\Gamma(\lambda j+a+b+1)\Gamma(\lambda(j+1)+1)}{\Gamma(\lambda j +a+1)\Gamma(\lambda j +b+1) \Gamma(1+\lambda)}.
\end{multline}Inserting this into the original expression for the Frame potential we can get:\begin{equation}
     F^k _{\rm gHaar}= \frac{1}{\mathcal{N}^{2k}}  \frac{M_L(k,k,1)}{M_L(0,0,1)}=\frac{1}{\mathcal{N}^{2k}} \prod^{L-1}_{j=0} \frac{\Gamma(j+2k+1)\Gamma(j+1)}{\Gamma(j+k+1)^2}.
\end{equation}Again we are interested in the large $L$ limit expansion:
\begin{equation}\label{Haarresferm}
     F^{(k)}_{\rm Haar} = \frac{\tilde{c}_k L^{k^2}}{\mathcal{N}^{2k}} + \ldots
\end{equation}
where we neglected subleading corrections in $L$,  and $\tilde{c}_k=\frac{{\rm sf}^2(k-1)}{{\rm sf}(2k-1)}$ with ${\rm sf}(n)=1!2!...n!$ a super factorial. Comparing this to the full Haar average, we again find 
\begin{equation}
     F^k _{\rm gHaar}\geq  F^k _{\rm Haar},
\end{equation}therefore the full Haar average is again smaller than the fermionic one.
\section{Dirac SYK saddle point with fixed charge density}\label{A:SP}

Let's start with replica-diagonal solution $Q^{+}_l=Q^-_l$, we use the following ansatz 

\begin{equation}
\hat{G}_d= \frac{1}{2} \begin{pmatrix}
f(\Theta(t_{12})(\hat 1-\hat n)-\Theta(-t_{12})\hat n)&-f\hat n\\
f(\hat 1-\hat n) & f(\Theta(-t_{12})(\hat 1-\hat n)-\Theta(t_{12})\hat n)
\end{pmatrix},
\end{equation}
where  $f_{ll'}(t)$ is a time-dependent matrix in replica space with initial condition $f_{l l'}(0)=f(0) \delta_{l l'}$. First, we consider the equation 
\begin{equation}
    \Sigma^{s s'}_{l l'}(t)=\frac{c_{s s'}}{q} (2 G^{s s'}_{l l'})^{\frac{q}{2}-1}(t)(-2G^{s' s}_{l'l})^{\frac{q}{2}}(-t) \delta(t),
\end{equation}
which gives 
\begin{equation}
    \hat \Sigma_d=\frac{f^{q-1}(0)}{q}\begin{pmatrix}
        \frac{1}{2} \Gamma(\hat 1-2  \hat n)&-\Gamma(\hat 1-\hat n) \\
        \hat n\Gamma&\frac{1}{2} \Gamma(\hat 1-2 \hat n)
    \end{pmatrix}\delta(t_{12}),
\end{equation}
where  $\hat n={\rm diag}(n_1,n_2,..,n_k)$ is a diagonal matrix in replica space, which consists of the fermion densities from different replicas, and $\Gamma=(\hat n(\hat 1-\hat n))^{\frac{q}{2}-1}$ is also a diagonal matrix. Using the equation $-(\hat G^{-1})^{s's}_{l'l}(t)=  \hat\sigma_z \partial_t-\hat \Sigma^{ss'}_{l l'}(t)+i \sigma_z\mu^{s}_l$ and Fourier transformation, we can find 
\begin{equation}
    \hat G^{-1}_d=\begin{pmatrix}
        i \omega+\frac{1}{2} \tilde \Gamma (\hat 1-2   \hat n)-i \hat \mu^+& \tilde \Gamma \hat n \\
        -(\hat 1-\hat n)\tilde\Gamma  &-i \omega+\frac{1}{2}\tilde \Gamma(\hat 1-2 \hat n)+i \hat \mu^-,
    \end{pmatrix}
\end{equation}
where $\tilde \Gamma =\frac{f^{q-1}(0)}{q} \Gamma$. Here we can use the formula for the block matrix and assume $\mu^+_l=\mu^-_l=\mu_l$ (since $Q^{+}_l=Q^-_l$), therefore we find
\begin{equation}
    \begin{pmatrix}
        A&B\\
        C&D
    \end{pmatrix}^{-1}=\begin{pmatrix}
        A^{-1}+A^{-1}B(D-CA^{-1}B)^{-1}CA^{-1}&-A^{-1}B(D-CA^{-1}B)^{-1}\\
        -(D-CA^{-1}B)^{-1}CA^{-1}&(D-CA^{-1}B)^{-1}
    \end{pmatrix},
\end{equation}

which gives 
\begin{equation}
    \hat G_d=\frac{1}{ (\omega-\hat \mu)^2+\tilde \Gamma^2 \frac{1}{4}}\begin{pmatrix}
        -i \omega+\tilde\Gamma/2(1-2\hat n)+i \hat \mu  &-
     \tilde \Gamma \hat n\\
        (1-\hat n)\tilde \Gamma&i \omega+\frac{1}{2}\tilde \Gamma (1-2\hat n)-i \hat \mu
    \end{pmatrix}.
\end{equation}
The inverse Fourier transform of each element of the matrix gives 
\begin{equation}
\hat{G}_d=  \begin{pmatrix}
e^{-\frac{\Gamma |t_{12}|}{2}-i \hat{\mu} t_{12}}(\Theta(t_{12})(\hat 1-\hat n)-\Theta(-t_{12})\hat n)&-e^{-\frac{\Gamma |t_{12}|}{2}-i \hat{\mu} t_{12}}\hat n\\
e^{-\frac{\Gamma |t_{12}|}{2}-i \hat{\mu} t_{12}}(\hat 1-\hat n) & e^{-\frac{\Gamma |t_{12}|}{2}-i \hat{\mu} t_{12}}(\Theta(-t_{12})(\hat 1-\hat n)-\Theta(t_{12})\hat n)
\end{pmatrix}.
\end{equation}

    \chapter{Pfaffian Calculation}\label{App:pfaffian}
\section{Majorana fermions, q>2,q=2}
For the case $q>2$ let us proceed with the calculation of the first term in the action. In fact, the Pfaffian can be understood as a fermionic (quadratic) thermal partition function with the  real inverse temperature equal to $T$:
\begin{equation} \label{pf}
    e^{S_1}=({\rm Pf}[\hat \sigma_z \partial_t-\hat\Sigma])^L={\rm Tr}(e^{-T \hat H})^{L},
\end{equation}
where 
\begin{equation}
    \hat H= -\frac{1}{2}\begin{pmatrix}\hat \chi^+&i \hat \chi^-
        
    \end{pmatrix} \hat \Sigma_0 \begin{pmatrix}
        \hat \chi^+\\
        i \hat \chi^-
    \end{pmatrix},
\end{equation}
with $\hat \Sigma_0$ time-idependent part of Eq.~\eqref{eq:sol_sp}:
\begin{equation}
\hat{\Sigma}_0= \frac{1}{q} \begin{pmatrix}
0&-\hat{\tau}^T\\
\hat{\tau} &0
\end{pmatrix}.
\end{equation}
Let us explain the form of the Hamiltonian.  The derivative in the $--$ sector eq.~(\ref{pf}) comes with an extra minus, therefore the Majorana fields in the partition function should come with an extra $i$.  Finally, we obtain
\begin{equation}
    ({\rm Pf}[\hat \sigma_z \partial_t-\hat\Sigma])^L={\rm Tr}(e^{i \frac{T}{q} \hat\chi^- \hat{\tau} \hat\chi^+})^{L}.
\end{equation}
Notice that the left part is just the result of path integral formulation of the right part in terms of Majorana fermions and integration over them (see \cite{Nakahara2003}). A similar expression was calculated in \cite{Jian2021Rep}. Let us  observe that any arbitrary permutation can be expressed as a direct sum of disjoint cycles, $\hat{\tau}=\bigoplus_i \hat{\tau}^c_i$ which implies, 
\begin{equation}
    \hat{\tau}=
\left(\begin{array}{c|c|c|c}
  \hat{\tau}^c_1
  & 0 & 0&0 \\
\hline
  0 &
  \hat{\tau}^c_2& 0&0\\
  \hline
  0 &
0 &  ...&0
\\
  \hline
  0 &
0 & 0& \hat{\tau}^c_m
\end{array}\right),
\end{equation}then ${\rm Tr}(e^{i \frac{T}{q}  \hat\chi^- \hat{\tau} \hat\chi^+})={\rm Tr}(e^{i \frac{T}{q} \sum_i \hat\chi_i^- \hat{\tau}^c_i \hat\chi_i^+}) $ , where the sum goes over the cycles.  Also, we notice that each cycle can be represented in the canonical form, therefore we have 
\begin{equation}\label{cyc}
    \hat{\tau}^c_i= \begin{cases}
    \delta^{\alpha+1, \beta},~~~n_c ~~~\text{odd}\\
    {\rm {\rm sgn}}(\beta-\alpha)\delta^{\alpha+1,\beta},~~~n_c ~~~\text{even},
    \end{cases}
\end{equation} where $n_c$ represents the length of a cycle, and $\delta^{n_c+1,\beta} = \delta^{1,\beta}$. Here, ${\rm sgn}(\beta-\alpha)$ merely affects the permutation on the boundaries. The boundary conditions have been chosen in a way to keep the parity $P$ equal to $1$. It can be observed that for any $\hat{\tau}$, this solution does not respect the parity of the initial state (the initial state is a thermofield double state between, where for each site, the Fermi parity is $P = 1$). Therefore, the boundary conditions eq.~(\ref{cyc}) arise from this fact. Then, for each cycle, we have:
\begin{equation}
    -T \hat H_c=i \frac{T}{q} \left(\hat \chi^-_{\alpha} \delta^{\alpha+1,\beta} \hat \chi^+_{\beta}+(-1)^{n_c+1} \hat \chi^-_{n_c}\hat \chi^+_1 \right).
\end{equation}
Introducing complex fermions \begin{equation}\label{ferm}
    c_{\alpha}=\frac{\hat \chi^+_{\alpha}+i \hat \chi^-_{\alpha}}{\sqrt{2}}, ~~~~  c^{\dagger}_{\alpha}=\frac{\hat \chi^+_{\alpha}-i \hat \chi^-_{\alpha}}{\sqrt{2}},
\end{equation}
and inserting this into the Hamiltonian we obtain
\begin{equation}
    -T \hat H(\hat{\tau}_c)=\frac{T }{2q} \sum_{\alpha} (c^{\dagger}_{\alpha} c_{\alpha+1}+c_{\alpha} c_{\alpha+1}+h.c.)+\frac{T }{2q}  ((-1)^{n_c+1}c^{\dagger}_n c_1 +h.c.), 
\end{equation} this is the Kitaev chain \cite{Kitaev2000} with periodic/antiperiodic boundary conditions, eq.~(\ref{cyc}) and odd/even length of chain respectively. The diagonalisation of the chain is readerly done, giving 
\begin{equation}
    {\rm Tr}(e^{-T \hat H(\hat{\tau}_c)})\sim e^{\frac{n_c T }{2 q}},~~~ T \rightarrow \infty.
\end{equation}
Now we can finally sum over all the cycles, noticing that the size of the permutation matrix $k$ should be equal to the sum of the length of cycles of this permutation $k=\sum_c n_c$ 
\begin{equation}
    {\rm Tr}(e^{-T \hat H})=\prod_c e^{\frac{n_c T }{2 q}}= e^{\frac{k T }{2 q}}.
\end{equation}Consequently we find the following expression for the Pfaffian:
\begin{equation}
    e^{S_1}=({\rm Pf}[\hat \sigma_z \partial_t-\Sigma])^L= e^{\frac{L k T }{2 q}},
\end{equation}

For the case $q=2$ let us  consider the Hamiltonian for the two-dimensional orthogonal sub-matrix:
\begin{equation}
     - T \hat H_R=i \frac{T}{q}\left( \cos(\phi) \hat \chi^-_1 \hat \chi^+_1+\sin(\phi) \hat\chi^-_1 \hat\chi^+_2-\sin(\phi)\hat \chi^-_2 \hat \chi^+_1+\cos(\phi) \hat \chi^-_2 \hat \chi^+_2\right).
 \end{equation}Here $\phi$ is an angle that parametrizes the rotational matrix, and we can again move to the complex fermions, see eq.~(\ref{ferm}), which gives a simple two-fermions Hamiltonian:
\begin{multline}
        - T \hat H_R=\frac{T}{2 q}(2 \cos(\phi)(c_1^{\dagger}c_1+c_2^{\dagger}c_2-1)+ 2 \sin(\phi)(c_1 c_2-c_1^{\dagger} c_2^{\dagger}))=\\=\frac{T}{ q}\left(c_1^{\dagger}~~c_2 \right)\begin{pmatrix}
\cos{\phi} & - \sin{\phi} \\
- \sin{\phi} & -\cos{\phi}
\end{pmatrix} \begin{pmatrix}
c_1 \\
c^{\dagger}_2 
\end{pmatrix},\end{multline}after the diagonalisation the trace can be easily calculated. Using the well-known formula for the trace of quadratic density matrices $\rm Tr(e^{c^{\dagger}_i \Gamma_{i j} c_j})=\det(1+e^{\Gamma_{ij}})$, we obtain:
\begin{multline}
        {\rm Tr}(e^{-T \hat H_R})=\det\left(1+\exp\left(\frac{T}{q}\begin{pmatrix}
1 & 0 \\
0 & -1 
\end{pmatrix}  \right) \right)=2+e^{\frac{T}{q}}+e^{-\frac{T}{q}}    \rightarrow e^{\frac{T}{q}}, ~~T\rightarrow \infty.
\end{multline}

    \chapter{Calculation on on shell action for Dirac fermions}\label{App:act_sp_dir}
In this appendix we calculate an on shell action using the saddle poiunt that we found.
\section{Dirac fermions,q>2}
Here we start with the calculation of the determinant part of the action Eq.~\eqref{det}. The Hamiltonian on the right hand side of Eq.~\eqref{det} has the form: 
\begin{gather}
    H=(c_+^{\dagger}~~~ c_-^{\dagger}) \begin{pmatrix}
        \hat 1~&0\\
        0~&\hat i
    \end{pmatrix} H_0 \begin{pmatrix}
        \hat 1~&0\\
        0~&\hat i
    \end{pmatrix} \begin{pmatrix}
        c_+\\
        c_-
    \end{pmatrix},\\
    H_0=\begin{pmatrix}
        \frac{2^{q-1}}{2q}\Gamma(1-2  \hat n)-i \hat \mu&-\frac{2^{q-1}}{q}\Gamma(1-\hat n) \\
        \frac{2^{q-1}}{q}\hat n\Gamma&\frac{2^{q-1}}{2q} \Gamma(1-2 \hat n)+i \hat{\mu}
    \end{pmatrix}. 
\end{gather}Then we use the determinant formula 
\begin{equation}
    {\rm Tr}(e^{-TH})=\det \left(1+{\rm exp}\left(
    {-T \begin{pmatrix}
        \hat 1~&0\\
        0~&\hat i
    \end{pmatrix} H_0 \begin{pmatrix}
        \hat 1~&0\\
        0~&\hat i
    \end{pmatrix}}\right)\right).
\end{equation} 
The diagonalisation of the matrix exponent gives eigenvalues $ e^{\pm\frac{T \Gamma_l}{2q}-i T \mu_l}=e^{\pm\frac{T (n_l(1-n_l))^{\frac{q}{2}}}{2q}- i T \mu_l}$, where $\Gamma_l$ is a matrix element of the diagonal matrix $\Gamma$.
\begin{equation}
    {\rm det}(\sigma_z \partial_t-\Sigma_d+i\sigma_z \hat\mu \delta(t_{12}))^L\rightarrow e^{\frac{ 2^{q-1} LT\sum_{l=1}^k \Gamma_l}{2q}-iL T \sum_l \mu_l}.
\end{equation}
Now let us  move to the second term in the action:\begin{multline}
    S_2=\frac{L}{2q^2}\sum\limits_{l,l's,s' } \int dt_1 dt_2 \delta(t_{12}) c_{s s'} (2 G^{s s'}_{l l'})^{q/2} (-2 G^{s' s}_{l' l})^{q/2}=\\
    =\frac{L}{q^2}\sum_l\left(n_l^{\frac{q}{2}}(1-n_l)^{\frac{q}{2}}-\int dt (\Theta^q (t)(1-n_l)^{\frac{q}{2}}n_\textit{}l^{\frac{q}{2}}+\Theta(-t)^q n_l^{\frac{q}{2}}(1-n_l)^{\frac{q}{2}})\delta(t)
    \right),
\end{multline}
We adopt the symmetric prescription $\Theta(0)=\tfrac12$, which implies, as distributions,
\begin{equation}
\Theta(t)\,\delta(t)=\Theta(-t)\,\delta(t)=\frac{1}{2}\,\delta(t).
\end{equation}
Equivalently, for constants $a,b$,
\begin{equation}
\int_{-\infty}^{\infty}\!\! dt\,\delta(t)\big[\Theta(t)\,a+\Theta(-t)\,b\big]
=\frac{a+b}{2}.
\end{equation}

\paragraph{Application to $S_2$.}
With $q$ even we have $\Theta(t)^q=\Theta(t)$ and $\Theta(-t)^q=\Theta(-t)$. Writing
$A_l \equiv n_l^{\frac{q}{2}}(1-n_l)^{\frac{q}{2}}$, the second term becomes
\begin{align}
S_2
&=\frac{L}{2q^2}\sum_{l,l',s,s'}\!\int dt_1 dt_2\,\delta(t_{12})\,c_{ss'}\,
\big(2G^{ss'}_{ll'}\big)^{\frac{q}{2}}
\big(-2G^{s's}_{l'l}\big)^{\frac{q}{2}} \notag\\
&=\frac{L}{q^2}\sum_{l}\!\left[
A_l-\int_{-\infty}^{\infty}\! dt\,\delta(t)\big(\Theta(t)\,A_l+\Theta(-t)\,A_l\big)
\right] \notag\\
&=\frac{L}{q^2}\sum_{l}\!\left[A_l-\frac{A_l+A_l}{2}\right]=0.
\end{align}
Hence,
\begin{equation}
\boxed{\,S_2=0\,}.
\end{equation}
notice that both side limit for the sum of Heaviside functions is not well-defined, therefore we assume: 
\begin{align}
    \delta(t)& (\Theta(t)a+\Theta(-t)b)=\frac{1}{2}\left(\lim_{t \rightarrow0^+}(\Theta(t)a+\Theta(-t)b)+\lim_{t \rightarrow0^-}(\Theta(t)a+\Theta(-t)b)\right) \nonumber \\& =\frac{1}{2}(a+b).
\end{align}
Therefore, as in the previous case, this term is equal to zero on the solution due to the forward and backward evolution, which comes with different signs.  The third term in the action gives,
\begin{multline}
       S_3=-L\sum_{l,l',s,s'} \int d t_1 d t_2 \Sigma^{s s'}_{ll'}(t_1,t_2)G^{s s'}_{ll'}(t_1,t_2)=\\=-L \int dt_1 dt_2 \left(\Sigma^{++}_{l l'}G^{++}_{l l'}+\Sigma^{--}_{l l'}G^{--}_{l l'}+\Sigma^{+-}_{l l'}G^{+-}_{l l'}+\Sigma^{-+}_{l l'}G^{-+}_{l l'} \right)=\\=-\frac{2^{q-1}}{2q}\sum_l(1-2n_l)^2 \Gamma_l- \frac{2^{q-1}}{q}\sum_l \Gamma_l(1-n_l)n_l=-\frac{ 2^{q-1} LT\sum_{l=1}^k \Gamma_l}{2q}.
\end{multline}
This formula finalises the calculation of the on shell action.
    \chapter{Ito derivation of the diffusive SSE from a coupled ancilla}\label{App:SSE}

\paragraph{Setup .}
At each time step $dt$ we prepare $N$ independent ancilla qubits in $|0_z\rangle^{\otimes N}$ and couple them to the system via
\begin{equation}
U_{\rm int}=\exp\!\Big[-i\sum_{j=1}^N \varepsilon_j\, M_j\!\otimes\!\sigma_x^{(j)}\Big],
\qquad 
\varepsilon_j\equiv\sqrt{\Gamma_j\,dt},
\end{equation}
while the system undergoes its unitary drive $U_{\rm sys}=\exp(-iH^{\rm uni}dt)$.
After the interaction, each ancilla is measured in the $\sigma_y$ basis 
$|m_j\rangle\equiv|\pm_y\rangle=(|0\rangle\pm i|1\rangle)/\sqrt2$ with outcomes $m_j=\pm1$.
The unnormalized conditional update for the system is
\begin{equation}
|\tilde\psi_{t+dt}\rangle
=K_{\{m\}}\,e^{-iH^{\rm uni}dt}\,|\psi_t\rangle,
\qquad
K_{\{m\}} \equiv \big\langle \{m\}\big|U_{\rm int}\big|0_z^{\otimes N}\big\rangle,
\end{equation}
where $|\{m\}\rangle\equiv\bigotimes_j|m_j\rangle$.

\paragraph{Joint Kraus operator to quadratic order (with cross terms).}
Expand $U_{\rm int}$ to Ito order ($\varepsilon_j\sim dt^{1/2}$, $\varepsilon_j^2\sim dt$):
\begin{align}
U_{\rm int}
&\simeq \mathbb{I}
- i\sum_j \varepsilon_j\, M_j\!\otimes\!\sigma_x^{(j)}
-\frac12\sum_j \varepsilon_j^2 M_j^2\!\otimes\!\mathbb{I}
-\frac12\sum_{j\neq k}\varepsilon_j\varepsilon_k\, M_jM_k\!\otimes\!\sigma_x^{(j)}\sigma_x^{(k)}.
\end{align}
Using $\langle m_j|0\rangle=2^{-1/2}$ and $\langle m_j|\sigma_x|0\rangle=-\,im_j\,2^{-1/2}$, one finds
\begin{equation}
\boxed{\
K_{\{m\}}=\frac{1}{2^{N/2}}\!\left[
\mathbb{I}
+ \sum_j m_j\varepsilon_j M_j
- \frac12\sum_j \varepsilon_j^2 M_j^2
- \frac12\sum_{j\neq k} m_j m_k\,\varepsilon_j\varepsilon_k\, M_j M_k
\right] + O(dt^{3/2}).}
\label{eq:K_joint}
\end{equation}
The last term is the explicit \emph{cross} contribution from squaring the sum in the exponent.

\paragraph{Outcome probabilities (normalization to Ito order).}
Let $p_{\{m\}}=\langle\psi_t|K_{\{m\}}^\dagger K_{\{m\}}|\psi_t\rangle$. 
Keeping terms up to $O(dt)$ and noting $M_j=M_j^\dagger$,
\begin{equation}
\boxed{\
K_{\{m\}}^\dagger K_{\{m\}}
=\frac{1}{2^N}\Big[
\mathbb{I} + 2\sum_j m_j\varepsilon_j M_j
\Big] + O(dt^{3/2}).
}
\end{equation}
Here the $O(dt)$ \emph{diagonal} pieces $\propto \varepsilon_j^2 M_j^2$ cancel between $-{\textstyle\frac12}\sum_j\varepsilon_j^2 M_j^2$ and $+\sum_j (m_j\varepsilon_j M_j)^2$, and the \emph{cross} pieces 
$\propto m_j m_k \varepsilon_j\varepsilon_k M_jM_k$ cancel between $-{\textstyle\frac12}\sum_{j\neq k}$ and $+\sum_{j\neq k}$ from $A^\dagger A$.
Therefore
\begin{equation}
\boxed{\
p_{\{m\}}=\frac{1}{2^N}\!\left(1+2\sum_j m_j\varepsilon_j\langle M_j\rangle_t\right) + O(dt^{3/2}).}
\end{equation}

\paragraph{Continuous records and innovations.}
Define for each channel the record increment and innovation
\begin{equation}
dy_{j,t}\equiv m_j\sqrt{dt},\qquad
dW_{j,t}\equiv dy_{j,t}-2\sqrt{\Gamma_j}\,\langle M_j\rangle_t\,dt,
\end{equation}
which obey the Ito table for independent meters:
\begin{equation}
dy_{j,t}\,dy_{k,t}=\delta_{jk}\,dt,\qquad dW_{j,t}\,dW_{k,t}=\delta_{jk}\,dt,\qquad dy_{j,t}\,dt=0.
\end{equation}

\paragraph{Normalized update (keep all $O(dt)$ terms).}
Write $K_{\{m\}}$ from \eqref{eq:K_joint} in $dy$ variables:
\begin{equation}
K_{\{m\}}=\frac{1}{2^{N/2}}\!\left[
\mathbb{I}
+ \sum_j \sqrt{\Gamma_j}\,M_j\,dy_{j,t}
- \frac12\sum_j \Gamma_j M_j^2\,dt
- \frac12\sum_{j\neq k}\sqrt{\Gamma_j\Gamma_k}\,M_j M_k\,dy_{j,t}\,dy_{k,t}
\right].
\end{equation}
Using $dy_{j,t}\,dy_{k,t}=\delta_{jk}\,dt$, the cross term $j\neq k$ vanishes at Ito order, so
\begin{equation}
K_{\{m\}}=\frac{1}{2^{N/2}}\!\left[
\mathbb{I}
+ \sum_j \sqrt{\Gamma_j}\,M_j\,dy_{j,t}
- \frac12\sum_j \Gamma_j M_j^2\,dt
\right]+O(dt^{3/2}).
\end{equation}
The normalization factor (to multiply $K_{\{m\}}$) is $p_{\{m\}}^{-1/2}$ with
\(
p_{\{m\}}=\frac{1}{2^N}\left(1+2\sum_j \sqrt{\Gamma_j}\,\langle M_j\rangle_t\,dy_{j,t}\right)+O(dt).
\)
Expanding $(1+\epsilon)^{-1/2}=1-\tfrac12\epsilon+\tfrac38\epsilon^2+O(\epsilon^3)$ with
$\epsilon=2\sum_j \sqrt{\Gamma_j}\,\langle M_j\rangle_t\,dy_{j,t}$ and using $dy_{j,t}\,dy_{k,t}=\delta_{jk}\,dt$ gives
\begin{equation}
\boxed{\
p_{\{m\}}^{-1/2}
=2^{N/2}\!\left[
1-\sum_j \sqrt{\Gamma_j}\,\langle M_j\rangle_t\,dy_{j,t}
+\frac{3}{2}\sum_j \Gamma_j \langle M_j\rangle_t^2\,dt
\right]+O(dt^{3/2}).}
\end{equation}
Multiplying $K_{\{m\}}$ by $p_{\{m\}}^{-1/2}$ and keeping $O(dt)$ terms:
\begin{align}
|\psi_{t+dt}\rangle
&=\frac{K_{\{m\}}}{\sqrt{p_{\{m\}}}}\,e^{-iH^{\rm uni}dt}\,|\psi_t\rangle \nonumber\\
&=\Bigg\{
\mathbb{I}
+\sum_j \sqrt{\Gamma_j}\,\big(M_j-\langle M_j\rangle_t\big)\,dy_{j,t}
-\sum_j\Gamma_j\Big[\tfrac12 M_j^2+\langle M_j\rangle_t M_j-\tfrac{3}{2}\langle M_j\rangle_t^2\Big]dt
-iH^{\rm uni}dt
\Bigg\}|\psi_t\rangle.
\label{eq:norm_dy_form}
\end{align}
This is the \emph{normalized} update in terms of the raw records $dy_{j,t}$.

\paragraph{Canonical Ito SSE (innovation form; cross terms gone).}
Insert $dy_{j,t}=dW_{j,t}+2\sqrt{\Gamma_j}\,\langle M_j\rangle_t\,dt$ into \eqref{eq:norm_dy_form}. 
All inter-channel quadratic terms vanish by $dW_{j,t}\,dW_{k,t}=\delta_{jk}\,dt$, and the drift reshuffles into the familiar “quadratic-in-fluctuations” form:
\begin{equation}
\boxed{\
\begin{aligned}
d|\psi_t\rangle
&=\Big[-iH^{\rm uni}\,dt
-\frac12\sum_j \Gamma_j \big(M_j-\langle M_j\rangle_t\big)^2 dt\Big]|\psi_t\rangle\\
&\quad+\sum_j \sqrt{\Gamma_j}\,\big(M_j-\langle M_j\rangle_t\big)\,|\psi_t\rangle\, dW_{j,t},
\end{aligned}}
\end{equation}
with independent Wiener increments $dW_{j,t}\,dW_{k,t}=\delta_{jk}\,dt$ and measurement currents
\begin{equation}
\boxed{\
dy_{j,t}=2\sqrt{\Gamma_j}\,\langle M_j\rangle_t\,dt + dW_{j,t}.}
\end{equation}
This finalises the derivation of the SSE for multiple channels. In the Appendix \ref{App:linearSSE} we connect this approach to non normalised density matrix evolution, and Hamiltonian representation.

\paragraph{Consistency checks.}
Averaging the innovation (i.e. $\mathbb{E}[dW_{j,t}]=0$) yields the Lindblad equation
\begin{equation}
\dot\rho=-i[H^{\rm uni},\rho]+\sum_j \Gamma_j\Big(M_j\rho M_j-\tfrac12\{M_j^2,\rho\}\Big).
\end{equation}
The conditional SME (Ito) is
\begin{equation}
d\rho_c=-i[H^{\rm uni},\rho_c]\,dt
+\sum_j \Gamma_j\,\mathcal D[M_j]\rho_c\,dt
+\sum_j \sqrt{\eta_j\Gamma_j}\,\mathcal H[M_j]\rho_c\,dW_{j,t},
\end{equation}
with $\mathcal D[M]\rho=M\rho M-\tfrac12\{M^2,\rho\}$ and $\mathcal H[M]\rho=M\rho+\rho M-2\langle M\rangle_\rho\,\rho$.

    \chapter{From normalized SSE to linearized form: Itô versus Stratonovich}
\label{App:linearSSE}

In the main text we emphasized that the normalized stochastic Schrödinger equation (SSE) is nonlinear, which obstructs analytic access to nonlinear observables such as entanglement entropies. In this appendix we show how the introduction of the unnormalized operator $\check\rho$ corresponds to a linearized version of the SSE and explain the relation between the Itô and Stratonovich conventions.

\section*{From normalized to unnormalized trajectories}

Consider continuous monitoring of a Hermitian operator $M$ at rate $\Gamma$, with Hamiltonian $H^{\rm uni}$. In the Itô convention, the normalized state $|\psi_t\rangle$ evolves according to the diffusive SSE
\begin{equation}
d|\psi_t\rangle
=\Big[-i H^{\rm uni}\,dt - \Gamma (M-\langle M\rangle_t)^2 dt
+\sqrt{2\Gamma}\,(M-\langle M\rangle_t)\,dW_t\Big]|\psi_t\rangle ,
\label{eq:SSE_norm}
\end{equation}
where $dW_t$ is a Wiener increment with $\mathbb{E}[dW_t]=0$, $dW_t^2=dt$, and $\langle M\rangle_t=\langle\psi_t|M|\psi_t\rangle$. 
This equation is explicitly nonlinear through the appearance of $\langle M\rangle_t$, and guarantees $\langle\psi_t|\psi_t\rangle=1$ along each trajectory.

To obtain a linear formulation, we remove the normalization and define an \emph{unnormalized} state $|\tilde\psi_t\rangle$, which evolves in Itô form as
\begin{equation}
d|\tilde\psi_t\rangle
=\Big[-i H^{\rm uni}\,dt - \Gamma M^2 dt + \sqrt{2\Gamma}\,M\,dW_t\Big]|\tilde\psi_t\rangle .
\label{eq:SSE_lin}
\end{equation}
This equation is linear in $|\tilde\psi_t\rangle$ and contains no expectation values. The norm is not preserved; instead $\langle\tilde\psi_t|\tilde\psi_t\rangle$ encodes the likelihood of the measurement record. The normalized state is recovered by
\[
|\psi_t\rangle = \frac{|\tilde\psi_t\rangle}{\sqrt{\langle\tilde\psi_t|\tilde\psi_t\rangle}} ,
\]
and substituting this relation into Eq.~\eqref{eq:SSE_norm} reproduces the nonlinear SSE. Thus the nonlinear, normalized SSE and the linear, unnormalized SSE are equivalent descriptions.

\section*{Density matrix formulation}

From the unnormalized state we construct the operator
\[
\check\rho(t) = |\tilde\psi_t\rangle \langle \tilde\psi_t| .
\]
Its evolution follows directly from Eq.~\eqref{eq:SSE_lin}, yielding
\begin{equation}
d\check\rho = -i [H^{\rm uni},\check\rho]\,dt - \Gamma \{M^2,\check\rho\}\,dt
+ \sqrt{2\Gamma}\,\big(M\check\rho\,dW_t + \check\rho M\,dW_t\big).
\end{equation}
This equation is \emph{linear} in $\check\rho$. 
The normalized density matrix is obtained by
\[
\rho(t) = \frac{\check\rho(t)}{\Tr \check\rho(t)} ,
\]
which then evolves according to the nonlinear SSE.

\section*{Itô versus Stratonovich conventions}

The equations above were written in the Itô convention. In this setting the stochastic increments satisfy $dW_t^2=dt$, and quadratic fluctuations generate explicit drift terms such as $-\Gamma M^2 dt$ in Eq.~\eqref{eq:SSE_lin}.  

In the Stratonovich convention, stochastic integrals are defined symmetrically (midpoint rule). One may then treat $dW_t$ as if it were generated by a formal white-noise process $\omega(t)=\dot W_t$ with $\langle \omega(t)\omega(t')\rangle=\delta(t-t')$, and write
\[
d|\tilde\psi_t\rangle = \Big(-i H^{\rm uni}\,dt + \sqrt{2\Gamma}\,M \circ dW_t\Big)|\tilde\psi_t\rangle
\;\;\Rightarrow\;\;
\frac{d}{dt}|\tilde\psi_t\rangle = \Big(-iH^{\rm uni} + \sqrt{2\Gamma}\,M\,w(t)\Big)|\tilde\psi_t\rangle.
\]
This compact form resembles an ordinary Schrödinger equation with a noisy Hamiltonian. Converting back to Itô recovers the explicit drift term $-\Gamma M^2 dt$.

\section*{Effective non-Hermitian Hamiltonian}

In Stratonovich form, the unnormalized state evolves under an effective non-Hermitian Hamiltonian,
\[
H_{\rm eff}(t) = H^{\rm uni}(t) - i \sqrt{2\Gamma}\,M\,w(t),
\]
so that
\[
|\tilde\psi_t\rangle = K(t)|\tilde\psi_0\rangle, \qquad
K(t) = \mathcal{T}\exp\!\Big(-i\int_0^t H_{\rm eff}(s)\,ds\Big).
\]
Here $K(t)$ is non-unitary since $H_{\rm eff}$ is non-Hermitian. The unnormalized density matrix can then be written as
\[
\check\rho(t) = K(t)\,\rho(0)\,K^\dagger(t).
\]
Normalization (and the Born rule) can be reinstated at the end by dividing by $\Tr \check\rho(t)$.

\subsection*{Summary}

In summary:
\begin{enumerate}
    \item  The nonlinear, norm-preserving SSE (Itô form) is equivalent to a linear, unnormalized SSE. 
    \item In the \emph{Itô convention}, explicit drift terms appear from the rule $dW_t^2=dt$.
    \item In the \emph{Stratonovich convention}, the same dynamics may be written with a white-noise Hamiltonian $H_{\rm eff}(t)=H^{\rm uni}-i\sqrt{2\Gamma}M\,w(t)$, with ordinary calculus rules.
    \item The linear formulation introduces a non-unitary propagator $K(t)$, and working with $\check\rho$ avoids explicit nonlinearities. This is the starting point for replica and path-integral methods developed in the main text.
\end{enumerate}
 
  \chapter{More on the replicated Hamiltonian}\label{App:rep}

For a given realization of the noises $h, w$, the density matrix of the system will evolve as
\begin{equation}
    \rho(t) = K(t) \rho(0) K^{\dag}(t) , \qquad K(t) = T \exp (- i \int_0^t \; dt' H(t')) .
\end{equation}
We want to study the evolution of $n$ replicas of the density matrix $\rho(t)$. In order to do so, we represent the dynamics in the vectorised and replicated Hilbert space. 
We introduce the replica indices $\alpha = (\sigma, a)$ with $\sigma = \pm$ and $a = 1,\ldots,n$. 
More specifically, we recall that within the standard vectorization construction, we use the $+$ for operators multiplying on the left and the $-$ for operators multiplying on the right, i.e.
\begin{equation}
    A \rho B = A^+ (B^-)^t \ket{\rho},
\end{equation}
where $t$ denotes the transpose.
We can thus define the replicated Hamiltonian as
\begin{equation}
    H^{(n)} = \sum_{a}  H^{(+,a)} -
    (H^{(-,a)})^\ast ,
\end{equation}
where we have replicated the system with a standard tensor product of the original Hamiltonian with identity on the remaining replicas so that $[H^{\alpha}, H^{\alpha'}] = 0$ whenever $\alpha \neq \alpha'$. 
More explicitly, our Majorana operators are transformed as
\begin{equation}
    \hat{\gamma}_{k}^{(+,a)} = \mathds{1}^{2(a-1)} \otimes \hat\chi_k \otimes \mathds{1}^{2n - 2a + 1} , \qquad \hat{\gamma}_{k}^{(-,a)} = \mathds{1}^{2a-1} \otimes \hat\chi_k^\ast \otimes \mathds{1}^{2n - 2a} ,
\end{equation}
so that we ordered the replicas as $((+,1), (-,1), (+,2), (-,2), \ldots, (+,n), (-,n))$.
The replicated Hamiltonians are obtained by substituting the original Majorana operators with the replicated ones $ H^\alpha = H_{\hat{\chi}_k \rightarrow \hat{\gamma}^\alpha_k}$.
However, the $\hat{\gamma}$ operators are not genuine Majorana fields since they commute (instead of anticommuting) on different replicas. 
To adjust this, we can introduce the following Klein factors 
\begin{equation}
    F^{(\sigma, a)} = \prod_{i,\mu}
    \chi_{i,\mu}^{(\sigma, a)},
\end{equation}
which allows us to define the operators
\begin{equation}
\begin{aligned}
    & \hchi^{(+,a)}_{i,\mu} = i (\prod_{a'<a} F^{+,a'} F^{-,a'}) F^{+, a} \gamma_{i,\mu}^{+, a} = i \mathcal{S}^a \gamma_{i,\mu}^{+,a} 
    ,
    \\
    & \hchi^{(-,a)}_{i,\mu} = (\prod_{a'<a} F^{+,a'} F^{-,a'}) F^{+, a} \gamma_{i,\mu}^{-, a} =  \mathcal{S}^a \gamma_{i,\mu}^{-,a} 
    ,
\end{aligned}
\end{equation}
from which we recover the standard anticommutation relations, up to a constant normalization,
\begin{equation}
    \{ \hchi_{k}^{\alpha}, 
    \hchi_{k'}^{\alpha'}\} = \delta_{\alpha, \alpha'} \delta_{k,k'} .
\end{equation}
One can verify that, for even $M$,
\begin{equation}
\hgamma_{k_1}^{\alpha}\ldots\hgamma_{k_M}^{\alpha} = \hchi_{k_1}^{\alpha}\ldots\hchi_{k_M}^{\alpha} .
\end{equation}
Thus the replicated Hamiltonians are easily expressed in terms of the proper Majorana operators with a simple substitution $\hgamma_k^\alpha \rightarrow \hchi_k^\alpha$, resulting in
\begin{equation}
  \begin{split}
    H^{(n)}(t)   = \sum_{\sigma,a,i} \left\{\rule{0cm}{20pt}\right. & (i)^{\frac{q_J}{2}}f_\sigma^{\rm uni} \sqrt{J} \left[\sum_{{\boldsymbol{\tilde \mu}}}  h^i_{ \boldsymbol{\tilde \mu}}(t) \prod^{q_J}_{k=1} \hchi^{(\sigma, a)}_{i \tilde \mu_{k}} +    \sum_{{\boldsymbol{\nu}}} \sum_{{\boldsymbol{\mu}}}  h^i_{ \boldsymbol{\mu}\boldsymbol{\nu}}(t) \prod^{q_J/2}_{k=1} \hchi^{(\sigma, a)}_{i \mu_{k}} \prod_{j=1}^{q_J/2} \hchi^{(\sigma, a)}_{i+1 \nu_{j}} \right]  \\
    + \ &  (i)^{\frac{q}{2} + 1}\sigma f_\sigma^{\rm mon} \sqrt{\Gamma} \sum_{\boldsymbol{\tilde \nu}} w^i_{\boldsymbol{\tilde \nu}} \prod_{j=1}^{q}\hchi^{(\sigma, a)}_{i\tilde \nu_j}  \left.\rule{0cm}{20pt}\right\}
,
\end{split}
\end{equation}
where $f^{\rm uni}_{+}=(-1)^{q_J/2+1}$, $f^{\rm uni}_-=1$, $f^{\rm mon}_+=-(-1)^{q/2+1}$ and $f^{\rm mon}_-=-1$,
and extra signs (global for either the unitary or monitoring parts) have been absorbed in the definition of our random variables $h, w$.

  \chapter{Boundary states}\label{App:bound_state}

The operators we calculate in Eq.~\eqref{eq_purity_final}, get vectorized into states according to Eq.~\eqref{eq_overlap_rho}.  
These are the boundary states, and it is important to understand how the replicated Majorana operators act on them.
We will start with the identity boundary state $\ket{\mathbb{I}}$. 
Before vectorization,
\begin{equation}
  \hgamma_{k}^{a} \mathbb{I} \hgamma_{k}^{a} = \mathbb{I} \, ,
\end{equation}
therefore 
\begin{equation}
  \hgamma_{k}^{(+,a)} \hgamma_{k}^{(-,a)} \ket{\mathbb{I}} = \ket{\mathbb{I}} \, .
\end{equation}
Now, we want to express this relation in terms of the new Majorana fields $\hchi^{(\pm,a)}_{k}$.
We have that
\begin{equation}
   i \hchi^{(+,a)}_{k} \hchi^{(-,a)}_{k} = - \mathcal{S}^a \hgamma_{k}^{(+,a)} \mathcal{S}^a \hgamma_{k}^{(-,a)} 
   = - (-1)^{M-1} \mathcal{S}^a \mathcal{S}^a \hgamma_{k}^{(+,a)} \hgamma_{k}^{(-,a)} ,
\end{equation}
where we anticommute $\hgamma_{k}^{(+,a)}$, $M-1$ times to move  through the string $F^{(+, a)}$ (here $M$ is the total number of Majorana operators $M = L N_F$ in each replica).
Computing the square of operator $\mathcal{S}^a$ results in $(\mathcal{S}^a)^2  = (-1)^{M(M-1)/2}$ such that, if we consider $M$ to be a multiple of four, then
\begin{equation}\label{eq_chi_boundary_id}
  i \hchi^{(+,a)}_{k} \hchi^{(-,a)}_{k} \ket{\mathbb{I}} = \ket{\mathbb{I}} .
\end{equation}
Next, we consider the state $\ket{\mathcal{C}_{A,2}}$, where the tensor product with identity is implicit. 
If $k \in \overline{A}$, 
\begin{equation}
  \hgamma_{k}^{a} \big( \mathcal{C}_{A,2} \otimes \mathbb{I} \big) \hgamma_{k}^{a} = \mathcal{C}_{A,2} \otimes \mathbb{I} 
  \ \Leftrightarrow \
  i \hchi^{(+,a)}_{k} \hchi^{(-,a)}_{k} \ket{\mathcal{C}_{A,2}} = \ket{\mathcal{C}_{A,2}} 
  ,
\end{equation}
acting just as the identity, as was previously derived. 
Instead for $k \in A$ we have 
\begin{align} \label{eq:bound}
  \begin{split}
    \hgamma_{k}^{1} \big( \mathcal{C}_{A,2} \otimes \mathbb{I} \big) \hgamma_{k}^{2} & = \mathcal{C}_{A,2} \otimes \mathbb{I} , \\
    \hgamma_{k}^{2} \big( \mathcal{C}_{A,2} \otimes \mathbb{I} \big) \hgamma_{k}^{1} & = \mathcal{C}_{A,2} \otimes \mathbb{I} .
  \end{split} 
\end{align}
With careful consideration of Majorana commutations and Klein factors, analogously to what was done for the identity boundary state, these relations can be vectorized and written in terms of $\hchi$ as
\begin{equation} \label{eq_cycle_bound_chi}
  \begin{aligned}
     i \hchi_{k}^{(+, 1)} \hchi_{k}^{(-, 2)} \ket{ \mathcal{C}_{A,2} } & = \ket{ \mathcal{C}_{A,2} } ,\\
     -i \hchi_{k}^{(+, 2)} \hchi_{k}^{(-, 1)} \ket{ \mathcal{C}_{A,2} }  & = \ket{ \mathcal{C}_{A,2}  } .
  \end{aligned} 
\end{equation}

Finally, we can calculate the expectation value of the fields $\hat \Phi$ on the boundary states,
\begin{equation}\label{eq_bound_state_phi_exp}
  \bra{\mathbb{I}} \Phi_i \ket{\mathbb{I}} = \bra{\mathcal{C}_{A,2} }\hat \Phi_{i \neq A} \ket{\mathcal{C}_{A,2}} = \frac{\Sigma}{2}  
  \ , \
  \bra{\mathcal{C}_{A,2} }\hat \Phi_A \ket{\mathcal{C}_{A,2}} = \frac{1}{2}
  \begin{bmatrix}
    0 & 0 & 0 & 0 & 1 & 0 \\
    0 & 0 & 0 & -1 & 0 & 0 \\
    0 & 0 & 0 & 0 & 0 & \mathbb{I}_{n-2} \\
    0 & 1 & 0 & 0 & 0 & 0 \\
    -1 & 0 & 0 & 0 & 0  & 0 \\
    0 & 0 & - \mathbb{I}_{n-2} & 0 & 0 & 0 \\
  \end{bmatrix}
  ,
\end{equation}
where $\Sigma$ is the symplectic matrix
\begin{equation}
\Sigma = \begin{pmatrix}
0 & \mathbb{I}_n \\ - \mathbb{I}_n & 0 
\end{pmatrix}
.
\end{equation}

Here we can also derive the form of the boundary state $\left|\mathcal{C}_{A,2} \right>$ as a coherent states in the case of two clusters. So we want to find the corresponding orthogonal matrices that define the boundary state. We suppose 
\begin{equation}
    \ket{\mathcal{C}_{A,2}}=e^{\frac{i N_F}{2} {\rm tr} (\omega^t_A \hat \Phi_A)}\left| \mathbb{I} \right>_A e^{\frac{i N_F}{2} {\rm tr} (\omega^t_B \hat \Phi_B)}\left| \mathbb{I} \right>_B,
\end{equation}
where $e^{\omega_A}=\Omega_A$ and $e^{\omega_B}=\Omega_B$ we are looking for. Finally, it is straightforward to confirm
\begin{equation}\begin{split}
    & \bra{\mathcal{C}_{A,2}} \hat \Phi_A  \ket{\mathcal{C}_{A,2}}=\frac{\Omega_A^t \Sigma \Omega_A}{2},\\
    & \bra{\mathcal{C}_{A,2}} \hat \Phi_B \ket{\mathcal{C}_{A,2}}=\frac{\Omega_B^t \Sigma \Omega_B}{2}.
     \end{split}
\end{equation}
By comparing these equations to  Eq.~\eqref{eq_bound_state_phi_exp} we find:
\begin{equation}\label{boundstate}
\begin{split}
    \ket{\mathcal{C}_{A,2}} = \ket{\Omega_A \Omega_B} \quad \Rightarrow \quad &\Omega_A = 
    \begin{bmatrix}
        0 & -1 & 0 \\
        1 & 0 & 0 \\
        0 & 0 & \mathds{1} \\
    \end{bmatrix}
    , \
    \Omega_B = \mathds{1}.
    \end{split}
\end{equation}
  \chapter{Averaging of replicated evolution operators}\label{App:aver}
In this appendix, we perform the averaging of the evolution operator in Eq.~\eqref{eq:evol} and find the averaged replicated Hamiltonian.
The Gaussian average of the evolution operator is given by
\begin{equation}\label{eq_avg_rep_ham_def}
   \mathbb{E}_{\rm G}\left[ \mathcal{T} \exp(-i \int_0^t H^{(n)}(s) ds ) \right] \cong \mathbb{E}_{\rm G}\left[ \prod_j \mathcal{T} \exp(-i \int_{t_j}^{t_j + \Delta t} H^{(n)}(s) ds ) \right] ,
\end{equation}
where in the last step we divided the evolution into segments of length $\Delta t \ll 1 $, with $t_j = j\Delta t$.
To perform the averaging, we expand the exponential into its Taylor series.
Since the white noise terms have zero mean, the leading term in Eq.~\eqref{eq_avg_rep_ham_def} is the second-order one, 
\begin{equation}
  \begin{aligned}
    & \mathbb{E}_{\rm G}\left[ \prod_j \mathcal{T} \exp(-i \int_{t_j}^{t_j + \Delta t} H^{(n)}(s) ds ) \right] \\
    = \ & \mathbb{E}_{\rm G}\left[\prod_j \left( 1 -i \int_{t_j}^{t_j + \Delta t} H^{(n)}(s) ds - \frac{1}{2} \iint_{t_j}^{t_j + \Delta t} \mathcal{T} H^{(n)}(s_1) H^{(n)}(s_2)ds_1 ds_2  + \dots \right) \right]  \\
    = \ & 1 - \frac{1}{2} \sum_j \int_{t_j}^{t_j + \Delta t} \int_{t_j}^{t_j + \Delta t} \mathcal{T} \mathbb{E}_{\rm G}\left[ H^{(n)}(s_1) H^{(n)} (s_2)\right]ds_1 ds_2  + \dots \\ 
    = \ & 1 - \sum_j \int_{t_j}^{t_j + \Delta t} \int_{t_j}^{t_j + \Delta t} \delta(s_1 - s_2) \mathcal{H}^{(n)} ds_1 ds_2  + \dots  = 1 - \mathcal{H}^{(n)}t + \dots = \exp(-t \mathcal{H}^{(n)}) 
  .
  \end{aligned} 
\end{equation}
Therefore we find that the averaged evolution operator gives the imaginary time evolution of the averaged replicated Hamiltonian 
\begin{equation}
  \mathcal{H}^{(n)} = \frac{1}{2} \left( J \sum_{i\mmu {\boldsymbol{\nu}}} \mathcal{H}^\uni_{i i+1, \boldsymbol{\mu}{\boldsymbol{\nu}}} \right. \\ \left.+J\sum_{i\tilde \mmu}\mathcal{H}^\uni_{i, \boldsymbol{\tilde \mu}}- \Gamma \sum_{i \boldsymbol{\tilde \nu} } \mathcal{H}^\mon_{i,\boldsymbol{\tilde \nu} }\right),
\end{equation}
with
\begin{equation}
\begin{split}
  & \mathcal{H}^\uni_{ii+1,\mmu {\boldsymbol{\nu}}} =\left(i^{\frac{q_J}{2}}  \sum_{\sigma, a}  f_\sigma^{\rm uni}\prod^{q_J/2}_{k=1} \hchi^{(\sigma, a)}_{i \mu_{k}} \prod_{j=1}^{q_J/2} \hchi^{(\sigma ,a)}_{i+1 \nu_{j}} \right)^2,
  \end{split}
\end{equation}
\begin{equation}
\begin{split}
  & \mathcal{H}^\uni_{i,\tilde \mmu} =\left(i^{\frac{q_J}{2}}  \sum_{\sigma, a}  f_\sigma^{\rm uni}\prod^{q_J}_{k=1} \hchi^{(\sigma ,a)}_{i \tilde \mu_{k}}\right)^2,
  \end{split}
\end{equation}
and
\begin{equation}
    \mathcal{H}^\mon_{\tilde \mnu} = \left(i^{\frac{q}{2}}  \sum_{\sigma, a} \sigma f_\sigma^{\rm mon}\prod_{j=1}^{q}\hchi^{(\sigma, a)}_{i\tilde \nu_j}   \right)^2 .
\end{equation}
Thus the replicated state $\rho^{(n)}(t)$ will evolve in imaginary time with the Hamiltonian $\mathcal{H}^{(n)}$, meaning that, at $t\rightarrow \infty$, it will act as a projector onto the groundstate $\ket{\mathrm{GS}}$ of $\mathcal{H}^{(n)}$.

  \chapter{Spin coherent states and two–cluster calculations}\label{App:SpinCoherent}

This appendix supplies the spin–coherent–state path integral used in Sec.~\ref{sec:n2q2}, and the detailed two–cluster purity calculations for $n=2$ with $q_J=2$ and $q_J=4$.

\section{Spin coherent states and the path integral}\label{app:spin_path}

We choose the $+\hat y$ direction as reference. A spin-$S$ coherent state pointing along
\[
  \hat{\mathbf n}=(n_x,n_y,n_z)=(\sin\theta\sin\phi,\ \cos\theta,\ \sin\theta\cos\phi)
\]
is obtained by rotating the fully $+y$–polarized state $\ket{\Uparrow^y}$:
\begin{equation}
  \ket{n}:= \ket{\Omega}
  := \exp\!\Big[-i\, \Theta(\hat n)\, \hat{\mathbf u}(\hat n)\!\cdot\!\mathbf S \Big]\ket{\Uparrow^y},
  ~
  \Theta(\hat n)=\arccos(n_y),~
  \hat{\mathbf u}=\frac{\hat{\mathbf n}\times \hat{\mathbf y}}{\|\hat{\mathbf n}\times \hat{\mathbf y}\|}.
  \label{eq:spin_coh_def}
\end{equation}
(Equivalently, use any representative of the coset $SU(2)/U(1)$ consistent with the $y$–axis gauge.)

The overlap of two spin coherent states is
\begin{equation}
  \braket{\Omega'}{\Omega}
  =\Big(e^{i(\phi-\phi')}\sin\tfrac{\theta}{2}\sin\tfrac{\theta'}{2}
          +\cos\tfrac{\theta}{2}\cos\tfrac{\theta'}{2}\Big)^{2S}.
  \label{eq:spin_overlap}
\end{equation}
Time–slicing the propagator, inserting $\mathbb I=\int d\Omega\,\ket{\Omega}\bra{\Omega}$ at each step, and expanding to $O(\delta t)$ gives the usual coherent–state path integral
\begin{equation}
  \langle \Omega_f| e^{-T \mathcal H}|\Omega_i\rangle
  = \int \mathcal D\Omega\,
  \exp\!\Big\{-\int_0^T\! dt\ \big[K(\Omega)-\langle \Omega|\mathcal H|\Omega\rangle\big]\Big\},
  \qquad
  K(\Omega)=-i\,\langle \Omega|\partial_t \Omega\rangle .
  \label{eq:spin_path_integral}
\end{equation}
Using Eq.~\eqref{eq:spin_overlap} at consecutive times, one finds
\begin{equation}
  K(\Omega)= -\,iS\,(1-\cos\theta)\,\dot\phi
            = -\,iS\,(1-n_y)\,\dot\phi ,
  \label{eq:Berry_yaxis}
\end{equation}
which is the Berry term in the $y$–axis gauge. In cylindrical variables we will often set \(y\equiv n_y=\cos\theta\).

For a bipartition \(A\cup B\) (two clusters in Sec.~\ref{sec:n2q2}), the full measure and Berry phase factorize over clusters (and over flavours inside each cluster when we restrict to the fully symmetric spin sector). The diagonal Hamiltonian matrix element is obtained by replacing spin operators by their classical expectation values on \(\ket{\Omega}\), i.e., \(\mathbf S\mapsto S\,\hat{\mathbf n}\).

\paragraph{Compact expression used in Sec.~\ref{sec:n2q2}.}
Specializing to the $y$–axis gauge and denoting by \((\phi_j,y_j)\) the cylindrical coordinates of cluster \(j\),
\begin{equation}
  \int \mathcal D\Omega\;
  \exp\!\Big\{-\int_0^T\! dt\ \sum_{j\in A,B}\Big[-\,iS(1-y_j)\dot\phi_j \;+\; \langle \Omega_j|\mathcal H|\Omega_j\rangle\Big]\Big\}.
  \label{eq:spin_PI_used}
\end{equation}
This is the starting point for Eqs.~\eqref{eq:act}–\eqref{eq:Huni2} in the main text.

\section{Two-Cluster Purity for \texorpdfstring{$n=2$}{n=2}}
\label{app:two_cluster}

This appendix derives the late-time purity for two clusters (\(L=2\)) in the \(n=2\) replica warm-up, complementing Sec.~\ref{sec:n2q2}. We introduce relative and center-of-mass angles
\[
  \tilde\phi := \frac{\phi_2 - \phi_1}{2}, \qquad
  \tilde\theta := \frac{\phi_1 + \phi_2}{2},
\]
and work at large \(S=N_F/2\).

\subsection{Quadratic unitary: \texorpdfstring{$q_J=2$}{qJ=2}}
\label{app:two_cluster_q2}

Next we consider the case $L=2$ of just two clusters and calculate the purity.
In this case, we introduce the relative coordinate and the centre of mass coordinate 
\begin{equation}
\begin{split}
\label{eq:actionn2q2}
     \mathcal{S} & = \mathcal{S}[\theta] + \mathcal{S}[\phi]   = \int^T_0 dt \left[ \frac{\dot{\tilde \theta}^2}{4 \Gamma} + \frac{\dot{\tilde \phi}^2}{4 \Gamma} -2 JS^2 \cos 2 \tilde \phi \right].
\end{split}
\end{equation}
Notice that, we can compute the overlap exactly, taking into account the boundary state $\left|X_A,Z_B \right> = \left|X_1,Z_2 \right>$. 
First, for the large times we can interpret the overlap as the transition amplitude between the boundary state and the ground state of a Hamiltonian of a particle on a circle with a pendulum potential, namely
\begin{equation}
    \mathcal{H} =  \frac{\dot{\tilde{\phi}}^2}{4 \Gamma} - 2 J S^2 \cos 2 \tilde \phi .
\end{equation}
For large $S$, the ground state can be approximated to be same as the ground state of a harmonic oscillator with $\tilde \phi = 0$,
\begin{equation}
    \mathcal{H} = \frac{p^2}{2m} + 4 J S^2\tilde \phi^2 ,
\end{equation}
with $m \omega^2/2 = 4 J S^2$, explicitly
\begin{equation}
    \Psi_{\overline{GS}} (\tilde \phi) \sim e^{-m \omega \tilde \phi^2/2}.
\end{equation}
So for the overlap we have
\begin{equation}
    \langle X_A, Z_B |GS  \rangle = \int d \tilde \phi d \tilde \theta e^{-m \omega \tilde \phi^2/2} \left<X_A,Z_B|\tilde \phi,\tilde \theta \right> .
\end{equation} Here the boundary state $\left|X_A,Z_B \right>$ corresponds to the coherent state with $\tilde \phi_b=\frac{\pi}{2}$ and $\tilde \theta_b=0$.
Using the overlaps between two coherent states (see Appendix \ref{App:SpinCoherent}) we obtain 
the following integral
\begin{equation}
\begin{split}
     \langle X_A, Z_B   & |GS \rangle= \frac{1}{2^{4 S}} \int_{- \pi}^{ \pi} d \tilde \phi \int_{0}^{2 \pi} d \tilde \theta  e^{-m \omega \tilde \phi^2/2} \left(e^{-i(\frac{\pi}{2} - \tilde \phi)} + 1\right)^{2S} \left(e^{i \tilde \theta} + 1\right)^{2S},
\end{split}
\end{equation}
with $m \omega = 2 S \sqrt{\frac{ J}{ \Gamma}}$. The leading order of this expression is given by $\tilde \phi =\frac{\pi}{2}$ and $\tilde \theta = 0$ or $\tilde \theta = 2\pi$, and the latter gives the replicated purity as
\begin{equation}\begin{split}
    \lim_{t \rightarrow \infty} \frac{\Tr \left[ \rho_A^{(2)}(t) \left(\mathcal{C}_2 \otimes \mathbb{I} \right)\right]}{\Tr \rho^{(2)}(t)} \sim e^{-2 S \sqrt{\frac{ J}{\Gamma}}\frac{\pi^2}{8}} ,
    \end{split}
\end{equation}
where $J\sim S^{\alpha}$, with $\alpha \in (-2,0)$.
In the next subsection, we repeat the analysis for the interacting case $q_J=4$. We show that the replicated purity can be again approximated as an overlap of the ground state of an oscillator with the boundary state. 
\subsection{The case with \texorpdfstring{$q_J = 4$}{qJ = 4}}

We now consider the case where the unitary part is interacting, but measurements remain quadratic. In this subsection, we derive a field theory action and calculate the purity in the case of two clusters. Using the mapping between the operator $\hat \Phi$ and the spin operator in Eq.~\eqref{eq:paramPhi}, we can write the replicated Hamiltonian:
\begin{equation}
\begin{split}
    & \mathcal{H}  = \frac{J}{2} \sum_j  \left[ (S^y_j)^{2} (S^y_{j+1})^{2} - ((S^x_j)^{2} (S^x_{j+1})^{2} \right.  \left.+(S^z_j)^{2} (S^z_{j+1})^{2}) \right]  + \frac{J }{12}\sum_j\left[ (S^y_j)^{4} -((S^x_j)^4 \right.\\
    & \left. + (S^z_j)^4) \right] +  2 { \Gamma }\sum_j (S^y_j)^2.
\end{split}
\end{equation}
We can then repeat the procedure of the previous sections, calculating the effective action for the path integral and integrating out the massive degrees of freedom as we did in the $q_J=2$ case, we find the following effective action 
\begin{equation}
\begin{split}
    \mathcal{S} &  = \int^t_0 d \tau \ \left[ \frac{\dot{\tilde \theta}_j^2}{4 \Gamma}  + \frac{\dot{\tilde \phi}_j^2}{4 \Gamma} \right.  \left.- \frac{J S^4}{24}(3+\cos  4 \tilde\phi_j)(3+ \cos 4 \tilde  \theta_j) \right],
\end{split}
\end{equation}
with $\tilde \phi_j = (\phi_{j+1}-\phi_j)/2$ and $\tilde \theta_j =(\phi_{j+1}+\phi_j)/2$. In the case of two clusters, for the large times, we can again compute the overlap explicitly, mapping the partition function into the Hamiltonian evolution.
This problem reduces to finding the ground state of the following Hamiltonian
\begin{equation}
\begin{split}
   H  = \frac{p_{\tilde \phi}^2}{2m} + \frac{p_{\tilde \theta}^2}{2m} - \frac{J S^4}{24}(3+\cos 4\tilde \phi)(3+ \cos  4\tilde \theta),
\end{split}
\end{equation}
where $m=\frac{1}{2 \Gamma}$ and $p_{\sigma}=\dot\sigma/(2\Gamma)$.
In order to find the ground states, we maximize the potential, applying the standard variational method
\begin{equation}
    \partial_{\tilde \phi} V = 0 \quad , \quad \partial_{\tilde \theta} V = 0 ,
\end{equation}
which corresponds to the condition:
\begin{gather}
     \sin 4\tilde \phi = 0 , \\ 
     \sin  4\tilde \theta= 0 ,
\end{gather}
where $\tilde \theta \in (0, 2 \pi)$ and $\tilde \phi \in (- \pi, \pi)$. This condition gives us a set of solutions for maximas
\begin{equation}\label{eq:min}
   \tilde \phi = n \frac{\pi}{2} \quad , \quad \tilde \theta = m \frac{\pi}{2} ,
\end{equation}
with $ n \in \{-2,-1,...,1,2\}$, $ m \in \{0,..,4\}$. We again expand around these minima (the consideration is the same for each of them), 
\begin{equation}
    \mathcal{H} = \frac{p^2_{\tilde \phi}}{2m} + \frac{p^2_{\tilde \theta}}{2m} + \frac{4 J S^4}{3}(\tilde \phi^2 + \tilde \theta^2) ,
\end{equation}
which is just a sum of two uncoupled oscillators. Thus, for the $\overline{GS}$ of the Hamiltonian we can write
\begin{equation}
    \Psi_{\overline{GS}}(\phi,\theta) \sim e^{- m \omega (\tilde \phi^2 + \tilde \theta^2)/2} ,
\end{equation}
with $m = \frac{1}{2 \Gamma}$, and $m \omega^2 = \frac{8 J S^4}{3}$. Finally, the overlap takes the form
\begin{equation}
\begin{split}
   &  \left<X_A,Z_B|GS\right> \sim \left< X_A, Z_B \right| \int d \tilde \phi d \tilde \theta \\
   & e^{- m \omega( \tilde \phi^2 + \tilde \theta^2)/2} \sum_{\rm min} \left|\tilde \phi - \tilde \phi_{\rm min}, \tilde \theta - \tilde \theta_{\rm min}\right>.
\end{split}
\end{equation}
Once again, using the overlaps of coherent states (Appendix \ref{App:SpinCoherent}) and the connection between the replicated purity and the overlap Eq.~\eqref{eq:puroverlSpin}, for $1/S^4<J \leq 1/S^2$ and dominating maximas Eq.~\eqref{eq:min} we obtain the result for the late-time purity 

\begin{equation}
    \lim_{t \rightarrow \infty} \frac{\Tr \left[ \rho_A^{(2)}(t) \left(\mathcal{C}_2 \otimes \mathbb{I} \right)\right]}{\Tr \rho^{(2)}(t)} \sim e^{- S^2 \sqrt{\frac{J}{3 \Gamma}} \left(\frac{\pi}{2}\right)^2},
\end{equation}
while in the case $1/S^2<J $, we can neglect the overlap, so we just have the Gaussian integral,
\begin{equation}
\begin{split}
  &   \lim_{t \rightarrow \infty} \frac{\Tr \left[ \rho_A^{(2)}(t) \left(\mathcal{C}_2 \otimes \mathbb{I} \right)\right]}{\Tr\rho^{(2)}(t)} 
  \\& = \frac{1}{2^{4 S} Z} \int d \tilde \phi d \tilde \theta e^{- m \omega (\tilde\phi^2 + \tilde\theta^2)/2} \sim e^{-4 S \log(2)},
\end{split}
\end{equation}
which gives a volume law scaling of the entanglement with respect to $S$. Here, $S=\frac{N_F}{2}$ effective spin and $N_F$ number of Majorana fermions in each cluster. Since we consider a system of two clusters, the entanglement entropy scales with cluster size $N_F$.
Thus, we have calculated the replicated purity for $n=2$ for two clusters in both cases $q_J=2$ and $q_J=4$ and derived the NLSM for $q_J=2$. 
In the next sections, we repeat this calculation for the generic number of replicas using the formalism of the coherent states on the initial $SO(n)$ group. We also discuss the scaling of the entanglement with the number of clusters $L$ in Sec(\ref{ssec:NLSM}). 
Next, we take the limit $n \rightarrow 1$ and arrive at the desired expression for the purity in the correct replica limit.

  \chapter{Cartan basis \label{sec:cartan}}
Here we explain how to move to the Cartan basis in the case of $so(2n)$ algebra, which we use to define the coherent states for the generic number of replicas.
Since $so(2n)$ is a semisimple Lie algebra, it is possible to transform it into the standard Cartan basis. Let us first recall that $so(2n)$ corresponds to a Dynkin diagram within the $D_n$ series with shape,
\begin{equation}
\scalebox{0.6}{
\begin{tikzpicture}
    \tikzset{every node/.style={circle, draw, minimum size=1cm, inner sep=0pt}} 
    \node (a1) at (0,0) [circle, draw] {$\vec\pi_1$};
    \node (a2) at (2,0) [circle, draw] {$\vec\pi_2$};
    \node (a3) at (4,0) [circle, draw] {$\vec\pi_3$};
    \node (an1) at (8,0) [circle, draw] {$\vec\pi_{n-2}$};
    \node (an2) at (10,0) [circle, draw] {$\vec\pi_{n-1}$};    
    \draw (a1) -- (a2);
    \draw (a2) -- (a3);
    \draw[dotted] (a3) -- (an1); 
    \draw (an1) -- (an2);

    \node (an) at (8,2) [circle, draw] {$\pi_n$};
    \draw (an1) -- (an); 
\end{tikzpicture}
}
\end{equation}
where the $\vec\pi_a$ with $a = 1,\ldots, n$ are the set of simple roots. All roots can be embedded in a $n$-dimensional real vector space endowed with a scalar product such that $\langle \vec\pi_a, \vec\pi_a \rangle = 2$ and $\langle \vec\pi_a, \vec\pi_{a'} \rangle = -1$ for all connected nodes in the diagram (and zero otherwise). Explicitly considering the orthogonal basis $[\vec{e}_a]^b = \delta_{a}^b $ in $\mathbb{R}^n$, we can set 
\begin{align}
    &\vec\pi_a = \vec{e}_a - \vec{e}_{a+1} , \quad a = 1,\ldots, n-1 , \\
    &\vec{\pi}_n = \vec{e}_{n-1} + \vec{e}_{n}.
\end{align}
Then, the full set of roots of $so(2n)$ can be obtained via the action of the Weyl group over the set of simple roots.
This leads to the full set of roots $\mathfrak{R}$ made of all vectors with two non-vanishing components of modulus $1$.
Explicitly, we decompose them into positive and negative roots as $\mathfrak{R} = \mathfrak{R}^+ \cup \mathfrak{R}^-$, with  $\mathfrak{R}^+ = \{ \vec{e}_{|a|} + \operatorname{sgn}(b) \vec{e}_{|b|} | a\in \mathbb{N}, b \in \mathbb{Z}, 1\leq a<|b|\leq n\}$ and $\mathfrak{R}^- = - \mathfrak{R}^+$.
We can then define the Cartan-Weyl basis of generators of $so(2n)$ as $\{\hhc_{a}, \eec_{\vec\rho}, \eec_{- \vec\rho} \}$.
The $\hhc_a$ are the commuting generators within the Cartan subalgebra, while the $\eec_{\vec{\rho}}$ are raising/lowering operators associated to each positive root $\vec\rho \in \mathfrak{R}^+$.
In this basis, the algebra of $so(2n)$ takes the form
\begin{equation}\label{eq:comm}
\begin{split}
    &[\hhc_a, \hhc_{a'}] = 0 ,  \\
    &[\hhc_a, \eec_{\vec{\rho}}] = [\vec\rho]_a  \eec_{\vec\rho} ,\\
    &[\eec_{\vec{\rho}}, \eec_{-\vec\rho}] = \sum_a [\vec\rho]_a \hhc_a ,\\
    &[\eec_{\vec\rho}, \eec_{\vec{\rho}^{\ \prime}}] = N_{\vec\rho, \vec{\rho}^{\ \prime}} \eec_{\vec\rho + \vec{\rho}^{\ \prime}} .
\end{split}
\end{equation}
where 
$a,a' = 1,\ldots, n$, $\vec\rho, \vec{\rho}^{\ \prime} \in \mathcal{R}$
the structure constants $N_{\vec\rho, \vec{\rho}^{\ \prime}}$ vanish if $\vec\rho + \vec{\rho}^{\ \prime} \notin \mathcal{R}$, while their explicit value depends on the normalisation convention and will be clarified afterward. 
We can now see how the Cartan-Weyl basis can be expressed realised in terms of fermionic operators. It is convenient to first move to the Dirac fermions representation,
\begin{equation}\label{eq:dirac}
\begin{split}
   & c_{a j \nu}^{\dagger} = \frac{\chi_{j\nu}^{(+,a)} - i \chi_{j \nu}^{(-,a)}}{\sqrt{2}} , \\
   & c_{a j \nu} =  \frac{\chi_{i\nu}^{(+,a)} + i \chi_{j \nu}^{(-,a)}}{\sqrt{2}},
\end{split}
\end{equation}
with the corresponding inverse relations
\begin{equation}
\begin{split}
&\chi_{j\nu}^{(+,a)} = \frac{c_{a j \nu}^{\dagger} + c_{a j \nu}}{\sqrt{2}},\\
&\chi_{j\nu}^{(-,a)} = i \frac{c_{a j \nu}^{\dagger} - c_{a j \nu}}{\sqrt{2}}.
\end{split}
\end{equation}
Since we have a $so(2n)$ algebra at each site, for simplicity of notation, we drop the chain index $j$ in the definition of the ladder operators and recover it later.
First, the generator within the Cartan subalgebra are direcly associated with occupation numbers setting 
\begin{equation}
\label{eq:cartangen}
\hhc_{a} = \frac1 {N_F} \sum_\nu (c^\dag_{a \nu} c_{a \nu} - 1/2)
=  \frac{i}{ N_F} \sum_{\nu} \chi^{(+,a)}_{\nu} \chi^{(-,a)}_{\nu}  =  \Phi^{(+,a),(-,a)} .
\end{equation}
Then the raising/lowering operators can then be defined for $\vec\rho \in \mathfrak{R}^+$ as
\begin{equation}
\label{eq:eecpos}
    \eec_{\vec\rho} = 
    \begin{cases}
    \frac 1 {N_F}\sum_\nu c^\dag_{a,\nu} c_{|b|,\nu} , & b<0, \\
    \frac 1 {N_F}\sum_{\nu} c^\dag_{a,\nu} c_{|b|,\nu}^\dag , & b>0, 
    \end{cases}
\end{equation}
and extended to the negative roots using $\eec_{-\vec\rho} = \eec_{\vec\rho}^\dag$.

  \chapter{Generic n: Matrix elements of the Hamiltonian }\label{App:matel}
In this appendix, we provide calculation of the matrix elements of the Hamiltonian needed for the formulation of the path integral.
As we already discussed in the section \ref{sec:coherent} in order to find the expectation value of the Hamiltonian on the coherent states $\left< O_\eta \right|
\mathcal{H}^{(n)}\left| O_\eta \right>$, we need to find the expectation values of operators $\left< O_\eta \right|
\hat \Phi\left| O_\eta \right>$. Notice that \begin{equation}
    \left< O_\eta \right|
 (\hat\Phi^{\alpha \alpha'}_i)^{\frac{q_J}{2}} (\hat\Phi^{\alpha \alpha'}_{i+1})^{\frac{q_J}{2}}\left| O_\eta \right>= \left< O_\eta \right|
\hat \Phi^{\alpha \alpha'}_i\left| O_\eta \right>^{\frac{q_J}{2}}\left< O_\eta \right|
\hat \Phi^{\alpha \alpha'}_{i+1}\left| O_\eta \right>^{\frac{q_J}{2}}\left( 1+ O\left(\frac{1}{N_F}\right) \right),
\end{equation}
due to the fact that Majorana operators in the definition of the coherent states are uncorrelated for different flavours and clusters.
Then moving to the representation in terms of Dirac fermions, we arrive to the calculation of the  matrix elements of quadratic operators in $c^{\dagger},c$. 
For instance, let us consider $\hat \Phi^{++}$:
\begin{equation}\label{eq:Phipp}
\begin{split}
   \left<O_\eta \right| \hat \Phi^{++} \left| O_\eta \right>= \frac{i}{N_F}(1-\delta_{a a'}) \sum_\nu \left<O_\eta \right| \hat \chi^{+}_{a \nu} \hat \chi^{+}_{a' \nu}\left| O_\eta \right>\\
   = \frac{i}{2N_F}\left<O_\eta \right|  \sum_\nu (c^{\dagger}_{a \nu} + c_{a \nu})(c^{\dagger}_{a'\nu} + c_{a' \nu})\left| O_\eta \right>  - \frac{i}{2} \delta_{aa'}.
\end{split}
\end{equation}
 Using Eq.~\eqref{Ttr} for each quadratic operator in fermions we obtain:
\begin{equation}\label{ccd}
\begin{split}
    &\frac{1}{N_F} \sum_{\nu}\left<O_\eta \right| c^{\dagger}_{a\nu}c_{a' \nu}\left| O_\eta \right> =(U^{t}U^t)_{a a'},\\
    &\frac{1}{N_F} \sum_{\nu} \left<O_\eta \right| c^{\dagger}_{a\nu}c^{\dagger}_{a' \nu}\left| O_\eta \right>=(U^t V^t)_{aa'},\\
     &\frac{1}{N_F} \sum_{\nu}\left<O_\eta \right| c_{a\nu}c_{a' \nu}\left| O_\eta \right>=-(V^{\dagger} U^t)_{aa'},\\
     &\frac{1}{N_F} \sum_{\nu}\left<O_\eta \right| c_{a\nu}c^{\dagger}_{a' \nu}\left| O_\eta \right>=-( V^{\dagger} V^t)_{aa'}.
     \end{split}
\end{equation}
Now we need to specify the form of the matrices $U$ and $V$. To do so we consider the operator $T$ acting on the fermionic operators, in particular:
\begin{equation}
    T_{\eta} c_{a\nu} T_{\eta}^{-1}= e^{\sum_{\nu,1 \leq a< a' \leq n} \eta_{a a'} c^{\dagger}_{a\nu} c^{\dagger}_{a'\nu}- h.c.} c_{a\nu} e^{-\sum_{\nu,1 \leq a< a' \leq n} \eta_{a a'} c^{\dagger}_{a\nu} c^{\dagger}_{a'\nu}+ h.c.},
\end{equation}
where we assume the summation over repeating indexes. Expanding to the second order in $\eta$ we can find:
\begin{equation}
     T_{\eta} c_{a\nu} T_{\eta}^{-1}=c_{a\nu}-\frac{1}{2} \eta_{ab} \eta^{\dagger}_{ba'} c_{a' \nu}-\eta_{aa'} c^{\dagger}_{a'}+O(\eta^3),
\end{equation}
recovering the orders we can obtain
\begin{equation}
     T_{\eta} c_{a\nu} T_{\eta}^{-1}=[\cos \sqrt{\eta \eta^{\dagger}}]_{a a'} c_{a' \nu}-\left(\frac{\sin \sqrt{\eta^{\dagger} \eta}}{\sqrt{\eta \eta^{\dagger}}} \eta \right)_{aa'} c^{\dagger}_{a' \nu}.
\end{equation}
We will use these relations to find the matrix elements of $\hat \Phi$, but first let us find the connection between variables $\eta$ that we were working with before and variables $\tau$ that we introduced in Eq.\eqref{eq:BCH1}. To do so, it is more convenient to work in the fundamental representation of $so(2n)$. This amounts to replacing the fermionic bilinears with the $2n \times 2n$ matrices 
\begin{equation}
\label{eq:fundrepr}
\begin{aligned}
    & \sum_{\nu} c_{a\nu}^{\dagger} c_{a'\nu} - \frac{1}{2} \delta_{aa'} \longrightarrow E^{a,a'} - E^{n+a', n+a} \\
    & \sum_{\nu} c_{a\nu}^{\dagger} c^{\dagger}_{a'\nu} \longrightarrow E^{a, n+a'} - E^{a', n+a} \\
    & \sum_{\nu} c_{a\nu} c_{a'\nu} \longrightarrow E^{n+a, a'} - E^{n+a', a},
    \end{aligned}
\end{equation}
where $E^{\alpha,\beta}$ is a $2n \times 2n$ matrix with $+1$ in the $i$-th column and $j$-th row, as introduced in Eq.~\eqref{eq:Erepmajo}. To clarify the notation, by the arrow $\rightarrow$ we mean that we replace the generators on the left with their matrix representation on the right. We also stress that this representation is unitarily equivalent to the one introduced for $\hat\Phi^{\alpha\beta} \to W^{\alpha\beta}$ below Eq.~\eqref{eq:Erepmajo}.

A useful fact is that that for any antisymmetric matrix $M$, i.e. $M^t = -M$, also $M f(M^\dag M) = f(M M^\dag) M$ is antisymmetric. First, notice that in this representation, the operator that generates the coherent states can be expressed as
\begin{equation}\label{eq:matrepEta}
    e^{\sum_{\nu, 1 \leq a < a' \leq n} \eta_{a a'} c^{\dagger}_{a\nu} c^{\dagger}_{a'\nu} - \text{h.c.}}=
    \begin{pmatrix}
      \sqrt{\mathbb{I}_n-z z^{\dagger}}& z\\
      -z^{\dagger}&\sqrt{\mathbb{I}_n- z^{\dagger}z}
    \end{pmatrix}, \quad 
    z:=\eta \frac{\sin{\sqrt{\eta^{\dagger}\eta}}}{\sqrt{\eta^{\dagger}\eta}}
\end{equation}
On the other hand, in this representation, the right-hand side of Eq.~\eqref{eq:BCH1} is
\begin{equation}\label{eq:matrepTau}
    \begin{pmatrix}
      \mathbb{I}_n& 0\\
      -\tau^{\dagger}& \mathbb{I}_n
    \end{pmatrix}
    \begin{pmatrix}
      e^{\xi} & 0\\
      0&  e^{-\xi^t  }  
    \end{pmatrix}
    \begin{pmatrix}
      \mathbb{I}_n& \tau\\
      0& \mathbb{I}_n
    \end{pmatrix}
     = 
     \begin{pmatrix}
      e^{\xi} & e^{\xi} \tau\\
      -\tau^\dag e^{\xi} & e^{-\xi^t} - \tau^\dag e^{\xi} \tau
    \end{pmatrix} \;.
\end{equation}
Therefore comparing Eqn.~\eqref{eq:matrepEta} and~\eqref{eq:matrepTau}, we find the connection between the matrices $\eta$ and $\tau$.
Explicitly one obtains the matrix relations
\begin{equation}
    e^{\gamma} = \sqrt{\mathbb{I}_n - z z^\dag} \;, \quad \tau = e^{-\gamma} z = (1 - z z^{\dagger})^{-1/2} z = z (1 - z^\dag z)^{-1/2} .
\end{equation}
These relations allow us to find the matrix elements of operators $\hat \Phi$ in terms of skew-symmetric matrices $\tau$ in a compact form.

\begin{equation}
\begin{aligned}
   & \Phi^{++}=\frac{i}{2}
       \left((\mathbb{I}+\tau^{\dagger} \tau)^{-1} \tau^{\dagger}+ \tau (\mathbb{I}+\tau^{\dagger} \tau)^{-1}+\tau(\mathbb{I}+\tau^{\dagger} \tau)^{-1} \tau^{\dagger}-\tau^{\dagger}  \tau(\mathbb{I}+\tau^{\dagger} \tau)^{-1}\right)\\
    &   \Phi^{+-}=\frac{1}{2}
       \left(-(\mathbb{I}+\tau^{\dagger} \tau)^{-1} \tau^{\dagger}+ \tau (\mathbb{I}+\tau^{\dagger} \tau)^{-1}-\tau(\mathbb{I}+\tau^{\dagger} \tau)^{-1} \tau^{\dagger}-\tau^{\dagger} \tau (\mathbb{I}+\tau^{\dagger} \tau)^{-1}+\mathbb{I}\right)\\
     &  \Phi^{-+}=\frac{1}{2}
       \left(-(\mathbb{I}+\tau^{\dagger} \tau)^{-1} \tau^{\dagger}+ \tau (\mathbb{I}+\tau^{\dagger} \tau)^{-1}+\tau(\mathbb{I}+\tau^{\dagger} \tau)^{-1} \tau^{\dagger}+\tau^{\dagger} \tau (\mathbb{I}+\tau^{\dagger} \tau)^{-1}-\mathbb{I}\right)\\
      & \Phi^{--}=-\frac{i}{2}
       \left((\mathbb{I}+\tau^{\dagger} \tau)^{-1} \tau^{\dagger}+ \tau (\mathbb{I}+\tau^{\dagger} \tau)^{-1}-\tau(\mathbb{I}+\tau^{\dagger} \tau)^{-1} \tau^{\dagger}+\tau^{\dagger}  \tau(\mathbb{I}+\tau^{\dagger} \tau)^{-1}\right)\;.
       \end{aligned}
       \end{equation}
 Using these expressions, one can easily recover the matrix elements of the Hamiltonian, as the matrices $\Phi$ serve as the fundamental building blocks of the Hamiltonian. Let us notice here that in order to recover cluster index one needs just to add an extra index $j$ to each $\tau \rightarrow \tau_j$.
  \chapter{Two-cluster stationary purity  }\label{App:purity2}
In this section, we redo the two-cluster calculation of the purity that we performed in section \ref{sec:n2q2}, but for a generic number of replicas $n$. 
Consider the action for the sigma model in the $q_J=2$ case, 
\begin{equation}
\begin{split}
    \mathcal{S}=\int d t \tr\Big(\frac{1}{16 \Gamma  }\sum_{i=A,B} \partial_t  Q_i^t \partial_t  Q_i-\frac{J N_F^2}{4}  Q_A  Q_{B}^t \Big) ,
    \end{split}
    \end{equation}
where  $Q \in SO(n)$.  
In the generic $n$ case, we introduce new variables $Q_+$ and $Q_-$, playing the role of center of mass and relative coordinates respectively,
\begin{equation} \label{basischange}
     Q_A=Q_+Q_-^t \ , \ Q_B=Q_+ Q_- .
\end{equation}
and applying the general procedure in Eqs.~(\ref{eq:pur},\ref{eq:transitAmpl}), in order to compute the stationary purity we need
\begin{equation}
     \braket{\mathcal{C}_{A,2}}{ GS} = \int dQ_+ dQ_- e^{- \mathcal{S}} \braket{\mathcal{C}_{A,2}}{Q_+, Q_-}\;.
\end{equation}
Due to the large weight of the unitary evolution term, we can assume that $Q_-=e^{\delta Q}$, where the cluster-to-cluster fluctuations $\delta Q$ are small. 
Doing the expansion with respect to $\delta Q$ and rescaling $\delta Q \rightarrow N_f^{-1/4} \delta Q$ , $t \rightarrow \tau/\sqrt{N_f}$  one finds
\begin{equation}
\begin{split}
    &\mathcal{S}=  \int d \tau \left[ \frac{1}{8 \Gamma} \left( \sqrt{N_F} \tr\Big( \dot{Q}_+^t\dot{Q}_+ \Big)+ \tr \Big( \delta \dot{Q}^t \delta \dot{Q} \Big) \right.\right.\\&\left. \left.+ \tr \Big( [\delta \dot{Q}, \delta Q^t]  \dot{Q}^t_+ Q_+\right)+ \frac{J N^{3/2}_F}{2} \tr \Big( \delta  Q^t \delta  Q \Big) \right],
    \end{split}
\end{equation}
Note that in contrast with the $n=2$ case, the $Q_+$ does not decouple from the rest (compare with $\tilde\theta$ in Eq.~\eqref{eq:actionn2q2}). However, since $\dot{Q}_+$ has a large weight $\sqrt{N_F}$, we can use saddle point to perform the integral. Specifically, we write $Q_+ = Q_0 \exp{N_F^{-1/4} \delta \Theta}$, where $Q_0$ is chosen to maximize the overlap $\braket{\mathcal{C}_{A,2}}{Q_+ = Q_0, Q_-}$. We postpone the determination of the value of $Q_0$ and expand the fluctuations in $\delta \Theta$, which we can integrate out the 
\begin{equation}
  \begin{split}
    \mathcal{S} & =\frac{1}{8 \Gamma}  \int d t \left[ \tr  \Big(\delta \dot{  Q}^t \delta \dot{ Q} \Big)  + 4 J N^2_f \Gamma \tr \Big( \delta  Q^t \delta  Q \Big) \right]\\
    & = \frac{1}{4\Gamma}  \int d t \sum_{i>k} \left( \delta \dot{Q}^2_{ik} + 4 J N^{2}_f \Gamma \delta Q^2_{ik} \right) \;.
    \end{split}
\end{equation}
Thus, at large $N_F$, we obtained the action for $n(n-1)/2$ independent harmonic oscillators.
As in the case $n=2$, for large $t$, the dominating configuration comes from the ground state of the harmonic oscillators
\begin{equation}
     \mathcal{H} = \frac{1}{4 \Gamma}\sum_{i>k}\delta \dot{Q}^2_{ik} +  J N^{2}_F\sum_{i>k} \delta Q^2_{i k} ,
\end{equation}
with $m =  1/(2 \Gamma) $, $m \omega^2/2 =  J N^2_F$, whose wave function reads
\begin{equation}
    \Psi_{\overline{GS}} (\delta Q) \sim e^{-m \omega \sum_{i>k} \delta Q_{i j}^2/2} .
\end{equation}
Therefore the integral we want to compute is the overlap of the ground state of the Hamiltonian and the boundary state $\left< \mathcal{C}_{A, 2}  | GS \right>$, resulting in
\begin{equation}\label{eq:overlosc}
    \left<\mathcal{C}_{A,2} | GS \right>=\int d \delta Q e^{- m \omega \sum_{i>k} \delta Q^2_{ik}/2} \left<\mathcal{C}_{A,2} | Q_-,Q_0 \right>,
\end{equation}
where $\left|Q_i\right>$ is a coherent state after taking the limit $ \tau_I \rightarrow 0$ .
The final ingredient of this calculation is understanding the overlap between an arbitrary coherent state and the boundary state $\left<\mathcal{C}_{A,2}| Q_-,Q_0\right>$.
This quantity can be calculated with the help of the formula Eq.~\eqref{eq:overlapPhi}, where we expressed the overlap of two arbitrary coherent states as a determinant of the corresponding $\Phi$'s. More explicitly, using $Q = (1 - \tau_R) / (1 + \tau_R)$ and Eq.~\eqref{eq:PhitauR}, we can relate $\Phi$ and $Q$ as
\begin{equation}
    \Phi =\frac{1}{2} \begin{pmatrix}
    0 & Q \\
    -Q^t & 0 
    \end{pmatrix} \;,
\end{equation}
where we ignore the sub-leading fluctuations in $\tau_I$.
Using the expression in Eq.~\eqref{eq:overlapPhi} and expression of the boundary state in terms of orthogonal matrices (see Appendix \ref{App:bound_state}), we get
\begin{equation}
\begin{split}
    &\left<\mathcal{C}_{A,2}| Q_0,Q_-  \right> = \braket{\mathcal{C}_A}{Q_-} \braket{\mathbb{I}}{Q_0}=
    \det[\frac{\mathbb{I}-2\Sigma \Phi(Q_0)}{2} \frac{\mathbb{I}-2\Omega^t_A \Sigma \Omega_A \Phi(Q_-)}{2}]^{N_F/4}
    \\=&\frac{1}{2^{N_f n }} \det\left[(\mathbb{I}_{n \times  n}+Q_0)(\mathbb{I}_{n \times  n}+Q_{-,1\leftrightarrow{}  2})\right]^{N_F/2},
    \end{split}
\end{equation}
where we used 
\begin{equation}
    \det[\frac{1-2\Sigma \Phi(Q_0)}{2} ]=\det\frac{1}{2}\begin{pmatrix}
        \mathbb{I}+Q^t_0&0\\
       0 & \mathbb{I}+Q_0
    \end{pmatrix},
\end{equation}
\begin{equation}\begin{split}
   &\det[ \frac{1-2\Omega_A^t \Sigma \Omega_A \Phi(Q_-)}{2} ]=\det \frac{1}{2}\begin{pmatrix}
        \mathbb{I}+Q^t_{-,1 \leftrightarrow{}2 }&0\\
       0 & \mathbb{I}+Q_{-,1 \leftrightarrow{}2 }
    \end{pmatrix},
    \end{split}
\end{equation}
and $Q_{-,1 \leftrightarrow{}2 }$ is $Q_-$ with the first and the second rows of the matrix swapped, and with the first row multiplied by $-1$. Therefore, the Eq.\eqref{eq:overlosc} can be rewritten as
\begin{equation}
    \begin{split}
    \left<\mathcal{C}_{A,2} | GS \right>=\frac{1}{2^{N_F n}}\int d \delta Q e^{- N_F \sqrt{\frac{2 J}{\Gamma}} \sum_{i>k} \delta Q^2_{ik}/2} \times e^{\frac{N_F}{2}\log \det\left[(\mathbb{I}_{n \times  n}+Q_0)(\mathbb{I}_{n \times  n}+Q_{-,1\leftrightarrow{}  2})\right]},
    \end{split}
\end{equation}the leading order of this expression is given by $Q_0=\mathbb{I}$ and $Q_{-,1\leftrightarrow{}  2}= \mathbb{I}$. Now we find $\delta Q= \log(Q_{-})$:
\begin{equation}
    \delta Q=\begin{bmatrix}
        0 & \frac{\pi}{2} & 0 \\
        -\frac{\pi}{2} & 0 & 0 \\
        0 & 0 & \mathbb{O} \\
    \end{bmatrix},
\end{equation}
then the replicated purity takes the form
\begin{equation}
     \lim_{t \rightarrow \infty}\frac{\Tr \left[ \rho^{(n)}(t) \left(\mathcal{C}_{A,2} \otimes \mathbb{I} \right)\right]}{\Tr \rho^{(2)}(t)}=\frac{1}{Z}\left<\mathcal{C}_{2,A}| GS  \right>\sim e^{- \frac{N_F}{2} \sqrt{\frac{ J}{\Gamma}}\left(\frac{\pi}{2}\right)^2} . 
    \end{equation}
Therefore, the limit $n \rightarrow 1$ becomes trivial and gives purity
\begin{equation}\label{eq_purity_gaussian}
    \left.\overline{\Tr\rho_A(t)^2} \right|_{t \rightarrow \infty}\sim e^{-\frac{N_F}{2}\sqrt{\frac{J}{ \Gamma} } \left(\frac{\pi}{2}\right)^2 },~~~1/N_F^2 < J \leq 1 ,
\end{equation}
which interpolates between area-law for $J \leqslant 1/N_F^2$ and volume-law for $J \geqslant 1$. One can notice that this result is consistent with $n=2$ case.

\begin{figure*}[t]
  \centering
  \begin{subfigure}[t]{0.473\linewidth}
    \includegraphics[trim=0 0 0 0, clip, width=\columnwidth]{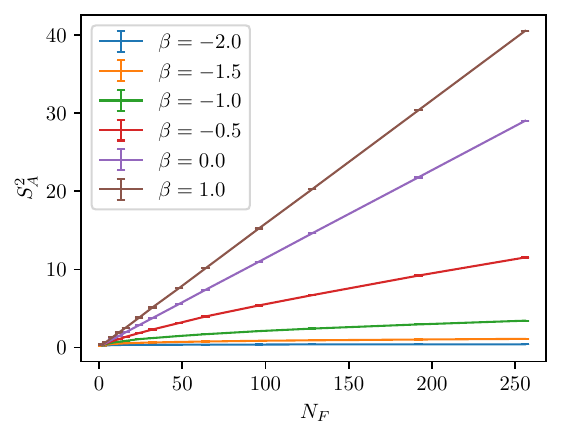}
    \caption{$S_A^2$ as a function of $N_F$, for different $\beta$.}
    \label{fig_2cluster_gaussian_entropies_NF_S2}
  \end{subfigure}
  \begin{subfigure}[t]{0.487\linewidth}
    \includegraphics[trim=0 0 0 0, clip, width=\columnwidth]{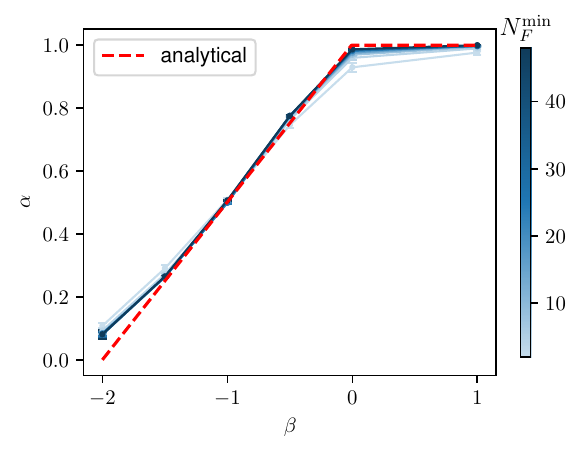}
    \caption{$\alpha$ from a fit with $N_F \geqslant N_F^{\textnormal{min}}$, as a function of $\beta$.}
    \label{fig_2cluster_gaussian_entropies_beta_alpha}
  \end{subfigure}
  \caption{ED results for the saturated ($t \rightarrow \infty$) second Rényi entropy $S_A^2$, with $q = q_J = 2$, $J = N_F^\beta$ and $\Gamma = 1$. 
  The entropy is expected to follow a law $S_A^2 \propto N_F^\alpha$, where the exponent $\alpha$ can be found with a fit.}
  \label{fig_2cluster_gaussian_entropies}
\end{figure*}

We compare our predictions for the purity with numerical exact diagonalization (ED) simulations. 
We compute the saturated second Rényi entropy, given by $S_A^2 = - \log \Tr \rho_A^2$, and approximate $\overline{S_A^2} \approx - \log \overline{\Tr \rho_A^2}$, assuming that the fluctuations are suppressed at large $N_F$. 
Setting a power law dependency of $J$ with $N_F$, $J = N_F^\beta$, different entropy growths are observed in Fig. \ref{fig_2cluster_gaussian_entropies_NF_S2}.
Assuming a power law $S_A^2 \propto N_F^\alpha$, the exponent $\alpha$ can be found with a fit.
Our theretical prediction for $\alpha$ can be obtained from the expression of Eq.~\eqref{eq_purity_gaussian}, with $\alpha = 0$ corresponding to an area-law state and $\alpha = 1$ to a volume-law state, giving this way
\begin{equation}
    \alpha = 
    \left\{
    \begin{array}{lll}
        1 + \beta/2 & , &  -2 < \beta \leqslant 0 \\
        1 & , & \beta > 0
    \end{array}
    \right.
    \, .
\end{equation}
We find that the numerically obtained $\alpha$ agrees with the predictions of Eq.~\eqref{eq_purity_gaussian}, as revealed by the convergence to the dashed line in Fig. \ref{fig_2cluster_gaussian_entropies_beta_alpha}.
For $\beta = -2$, the numerics and the analysis agree less well, but that is precisely the point where we expect $\alpha = 0$ and the numerically obtained fit is affected by strong finite-size effects. 

Now we move to the case $q_J=4$. 
We will follow the same steps as for the case $q=2$. 
Let us again start with the action,
\begin{equation}
   -\mathcal{S}=-\frac{1}{16 \Gamma} \int dt~~\sum_{j=A,B }{\rm tr} \Big( \partial_t Q_j \partial_t Q_j^t \Big)  +\frac{JN_f^4}{64}\int d t\left[  \sum_{\alpha \alpha'} (Q^{\alpha \alpha'}_A)^2 (Q^{\alpha \alpha'}_B)^2\right.   \left. +\frac{1}{6} \sum_{j \alpha \alpha'} (Q^{\alpha \alpha'}_j)^4 \right] .
\end{equation}
The minima of the action are given by the permutation matrices with determinant equal to one.
We then introduce new variables
\begin{equation} 
    Q_A \rightarrow D_A Q_A, ~~~Q_B \rightarrow D_B Q_B ,
\end{equation}
where $D_i$ is a permutation matrix, and $ Q_i=e^{\delta  Q_i}$ is a relative coordinate close to the identity matrix, with $\delta  Q$ a skew-symmetric matrix. 
We expand the exponent concerning $\delta Q$ and the expansion reveals once again the action of the quadratic oscillators,
\begin{equation}
\begin{split}
   -\mathcal{S}=-\frac{1}{16 \Gamma} \sum_{j=A,B }\int dt~~{\rm tr} \left(\delta \dot Q_j ( \delta \dot Q_j)^t\right)+\frac{JN_F^4}{48}\int dt ~({\rm tr}(\delta Q_A\delta Q_A)+{\rm tr}(\delta Q_B\delta Q_B)).
   \end{split}
  \end{equation}
Taking into account the boundary states we get the following expression for the overlap,
\begin{equation}
\begin{split}
    \left<  \mathcal{C}_{2,A}| GS\right>&=\frac{1}{2^{N_F n}}\int d \delta Q_A d \delta Q_B    e^{- m \omega \sum_{\alpha \alpha' i} (\delta Q^{\alpha \alpha'}_{i})^2/2}\\
    \times & e^{N_F {\rm tr} \log(\mathbb{I}+e^{\delta Q_{B, 1 \leftrightarrow{}2}})+N_F{\rm tr} \log(\mathbb{I}+e^{\delta Q_{A}})}.
    \end{split}
\end{equation}
where $m\omega= \frac{N_F^2}{2} \sqrt{\frac{J}{3 \Gamma}}$. 
Once again, using the overlaps of coherent states  for $1/N_F^4<J \leq 1/N_F^2$ and dominating maximas $e^{\delta Q_A}=\mathbb{I}$ and $e^{\delta Q_{B,1\leftrightarrow{}  2}}= \mathbb{I}$  we obtain the result for the late-time purity 
\begin{equation}
    \left< \mathcal{C}_{2,A}|GS  \right>\sim e^{-\frac{N_F^2}{4} \sqrt{\frac{J}{3 \Gamma}} \left(\frac{\pi}{2}\right)^2} .
\end{equation}
Finally, in the case, $1/N_F^2<J<1$ we neglect the overlap and get the Gaussian integral
\begin{equation}
\begin{split}
   & \frac{1}{2^{N_F n}}\int d \delta Q_A d \delta Q_B e^{- m \omega \sum_{\alpha \alpha' i} (\delta Q^{\alpha \alpha'}_{i})^2/2} \\
   &\sim e^{-N_F n \log(2)}.
\end{split}
\end{equation}
In both cases, the replica limit $n \rightarrow 1$ can be obtained trivially, resulting in the infinite-time purities
\begin{equation}\label{eq_purity_qJ4}
    \begin{split}
        &\left.\overline{\Tr\rho_A(t)^2}\right|_{t\rightarrow \infty} \sim e^{-\frac{N_F^2}{4}\sqrt{\frac{J}{3\Gamma}} \left(\frac{\pi}{2}\right)^2},~\frac{1}{N_F^4} < J \leq \frac{1}{N_F^2}\\&
        \left.\overline{\Tr\rho_A(t)^2}\right|_{t \rightarrow \infty}\sim e^{-N_F \log(2)},~\frac{1}{N_F^2}<J<1 \,.
    \end{split}
\end{equation}

\begin{figure*}[t]
  \centering
  \begin{subfigure}[t]{0.48\linewidth}
    \includegraphics[trim=0 0 0 0, clip, width=\columnwidth]{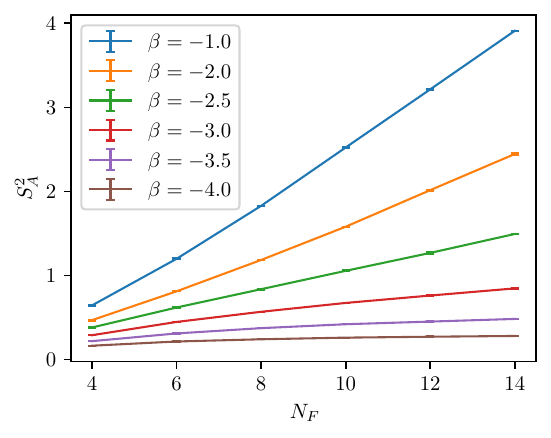}
    \caption{$S_A^2$ as a function of $N_F$, for different $\beta$.}
    \label{fig_2cluster_trotterv2_entropies_NF_S2}
  \end{subfigure}
  \begin{subfigure}[t]{0.50\linewidth}
    \includegraphics[trim=0 0 0 0, clip, width=\columnwidth]{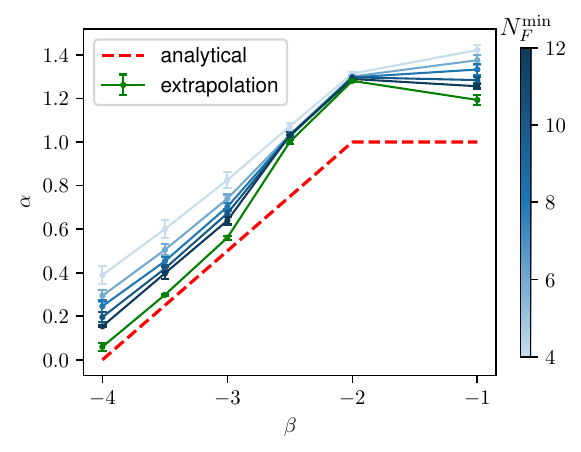}
    \caption{$\alpha$ from a fit with $N_F \geqslant N_F^{\textnormal{min}}$, as a function of $\beta$.}
    \label{fig_2cluster_trotterv2_entropies_beta_alpha}
  \end{subfigure}
  \caption{ED results for the saturated ($t \rightarrow \infty$) second Rényi entropy $S_A^2$, with $q = 2$, $q_J = 4$, $J = N_F^\beta$ and $\Gamma = 1$. 
  The entropy is expected to follow a law $S_A^2 \propto N_F^\alpha$, where the exponent $\alpha$ can be found with a fit.}
  \label{fig_2cluster_trotterv2_entropies}
\end{figure*}

Analogously to the Gaussian case, we compare the predictions to numerical ED simulations.
However, now the system is interacting (exponentially hard to simulate in $N_F$) and there is a larger number of terms in the unitary part of the Hamiltonian, so results were only obtained for $N_F \leqslant 16$.
Setting $J = N_F^\beta$, different entropy growths are observed in Fig. \ref{fig_2cluster_trotterv2_entropies_NF_S2}.
The exponent $\alpha$ is obtained by fitting $S_A^2 \propto N_F^\alpha$ in Fig. \ref{fig_2cluster_trotterv2_entropies_beta_alpha}, progressively increasing the minimum number of flavours $N_F^{\min}$.
We observe a qualitative similarity of the dependence of $\alpha$ with $\beta$ between the ED results and the analytical prediction of Eq.~\eqref{eq_purity_qJ4}, which corresponds to
\begin{equation}
    \alpha = 
    \left\{
    \begin{array}{lll}
        2 + \beta/2 & , &  -4 < \beta \leqslant -2 \\
        1 & , & \beta > -2
    \end{array}
    \right.
    \, .
\end{equation}
Although this generally corroborates our analytical results, due to large finite-size effects, the two do not quantitively match, even though increasing $N_F^{\min}$ generally brings us closer to the analytical prediction. 
Finite-size effects are also evidently present, since regimes where $\alpha > 1$ are not possible in the thermodynamic limit (the maximum is set by the volume-law at $\alpha = 1$).
An extrapolation to $N_F^{\min} \rightarrow \infty$ is performed by linearly fitting $\alpha$ as a function of $1/N_F^{\min}$, which takes us even closer to the predicted values, but still suffers from finite-size effects.
{
Finite-size effects also progressively increase as we approach the volume-law regime in the range $-4 \leqslant \beta \leqslant -2$. 
Going deeper into the volume-law regime, finite-size effects seem to reduce, possibly since measurements become irrelevant, as Eq.~\eqref{eq_purity_qJ4} suggests.
}

In this section, we calculated the purity for a system with two clusters ($A$ and $B$) in the case $q_J=2$ and $q_J=4$. 
{
In both cases, we find that the purity interpolates between area-law and volume-law regimes, which we confirm numerically.
However, in the $q_J = 2$ case, this interpolation occurs when $J$ scales as $1/N_F^2 < J \leq 1$, whereas for $q_J = 4$, due to the larger strength of unitary evolution, it occurs for the scaling $1/N_F^4 < J \leq 1/N_F^2$.
}
  \chapter{Details on the measurement-only action}\label{App:mon}
In this appendix, we first provide the derivation of the expansion for the kinetic term and the expansion of the Hamiltonian part of the action. Here we start with the formula we derived in the section \ref{sec:kin}:

\begin{equation}
K[\tau,\tau^\ast] =  \frac{1}{2} \sum_{\alpha \beta} \left[\partial_{\tau_{ij}}(\ln \mathcal{N}) \dot \tau_{ij} - \partial_{\tau_{ij}^\ast}(\ln \mathcal{N}) \dot \tau^\ast_{ij} \right] d t.
\end{equation}
Now by computing the metric in $\tau + \Delta x \delta \tau$ at the first order
\begin{equation}
\begin{split}
    &K[\tau + \Delta x \delta \tau, \tau + \Delta x \delta \tau^\ast] - K[\tau, \tau^\ast] = \\ &\frac{\Delta x}{2} \big( \partial_{\tau_{ij},\tau_{kl}}(\ln \mathcal{N}) \dot \tau_{ij} \delta \tau_{kl} + \partial_{\tau_{ij},\tau_{kl}^\ast}(\ln \mathcal{N}) \dot \tau_{ij} \delta \tau_{kl}^\ast + \partial_{\tau_{ij}}(\ln \mathcal{N}) \delta \dot \tau_{ij} \\ 
    &- \partial_{\tau_{ij}^\ast \tau_{lk}} (\ln \mathcal{N}) \dot \tau^\ast_{ij} \delta \tau_{lk} - \partial_{\tau_{ij}^\ast \tau_{lk}^\ast} (\ln \mathcal{N}) \dot \tau^\ast_{ij} \delta \tau_{lk}^\ast - \partial_{\tau_{ij}^\ast}(\ln \mathcal{N}) \delta \dot \tau^\ast_{ij}  \big) + O(\delta \tau^2).
\end{split}
\end{equation}
By integrating by parts the terms containing $\delta \dot \tau$ we have: 
\begin{equation}
    \begin{split}
         \partial_{\tau_{ij}}(\ln \mathcal{N}) \delta \dot \tau_{ij} &=  - \partial_{\tau_{ij} \tau_{lk}}(\ln \mathcal{N}) \dot \tau_{lk}\delta \tau_{ij} - \partial_{\tau_{ij} \tau_{lk}^\ast}(\ln \mathcal{N}) \dot \tau_{lk}^\ast \delta \tau_{ij} \\
         \partial_{\tau_{ij}^\ast}(\ln \mathcal{N}) \delta \dot \tau_{ij}^\ast &=  - \partial_{\tau_{ij}^\ast \tau_{lk}}(\ln \mathcal{N}) \dot \tau_{lk} \delta \tau_{ij}^\ast - \partial_{\tau_{ij}^\ast \tau_{lk}^\ast}(\ln \mathcal{N}) \dot \tau_{lk}^\ast \delta \tau_{ij}^\ast, \\
    \end{split}
\end{equation}
so that we get 
\begin{equation}
\begin{split}
   & K[\tau + \delta \tau, \tau + \delta \tau^\ast] - K[\tau, \tau^\ast] =  \frac{\Delta x}{2} \big(   \partial_{\tau_{ij},\tau_{kl}^\ast}(\ln \mathcal{N}) \dot \tau_{ij} \delta \tau_{kl}^\ast - \partial_{\tau_{ij} \tau_{lk}^\ast}(\ln \mathcal{N}) \dot \tau_{lk}^\ast \delta \tau_{ij} \\ 
    &- \partial_{\tau_{ij}^\ast \tau_{lk}} (\ln \mathcal{N}) \dot \tau^\ast_{ij} \delta \tau_{lk} + \partial_{\tau_{ij}^\ast \tau_{lk}}(\ln \mathcal{N}) \dot \tau_{lk} \delta \tau_{ij}^\ast  \big) + O(\delta \tau^2) = \mathtt{g}_{\alpha \beta} \dot \tau_\alpha \delta \tau_\beta^\ast - \mathtt{g}_{\beta\alpha} \dot \tau_\alpha^\ast  \delta \tau_\beta
\end{split}\;,
\end{equation} where we defined $\mathtt{g}_{\alpha \beta}= \frac{\partial^2 \ln \mathcal{N}}{\partial \tau_{\alpha } \partial \tau^*_{\beta}}$ is a metric tensor see Eq.~\eqref{eq:kahler}
(notice that $\mathtt{g}_{\beta \alpha} = \mathtt{g}_{\alpha \beta}^{*}$). Setting $\tau_{2i-1} = \tau + \Delta x\delta \tau$ and $\tau_{2i}= \tau -\Delta x \delta \tau$, one can expand the difference of kinetic terms in terms of the metric
\begin{equation}
    K_{2i-1} - K_{2i} = 2N_F \Delta x \mathtt{g}_{\alpha \beta}(\tau^{2i-1}) \dot \tau^{2i-1}_{\alpha} (\delta\tau^{2i-1}_{\beta})^*-c.c.
\end{equation}
Similarly, the interaction term expanded for small $\delta \tau$ up to second order must be invariant, and so we have:
\begin{equation}
    \sum_i H_{i ,i+1}=\sum_j H_{2j-1, 2j}+H_{2j,2j+1},
\end{equation}
where $H_{2j-1,2j}$ is a Hamiltonian defined inside each dimer and $H_{2j,2j+1}$ is a Hamiltonian defined in between different dimers. Therefore, for the interaction inside each dimer we write:
\begin{equation}
    H_{2i-1,2i}=- 4 N_F^2 \Delta x^2 \Gamma(1-\Delta) \  \mathtt{g}_{\alpha\beta}(\tau^{2i-1}) \delta \tau^{2i-1}_\alpha (\delta \tau^{2i-1}_\beta)^\ast,
\end{equation}
and for the interaction term between each dimer  we do the expansion:
\begin{equation}
    H_{2i,2i+1}= - 4 N_F^2 \Gamma(1+\Delta) \mathtt{g}_{\alpha \beta}(\tau^{2i-1})(\Delta x \delta \tau_{\alpha}^{2i-1}+\Delta x \partial_x \tau_{\alpha}^{2i-1})(\Delta x (\delta \tau_{\beta}^{2i-1})^*+\Delta x (\partial_x \tau_{\beta}^{2i-1})^*).
\end{equation}
Here we also provide the calculation of the Gaussian integral in $\delta \tau$. Notice that it can be presented as
\begin{equation}
    \int \delta \tau e^{-8 N_F^2 \Delta x \Gamma \int d^2 x \delta \tau_{\alpha} \mathtt{g}_{\alpha \beta} \delta \tau^*_{\beta}+\int d^2 x A_{\alpha} \delta \tau^*_{\alpha}+\int d^2 x \delta \tau_{\alpha} B_{\alpha}}\sim e^{\frac{A_{\alpha} \mathtt{g}^{-1}_{\alpha \beta }B_{\beta}}{8 N_F^2 \Delta x \Gamma}},
\end{equation}
where we assume summation over repeating indexes, here 
\begin{equation}
    A_{\alpha}=2 N_F \dot{\tau}_{\gamma} \mathtt{g}_{\gamma \alpha} -4 N_F^2\Gamma (1+\Delta) \Delta x \partial_x \tau_{\gamma} \mathtt{g}_{\gamma \alpha},
\end{equation}
\begin{equation}
    B_{\alpha}=-2 N_F \mathtt{g}_{\alpha \gamma} \dot{\tau}^*_{\gamma} -4 N_F^2\Gamma(1+\Delta) \Delta x \mathtt{g}_{\alpha \gamma} \partial_x \tau^*_{\gamma}.
\end{equation}
  \chapter{Details on the expansion of the action for small \texorpdfstring{$\tau_I$}{tau	extunderscore I}}\label{App:tauexp}
In this appendix we perform the expansion of the kinetic term and the Hamilotnian discussed in the subsection \ref{ssec:integ_tau}, with respect to $\tau_I$. Remind that $\tau_I$ is the imaginary part of the matrix $\tau$. As we are going to show, monitoring part of the Hamiltonian has only second order in this expansion, while unitary part has also zero order. So the action has the following form:
\begin{equation}
    -\mathcal{S}=\int dt \left(N_F \tr (A \tau_I)+ J N_F^2 (\tr(B)+\tr(C\tau_IC\tau_I))+\Gamma N_F^2 \tr(D\tau_ID\tau_I) \right),
\end{equation}
where $A,B,C,D$ matrices that depend on the real part of $\tau$. Rescaling $\tau_I \rightarrow N_F \tau_I$ as we did with the parameter $y$ in the section \ref{sec:n2q2} we can neglect the quadratic order in the unitary part of the action, since it is subleading due to the scaling of the unitary  coupling $J \sim 1/N_F^{\alpha}$.
Therefore we proceed with the expansion of the kinetic term. For the first matrix under the trace in Eq.~\eqref{eq:kin} we get in the first order:
\begin{equation}
   (\mathbb{I}+\tau^{\dagger} \tau)^{-1}= (\mathbb{I}-( \tau_R -i \tau_I)(\tau_R+i \tau_I))^{-1}=((\mathbb{I}-\tau_R^2)^{-1}-i(\mathbb{I}-\tau_R^2)^{-1}[\tau_I, \tau_R](\mathbb{I}-\tau_R^2)^{-1}),
\end{equation}
where $[.,.]$ is a commutator, and
\begin{equation}
    \tau^{\dagger} \dot{\tau}-\dot{\tau}^{\dagger}\tau=-\tau_R \dot{\tau}_R+\dot \tau_R {\tau}_R-i\{\tau_R, \dot \tau_I\}+i\{\tau_I,  \dot\tau_R \},
\end{equation}
all together it gives:
\begin{equation}
\begin{split}
     &(\mathbb{I}+\tau^{\dagger} \tau)^{-1}(\tau^{\dagger} \dot{\tau}-\dot{\tau}^{\dagger}\tau)=\\&=((\mathbb{I}-\tau_R^2)^{-1}-i(\mathbb{I}-\tau_R^2)^{-1}[\tau_I, \tau_R](\mathbb{I}-\tau_R^2)^{-1}) ([\dot \tau_R,\tau_R]+i \{\tau_I,  \dot\tau_R \}-i \{\tau_R, \dot \tau_I\}),
     \end{split}
\end{equation}
and for the trace we naturally have the expansion
\begin{equation}\label{eq:kin_tau}
\begin{split}
    {\rm tr} (\mathbb{I}+\tau^{\dagger} \tau)^{-1}(\tau^{\dagger} \dot{\tau}-\dot{\tau}^{\dagger}\tau)=- i {\rm tr} (\mathbb{I}-\tau_R^2)^{-1} [\tau_I,\tau_R](\mathbb{I}-\tau_R^2)^{-1} [\dot \tau_R,\tau_R]+i {\rm tr}(\dot \tau_R (\mathbb{I}-\tau_R^2)^{-1} \tau_I)\\+i {\rm tr}( (\mathbb{I}-\tau_R^2)^{-1} \dot \tau_R \tau_I)-i {\rm tr}( (\mathbb{I}-\tau_R^2)^{-1} \tau_R \dot \tau_I)-i {\rm tr}( \tau_R (\mathbb{I}-\tau_R^2)^{-1}\dot \tau_I ).\end{split}\end{equation}
As in the case $n=2$ we can perform integration by parts for the last two terms
\begin{equation}
\begin{split}
    &-i \int dt ~~{\rm tr}( (\mathbb{I}-\tau_R^2)^{-1} \tau_R \dot \tau_I)-i \int dt ~~ {\rm tr}( \tau_R (\mathbb{I}-\tau_R^2)^{-1}\dot \tau_I  )=\\&=i \int dt ~~ {\rm tr}~\left( \tau_I \frac{\partial}{\partial t} (\mathbb{I}-\tau_R^2)^{-1} \tau_R \right)+ i \int dt ~~ {\rm tr}~\left( \tau_I \frac{\partial}{\partial t} \tau_R (1-\tau_R^2)^{-1}  \right)=\\&=
 i \int dt ~~ {\rm tr} \left( \tau_I(\mathbb{I}-\tau_R^2)^{-1}(\dot \tau_R \tau_R+\tau_R \dot \tau_R) \tau_R (\mathbb{I}-\tau_R^2)^{-1} \right) +  i \int dt ~~ {\rm tr} (\tau_I (\mathbb{I}-\tau_R^2)^{-1} \dot \tau_R)+\\&+
    i \int dt ~~ {\rm tr} (\tau_I \dot \tau_R (\mathbb{I}-\tau_R^2)^{-1} )+ i \int dt ~~ {\rm tr} \left( \tau_I (\mathbb{I}-\tau_R^2)^{-1} \tau_R (\dot \tau_R \tau_R+\tau_R \dot \tau_R)  (\mathbb{I}-\tau_R^2)^{-1} \right),
    \end{split}
\end{equation}
where we used $\frac{\partial }{\partial t} A^2= A' A+A A'$. Together with the first  terms from Eq.~\eqref{eq:kin_tau}
\begin{equation}
\begin{split}
    &-2 i \int dt~~ {\rm tr}~~ \frac{\tau_R}{\mathbb{I}-\tau_R^2} \tau_I \frac{\tau_R}{\mathbb{I}-\tau_R^2} \dot \tau_R+i \int d t~~{\rm tr}~~ \frac{1}{\mathbb{I}-\tau_R^2} \tau_I \frac{\tau^2_R}{1-\tau_R^2} \dot \tau_R+i \int d t~~{\rm tr}~~ \frac{\tau^2_R}{\mathbb{I}-\tau_R^2} \tau_I \frac{1}{\mathbb{I}-\tau_R^2} \dot \tau_R
    \\&+4 i \int dt ~~ {\rm tr} \left((\mathbb{I}-\tau_R^2)^{-1} \dot \tau_R (\mathbb{I}-\tau_R^2)^{-1}  \tau_I \right)-i\int dt~~ {\rm tr} \left((\mathbb{I}-\tau_R^2)^{-1} \tau_R^2 \dot \tau_R (\mathbb{I}-\tau_R^2)^{-1} \tau_I \right)-\\&-i\int dt~~ {\rm tr} \left((\mathbb{I}-\tau_R^2)^{-1}  \dot \tau_R \tau_R^2 (\mathbb{I}-\tau_R^2)^{-1} \tau_I \right)+2 i\int dt~~ {\rm tr} \left((\mathbb{I}-\tau_R^2)^{-1} \tau_R \dot \tau_R \tau_R (\mathbb{I}-\tau_R^2)^{-1} \tau_I \right)=\\&=
    i \int dt ~~{\rm tr}\left( 4 \dot \tau_R  \frac{1}{\mathbb{I}-\tau_R^2} \tau_I \frac{1}{\mathbb{I}-\tau_R^2}\right),
    \end{split}
\end{equation}
which gives us the expansion of the kinetic term, notice that it is in complete analogy with the case $n=2$.
For the Hamiltonian part, we first proceed with the expansion of the matrix $\Phi$. Starting from the expression we derived in the appendix \ref{App:matel} for $\Phi$  in terms of $\tau$,
we find the following expansion at linear order in $\tau_I$
\begin{equation}
\begin{split}
    \Phi^{++} =
    -\frac{1}{\mathbb{I}-\tau_R^2} \tau_I \frac{1}{\mathbb{I}-\tau_R^2}+\frac{1}{\mathbb{I}-\tau_R^2} \tau_I \frac{\tau_R}{\mathbb{I}-\tau_R^2}-\frac{\tau_R}{\mathbb{I}-\tau_R^2} \tau_I \frac{1}{\mathbb{I}-\tau_R^2}+\frac{\tau_R}{\mathbb{I}-\tau_R^2} \tau_I \frac{\tau_R}{\mathbb{I}-\tau_R^2},
    \end{split}
\end{equation}
\begin{equation}
\begin{split}
    \Phi^{--} 
    =
    \frac{1}{\mathbb{I}-\tau_R^2} \tau_I \frac{1}{\mathbb{I}-\tau_R^2}+\frac{1}{\mathbb{I}-\tau_R^2} \tau_I \frac{\tau_R}{\mathbb{I}-\tau_R^2}-\frac{\tau_R}{\mathbb{I}-\tau_R^2} \tau_I \frac{1}{\mathbb{I}-\tau_R^2}-\frac{\tau_R}{\mathbb{I}-\tau_R^2} \tau_I \frac{\tau_R}{\mathbb{I}-\tau_R^2}\;;
    \end{split}
\end{equation}
and 
\begin{equation}
    \Phi^{+-}=\frac{1}{2}\frac{\mathbb{I}-\tau_R}{\mathbb{I}+\tau_R}+O(\tau_I^2).
\end{equation}
So for the monitoring part of the Hamiltonian we arrive at 
\begin{equation}
    \mathcal{ H}_{\rm mon}= \frac{N_F^2 \Gamma }{4}\sum_{i=1}^L \left(2 {\rm tr} \left( \Phi^{++}_i(\Phi^{++}_i)^t \right) +2{\rm tr} \left( \Phi^{--}_i(\Phi^{--}_i)^t \right)\right).
\end{equation}
where we used $\tr \Phi \Phi^t=\frac{n}{4}={\rm tr} \Phi^{++}_i(\Phi^{++}_i)^t+{\rm tr} \Phi^{+-}_i(\Phi^{+-}_i)^t+{\rm tr} \Phi^{-+}_i(\Phi^{-+}_i)^t+{\rm tr} \Phi^{--}_i(\Phi^{--}_i)^t $, and  ${\rm tr} \Phi^{+-}_i(\Phi^{+-}_i)^t={\rm tr} \Phi^{-+}_i(\Phi^{-+}_i)^t$. And in terms of $\tau$ we find a full square under the trace: 
\begin{equation}
    \mathcal{ H}_{\rm mon}=- \Gamma N_f^2 \sum_{i=1}^L {\rm tr} \left(\frac{1}{(\mathbb{I}-\tau_{R,i}^2)^2} \tau_{I,i}-\frac{\tau_{R,i}^2}{(\mathbb{I}-\tau_{R,i}^2)^2} \tau_{I,i} \right)^2=- \Gamma N_f^2\sum_{i=1}^L {\rm tr}\left( \frac{1}{\mathbb{I}-\tau_{R,i}^2} \tau_{I,i}\frac{1}{\mathbb{I}-\tau_{R,i}^2} \tau_{I,i} \right).
\end{equation}
We also find the unitary part in the relevant zeroth order:

\begin{equation}
\begin{split}
    \mathcal{H}_{\rm uni}\Big|_{q_J=2} & = \frac{J}{4} N_F^2 \sum_{i=1}^L {\rm tr}\left(\frac{\mathbb{I} - \tau_{R,i}}{\mathbb{I} + \tau_{R,i}} \frac{\mathbb{I} - \tau^t_{R,i+1}}{\mathbb{I} + \tau^t_{R,i+1}}\right)+O(\tau^2_I)\\ 
    \mathcal{H}_{\rm uni}\Big|_{q_J=4} & = -\frac{J N_F^4}{64}\sum_{a,a'=1}^n\sum^L_{i=1}   \left( \left(\frac{\delta_{aa'} - \tau^{aa'}_{R,i}}{\delta_{aa'} + \tau^{aa'}_{R,i}}\right)^2 \left(\frac{\delta_{aa'} - \tau^{aa'}_{R,i+1}}{\delta_{aa'} + \tau_{R,i+1}^{aa'}}\right)^2 
        +\frac{1}{6}  \left(\frac{\delta_{aa'} - \tau^{aa'}_{R,i}}{\delta_{aa'} + \tau^{aa'}_{R,i}}\right)^4\right)+O(\tau^2_I).
    \end{split}
\end{equation}
In this appendix, we have presented the detailed derivation of the expansion of the kinetic term and the Hamiltonian with respect to $\tau_I$ modes. This expansion forms the basis for integrating out $\tau_I$ (see subsection \ref{ssec:integ_tau}), ultimately leading to the field theory descriptions for both interacting and non-interacting models, as discussed in the subsection \ref{ssec:NLSM}.

  \chapter{Cayley transform}\label{App:Caley}In this appendix we derive the integration measure for the matrices formed by Cayle transformation of a skew-symmetric matrices.
Consider an antisymmetric real matrix $A$ of size $n$. The Cayley transform is defined as
\begin{equation}
    O = \frac{\mathbb{I} - A}{\mathbb{I} + A}.
\end{equation}
It is immediate to check that $O$ is an orthogonal matrix, i.e. $O^t O = \mathbb{I}$. Also, the transformation is an involution and therefore
\begin{equation}
    A = (\mathbb{I} - O)(\mathbb{I} + O)^{-1}.
\end{equation}
The matrix $O$ cannot have the eigenvalue $-1$ and therefore $\det O = 1$. So, the Cayley transform maps antisymmetric matrices to $SO(n)$.

We can use this transformation to express the Haar measure over $SO(n)$ by means of the simpler measure over the entries of $A$. Let us define the standard measure on real antisymmetric matrices as,
\begin{equation}
    dA = \prod_{1\leq i<j\leq n} dA_{ij}.
\end{equation}
We denote the Haar measure on $SO(n)$ as $dO_{\rm Haar}$. The Haar measure is invariant under left and right multiplication, $O \to W_1 O W_2$. We want to find a function $f$ such that
\begin{equation}
\label{eq:maphaar0}
    f(A) dA = dO_{\rm Haar}.
\end{equation}
To do so, we diagonalise $A$ writing $A = W X W^t$, with $W$ an orthogonal matrix and $X$ a diagonal matrix of eigenvalues. Since $A$ in antisymmetric, the spectrum is organized in pairs of imaginary numbers. 
We have to distinguish the two cases
\begin{align}
    &X = \mbox{diag}(+i x_1, -i x_1, +i x_2, -i x_2, \ldots, + i x_k, -i x_k) , \qquad n = 2k ,\\
    &X = \mbox{diag}(0, +i x_1, -i x_1, +i x_2, -i x_2, \ldots, + i x_k, -i x_k) , \qquad n = 2k + 1,
\end{align}
using standard measure, one can ttransform the measure $dA$ into a measure over the rotation $W$ and one over the eigenvalues. Following \cite{Mehta}, we obtain,
\begin{align}
    &dA = \frac{1}{Z_A} dW_{\rm Haar} \prod_{1\leq i<j \leq k} (x_i^2 - x_j^2)^2 \prod_{i=1}^k dx_i \qquad n = 2k \\
    &dA = \frac{1}{Z_A} dW_{\rm Haar} \prod_{1\leq i<j \leq k} (x_i^2 - x_j^2)^2 \prod_{i=1}^k x_i^2 dx_i \qquad n = 2k  + 1.
\end{align}
We can then apply the same procedure to the matrix $O$. In this case, the eigenvalues of $O$ come in complex conjugate pairs of the form $e^{\pm i \theta_k}$, with an additional $1$ in case of odd  and express the Haar measure as
\begin{align}
    &dO_{\rm Haar} = \frac{dW_{\rm Haar}}{Z_O} \prod_{1 \leq i < j \leq k} (\cos(\theta_i) - \cos(\theta_j))^2 \prod_{i=1}^k d\theta_k \qquad n = 2k \\
    &dO_{\rm Haar} = \frac{dW_{\rm Haar}}{Z_O} \prod_{1 \leq i < j \leq k} (\cos(\theta_i) - \cos(\theta_j))^2 \prod_{i=1}^k (1 - \cos(\theta_k)) d\theta_k \qquad n = 2k + 1.
\end{align}
We now use the Cayley transform to determine the function $f$ in Eq.~\eqref{eq:maphaar0} which produces the right correspondence.
We see that the Haar matrix over $W$ is common to both, so we can focus on the mapping of the eigenvalue distribution. 
Under the Cayley transformation, we have the mapping of the eigenvalues
\begin{equation}
    e^{i \theta} = \frac{1 + i x}{1 - i x} \quad \Rightarrow \quad \theta = 2 \arctan(x)  , \; x = \tan(\theta/2).
\end{equation}
With this transformation, we have
\begin{multline}
        \prod_{1 \leq i < j \leq k} (\cos(\theta_i) - \cos(\theta_j))^2 =     \prod_{1 \leq i < j \leq k} (\frac{2}{1 + x_i^2} - \frac{2}{1 + x_j^2})^2  =\\= 2^{k(k-1)}\prod_{1 \leq i<j} (x_i^2 - x_j^2)^2  \prod_{i=1}^k (1 + x_i^2)^{2(1-k)},
\end{multline}
also notice that this formula can be presented as a determinant:
\begin{equation}
    \prod_{i=1}^k (1 + x_i^2)^2 = \det(1 - A^2).
\end{equation}
So for the even number of replicas $n = 2k$, we arrive at
\begin{equation}
    dO_{\rm Haar} = \frac{Z_A}{Z_0} 2^{k^2} \prod_{i=1}^k (1 + x^2_i)^{1-2k} dA = \frac{Z_A}{Z_O} 2^{k^2} \det(1 - A^2)^{(1 - 2k)/2} dA.
\end{equation}For the odd number of replicas, we similarly  have
\begin{equation}
    \prod_{i=1}^k ( 1 - \cos(\theta_k)) = 2^{k} \prod_{i=1}^k \frac{x^2_i}{1 + x_i^2},
\end{equation}
which gives us the Haar measure on the $SO(n)$ group
\begin{equation}
    dO_{\rm Haar} = \frac{Z_A}{Z_O} 2^{k(k+1)} \det(1 - A^2)^{-k} dA.
\end{equation}
So we see that except for normalisation constants, we can set in both cases, $f(A) = \det(\mathbb{I} - A^2)^{(1 -n)/2}$. This result confirms Theorem 3 of \cite{Toyama1948}. Therefore, in this appendix,1 we confirmed that the Caley transformation of a skew symmetric matrix gives a Haar measure on the special orthogonal group.
  \chapter{Cumulant generator and saddle inversion }
\label{app:cumulant-saddle}

\subsection*{From $\Pr(m=M)$ to the cumulant generator}

Starting from the Laplace–space expression for the probability of \emph{exactly} $M$ links 
\[
\mathcal{L}\{\Pr(m=M)\}(u)=\frac{u+3}{\big(1+u(u+3)\big)^{M+1}}
=\frac{u+3}{\big(u^2+3u+1\big)^{M+1}}.
\]
 The denominator $(u^2+3u+1)^{M+1}$ encodes the $M{+}1$ waiting intervals (the $M$ that occur plus the final survival), while the factor $(u+3)$ comes from the hazard/survival relation.

We now sum over $M$ with weight $e^{-zM}$:
\begin{equation}
\label{eq:Gz-u}
\sum_{M\ge0} \Pr(m=M)\,e^{-zM}
=\mathcal{L}^{-1}\!\left[\frac{(u+3)e^{z}}{\left(u^2+3u+1\right)e^{z}-1}\right](t)
=:e^{-t\,g(z)}.
\end{equation}
 Weighting by $e^{-zM}$ builds the generating function of the count $m$. In Laplace space this becomes a simple rational function; after inverting back in $t$ we \emph{define} $g(z)$ so that the whole object behaves as $e^{-t\,g(z)}$, i.e. $g(z)$ is the \emph{cumulant rate} (cumulants per unit $t$).

For finite $t$ the inverse Laplace can be computed explicitly; for large $t$ the dominant contribution comes from the simple pole in $u$ that solves
\[
\big(u^2+3u+1\big)e^z=1.
\]
 At large $t$, residues/poles of the $u$–integrand control the asymptotics. The leading pole satisfies the above algebraic equation.

This gives the scaled cumulant generator
\begin{equation}
\label{eq:g-of-z}
g(z)=\frac{1}{2}\left(3-e^{-z/2}\sqrt{5e^{z}+4}\right).
\end{equation}
 Solving the pole condition for $u$ and plugging into the prefactor yields a closed form for $g(z)$. All cumulants of $m$ at large $t$ follow from derivatives of this function at $z=0$.

Equivalently, approximating the discrete sum by an integral,
\[
\sum_M \Pr(m(t)=M)\,e^{-zM}\;\sim\;t\int_0^\infty dq\,\Pr(m=qt)\,e^{-zqt}\;\sim\;e^{-t\,g(z)},
\]
 Replacing the sum over $M$ with an integral over the intensive variable $q=M/t$ and applying Laplace’s method leads to the same exponential rate $g(z)$: both routes are consistent.

\subsection*{Back to $\Pr(m=qt)$ by saddle point in $z$}

We invert the $z$–transform by steepest descent:
\begin{equation}
\label{eq:pt-ml}
\Pr(m=qt)\;\sim\;\frac{1}{2\pi i}\int dz\,\exp\!\big\{t\,[\,q z-g(z)\,]\big\}
\;\asymp\;e^{t\,[\,q z_*-g(z_*)\,]}\,
\sqrt{\frac{1}{2\pi t\,g''(z_*)}},
\end{equation}
where \(z_*\) solves \(g'(z_*)=m\).
 The probability of a specific density $q$ is dominated by the saddle point where the exponent $q z-g(z)$ is stationary; the Gaussian factor comes from the local curvature $g''$ at that saddle.

Evaluating with the explicit \(g(z)\) in \eqref{eq:g-of-z} gives
\begin{equation}
\label{eq:pt-saddle-final}
\begin{split}
&\Pr(m=qt)\;\asymp\;
\exp\!\left\{\frac{t}{2}\Big[-3+g(q)-2q\ln\!\big(q\,g(q)\big)\Big]\right\}
\frac{\sqrt{g(q)}}{(5+4q^2)^{1/4}\,\sqrt{2\pi\,t\,q}},\\
&g(q)=2q+\sqrt{5+4q^2}.
\end{split}
\end{equation}
 The exponential part is the large–deviation rate (how rare a given $q$ is); the prefactor gives the leading Gaussian fluctuation around the saddle. The function $g(q)$ is a convenient shorthand that simplifies both the exponent and prefactor.

  \chapter{Renewal counting: mean, variance, skewness, and kurtosis}
\label{app:renewal-moments}
\subsection{Distribution of the number of singlets}
Since we want to find the probability distribution of a number of singlets over a bond we first fix a bond $B$. Along the SDRG, a singlet across $B$ forms only when the coupling on $B$ reaches the running cutoff $\Omega$. Right after such a decimation (at cutoff $\Omega_0$) the bond $B$ is ``reset'' to a new effective coupling. From that moment on we can ask:

\medskip
 As the cutoff is lowered from $\Omega_0$ to $\Omega<\Omega_0$, what is the probability that $B$ has \emph{not} been decimated again? This is the \emph{survival probability}
\begin{equation}
\label{eq:def-surv}
p(\Omega\,|\,\Omega_0):=\Pr\!\big(\text{no new singlet across $B$ as }\Omega_0\downarrow \Omega\big).
\end{equation}
It is a standard survival function for a single bond viewed as a stochastic process under the RG, and it will be the key quantity that controls the counting statistics below.

We track (i) the bulk bond distribution $P(\Gamma,\Omega)$ for a generic bond, and (ii) the \emph{conditioned} distribution $P^{(B)}(\Gamma,\Omega)$ for our marked bond $B$, \emph{given that it has survived} down to $\Omega$. They are related to their logarithmic versions by
\[
\beta=\ln\frac{\Omega}{\Gamma},\qquad
P_\ell(\beta)\,d\beta=P(\Gamma,\Omega)\,d\Gamma,\quad
P^{(B)}_\ell(\beta)\,d\beta=P^{(B)}(\Gamma,\Omega)\,d\Gamma,\qquad
\ell=\ln\frac{\Omega_0}{\Omega}.
\]
So we proceed to derive the flow equation for the distribution $P^{(B)}(\Gamma,\Omega)$.
Over $\Omega\to\Omega-d\Omega$ the link can be formed over $B$ if it hits the cutoff \(\Gamma=\Omega\). Therefore we can write for the survival probability
\begin{equation}
\label{eq:surv-hazard}
p(\Omega-d\Omega\,|\,\Omega_0)
= p(\Omega\,|\,\Omega_0)\,\big[\,1-P^{(B)}(\Omega,\Omega)\,d\Omega\,\big],
\end{equation}
i.e.\ $P^{(B)}(\Omega,\Omega)$ is the \emph{instantaneous decimation rate } for $B$ . 
Two things happen when we lower the cutoff a little:
\begin{itemize}
\item[(i)] \textbf{Renormalization by conditioning.} Because we condition on survival, we must re-normalize $P^{(B)}$ on the smaller domain $\Gamma\in[0,\Omega-d\Omega]$:
\begin{equation}
\label{eq:cond-norm}
P^{(B)}(\Gamma,\Omega)\ \longrightarrow\
\frac{P^{(B)}(\Gamma,\Omega)}{1-P^{(B)}(\Omega,\Omega)\,d\Omega}.
\end{equation}
\item[(ii)] \textbf{Shape change from neighbor decimations.}
If the left or right neighbor hits the cutoff (probability $P(\Omega,\Omega)\,d\Omega$ each), the bond across $B$ is \emph{updated} by the SDRG rule
\[
\Gamma_{\rm eff}=\frac{\Gamma_B\,\Gamma'}{2\Omega}\qquad(\text{the decimated neighbor has strength } \Omega).
\]
This injects weight at the new value $\Gamma_{\rm eff}$ and removes it at $\Gamma_B$.
\end{itemize}
Putting (i) and (ii) together and keeping $O(d\Omega)$ terms gives the continuum equation
\begin{equation}
\label{eq:PBevoGamma-cont}
\begin{split}
&-\partial_\Omega P^{(B)}(\Gamma,\Omega)
= P^{(B)}(\Gamma,\Omega)\,P^{(B)}(\Omega,\Omega)
-2P(\Omega,\Omega)\,P^{(B)}(\Gamma,\Omega)
\\&+2P(\Omega,\Omega)\!\int d\Gamma_B\,d\Gamma'\,P^{(B)}(\Gamma_B,\Omega)\,P(\Gamma',\Omega)\,
\delta\!\big(\Gamma-\tfrac{\Gamma_B\Gamma'}{2\Omega}\big).
\end{split}
\end{equation}
Lowering $\Omega$ increases \(\beta=\ln(\Omega/\Gamma)\) uniformly by $d\beta=d\ell$, which produces a convective ``drift'' term $-\partial_\beta P^{(B)}_\ell$. Transforming Eq.~\eqref{eq:PBevoGamma-cont} we obtain
\begin{equation}
\label{eq:PBevoBeta}
\begin{split}
&\partial_\ell P^{(B)}_\ell(\beta)-\partial_\beta P^{(B)}_\ell(\beta)
= P^{(B)}_\ell(0)\,P^{(B)}_\ell(\beta)-2P_\ell(0)\,P^{(B)}_\ell(\beta)
\\&+2P_\ell(0)\!\int d\beta_B\,d\beta'\,P^{(B)}_\ell(\beta_B)\,P_\ell(\beta')\,
\delta\!\big(\beta-\beta_B-\beta'-\ln 2\big),
\end{split}
\end{equation}
i.e.\ when a neighbor is decimated, the new logarithmic coupling is the sum of the two neighbors’ $\beta$’s plus the fixed shift $\ln 2$.

It is also convenient to rewrite this variable in terms of
\begin{equation}
Q_\ell(\beta) := p(\ell\,|\,\ell_0)\,P^{(B)}_\ell(\beta),\qquad
\int_0^\infty d\beta\,Q_\ell(\beta)=p(\ell\,|\,\ell_0),
\end{equation}
so that the integral over $\beta$ of the function $Q_{\ell}(\beta)$ gives back the survival probability $p(\ell | \ell_0)$.
Differentiating the last identity and using Eq.~\eqref{eq:surv-hazard} gives the relation
\begin{equation}
\label{eq:hazard-Q}
\partial_\ell p(\ell\,|\,\ell_0) = -\,Q_\ell(0).
\end{equation}
The flow equation \eqref{eq:PBevoBeta} in terms of $Q$ takes the form
\begin{equation}
\label{eq:Q-eq}
\begin{split}
&\partial_\ell Q_\ell(\beta)-\partial_\beta Q_\ell(\beta)
=-2P_\ell(0)\,Q_\ell(\beta)
\\&+2P_\ell(0)\!\int d\beta_B\,d\beta'\,Q_\ell(\beta_B)\,P_\ell(\beta')\,
\delta\!\big(\beta-\beta_B-\beta'-\ln 2\big).
\end{split}
\end{equation}
Now we are going to solve this equation using the solution for $P_{\ell}(\beta)$ in infinite randomness fixed point Eq.~\eqref{eq:fixedpoint}.
Notice that immediately after $B$ is decimated (at $\ell=\ell_0$), $B$ is the product of its two neighbors, so
\begin{equation}
\label{eq:Q-init}
Q_{\ell_0}(\beta)=P^{(B)}_{\ell_0}(\beta)
=\int d\beta_1\,d\beta_2\,P_{\ell_0}(\beta_1)\,P_{\ell_0}(\beta_2)\,
\delta\!\big(\beta-\beta_1-\beta_2-\ln 2\big).
\end{equation}
And at the infinite-randomness fixed point, \(P_\ell(\beta)=\frac{1}{\ell}e^{-\beta/\ell}\) and \(P_\ell(0)=1/\ell\) (cf.\ Eq.~\eqref{eq:PBevoBeta}); neglecting the small $\ln 2$ shift in Eq.~\eqref{eq:Q-init}, we get \(Q_{\ell_0}(\beta)\sim(\beta/\ell_0^2)\,e^{-\beta/\ell_0}\).

Guided by scaling (only the ratio $\ell/\ell_0$ matters), we use the ansatz to solve Eq.~\eqref{eq:Q-eq}
\begin{equation}
\label{eq:ansatz_ch4}
Q_\ell(\beta)=\Big(a(t)+b(t)\,\frac{\beta}{\ell}\Big)\,P_\ell(\beta),\qquad t:=\ln\frac{\ell}{\ell_0}.
\end{equation}
Plugging Eq.~\eqref{eq:ansatz_ch4} into Eq.~\eqref{eq:Q-eq} gives the simple ODE system
\begin{equation}
\label{eq:ab-odes-t}
\frac{da}{dt}=b-2a,\qquad \frac{db}{dt}=-\,b+a,\qquad a(0)=0,\ \ b(0)=1,
\end{equation}
whose solution is
\begin{equation}
\label{eq:a-b-sol}
a(t)=\frac{2e^{-3t/2}}{\sqrt{5}}\sinh\!\Big(\frac{\sqrt{5}\,t}{2}\Big),~~
b(t)=\frac{e^{-3t/2}}{5}\Big[\sqrt{5}\,\sinh\!\Big(\frac{\sqrt{5}\,t}{2}\Big)
+5\,\cosh\!\Big(\frac{\sqrt{5}\,t}{2}\Big)\Big].
\end{equation}
Therefore the survival probability depends only on the scale ratio:
\begin{equation}
\label{eq:rho}
\begin{split}
p(\ell\,|\,\ell_0)=\rho(t):=\int_0^\infty d\beta\,Q_\ell(\beta)=a(t)+b(t)
\\=\frac{e^{-3t/2}}{5}\Big[3\sqrt{5}\,\sinh\!\Big(\frac{\sqrt{5}\,t}{2}\Big)
+5\,\cosh\!\Big(\frac{\sqrt{5}\,t}{2}\Big)\Big].
\end{split}
\end{equation}
Further we move to the calculation of the probability distribution of the number of singlets across the bond.

\subsubsection{Counting singlets as a renewal process (from survival to counts).}
We now explain how the \emph{survival probability} produces the full
counting statistics of singlets across the marked bond $B$.
Throughout we use couplings $\Gamma$, RG time $\ell=\ln(\Omega_0/\Omega)$, and the
scale-invariant clock $t:=\ln(\ell/\ell_0)$, so that only the ratio $\ell/\ell_0$ matters at the fixed point (as in Eq.~\eqref{eq:rho}).

\medskip
\noindent\textbf{1) Survival $\boldsymbol{\rho}$ and waiting time.}
Right after a decimation across $B$ at $\ell_0$, start a stopwatch.
Let $t_1$ be the (random) time in $t$-units until the \emph{next} singlet across $B$.
The survival probability is
\[
\rho(t)\equiv p(\ell\,|\,\ell_0)=\Pr(t_1>t)\,,
\]
i.e.\ the probability that no new singlet has formed across $B$ up to time $t$
(see Eq.~\eqref{eq:rho}).
Therefore the CDF and PDF (density) of random variable $t_1$ are
\[
F(t)=\Pr(t_1\le t)=1-\rho(t),\qquad
f(t)=\frac{d}{dt}F(t)=-\rho'(t).
\]
From the RG solution at fixed point Eq.\eqref{eq:rho} we already have
\begin{equation}
\label{eq:waiting}
f(t)=-\rho'(t)=a(t).
\end{equation}
Let us notice that at the infinite-randomness fixed point, each time a singlet forms across $B$ the marked
bond is \emph{re-initialized} to the same fixed-point law in the logarithmic coupling
$\beta=\ln(\Omega/\Gamma)$ (up to the current $\ell$). Measuring time by $t=\ln(\ell/\ell_0)$ removes the absolute scale,
so each inter-decimation interval has the same distribution and is independent of the past.
Hence
\[
t_1,t_2,t_3,\dots\quad\text{are i.i.d.\ with density } f(t)=-\rho'(t)
\ \text{from Eq.~ \eqref{eq:waiting}}.
\]
This makes the singlet events across $B$ a classical \emph{renewal process}.

\medskip
\noindent\textbf{2) Event times and the counting process.}
So let $\{t_j\}_{j\ge1}$ be the i.i.d.\ inter-decimation waiting times with density $f(t)$ from Eq.~\eqref{eq:waiting}.
Define the \emph{arrival times} $S_M:=\sum_{j=1}^M t_j$ and the \emph{counting process}
$m(t):=\max\{M\ge 0:\ S_M\le t\}$.
Then "having at least M singlets by time t" is exactly the same event as "the M-th singlet happened no later than t":
\[
\{m(t)\ge M\}=\{S_M\le t\}\quad\Longrightarrow\quad
\Pr(m(t)\ge M)=\Pr(S_M\le t).
\]
To have at least $M$ events by time $t$, choose a first waiting time $t_1\in[0,t]$; given that choice, choose a second waiting time $t_2\in[0,\,t-t_1]$; continue this way so that the $M$-th waiting time $t_M$ lies in $[0,\,t-\sum_{j=1}^{M-1}t_j]$.
Because the waiting times $(t_j)$ are independent, the joint probability density for a particular tuple $(t_1,\dots,t_M)$ is the product $f(t_1)\cdots f(t_M)\,dt_1\cdots dt_M$.
Summing over all such tuples—i.e.\ integrating over the simplex $\{t_1,\dots,t_M\ge 0:\ t_1+\cdots+t_M\le t\}$—gives
\begin{equation}\label{eq:convol}
\begin{split}
&F_M(t)=\Pr(m(t)\ge M)=\Pr(S_M\le t)\\=&\int_{0}^{t} dt_1 f(t_1) \int_{0}^{t - t_1} dt_2 f(t_2) 
  \ldots \int_{0}^{t - \sum_{j=1}^{M-1} t_j} dt_M f(t_M) \;,
  \end{split}
\end{equation}

with $F_0(t)\equiv 1$.
Hence the probability of \emph{exactly} $M$ links over $B$ is "probability of at least $M$" minus "probability of at least $M+1$,” i.e.
\[
\Pr(m(t)=M)=F_M(t)-F_{M+1}(t).
\]
Equivalently, conditioning on the time $S_M=s$ of the $M$-th event and then requiring \emph{survival} over the remainder $t-s$ gives
\begin{equation}
\label{eq:exactM}
\Pr(m(t)=M)=\int_0^{t} f^{*M}(s)\,\rho(t-s)\,ds=(f^{*M}*\rho)(t),
\end{equation}
where $f^{*M}$ is the $M$-fold convolution and $\rho$ is from Eq.~\eqref{eq:rho}.

\medskip
\noindent\textbf{3) Laplace-transform shortcuts.}
As we saw adding independent waiting times Eq.~\eqref{eq:convol} corresponds to a \emph{convolution} in time.
The Laplace transform turns convolutions into simple products, and turning a
time–integral $\int_0^t(\cdots)ds$ into division by $u$:
\[
\mathcal{L}\{f^{*M}\}(u)=[\tilde f(u)]^M,\qquad
\mathcal{L}\Big\{\int_0^t g(s)\,ds\Big\}(u)=\frac{\tilde g(u)}{u}.
\]
So, since “at least $M$ events by time $t$” is
$F_M(t)=\Pr(S_M\le t)=\int_0^t f^{*M}(s)\,ds$, its Laplace transform is
\begin{equation}
\label{eq:FM-laplace}
\tilde F_M(u)=\mathcal{L}\{F_M\}(u)=\frac{[\tilde f(u)]^M}{u}
=\frac{1}{u\,(u^2+3u+1)^M},
\end{equation}
because here $\tilde f(u)=\mathcal{L}\{f\}(u)=1/(u^2+3u+1)$.

\smallskip
“\emph{Exactly $M$} events by time $t$” means “the $M$-th arrival happens at some
time $s\le t$ and then the process \emph{survives} (no new event) for the remaining
interval $t-s$.” That is a convolution in time:
\[
\Pr(m(t)=M)=(f^{*M} * \rho)(t),
\]
so in Laplace space it becomes a product:
\begin{equation}\label{eq:conv}
\mathcal{L}\{\Pr(m(t)=M)\}(u)=\tilde f(u)^M\,\tilde\rho(u).
\end{equation}
The factor $[\tilde f(u)]^M$ accounts for the $M$ inter-arrival waits, while
$\tilde\rho(u)$ enforces survival (no additional link) after the $M$-th arrival.
To eliminate $\tilde\rho$, use $f=-\rho'$ and $\rho(0)=1$, which implies
$\tilde f(u)=1-u\tilde\rho(u)$.
Substituting $\tilde\rho(u)=(1-\tilde f(u))/u$ into the convolution formula Eq.~\eqref{eq:conv} yields the
Laplace–space expression for the probability of observing \emph{exactly} $M$ links across $B$:
\begin{equation}
\label{eq:exactM-laplace}
\begin{split}
\mathcal{L}\{\Pr(m(t)=M)\}(u)
= \frac{1-\tilde f(u)}{u}\,[\tilde f(u)]^M
\\= \frac{u+3}{\big(u^2+3u+1\big)^{M+1}}
= \frac{u+3}{\big(1+u(u+3)\big)^{M+1}}.
\end{split}
\end{equation}
 Inverting Eq.~\eqref{eq:exactM-laplace} gives $\Pr(m(t)=M)$ in the time domain and, via
Eq.~\eqref{eq:links_ent}, the full distribution of the second Rényi entropy.
However a closed-form inverse is cumbersome, so in the large-$t$ (or large-$M$) regime we evaluate
the Bromwich integral by the \emph{saddle-point} (steepest-descent) method further.
 \smallskip

\subsection{Second, third, and fourth raw moments}
Apply the same summation with $M^k$ weights.  
With $r=\tilde f(u)$ and the identities
\begin{equation}
\begin{split}
   &\sum_{M\ge0}M r^M=\frac{r}{(1-r)^2},\quad
\sum_{M\ge0}M^2 r^M=\frac{r(1+r)}{(1-r)^3},\\&
\sum_{M\ge0}M^3 r^M=\frac{r(1+4r+r^2)}{(1-r)^4},\quad
\sum_{M\ge0}M^4 r^M=\frac{r(1+11r+11r^2+r^3)}{(1-r)^5}, 
\end{split}
\end{equation}
these are the standard generating–series formulas for $M^k$-weighted geometric sums, used to do the $M$-sums in closed form. We obtain
\begin{equation}
\begin{split}
\mathcal{L}\{\langle m^{2}\rangle\}(u)
&=\frac{(u+1)(u+2)}{u^3(u+3)^2},\qquad
\mathcal{L}\{\langle m^{3}\rangle\}(u)
=\frac{u^2(u+3)^2+6u(u+3)+6}{u^4(u+3)^3},\\
\mathcal{L}\{\langle m^{4}\rangle\}(u)
&=\frac{u^3(u+3)^3+14u^2(u+3)^2+36u(u+3)+24}{u^5(u+3)^4}.
\end{split}
\end{equation}
after inserting the $M^k$ sums and the ladder prefactor, each moment’s Laplace transform becomes a simple rational function of $u$ and $(u+3)$, ready for inversion.

Inverse Laplace transforms give the exact time–domain forms (exponentially small pieces explicit):
\begin{equation}
\begin{split}
\langle m^2\rangle(t)
&=\frac{t^2}{9}+\frac{5}{27}t-\frac{1}{27}
+e^{-3t}\!\left(\frac{1}{27}-\frac{2}{27}\,t\right),\\
\langle m^3\rangle(t)
&=\frac{t^3}{27}+\frac{2}{9}t^2+\frac{1}{27}t+\frac{7}{243}
+e^{-3t}\!\left(-\frac{7}{243}-\frac{10}{81}\,t+\frac{1}{27}\,t^2\right),\\
\langle m^4\rangle(t)
&=\frac{t^4}{81}+\frac{38}{243}t^3+\frac{67}{243}t^2-\frac{25}{729}t+\frac{91}{2187}
\\+e^{-3t}\!&\left(-\frac{91}{2187}-\frac{22}{243}\,t+\frac{34}{243}\,t^2-\frac{4}{243}\,t^3\right).
\end{split}
\end{equation}
 each raw moment is a polynomial in $t$ (dominant growth) plus a constant and exponentially decaying corrections from the pole at $u=-3$.
From Eq.~\eqref{eq:exactM-laplace} (or via the PGF), the mean count is
\begin{equation}
\label{eq:mean-exact}
\langle m(t)\rangle=\frac{t}{3}-\frac{1}{9}+\frac{1}{9}e^{-3t}.
\end{equation}

\subsection{Variance, skewness, and kurtosis}
The variance follows from $\mathrm{Var}\,m=\langle m^2\rangle-\langle m\rangle^2$:
\begin{equation}
\label{eq:var-time}
\mathrm{Var}\,m(t)=\frac{7}{27}\,t-\frac{4}{81}+\mathcal{O}(e^{-3t}).
\end{equation}
subtracting the square of the mean removes the $t^2$ term, leaving linear growth with coefficient $7/27$ and small corrections.

The third and fourth \emph{central} moments are
\begin{equation}
\begin{split}
\mu_3(t)&:=\langle(m-\langle m\rangle)^3\rangle
=\langle m^3\rangle-3\langle m^2\rangle\langle m\rangle+2\langle m\rangle^3
=\frac{13}{81}\,t+\frac{10}{729}+\mathcal{O}(e^{-3t}),\\
\mu_4(t)&:=\langle(m-\langle m\rangle)^4\rangle
=\langle m^4\rangle-4\langle m^3\rangle\langle m\rangle+6\langle m^2\rangle\langle m\rangle^2-3\langle m\rangle^4
\\=&\frac{441\,t^2-45\,t+112}{2187}+\mathcal{O}(e^{-3t}).
\end{split}
\end{equation}
these are the standard polynomial relations converting raw moments to central moments; the leading behaviors are linear for $\mu_3$ and quadratic for $\mu_4$.

Hence the standardized skewness and (non-excess) kurtosis are
\begin{align}
\mathrm{Skew}(t)
&:=\frac{\mu_3(t)}{\mathrm{Var}(m(t))^{3/2}}
=\frac{13}{7}\sqrt{\frac{3}{7}}\;\frac{1}{\sqrt{t}}
+\mathcal{O}(t^{-3/2}),
\\[2pt]
\mathrm{Kur}(t)
&:=\frac{\mu_4(t)}{\mathrm{Var}(m(t))^{2}}
=3+\mathcal{O}\!\left(\frac{1}{t}\right).
\end{align}
dividing by the appropriate powers of the variance yields dimensionless shape parameters: skewness decays as $t^{-1/2}$ (symmetry emerges) and kurtosis tends to the Gaussian value $3$ with $1/t$ corrections.

\medskip
\noindent
Step by step: write $\langle m^k\rangle=\sum_M M^k\,p_t(n=M)$, take Laplace transforms using \eqref{eq:FM-laplace}, perform the $M$-sum via closed-form weighted geometric series, then invert the Laplace transform. Central moments are formed by the standard polynomial combinations of raw moments; skewness and kurtosis are the corresponding standardized ratios.

  \chapter{RG evolution of the distribution of couplings}
\label{app:RG-bulk-P}

Consider a chain of length $L$ with couplings $\Gamma_1,\ldots,\Gamma_L$ and a running cutoff $\Omega>0$ such that all $\Gamma\in[0,\Omega]$. Let $N(\Gamma)\,\Delta\Gamma$ be the number of couplings in $[\Gamma,\Gamma+\Delta\Gamma]$ and
\begin{equation}
N(\Gamma)=\frac{1}{\Delta\Gamma}\#\{\,i:\ \Gamma_i\in[\Gamma,\Gamma+\Delta\Gamma]\,\},
\qquad
P(\Gamma)=\frac{N(\Gamma)}{L},
\end{equation}
so $P(\Gamma)$ is a normalized empirical distribution on $[0,\Omega]$.

\paragraph{One RG step at fixed $\Omega$.}
Pick a thin shell of ``large'' couplings $\Gamma\in[\Omega-\Delta\Omega,\Omega]$. We assume such events are dilute (no two decimated bonds are adjacent). For a decimated bond of strength $\Gamma\approx\Omega$ with neighbors $\Gamma_\ell$ (left) and $\Gamma_r$ (right), the SDRG update generates a new bond
\begin{equation}
\label{eq:update-Gamma-prime}
\Gamma'=\frac{\Gamma_\ell\,\Gamma_r}{2\,\Gamma}\ \simeq\ \frac{\Gamma_\ell\,\Gamma_r}{2\,\Omega}.
\end{equation}
Each such move removes two sites; since there are $N(\Omega)\Delta\Omega$ decimations in the shell, the new system size is
\begin{equation}
\label{eq:L-tilde}
\tilde L = L - 2\,N(\Omega)\,\Delta\Omega.
\end{equation}

Let $\tilde N(\Gamma)\,\Delta\Gamma$ be the number of bonds in $[\Gamma,\Gamma+\Delta\Gamma]$ \emph{after} the RG step. Two effects contribute:
(i) removals when either neighbor $\Gamma_\ell$ or $\Gamma_r$ of a decimated triplet lies in $[\Gamma,\Gamma+\Delta\Gamma]$; and
(ii) insertions when the generated $\Gamma'$ lands in the bin. Denoting by $\mathbf{1}_{\Gamma,\Delta\Gamma}(\cdot)$ the indicator of $[\Gamma,\Gamma+\Delta\Gamma]$, we have
\begin{equation}
\begin{aligned}
\tilde N(\Gamma)\Delta\Gamma - N(\Gamma)\Delta\Gamma
=&\ -\,N(\Omega)\Delta\Omega
\!\int\! d\Gamma_\ell d\Gamma_r\,P(\Gamma_\ell)P(\Gamma_r)
\big[\mathbf{1}_{\Gamma,\Delta\Gamma}(\Gamma_\ell)+\mathbf{1}_{\Gamma,\Delta\Gamma}(\Gamma_r)\big]
\\
&\ +\,N(\Omega)\Delta\Omega
\!\int\! d\Gamma_\ell d\Gamma_r\,P(\Gamma_\ell)P(\Gamma_r)\,
\mathbf{1}_{\Gamma,\Delta\Gamma}\!\left(\frac{\Gamma_\ell\Gamma_r}{2\Omega}\right).
\end{aligned}
\end{equation}
In the limit $\Delta\Gamma\to0$, use $\mathbf{1}_{\Gamma,\Delta\Gamma}(x)=\delta(\Gamma-x)\,\Delta\Gamma$ and define the post-step density $\tilde P(\Gamma)=\tilde N(\Gamma)/\tilde L$. A short algebra (including the Jacobian from $\tilde L$ in \eqref{eq:L-tilde}) yields
\begin{equation}
\label{eq:P-tilde}
\begin{split}
\tilde P(\Gamma)
=(1-2P(\Omega)\Delta\Omega)^{-1}\!\left[
P(\Gamma)
+ P(\Omega)\Delta\Omega\!\int\! d\Gamma_\ell d\Gamma_r\,P(\Gamma_\ell)P(\Gamma_r)\times
\right.\\\left.\times \Big(\delta\!\left(\Gamma-\frac{\Gamma_\ell\Gamma_r}{2\Omega}\right)-\delta(\Gamma-\Gamma_\ell)-\delta(\Gamma-\Gamma_r)\Big)
\right].
\end{split}
\end{equation}

\paragraph{Continuum flow in $\Omega$.}
Treat $P(\Gamma,\Omega)$ as the running distribution at cutoff $\Omega$ and note that $P(\Gamma,\Omega-\Delta\Omega)=\tilde P(\Gamma)$. Expanding \eqref{eq:P-tilde} to first order in $\Delta\Omega$ gives the integro–differential flow
\begin{equation}
\label{eq:P-GammaOmega-flow}
\partial_\Omega P(\Gamma,\Omega)
= -\,P(\Omega,\Omega)\!\int\! d\Gamma_\ell d\Gamma_r\,
P(\Gamma_\ell,\Omega)P(\Gamma_r,\Omega)\,
\delta\!\left(\Gamma-\frac{\Gamma_\ell\Gamma_r}{2\Omega}\right).
\end{equation}
(The removal terms and the normalization change cancel, leaving only the source term from newly generated bonds.)

\paragraph{Logarithmic variables and RG ``time''.}
Introduce
\begin{equation}
\label{eq:beta-ell-defs}
\beta=\ln\frac{\Omega}{\Gamma},\qquad
\ell=\ln\frac{\Omega_0}{\Omega}
\quad\Rightarrow\quad
\Gamma=\Omega e^{-\beta},\ \ \Omega=\Omega_0 e^{-\ell}.
\end{equation}
Lowering $\Omega$ by $d\Omega$ increases both $\beta$ and $\ell$ by the same amount: $d\beta=d\ell$. Define the $\beta$–density at fixed $\ell$:
\begin{equation}
\label{eq:P-ell-beta}
P_\ell(\beta):=P(\Gamma,\Omega)\left|\frac{d\Gamma}{d\beta}\right|
= \Omega\,e^{-\beta}\,P(\Gamma,\Omega),\qquad
\int_0^\infty d\beta\,P_\ell(\beta)=1.
\end{equation}
Transforming \eqref{eq:P-GammaOmega-flow} to $(\beta,\ell)$ and adding the uniform drift $-\partial_\beta P_\ell$ from $d\beta=d\ell$, we obtain
\begin{equation}
\label{eq:P-ell-PDE}
\partial_\ell P_\ell(\beta)-\partial_\beta P_\ell(\beta)
= P_\ell(0)\!\int\! d\beta_\ell d\beta_r\,
P_\ell(\beta_\ell)P_\ell(\beta_r)\,
\delta\!\big(\beta-\beta_\ell-\beta_r-\ln 2\big).
\end{equation}
(When a neighbor is decimated, the new logarithmic coupling is the sum of the neighbors’ $\beta$’s plus the fixed shift $\ln 2$ from the factor $1/2$ in \eqref{eq:update-Gamma-prime}.)

\paragraph{IRFP and scaling form.}
At large $\ell$ we look for a scaling solution
\begin{equation}
\label{eq:scaling-ansatz}
P_\ell(\beta)=\frac{1}{\ell}\,g\!\left(\frac{\beta}{\ell}\right).
\end{equation}
Neglecting the subleading constant shift $\ln 2$ (irrelevant at large $\ell$), \eqref{eq:P-ell-PDE} reduces to
\begin{equation}
\label{eq:g-eq}
g(x)+(1+x)g'(x)
= -\,g(0)\!\int\! dx_\ell dx_r\,g(x_\ell)g(x_r)\,\delta(x-x_\ell-x_r).
\end{equation}
The unique normalized solution is
\begin{equation}
\label{eq:IRFP-solution}
g(x)=e^{-x}\quad\Longrightarrow\quad
P_\ell(\beta)=\frac{1}{\ell}\,e^{-\beta/\ell},\qquad P_\ell(0)=\frac{1}{\ell},
\end{equation}
the well-known infinite-randomness fixed point (IRFP).
The solution \eqref{eq:IRFP-solution} also solves the full equation \eqref{eq:P-ell-PDE}; the constant $\ln 2$ shift merely produces $O(1/\ell)$ corrections at large $\ell$.

  \chapter{Jordan--Wigner ordering and spin representation of \texorpdfstring{$\Phi$}{Phi}}
\label{app:JW-Phi}
In this appendix we provide numerical analysis of the Dagupta-Ma derivation done in Sec. \ref{sec:DM-Phi}. First, we express the generators $\hat \Phi$ in terms of spin operators, then calculate Eq.\eqref{eq:PertPhi} in mathematica for a fixed set of $n$ and $N_F$.
\section{Linear ordering and Jordan--Wigner map}
We work on a 1D chain with sites $j=1,\dots,L$. Each site carries $N_F$ ``flavors'' (labelled by $\nu=1,\dots,N_F$) and $n$ replica indices ($a=1,\dots,n$). We impose a linear order on the triple $(j,\nu,a)$ by taking the replica index fastest, then flavor, then site:
\begin{equation}
\label{eq:I-order}
I(j,\nu,a)\;=\; a\;+\; n\,(\nu-1)\;+\; nN_F\,(j-1).
\end{equation}
Following Fradkin~\cite{Fradkin2013}, we define the Jordan--Wigner (JW) map
\begin{equation}
\label{eq:JW-map}
c^\dagger_{\nu,a,j}\equiv c^\dagger_{I}
\;=\;e^{-i\pi \sum_{I'<I} S^+_{I'}S^-_{I'}}\,S^+_{I},
\qquad
c_{\nu,a,j}\equiv c_{I}
\;=\;e^{+i\pi \sum_{I'<I} S^+_{I'}S^-_{I'}}\,S^-_{I},
\end{equation}
where $S^{\pm}_{I}$ act on a spin-$\tfrac12$ at position $I$ in the linear order. The condition $I'<I$ fixes the total ordering of all spins.

\paragraph{Decomposing the JW string.}
Using \eqref{eq:I-order}, the prefix sum in the string factor separates as
\begin{equation}
\label{eq:string-ABC}
\sum_{I'<I} S^+_{I'}S^-_{I'}
=\sum_{j'<j}\ \sum_{\nu'=1}^{N_F}\ \sum_{a'=1}^{n} S^+_{j',\nu',a'}S^-_{j',\nu',a'}
+\sum_{\nu'<\nu}\ \sum_{a'=1}^{n} S^+_{j,\nu',a'}S^-_{j,\nu',a'}
+\sum_{a'=1}^{a-1} S^+_{j,\nu,a'}S^-_{j,\nu,a'}.
\end{equation}
For brevity we write
\[
\sum_{I'<I} S^+_{I'}S^-_{I'} \;=\; A(j;N_F,n)\;+\;B(j,\nu;n)\;+\;C(j,\nu;a-1).
\]
The inequalities ensuring the blocks do not overlap are immediate from the ordering:
\begin{equation}
\label{eq:I-inequalities}
\begin{aligned}
&I_{\max}(j-1)=nN_F(j-1)<I_{\min}(j)=nN_F(j-1)+1,\\
&I_{\max}(j,\nu-1)=n(\nu-1)+nN_F(j-1)<I_{\min}(j,\nu)=n(\nu-1)+nN_F(j-1)+1,\\
&I(j,\nu,a-1)<I(j,\nu,a).
\end{aligned}
\end{equation}

\section{Local parity strings along the replica axis}
Let $a'<a$ without loss of generality (the case $a'>a$ is obtained by swapping the labels). Consider the bilinear at fixed $(j,\nu)$:
\begin{equation}
\label{eq:ccdaggerJJ}
\begin{aligned}
c^\dagger_{a'\nu,j}\,c^\dagger_{a\nu,j}
&=e^{-i\pi\big(A+B+C(a'-1)\big)}\,S^+_{j,\nu,a'}\,
  e^{-i\pi\big(A+B+C(a-1)\big)}\,S^+_{j,\nu,a}\\
&=e^{-i2\pi(A+B)}\,e^{-i\pi\big(C(a'-1)+C(a-1)\big)}\,
  S^+_{j,\nu,a'}S^+_{j,\nu,a}.
\end{aligned}
\end{equation}
Because $e^{2i\pi S^+S^-}=\mathbb{I}$ on a spin-$\tfrac12$, the blocks $A$ and $B$ drop out:
\begin{equation}
\label{eq:AB-drop}
e^{2i\pi A}=\prod_{j'<j,\nu',a'} e^{2i\pi S^+_{j'\nu'a'}S^-_{j'\nu'a'}}=1,
\qquad
e^{2i\pi B}=\prod_{\nu'<\nu,a'} e^{2i\pi S^+_{j\nu'a'}S^-_{j\nu'a'}}=1.
\end{equation}
For the residual string,
\begin{equation}
\label{eq:C-sum-simplify}
\begin{aligned}
C(a'-1)+C(a-1)
&=\sum_{b=1}^{a'-1} S^+_{j,\nu,b}S^-_{j,\nu,b}+\sum_{b=1}^{a-1} S^+_{j,\nu,b}S^-_{j,\nu,b}\\
&=2\sum_{b=1}^{a-1} S^+_{j,\nu,b}S^-_{j,\nu,b}
+\sum_{b=a}^{a'-1} S^+_{j,\nu,b}S^-_{j,\nu,b}.
\end{aligned}
\end{equation}
The first term gives $e^{2i\pi(\cdots)}=\mathbb{I}$. Define the replica-string (parity) operator
\begin{equation}
\label{eq:S-string}
S_{[a:a'),\nu}(j):=\prod_{b=a}^{a'-1}\sigma^z_{j,\nu,b},
\end{equation}
so that
\begin{equation}
\label{eq:ccdagger-final}
c^\dagger_{a'\nu,j}\,c^\dagger_{a\nu,j}
= e^{\,i\pi\sum_{b=a}^{a'-1}\!S^+_{j,\nu,b}S^-_{j,\nu,b}}\ S^+_{j,\nu,a'}S^+_{j,\nu,a}
= \big[(-1)^{a'-a}\,S_{[a:a'),\nu}(j)\big]\ S^+_{j,\nu,a'}S^+_{j,\nu,a}.
\end{equation}
Proceeding analogously for all bilinears at fixed $(j,\nu)$, we obtain
\begin{equation}
\label{eq:all-bilinears}
\begin{aligned}
c^\dagger_{a'\nu,j}\,c_{a\nu,j}
&= (-1)^{a'-a}\,S_{[a:a'),\nu}(j)\ S^+_{j,\nu,a'}S^-_{j,\nu,a},\\
c_{a'\nu,j}\,c_{a\nu,j}
&= (-1)^{a'-a}\,S_{[a:a'),\nu}(j)\ S^-_{j,\nu,a'}S^-_{j,\nu,a},\\
c_{a'\nu,j}\,c^\dagger_{a\nu,j}
&= (-1)^{a'-a}\,S_{[a:a'),\nu}(j)\ S^-_{j,\nu,a'}S^+_{j,\nu,a}.
\end{aligned}
\end{equation}
For $a'>a$ swap $a\leftrightarrow a'$ in the right-hand sides.

\section{Spin representation of \texorpdfstring{$\Phi^{\pm\pm}$}{Phi}}
With $\chi^{\pm}_{a\nu}=\tfrac{1}{\sqrt2}(c_{a\nu}\pm c^\dagger_{a\nu})$ and suppressing $(j,\nu)$ where unambiguous, the replica bilinears read
\begin{equation}\label{eq:Phi-ppmm}
\begin{split}
   & \Phi^{++}_{a'a}=\frac{i}{N_F}(1-\delta_{a' a})\sum_{\nu }\chi^{+}_{a'\nu} \chi^{+}_{a \nu}=\frac{i}{2 N_F} \sum_{\nu}(c^{\dagger}_{a'\nu} c^{\dagger}_{a\nu}-c^{\dagger}_{a\nu} c_{a'\nu}+c^{\dagger}_{a'\nu} c_{a\nu}+c_{a'\nu} c_{a\nu}+\delta_{a'a})=\\&\frac{i}{2 N_F} \sum_{\nu}(-1)^{a'-a}\left[S_{aa',\nu} S^+_{a',\nu} S^+_{a,\nu}+S_{aa',\nu} S^-_{a',\nu} S^+_{a,\nu}+  S_{a,a',\nu} S^+_{a'} S^-_a+S_{aa',\nu} S^-_{a',\nu} S^-_{a,\nu}+\delta_{aa'}\right],\\& \Phi^{--}_{a'a}=-\frac{i}{N_F}(1-\delta_{a a'})\sum_{\nu }\chi^{-}_{a'\nu} \chi^{-}_{a \nu}=-\frac{i}{2 N_F} \sum_{\nu}(c^{\dagger}_{a'} c^{\dagger}_{a}+c^{\dagger}_{a} c_{a'}-c^{\dagger}_{a'} c_{a}+c_{a'} c_{a}+\delta_{aa'})=\\&-\frac{i}{2 N_F} \sum_{\nu}(-1)^{a'-a}\left[S_{aa',\nu} S^+_{a',\nu} S^+_{a,\nu}-S_{aa',\nu} S^-_{a',\nu} S^+_{a,\nu}-  S_{a,a',\nu} S^+_{a'} S^-_a+S_{aa',\nu} S^-_{a',\nu} S^-_{a,\nu}+\delta_{aa'}\right],\\&
    \Phi^{+-}_{a'a}=-\frac{1}{2 N_F} \sum_{\nu}(c^{\dagger}_{a'} c^{\dagger}_{a}-c^{\dagger}_{a} c_{a'}-c^{\dagger}_{a'} c_{a}-c_{a'} c_{a}+\delta_{aa'})=\\&-\frac{1}{2 N_F} \sum_{\nu}(-1)^{a'-a}\left[S_{aa',\nu} S^+_{a',\nu} S^+_{a,\nu}+S_{aa',\nu} S^-_{a',\nu} S^+_{a,\nu} \right.\left.- S_{a,a',\nu} S^+_{a'} S^-_a-S_{aa',\nu} S^-_{a',\nu} S^-_{a,\nu}+\delta_{aa'}\right],\\
    &\Phi^{-+}_{a'a}=-\frac{1}{2 N_F} \sum_{\nu}(c^{\dagger}_{a'} c^{\dagger}_{a}+c^{\dagger}_{a} c_{a'}+c^{\dagger}_{a'} c_{a}-c_{a'} c_{a}-\delta_{aa'})=\\&-\frac{1}{2 N_F} \sum_{\nu}(-1)^{a'-a}\left[S_{aa',\nu} S^+_{a',\nu} S^+_{a,\nu}-S_{aa',\nu} S^-_{a',\nu} S^+_{a,\nu}\right.\left.+  S_{a,a',\nu} S^+_{a'} S^-_a-S_{aa',\nu} S^-_{a',\nu} S^-_{a,\nu}-\delta_{aa'}\right].
    \end{split}
\end{equation}
(Here $\delta$-terms encode the normal-ordering convention consistent with the chosen JW map.)

\section{Bond-parity operators and a reference state}
Define bond parities on sites $2$ and $3$:
\begin{equation}
\label{eq:parities}
P_2=\prod_{a=1}^{n}\prod_{\nu=1}^{N_F}\bigl(-\sigma^z_{2,a,\nu}\bigr),
\qquad
P_3=\prod_{a=1}^{n}\prod_{\nu=1}^{N_F}\bigl(-\sigma^z_{3,a,\nu}\bigr),
\end{equation}
so $P_2P_3=\prod_{a,\nu}(\sigma^z_{2,a,\nu}\sigma^z_{3,a,\nu})$ since $(-1)^{2nN_F}=+1$.
Let $\lvert\mathbb{I}\rangle$ be the reference state fixed by our convention for occupancies. With $S^z_{a\nu j}
=\big(c^\dagger_{a\nu j}c_{a\nu j}-\tfrac12\big)$ we have (by construction)
\begin{equation}
\label{eq:P2P3-on-1}
P_2P_3\,\lvert\mathbb{I}\rangle = (-1)^{nN_F}\,\lvert\mathbb{I}\rangle.
\end{equation}
We project $\Phi_j$ onto the parity-$+1$ sector using $P_j=\exp\!\big[\tfrac{i\pi}{2}(\sum_{\nu,a}\sum_{j'<j}\sigma^z_{a\nu j'}-nN_F j)\big]$ when needed.

\section{Second-order decimation and effective coupling}
In second-order perturbation theory around the strong central bond (site $2$), the effective interaction between blocks $1$ and $4$ reads
\begin{equation}
\label{eq:Heff-general}
H_{\rm eff}
=-\,\frac{1}{\Delta}\,\langle s\lvert V_{12}\,\mathcal{P}_t\,V_{34}\rvert s\rangle
=\frac{N_F^4}{16}\,\frac{2\,\Gamma_1\Gamma_2}{-\Delta}\;
\sum_{a,a',b,b'}\sum_{\sigma,\sigma'}\Phi^{\sigma\sigma'}_{1,aa'}\,
\langle s\lvert \Phi^{\sigma\sigma'}_{2,aa'}\Phi^{\delta\delta'}_{2,bb'}\rvert s\rangle\,
\Phi^{\delta\delta'}_{4,bb'},
\end{equation}
where $\Delta=E_s-E_t<0$ is the singlet--triplet gap of the central pair, $\mathcal{P}_t$ projects onto the excited manifold, and $\Gamma_{1,2}$ are the couplings to the central sites. Writing the singlet contraction as
\[
\sum_{aa'bb',\sigma,\delta}\Phi^{\sigma\sigma'}_{1,aa'}\,
\langle s\lvert \Phi^{\sigma\sigma'}_{2,aa'}\Phi^{\delta\delta'}_{2,bb'}\rvert s\rangle\,
\Phi^{\delta\delta'}_{4,bb'}
\;=\; N_F^2\mathrm{Sum}\;\mathrm{tr}\!\big(\Phi_1\Phi_4\big),
\]
we define the dimensionless coefficient $\mathrm{Sum}=\mathrm{Sum}(n,N_F)$ and tabulate it together with $\Delta$:
\begin{table}[h!]
\centering
\resizebox{\textwidth}{!}{%
\begin{tabular}{lcccc}
\toprule
$N_F \backslash n$ & $n=2$ & $n=3$ & $n=4$ & $n=5$ \\
\midrule
$N_F=1$ & $\mathrm{Sum}=4,\ \Delta=-1$ & $\mathrm{Sum}=2,\ \Delta=-2$ & $\mathrm{Sum}=2,\ \Delta=-3$ & $\mathrm{Sum}=2,\ \Delta=-4$\\
$N_F=2$ & $\mathrm{Sum}=0.6,\ \Delta=-1$ & $\mathrm{Sum}=0.3,\ \Delta=-2$ & $\mathrm{Sum}=2/7,\ \Delta=-3$ & \\
$N_F=3$ & $\mathrm{Sum}=0.246914,\ \Delta=-1$ & $\mathrm{Sum}=0.103703,\ \Delta=-2$ &  & \\
$N_F=4$ & $\mathrm{Sum}=0.125,\ \Delta=-1$ &  &  &  \\
\bottomrule
\end{tabular}}
\caption{Values of Sum and $\Delta$ for different $N_F$ and $n$.}
\end{table}

The combination $\mathrm{Sum}/\Delta$ admits a closed form:
\begin{equation}
\label{eq:Sum-over-Delta}
\frac{\mathrm{Sum}}{\Delta}=
\begin{cases}
\displaystyle \frac{4}{3}\,\frac{N_F+2}{N_F^3}, & n=2,\\[6pt]
\displaystyle \frac{2}{(n-1)(2n-1)}\,\frac{N_F+2(n-1)}{N_F^3}, & n\ge 3.
\end{cases}
\end{equation}
Thus the effective Hamiltonian can be written compactly as
\begin{equation}\label{eq:Ham_app}
H_{\rm eff}
=\frac{N_F^3}{4}\;\frac{\Gamma_1\Gamma_2}{\Gamma'}\;
\frac{N_F+2(n-1)}{(n-1)(2n-1)}\;
\mathrm{tr}\!\big(\Phi_1\Phi_4\big),
\qquad (n>2),
\end{equation}
with the $n=2$ case obtained by multiplying the bracket by $2$ (cf.\ \eqref{eq:Sum-over-Delta}). Here $\Gamma'$ denotes the decimated central coupling.

\bigskip
\noindent\textbf{Remark.} Equations \eqref{eq:Phi-ppmm}--\eqref{eq:Ham_app} provide the spin representation and the resulting second-order SDRG coefficient for the monitored, replica-extended problem with the JW ordering \eqref{eq:I-order}. All signs and string factors are fixed by the convention in \eqref{eq:JW-map}.

  \chapter{SSE numerics for the \(n{=}1\) check }
\label{app:gaussian-n1-code}

\subsection{Time evolution with the stochastic Schrödinger equation}

The time discretized evolution corresponding to the SSE of Eq.~\eqref{eq:SSE} and the time-dependent random Hamiltonian Eq.~\eqref{eq:mon_only} is given by 
\begin{equation}\label{eq_discretized_SSE}
  \begin{aligned}
    &\delta \ket{\psi} 
    =  \ i \sqrt{\Gamma} \sum_{\nu_1 \nu_2} \delta w_{\nu_1\nu_2}^j(t)   
    \left( \hchi_{j \nu_1} \hchi_{j+1 \nu_2} - \left\langle \hchi_{j \nu_1} \hchi_{j+1 \nu_2} \right\rangle_t \right)
    \\
    - & \left. \ i^2 \frac{\Gamma \delta t}{2} \sum_{\boldsymbol{\tilde{\nu}}}  
    \left( \hchi_{j \nu_1} \hchi_{j+1 \nu_2} - \left\langle \hchi_{j \nu_1} \hchi_{j+1 \nu_2} \right\rangle_t \right)^2
    \right\} \ket{\psi}
    ,
    \\
  \end{aligned}
\end{equation}
with $\delta t \ll 1$.
The random variables $\delta w$ have zero mean and variance

  \begin{equation}
    \mathbb{E}_{\rm G} \left[ \delta w_{\mnu_1}^i \delta w_{\mnu_2}^j \right] = \sqrt{\Gamma_i \Gamma_j}\delta_{ij} \delta_{\mnu_1 \mnu_2} \delta t .
  \end{equation}

Where $\Gamma_i$ are random variables as well from Cauchy distribution.
The discretized time evolution must be taken with a small enough $\delta t$, such that we see a convergence in the results in $\delta t$.
A proper $\delta t$ needs to be found empirically for each set of $\Gamma_i$ and $N_F$ .

We are interested in computing the entropies.
Since the entropy saturates, we can instead compute time-averaged values over an appropriate window, reducing the total amount of need sampling. 
Note that this window also needs to be found empirically, for just changing $N_F$ can alter the timescale of saturation of the dynamics.
\subsection{The Gaussian case}

If all terms in the Hamiltonian are quadratic, meaning $q = q_J = 2$, then simulations can be greatly simplified since, due to Wick's theorem, all observables can be computed from the covariance matrix
\begin{equation}
    M_{jk} = \frac{i}{2} \Tr \Big( \rho [\hchi_j, \hchi_k] \Big) = i \left( \langle \hchi_j \hchi_k \rangle - \delta_{jk} \right) ,
\end{equation}
where we have introduced a flattening of the Majorana operator indices (cluster and flavour) which will be generally useful for numerics.

The evolution is given by
\begin{equation}
\begin{aligned}
    \ket{\psi} \rightarrow &
    \exp{  \sum_{j}  \sum_{\nu_1 < \nu_2} \left(  \delta w_{\nu_1 \nu_2}^j \sqrt{\Gamma} + \Gamma \delta t M_{(j,\nu_1),(j+1,\nu_2)}   \right) i \hchi_{j,\nu_1} \hchi_{j+1,\nu_2} }  \ket{\psi} \\ 
    = &
    \exp{ i \sum_{j}    \delta w_{j,j+1} \hchi_{j} \hchi_{j+1} }  \ket{\psi}
    = \mathcal{D} \ket{\psi} ,
\end{aligned}
\end{equation}
where we have hidden the microscopical details in the skew-symmetric real matrix $\delta w$.
Since $\mathcal{D}$ is not a unitary operator, the state also needs to be normalized as
\begin{equation}\label{eq_monitoring_evol_step}
  \ket{\psi} \rightarrow \frac{\mathcal{D}\ket{\psi}}{\|\mathcal{D} \ket{\psi}\|} .
\end{equation}
Following the approach of \cite{Fava:2023tgg}, we can note that any Gaussian state can be defined by a set of Dirac fermion annihilation operators (with respect to the state)
\begin{equation}
  d_\mu = \sum_jU_{\mu j} \hat{c}_j + V_{\mu j} \hat{c}_j^\dagger ,
\end{equation}
where $\hat{c}_j = \left(\hchi_{2j-1} + i\hchi_{2j} \right)/2$  and $\hat{c}_j^\dagger = \left(\hchi_{2j-1} - i\hchi_{2j} \right)/2$.
The operators $d_i$ satisfy the usual fermionic commutation relations if and only if
\begin{equation}\label{eq:an_op_commutation_relations}
  U U^\dagger + V V^\dagger = \mathds{1} \quad \land \quad V U^t + U V^t = 0 .
\end{equation}
After applying the monitoring step of Eq.~\eqref{eq_monitoring_evol_step}, the annihilation operators evolve as 
\begin{equation}\label{eq:an_op_UV_evol}
  d_\mu' =  \mathcal{D} d_\mu \mathcal{D}^{-1} \quad \Rightarrow \quad
  \begin{bmatrix}
    U' & V' \\
  \end{bmatrix}
  =
  \begin{bmatrix}
    U & V \\
  \end{bmatrix}
  \frac{1}{2}
  \begin{bmatrix}
    \mathds{1} & -i \mathds{1} \\
    \mathds{1} & i \mathds{1} \\
  \end{bmatrix}
  P e^{-i 4 \delta w} P^{-1}
  \begin{bmatrix}
    \mathds{1} & \mathds{1} \\
    i\mathds{1} & - i \mathds{1} \\
  \end{bmatrix}
  ,
\end{equation}
where $P$ is a permutation that reorders the Majorana operator flattened indices, such that the odd ones come before the even ones, since
\begin{equation}
    \mathcal{D} \hchi_i \mathcal{D}^{-1} = \sum_j [e^{-4i \delta w}]_{ij} \hchi_j .
\end{equation}

$U'$ and $V'$ do not generally respect the commutation relations of Eq.~\eqref{eq:an_op_commutation_relations}, but that can be fixed by applying the following QR decomposition
\begin{equation}\label{eq:QRUV}
  \begin{bmatrix}
    {U'}^t \\ 
    {V'}^t \\ 
  \end{bmatrix}
  = 
  Q
  \begin{bmatrix}
    R \\ 
    0 \\ 
  \end{bmatrix}
  =
  \begin{bmatrix}
    Q_{11} & Q_{12} \\
    Q_{21} & Q_{22} \\
  \end{bmatrix}
  \begin{bmatrix}
    R \\ 
    0 \\ 
  \end{bmatrix}
  = 
  \begin{bmatrix}
    Q_{11}R \\
    Q_{21}R \\
  \end{bmatrix}
  ,
\end{equation}
where $Q$ is a unitary matrix and $R$ is an upper triangular matrix.
We now define new valid annihilation operators 
\begin{equation}
  \tilde{d}_\mu = \sum_\nu \left[ {R^t}^{-1}\right]_\nu d'_\nu = \sum_j \tilde{U}_j c_j +  \tilde{V}_j c^\dagger_j , 
\end{equation}
with $\tilde{U} = Q_{11}^t$ and $\tilde{V} = Q_{21}^t$, with the correct commutation relations.

Finally, we can show that the covariance matrix can be found for any set of annihilation operators $d_\mu$.
For a Gaussian state $\ket{\psi} = \mathcal{O} \ket{0}$ such that $\mathcal{O}^\dagger \hchi_i \mathcal{O} = \sum_j O_{ij} \hchi_j$, the state is annihilated by $d_\mu = \mathcal{O} c_\mu \mathcal{O}^\dagger $, which is equivalent to the following relation
\begin{equation}
  O = 
  \frac{1}{2} 
  P^{-1}
  \begin{bmatrix}
    \mathds{1} & \mathds{1} \\
    -i\mathds{1} & i \mathds{1} \\
  \end{bmatrix}
  \begin{bmatrix}
    U^t & V^\dagger \\
    V^t & U^\dagger \\
  \end{bmatrix}
  \begin{bmatrix}
    \mathds{1} & i \mathds{1} \\
    \mathds{1} & -i \mathds{1} \\
  \end{bmatrix}
  P
  .
\end{equation}
The covariance matrix is given by $M = O M_0 O^t$, where $M_0$ is the vacuum state covariance matrix
\begin{equation}\label{eq:vacuum_covariance}
  M_0 = 
  \begin{bmatrix}
    0 & 1 & & & \\
    -1 & 0 & & & \\
    & & 0 & 1 & \\
    & & -1 & 0 & \\
    & & & & \ddots \\
  \end{bmatrix} .
\end{equation}

  \bibliographystyle{unsrt}         
  \bibliography{bibliography/refs}  

\end{document}